\documentclass[12pt]{article}

\usepackage{jheppub, bm, wrapfig,float,array}
\usepackage[utf8]{inputenc}
\numberwithin{equation}{section}
\renewcommand\afterTocRuleSpace{\bigskip}

\usepackage{amsfonts}
\usepackage{amsmath}
\usepackage{color}
\usepackage{graphicx}

\usepackage{tikz}
\usetikzlibrary{arrows}
\usetikzlibrary{shapes.geometric,calc,arrows, positioning,shapes.misc,decorations.markings}
\tikzset{
  big arrow/.style={
    decoration={markings,mark=at position 1 with {\arrow[scale=2,#1]{>}}},
    postaction={decorate},
    shorten >=0.4pt},
  big arrow/.default=black}

\newcommand{\bea}{\begin{eqnarray}}
\newcommand{\eea}{\end{eqnarray}}
\newcommand{\be}{\begin{equation}}
\newcommand{\ee}{\end{equation}}
\newcommand{\bit}{\begin{itemize}}
\newcommand{\eit}{\end{itemize}}
\newcommand{\ben}{\begin{enumerate}}
\newcommand{\een}{\end{enumerate}}
\newcommand{\nn}{\nonumber}

\newcommand{\mbb}{\mathbb}

\newcommand{\R}{{\mathbb R}}

\newcommand{\F}{{\mathbb F}}
\renewcommand{\P}{{\mathbb P}}

\newcommand{\cF}{\mathcal{F}}

\newcommand{\cL}{\mathcal{L}}
\newcommand{\cM}{\mathcal{M}}
\newcommand{\cN}{\mathcal{N}}

\newcommand{\cT}{\mathcal{T}}

\newcommand{\cX}{\mathcal{X}}

\newcommand{\mf}{\mathfrak}
\newcommand{\fT}{\mathfrak{T}}

\newcommand{\fe}{\mathfrak{e}}
\newcommand{\ff}{\mathfrak{f}}
\newcommand{\fg}{\mathfrak{g}}
\newcommand{\fh}{\mathfrak{h}}
\newcommand{\su}{\mathfrak{su}}
\renewcommand{\sp}{\mathfrak{sp}}
\newcommand{\so}{\mathfrak{so}}

\newcommand{\ubf}[1]{\underline{\bf #1}}

\newcommand{\lra}{\longrightarrow}

\title{Classifying $5d$ SCFTs via $6d$ SCFTs: Arbitrary rank}

\author{Lakshya Bhardwaj\footnote{lbhardwaj@fas.harvard.edu}, Patrick Jefferson\footnote{patrickjefferson@fas.harvard.edu}}

\affiliation{Department of Physics, Harvard University, Cambridge, MA 02138, USA}

\abstract{According to a conjecture, all $5d$ SCFTs should be obtainable by rank-preserving RG flows of $6d$ SCFTs compactified on a circle possibly twisted by a background for the discrete global symmetries around the circle. For a $6d$ SCFT admitting an F-theory construction, its untwisted compactification admits a dual M-theory description in terms of a ``parent'' Calabi-Yau threefold which captures the Coulomb branch of the compactified $6d$ SCFT. The RG flows to $5d$ SCFTs can then be identified with a sequence of flop transitions and blowdowns of the parent Calabi-Yau leading to ``descendant'' Calabi-Yau threefolds which describe the Coulomb branches of the resulting $5d$ SCFTs. An explicit description of parent Calabi-Yaus is known for untwisted compactifications of rank one $6d$ SCFTs. In this paper, we provide a description of parent Calabi-Yaus for untwisted compactifications of arbitrary rank $6d$ SCFTs. Since 6d SCFTs of arbitrary rank can be viewed as being constructed out of rank one SCFTs, we accomplish the extension to arbitrary rank by identifying a prescription for gluing together Calabi-Yaus associated to rank one 6d SCFTs.}

\begin{document}

\maketitle

\section{Introduction} \label{intro}
Following a recent proposal \cite{Jefferson:2018irk} (see also \cite{Jefferson:2017ahm}), a strategy for classifying $5d$ SCFTs in terms of $6d$ SCFTs was spelled out in \cite{Bhardwaj:2018yhy}. It seeks to obtain Coulomb branches of $5d$ SCFTs starting from Coulomb branches of $5d$ KK theories. A $5d$ KK theory is defined to be a $6d$ SCFT $\fT$ compactifed on $S^1$ with some choice of background for the global discrete symmetry of $\fT$ around $S^1$. When the discrete symmetry background is trivial, we say that we have an \emph{untwisted} compactification of $\fT$. When the discrete symmetry background is not trivial, we say that we have a \emph{twisted} compactification of $\fT$. In this paper, we will only study untwisted compactifications. Moreover, we will always turn on generic mass parameters for $5d$ theories. Hence, the phrase ``Coulomb branch'' will always refer to the Coulomb branch with generic mass parameters turned on.

Let us consider a $\fT$ that admits an F-theory construction without frozen singularities\footnote{See \cite{Heckman:2015bfa} for a classification of such $6d$ SCFTs and \cite{Bhardwaj:2018jgp} for an F-theory construction (involving frozen singularities) of $6d$ SCFTs not obtainable by the methods of \cite{Heckman:2013pva,Heckman:2015bfa} but visible in the classification of \cite{Bhardwaj:2015xxa} and having a known string theory construction since the appearance of \cite{Hanany:1997gh}.}. The tensor branch of $\fT$ is described by F-theory compactified on a non-compact Calabi-Yau threefold $X_\fT$ which takes the form of an elliptic fibration over a smooth non-compact K\"ahler base $B$. An important point to note is that, in general, $X_\fT$ is a singular threefold even though $B$ is smooth. The loci in $B$ over which the singularities of $X_\fT$ live are then identified as the loci wrapped by 7-branes.

Let us consider an untwisted compactification of $\fT$ (on its tensor branch) on a circle $S^1_R$ of radius $R$ and let us turn on generic holonomies for the continuous gauge and global symmetries of $\fT$ around $S^1_R$. This corresponds to compactifying F-theory on $X_\fT\times S^1_R$ and turning on holonomies on the 7-branes around $S^1_R$. This F-theory setup is then dual to M-theory compactified on a fully smooth resolution $\tilde X_\fT$ of $X_\fT$ with the elliptic fiber in $\tilde X_\fT$ carrying volume $\frac{1}{R}$. 

M-theory compactified on $\tilde X_\fT$ describes the Coulomb branch of the corresponding $5d$ KK theory which we denote as $\fT_{KK}$. The objects of interest are the sets of holomorphic curves and holomorphic surfaces inside $\tilde X_\fT$. M2 branes wrapping holomorphic curves give rise to BPS particles in the low energy $5d$ theory and M5 branes wrapping holomorphic surfaces give rise to BPS strings in the low energy $5d$ theory. Each BPS string can be seen as a monopole for a $U(1)$ gauge group in the low energy $5d$ theory, and hence the number of surfaces equal the rank of the $5d$ theory. The intersection numbers between surfaces capture the full prepotential and hence the Coulomb branch metric of the $5d$ theory.

Reference \cite{Bhardwaj:2018yhy} provided a description of $\tilde X_\fT$ for all $\fT$ that are rank one $6d$ SCFTs. This was done by explicitly performing resolutions of $X_\fT$ using methods described in detail in \cite{Esole:2017kyr,Esole:2015xfa,Esole:2017qeh,Esole:2017rgz,Esole:2014hya,Esole:2014bka,Esole:2011sm}. In this paper, we extend the work of \cite{Bhardwaj:2018yhy} and provide a description of $\tilde X_\fT$ for $\fT$ that are $6d$ SCFTs of any arbitrary rank (see \cite{Esole:2018mqb,Esole:2018csl,Esole:2017hlw} for related analyses of F-theory models involving collisions of elliptic fibers in cases of semi-simple, as opposed to simple, gauge algebras.)

We can use the data of $\tilde X_\fT$ to perform RG flows to $5d$ SCFTs. There are many kinds of RG flows that are possible. The simplest and the most widely studied ones involve sending some surface inside $\tilde X_\fT$ to infinite volume. Under such an RG flow, the rank of the $5d$ theory is reduced. The suggestion in \cite{Bhardwaj:2018yhy,Jefferson:2018irk} was to instead study \emph{rank preserving} RG flows\footnote{Field theoretically, such RG flows correspond to integrating out BPS particles. When the $5d$ theory has an interpretation in terms of a $5d$ gauge theory, such BPS particles can either be perturbative or instantonic from the point of view of gauge theory.}, and it was conjectured there that all rank $n$ $5d$ SCFTs can be obtained by rank preserving RG flows starting from rank $n$ $5d$ KK theories\footnote{Evidence for this conjecture was provided in \cite{Jefferson:2018irk} for $5d$ SCFTs upto rank two that can be given a geometric construction in M-theory.}. For the KK theories $\fT_{KK}$ that we discussed above, such RG flows are described by a sequence of flop transitions and blowdowns on $\tilde X_\fT$ generating new smooth non-compact Calabi-Yau threefolds that do not admit an elliptic fibration. 

In this sense, one can regard $\tilde X_\fT$ as ``parent'' Calabi-Yaus and the Calabi-Yaus obtained after flops and blowdowns as their ``descendant'' Calabi-Yaus. M-theory compactified on the descendant Calabi-Yaus describes the Coulomb branch of $5d$ SCFTs. Different descendant Calabi-Yaus lead to different Coulomb branches and hence are identified with different $5d$ SCFTs\footnote{It is possible for multiple $5d$ KK theories to give rise to flow to the same $5d$ SCFT. Many such examples can be found in \cite{Jefferson:2018irk}. Thus, naively computing all possible blowdowns of parent Calabi-Yaus leads to an overcounting of $5d$ SCFTs. In practice, this overcounting is easy to eliminate since two descendant Calabi-Yaus (of two different parent Calabi-Yaus) describing the same $5d$ SCFT are related to each other flops.}. This is the basis for the classification scheme sketched in \cite{Bhardwaj:2018yhy}.

This paper is organized as follows. In Section \ref{gen}, we argue that the relevant data of $\tilde X_\fT$ can be captured in terms of a graph whose nodes label compact holomorphic surfaces in $\tilde X_\fT$ and the edges describe how the surfaces intersect. The nodes, edges and faces of the graph are decorated by some numbers. These numbers identify each surface in terms of some simple known complex surfaces, and describe the intersection numbers between curves and surfaces in the threefold $\tilde X_\fT$. In Section \ref{over}, we describe our methods of computation using which we assemble the graph associated to $\tilde X_\fT$ given a Weierstrass model for $X_\fT$. In Section \ref{flop-eq}, we assign a graph to each compact holomorphic curve $C$ in the base $B$ of $X_\fT$. In Section \ref{gluing}, we provide rules to join the graphs corresponding to different curves $C$ in $B$, hence resulting in a graph describing $\tilde X_\fT$. In Section \ref{conc}, we describe some possible extensions of this work in interesting directions. Lastly, in Appendix \ref{appendix}, we provide instructions on using the mathematica notebook ``Pushforward.nb'' attached as an ancillary file with this paper.

\section{General ideas}\label{gen}
In this section, we describe the structure of $\tilde X_\fT$ in as general terms as possible without invoking any detailed computations. We start by reviewing some relevant details of the F-theory construction of $\fT$ in Section \ref{F}. We proceed by reviewing some relevant aspects of M-theory compactifications on smooth non-compact Calabi-Yau threefolds in Section \ref{M}. We also list the relevant data that we need to track for each Calabi-Yau. After that, in Sections \ref{single} and \ref{multiple}, we take a closer look at the structure of the Calabi Yaus $\tilde X_\fT$ due to the fact that they admit an elliptic fibration. We find that $\tilde X_\fT$ can be described in terms of a collection of compact holomorphic surfaces glued to each other along some holomorphic curves inside these surfaces. We show that each surface in the collection can either be described in terms of a Hirzebruch surface, or in terms of the projective space $\P^2$. We also describe some general constraints that the gluing curves have to satisfy. In Sections \ref{mori} and \ref{num}, we show how we can recover each item in the list presented at the end of Section \ref{M} in terms of the data of this collection of surfaces along with the data of their gluings. In Section \ref{RG}, we describe how one can perform a sequence of flops and blowdowns on $\tilde X_\fT$ to reach Calabi-Yaus that are descendants of $\tilde X_\fT$. The data of a descendant $Y$ is also naturally packaged as a collection of surfaces with some gluing rules, and the data listed at the end of Section \ref{M} can be recovered by applying the ideas in Sections \ref{mori} and \ref{num} to $Y$. In Section \ref{hate}, we discuss the notion of ``decoupled states''. These are states that can be decoupled from the $5d$ theory when they are massless. Removing the decoupled states changes the Calabi-Yau, and the Calabi-Yau with no decoupled states provides the maximum number of RG flows. In Section \ref{rhate}, we argue that in order to remove decoupled states, we should replace some of the F-theory configurations with less singular and potentially inconsistent F-theory configurations. Even if the less singular F-theory configuration is inconsistent, we can still assign to it a consistent Calabi-Yau which fails to admit elliptic fibration structure due to some subtle reason and hence is inconsistent as an F-theory background. However, it makes sense as an M-theory background and allows access to the maximum number of RG flows. We capture these replacements of F-theory configurations by defining a notion of ``formal gauge algebra'' in Section \ref{rhate}. Finally, in Section \ref{close}, we describe an algorithm that sums up the discussion and allows one to construct $\tilde X_\fT$ starting from an F-theory configuration $X_\fT$ constructing a $6d$ SCFT $\fT$.

\subsection{F-theory construction of $\fT$} \label{F}
The elliptic fibration $X_\fT$ is defined by a Weierstrass equation
\be
y^2z=x^3+f xz^2+gz^3 \label{W}
\ee
where $[x:y:z]$ are homogeneous coordinates defining a $\P^2$. $z$ is a function on $B$ and $x,y,f,g$ are sections of $-2K_B,-3K_B,-4K_B,-6K_B$ respectively where $K_B$ is the canonical divisor class of $B$. The discriminant $\Delta$ of (\ref{W}) is
\be
\Delta = 4f^3+27g^2\label{dis}
\ee
which is a section of $-12K_B$. The points in $B$ over which $\Delta$ vanishes carry a singular elliptic fiber and the collection of such points is known as the \emph{discriminant locus} of the elliptic fibration. The discriminant locus can be decomposed into a set of irreducible holomorphic curves in $B$ which are referred to as the components of discriminant locus. To each component of discriminant locus, we can assign the orders of vanishing of $f$, $g$ and $\Delta$. In general, several components of discriminant locus can pass through a point $p$ in the discriminant locus. The orders of vanishing of $f$, $g$ and $\Delta$ at $p$ are then defined to be the sum of orders of vanishings of $f$, $g$ and $\Delta$ over all such components. One requires that the orders of vanishings of $f$, $g$ and $\Delta$ at any point or at any curve are neither greater than nor equal to $4$, $6$ and $12$ respectively.

The conformal point of $\fT$ corresponds to all the compact holomorphic curves in $B$ shrinking simultaneously to a point. Thus we require that we can contract all compact holomorphic curves in $B$ at a finite distance in the moduli space of the Calabi-Yau. This condition along with the requirement that the orders of vanishings of $f$, $g$ and $\Delta$ remain less than $4$, $6$ and $12$ has several consequences \cite{Heckman:2013pva}. First, all compact holomorphic curves in $B$ must be rational, i.e. they must have genus zero. Second, two distinct compact holomorphic curves can only intersect in at most one point. Third, the intersection pairing of compact holomorphic curves on $B$ must be negative definite.

\subsection{M-theory construction of $5d$ Coulomb branches} \label {M}
We will now like to discuss the structure of $\tilde X_\fT$. But before getting there, we review some aspects of M-theory compactified on a smooth non-compact Calabi-Yau threefold $Y$ which may or may not admit an elliptic fibration.

The dual of the K\"ahler form is a divisor $J$ on $Y$ can be written in terms of compact divisors $S^i$ and non-compact divisors $N^j$ in $Y$ as $J=\phi_i S^i + m_j N^j$. The coefficients $\phi_i$ in $J$ are the Coulomb branch moduli and the coefficients $m_j$ in $J$ are the mass parameters of the low energy $5d$ theory arising from M-theory compactified on $Y$. If $Y=\tilde X_\fT$, then the base $B$ of the elliptic fibration appears as a special non-compact divisor whose coefficient $m_B$ in $J$ can be identified with $\frac{1}{R}$ where $R$ is the radius of the circle used to compactify the $6d$ theory $\fT$.

The intersection number $-J\cdot C$ for a compact holomorphic curve $C$ computes the volume of $C$ which controls the mass of the BPS particle arising from an M2 brane wrapping $C$. Similarly, the intersection number $J\cdot J\cdot S$ for a compact holomorphic surface $S$ computes the volume of $S$ which controls the tension of the BPS string arising from an M5 brane wrapping $S$. Finally, the intersection number $J\cdot J\cdot J$ computes the prepotential $\cF$ of the low energy $5d$ theory. We see that $\cF$ is a cubic polynomial as it should be \cite{Seiberg:1996bd}. Taking the partial derivative of $\cF$ with respect to $\phi_i$, $\phi_j$ gives us the metric on the Coulomb branch.

The individual intersection numbers $S^i\cdot S^j\cdot S^k$ for three compact surfaces are also relevant physically as they compute the coefficient of the Chern-Simons term for the low-energy $U(1)$ gauge groups associated to the three surfaces. Similarly, the individual intersection numbers $S\cdot C$ for a compact surface $S$ and a compact curve $C$ capture the Dirac pairing between the BPS string associated to $S^i$ and the BPS particle associated to $C$, and so they are also physically relevant.

Thus, for the purpose of describing physics on the Coulomb branch, the data of the Calabi-Yau that we track in this paper is:
\bit
\item The set of compact holomorphic surfaces $S^i$.
\item The set of compact holomorphic curves.
\item The set of intersection numbers $S\cdot C$ between a compact holomorphic surface and a compact holomorphic curve.
\item The set of triple intersection numbers $S^i\cdot S^j\cdot S^k$ of three compact holomorphic surfaces.
\eit
Note that the data related to a non-compact holomorphic surface $N^j$ can be encoded in the above data once we know the compact curves $C^{ij}=S^i\cdot N^j$ for all $i$. We ignore the data of $N^j$ in this paper except for the base $B$ in the case when $Y=\tilde X_\fT$.

\subsection{Consequences of the structure of elliptic fibration: Single curve} \label{single}
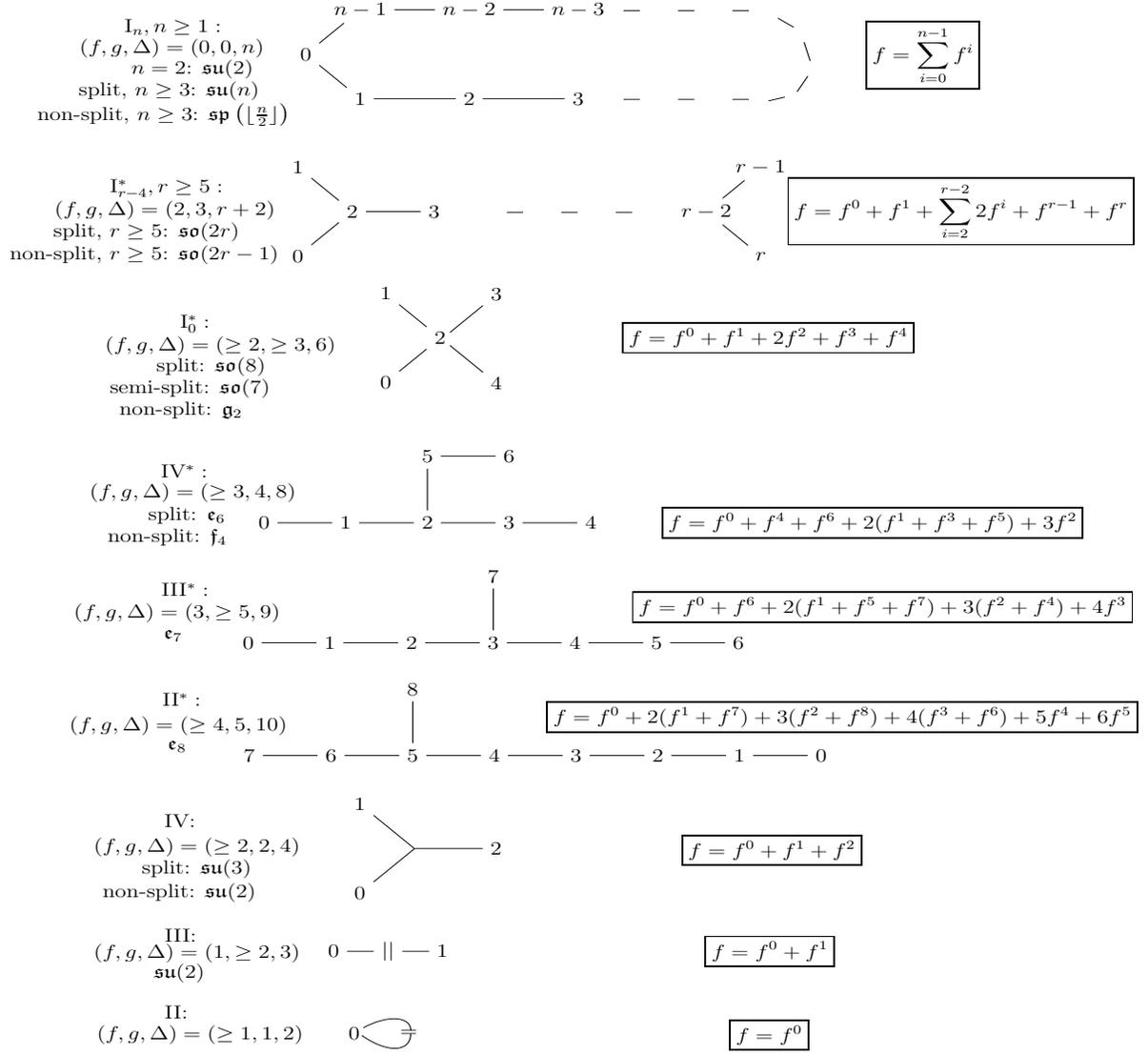
\begin{figure}
\begin{center}
 \scalebox{0.76}[0.62]{
\begin{tikzpicture}
\node at (-1.6015,2.543) {I$_n,n\geq1:$};
\node (v1) at (0.8369,1.9426) {0};
\node (v2) at (1.8369,2.9426) {$n-1$};
\node (v5) at (1.8369,0.9426) {1};
\node (v6) at (3.8369,0.9426) {2};
\node (v7) at (5.8369,0.9426) {3};
\node (v3) at (3.8369,2.9426) {$n-2$};
\node (v4) at (5.8369,2.9426) {$n-3$};
\draw  (v1) edge (v2);
\draw  (v2) edge (v3);
\draw  (v3) edge (v4);
\draw  (v1) edge (v5);
\draw  (v5) edge (v6);
\draw  (v6) edge (v7);
\draw (6.6716,2.9262) -- (6.9716,2.9262);
\begin{scope}[shift={(1,0)}]
\draw (6.6716,2.9262) -- (6.9716,2.9262);
\end{scope}
\begin{scope}[shift={(2,0)}]
\draw (6.6716,2.9262) -- (6.9716,2.9262);
\end{scope}
\begin{scope}[shift={(0,-2)}]
\draw (6.6716,2.9262) -- (6.9716,2.9262);
\end{scope}
\begin{scope}[shift={(1,-2)}]
\draw (6.6716,2.9262) -- (6.9716,2.9262);
\end{scope}
\begin{scope}[shift={(2,-2)}]
\draw (6.6716,2.9262) -- (6.9716,2.9262);
\end{scope}
\draw (9.5048,2.7545) -- (9.7967,2.5485);
\draw (9.9855,2.2136) -- (10.0971,1.793);
\draw (10.0542,1.4495) -- (9.8568,1.0975) (9.333,0.8915) -- (9.6335,0.9774);
\node at (12.1711,1.9184) {\boxed{f=\sum_{i=0}^{n-1}f^i}};
\node at (-1.624,2.0703) {$(f,g,\Delta)=(0,0,n)$};
\node at (-1.6718,1.1097) {split, $n\ge3$: $\mf{su}(n)$};
\begin{scope}[shift={(-0.1445,4.4582)}]
\node at (-1.5768,-5.5428) {I$_{r-4}^{*},r\geq5:$};
\node at (-1.6076,-6.0181) {$(f,g,\Delta)=(2,3,r+2)$};
\node (v1_1_2) at (0.8369,-7.0574) {0};
\node (v1_1_3) at (0.8369,-5.0574) {1};
\node (v1_1_4) at (1.8369,-6.0574) {2};
\node (v1_1_5) at (3.3369,-6.0574) {3};
\begin{scope}[shift={(-2,-7)}]
\begin{scope}[shift={(0,-2)}]
\draw (6.6716,2.9262) -- (6.9716,2.9262);
\end{scope}
\begin{scope}[shift={(1,-2)}]
\draw (6.6716,2.9262) -- (6.9716,2.9262);
\end{scope}
\begin{scope}[shift={(2,-2)}]
\draw (6.6716,2.9262) -- (6.9716,2.9262);
\end{scope}
\end{scope}
\node (v1_1_6) at (8.3369,-6.0574) {$r-2$};
\node (v1_1_7) at (9.3369,-7.0574) {$r$};
\node (v1_1_8) at (9.3369,-5.0574) {$r-1$};
\node at (13.011,-6.0574) {\boxed{f=f^0+f^1+\sum_{i=2}^{r-2}2f^i+f^{r-1}+f^r}};
\node at (-1.9418,-6.5256) {split, $r\ge5$: $\mf{so}(2r)$};
\node at (-1.9839,-7.0233) {non-split, $r\ge5$: $\mf{so}(2r-1)$};
\end{scope}
\begin{scope}[shift={(-0.9949,1.6926)}]
\node at (-0.2031,-5.816) {I$_{0}^{*}:$};
\node at (0.2642,-6.3063) {$(f,g,\Delta)=\left(\ge2,\ge3,6\right)$};
\node (v1_1_2_1) at (3.2986,-7.1392) {0};
\node (v1_1_3_1) at (3.2986,-5.1392) {1};
\node (v1_1_4_1) at (4.2986,-6.1392) {2};
\node (v1_1_6_1) at (4.3328,-6.1606) {$$};
\node (v1_1_7_1) at (5.3328,-7.1606) {$4$};
\node (v1_1_8_1) at (5.3328,-5.1606) {$3$};
\node at (10.3239,-6.1407) {\boxed{f=f^0+f^1+2f^2+f^3+f^4}};
\node at (0.0911,-6.7927) {split: $\mf{so}(8)$};
\node at (-0.2937,-7.2835) {semi-split: $\mf{so}(7)$};
\node at (-0.4588,-7.7583) {non-split: $\mf{g}_2$};
\end{scope}
\draw  (v1_1_3_1) edge (v1_1_4_1);
\draw  (v1_1_4_1) edge (v1_1_2_1);
\draw  (v1_1_6_1) edge (v1_1_8_1);
\draw  (v1_1_6_1) edge (v1_1_7_1);
\begin{scope}[shift={(-0.7678,-2.551)}]
\node at (-0.627,-4.894) {IV$^{*}:$};
\node at (-0.47,-5.3859) {$(f,g,\Delta)=\left(\ge3,4,8\right)$};
\node (v1_1_2_1_1) at (0.8369,-6.0574) {0};
\node (v1_1_2_1_2) at (2.3369,-6.0574) {1};
\node (v1_1_3_1_1) at (3.8369,-4.5574) {5};
\node (v1_1_4_1_1) at (3.8369,-6.0574) {2};
\node (v1_1_5_1_1) at (5.3369,-6.0574) {3};
\node (v1_1_5_1_2) at (6.8369,-6.0574) {4};
\node (v1_1_3_1_2) at (5.3369,-4.5574) {6};
\node at (12.011,-6.0574) {\boxed{f=f^0+f^4+f^6+2(f^1+f^3+f^5)+3f^2}};
\end{scope}
\draw  (v1_1_2_1_1) edge (v1_1_2_1_2);
\draw  (v1_1_2_1_2) edge (v1_1_4_1_1);
\draw  (v1_1_4_1_1) edge (v1_1_5_1_1);
\draw  (v1_1_5_1_1) edge (v1_1_5_1_2);
\draw  (v1_1_4_1_1) edge (v1_1_3_1_1);
\draw  (v1_1_3_1_1) edge (v1_1_3_1_2);
\begin{scope}[shift={(-1.0568,-5.2507)}]
\node at (-0.323,-4.8675) {III$^{*}:$};
\node at (-0.47,-5.3859) {$(f,g,\Delta)=\left(3,\ge5,9\right)$};
\node (v1_1_2_1_1_2) at (0.8369,-6.0574) {0};
\node (v1_1_2_1_2_2) at (2.3369,-6.0574) {1};
\node (v1_1_3_1_1_1) at (5.3369,-4.5574) {7};
\node (v1_1_4_1_1_2) at (3.8369,-6.0574) {2};
\node (v1_1_5_1_1_2) at (5.3369,-6.0574) {3};
\node (v1_1_5_1_2_2) at (6.8369,-6.0574) {4};
\node (v1_1_5_1_3) at (8.3369,-6.0574) {5};
\node (v1_1_5_1_4) at (9.8369,-6.0574) {6};
\node at (12.4816,-5.2574) {\boxed{f=f^0+f^6+2(f^1+f^5+f^7)+3(f^2+f^4)+4f^3}};
\end{scope}
\draw  (v1_1_2_1_1_2) edge (v1_1_2_1_2_2);
\draw  (v1_1_2_1_2_2) edge (v1_1_4_1_1_2);
\draw  (v1_1_4_1_1_2) edge (v1_1_5_1_1_2);
\draw  (v1_1_5_1_1_2) edge (v1_1_5_1_2_2);
\draw  (v1_1_5_1_1_2) edge (v1_1_3_1_1_1);
\draw  (v1_1_5_1_3) edge (v1_1_5_1_4);
\draw  (v1_1_5_1_2_2) edge (v1_1_5_1_3);
\begin{scope}[shift={(-1.0362,-7.7919)}]
\node at (-0.4059,-4.7966) {II$^{*}:$};
\node at (-0.47,-5.3859) {$(f,g,\Delta)=\left(\ge4,5,10\right)$};
\node (v1_1_2_1_1_3) at (0.8369,-6.0574) {7};
\node (v1_1_2_1_2_3) at (2.3369,-6.0574) {6};
\node (v1_1_3_1_1_2) at (3.8369,-4.5574) {8};
\node (v1_1_4_1_1_3) at (3.8369,-6.0574) {5};
\node (v1_1_5_1_1_3) at (5.3369,-6.0574) {4};
\node (v1_1_5_1_2_3) at (6.8369,-6.0574) {3};
\node (v1_1_5_1_3_1) at (8.3369,-6.0574) {2};
\node (v1_1_5_1_4_1) at (9.8369,-6.0574) {1};
\node (v1_1_5_1_5) at (11.3369,-6.0574) {0};
\node at (11.721,-5.1932) {\boxed{f=f^0+2(f^1+f^7)+3(f^2+f^8)+4(f^3+f^6)+5f^4+6f^5}};
\end{scope}
\draw  (v1_1_2_1_1_3) edge (v1_1_2_1_2_3);
\draw  (v1_1_2_1_2_3) edge (v1_1_4_1_1_3);
\draw  (v1_1_4_1_1_3) edge (v1_1_3_1_1_2);
\draw  (v1_1_4_1_1_3) edge (v1_1_5_1_1_3);
\draw  (v1_1_5_1_1_3) edge (v1_1_5_1_2_3);
\draw  (v1_1_5_1_2_3) edge (v1_1_5_1_3_1);
\draw  (v1_1_5_1_3_1) edge (v1_1_5_1_4_1);
\draw  (v1_1_5_1_4_1) edge (v1_1_5_1_5);
\begin{scope}[shift={(-0.8504,-9.7895)}]
\node at (-0.6228,-5.5321) {IV$:$};
\node at (-0.3124,-6.1174) {$(f,g,\Delta)=\left(\ge2,2,4\right)$};
\node (v1_1_2_1_3) at (2.6827,-7.1552) {0};
\node (v1_1_3_1_3) at (2.6827,-5.1552) {1};
\coordinate (v1_1_4_1_2) at (3.6827,-6.1552) {} {} {} {} {};
\node (v1_1_5_1) at (5.1827,-6.1552) {2};
\node at (10.2475,-6.1925) {\boxed{f=f^0+f^1+f^2}};
\end{scope}
\begin{scope}[shift={(-0.8317,-6.6746)}]
\node at (-0.6303,-11.2213) {III$:$};
\node at (-0.3282,-11.631) {$(f,g,\Delta)=\left(1,\ge2,3\right)$};
\node (v1_1_4_2) at (2.1785,-11.5818) {0};
\node (v1_1_4_3) at (3.1785,-11.5818) {$||$};
\node (v1_1_4_4) at (4.1785,-11.5818) {1};
\node at (10.1915,-11.5819) {\boxed{f=f^0+f^1}};
\end{scope}
\draw  (v1_1_4_2) -- (v1_1_4_3);
\draw  (v1_1_4_3) -- (v1_1_4_4);
\begin{scope}[shift={(-0.8679,-8.586)}]
\node at (-0.6732,-11.04) {II$:$};
\node at (-0.2382,-11.5522) {$(f,g,\Delta)=\left(\ge1,1,2\right)$};
\node (v1_1_4_2_1) at (2.5959,-11.5145) {0};
\node (v1_1_4_3_1) at (3.5959,-11.5145) {$=$};
\node at (10.2127,-11.5508) {\boxed{f=f^0}};
\end{scope}
\draw (1.864,-20.1494) .. controls (2.3361,-20.5606) and (2.763,-20.5304) .. (2.7626,-20.1124);
\draw (1.8934,-19.9795) .. controls (2.3161,-19.6897) and (2.7616,-19.7054) .. (2.764,-20.0377);
\node at (-1.2471,1.6074) {$n=2$: $\mf{su}(2)$};
\node at (-1.7724,0.5736) {non-split, $n\ge3$: $\mf{sp}\left(\lfloor \frac{n}{2}\rfloor\right)$};
\node at (-1.3475,-8.4389) {split: $\mf{e}_6$};
\node at (-1.7063,-8.9316) {non-split: $\mf{f}_4$};
\node at (-1.5981,-11.1506) {$\mf{e}_7$};
\node at (-1.5268,-13.6402) {$\mf{e}_8$};
\node at (-1.1666,-16.4051) {split: $\mf{su}(3)$};
\node at (-1.4841,-16.928) {non-split: $\mf{su}(2)$};
\node at (-1.482,-18.7465) {$\mf{su}(2)$};
\draw  (v1_1_3) edge (v1_1_4);
\draw  (v1_1_4) edge (v1_1_2);
\draw  (v1_1_4) edge (v1_1_5);
\draw  (v1_1_6) edge (v1_1_8);
\draw  (v1_1_6) edge (v1_1_7);
\draw  (v1_1_3_1_3) edge (v1_1_4_1_2);
\draw  (v1_1_4_1_2) edge (v1_1_5_1);
\draw  (v1_1_2_1_3) edge (v1_1_4_1_2);
\end{tikzpicture}
}
\end{center}
\caption{\ubf{Left}: Each Kodaira type is associated to an order of vanishing of $f$, $g$ and $\Delta$ which we denote as $(f,g,\Delta)$. We also display the corresponding $6d$ gauge algebras which are different for split, semi-split and non-split cases.
\ubf{Middle}: The graph displays the intersection pattern for the rational curves composing the elliptic fiber. The numbers in the nodes of the graph are labels and each edge between two nodes corresponds to a transverse intersection of the corresponding rational curves. The components of type IV fiber all intersect each other transversely at a common point. The $||$ in between the edges for types II and III denote the fact that those intersections are tangential rather than transverse.
\ubf{Right}: The elliptic fiber $f$ is written in terms of the rational curves $f^i$ where $i$ is the label appearing in the node.}
\label{hell}
\end{figure}

We now look more closely at the consequences of the fact that $\tilde X_\fT$ has the structure of an elliptic fibration. Consider a compact holomorphic curve $C$ in $B$ which is a component of the discriminant locus. Then over a generic point of $C$, the possible degenerations are classified by a Kodaira type which we list in Figure \ref{hell}. Over special points of $C$, we can see other types of degenerations. Now consider a generic point $p$ of $C$. As is visible from Figure \ref{hell}, the degenerate elliptic fiber over $p$ takes the form of a collection of rational curves $f^i_C$ intersecting each other transversely, and the elliptic fiber $f_C$ is recovered as
\be
f_C=\sum_i n_i f^i_C \label{fiber}
\ee
where $n_i$ are positive integers which can be read from Figure \ref{hell}. As one of the rational components $f^i_C$ moves over $C$, it sweeps out a holomorphic surface which is a $\P^1$ fibration over $C$. As we noted above, for $6d$ SCFTs, every compact holomorphic curve in $B$ must be a rational curve. Thus, the holomorphic surface swept out by $f^i_C$ is a $\P^1$ fibration over $\P^1$, or in other words a Hirzebruch surface. We label the surface as $S^i_C$. $f^i_C$ can then be recognized as the fiber of the Hirzebruch surface $S^i_C$. 

Sometimes the degenerate elliptic fiber on $C$ can have a monodromy such that $f^i_C$ is transported to $f^j_C$ as one starts at $p$, moves over $C$, and then returns back to $p$. In such situations $S^i_C=S^j_C$ and $f^i_C$ is in the same homotopy class as $f^j_C$ inside $S^i_C=S^j_C$. Such a monodromy is only possible for Kodaira types I$_{n\ge3}$, I$^*_n$, IV$^*$ and IV. In all of these cases except I$_0^*$, there is only one kind of non-trivial monodromy that is possible. So, in these cases, if there is monodromy, one attaches a superscript $ns$ to the Kodaira type and refers to the fiber as a \emph{non-split} fiber, and if there is no monodromy, one attaches a superscript $s$ to the Kodaira type and refers to the fiber as a \emph{split} fiber. In the case of I$_0^*$, there are two possible non-trivial monodromies. One of them is referred to as \emph{semi-split} and denoted by superscript $ss$, and the other is referred to as non-split and denoted by superscript $ns$. In literature, the data of Kodaira type along with the data of monodromy is known as the \emph{Kodaira-Tate} type of the degenerate fiber, and we will continue this terminology in the paper.

$C$ gives rise to a simple gauge algebra $\fg_C$ in the $6d$ theory. $\fg_C$ is completely determined by the Kodaira-Tate type of the elliptic fiber over $C$. Note that $\fg_C$ does not have to be non-trivial. It is trivial for Kodaira-Tate types I$_1$ and II which are singular fibers. It is also trivial for the smooth fiber which is denoted by Kodaira-Tate type I$_0$.

The surfaces $S^i_C$ living over $C$ intersect each other transversely in the way the rational curves $f^i_C$ intersect each other. The intersection locus $\ell^{ij}_C$ between $S^i_C$ and $S^j_C$ can be recognized as a holomorphic curve $\ell^{ij,i}_C$ inside $S^i_C$ and a holomorphic curve $\ell^{ij,j}_C$ inside $S^j_C$. So the intersection can also be thought of as a \emph{gluing} of surfaces $S^i_C$ and $S^j_C$ where we glue the curve $\ell^{ij,i}_C$ inside $S^i_C$ to the curve $\ell^{ij,j}_C$ inside $S^j_C$. In fact, any transverse intersection of any two holomorphic surfaces $S$ and $S'$ in any threefold $Y$ can be thought of as a transverse gluing of a holomorphic curve $\ell$ in $S$ to a holomorphic curve $\ell'$ in $S'$. For any such transverse gluing to be consistent with the Calabi-Yau condition on $Y$, the self-intersections of $\ell$ and $\ell'$ in $S$ and $S'$ respectively have to satisfy a condition which we now describe.

We can can break the tangent bundle $\cT Y$ of the threefold $Y$ into the tangent bundle $\cT L$ and the normal bundle $\cN L$ of the holomorphic curve $L=\ell\sim\ell'$ in $Y$. Furthermore, we can break $\cN L$ into a sum of the normal bundle $\cN\ell$ of $\ell$ in $S$ and the normal bundle $\cN\ell'$ of $\ell'$ in $S'$. The first Chern classes $c_1(\cT L)$, $c_1(\cN\ell)$ and $c_1(\cN\ell')$ are $2-2g$, $(\ell\cdot\ell)|_S$ and $(\ell'\cdot\ell')|_{S'}$ respectively where $g$ is the genus of $\ell$ and $\ell'$. Thus the vanishing of $c_1(\cT Y)$ implies the following condition which we will refer to as the \emph{Calabi-Yau condition}
\be
(\ell\cdot\ell)|_S+(\ell'\cdot\ell')|_{S'}=2g-2 \label{CY}
\ee
There can also be situations in which the intersection locus $L$ decomposes into a sum of intersection loci such that $L=\sum_\alpha L_\alpha$ with $L_\alpha$ mutually intersecting each other. Each $L_\alpha$ can be identified as a curve $\ell_\alpha$ in $S$ with the intersections of $\ell_\alpha$ descending from intersections of $L_\alpha$. Similarly, $L_\alpha$ can be identified with curves $\ell'_\alpha$ in $S'$ with the intersections of $\ell'_\alpha$ again descending from the intersections of $L_\alpha$. Then, the gluing of $S$ and $S'$ can be described as $\ell_\alpha\sim\ell'_\alpha$ for all $\alpha$ which implies that $\ell\sim\ell'$ where $\ell=\sum_\alpha\ell_\alpha$ and $\ell'=\sum_\alpha\ell'_\alpha$. In such a situation the Calabi-Yau condition ($\ref{CY}$) applies only to $\ell$ and $\ell'$, and not to their components $\ell_\alpha$ and $\ell'_\alpha$.  We will see examples of such gluings later in the paper. All of the situations described above are consistent with the assertion that the Calabi-Yau condition must be imposed locally on each geometrically irreducible curve along which a pair of surfaces intersects.

Going back to intersections among the surfaces $S^i_C$ living over $C$, we note that there can also be further gluings among $S^i_C$ not captured by the intersection pattern of rational curves in the Kodaira fiber over $C$. We will see such examples later in the paper and this fact was also noticed in \cite{Bhardwaj:2018yhy}, for example for a $6d$ $SO(n)$ gauge theory with $n-8$ hypers in vector representation. This has to do with the behavior of the elliptic fiber over some special points of $C$ as we will make clear when we discuss such examples.

At other special points over $C$, the elliptic fiber can degenerate further in the sense that the number of rational curves composing the elliptic fiber increase as compared to the elliptic fiber over a generic point of $C$. Such points can be identified as the locations where some other component of the discriminant locus intersects $C$. More precisely, by analyzing the Weierstrass equation, one can see that some of the individual $f^i_C$ split into multiple rational curves over such points. We interpret this to mean that the Hirzebruch surfaces $S^i_C$ carry some blowups which act  on the fiber $f^i_C$ splitting it into multiple pieces. For example, if we just do a single blowup at a generic point $p$ in $S^i_C$, then since some holomorphic representative of the homology class of $f^i_C$ passes through $p$, the blowup at least breaks the fiber into two pieces $f^i_C-x$, $x$ where $x$ is the exceptional curve created by the blowup. We note that, in general, the blowups \emph{do not} occur at generic points. For instance, $f^i_C$ can split into $f^i_C-x_1-x_2$, $x_1$ $x_2$ where $x_1$ and $x_2$ are exceptional curves produced by two blowups. These blowups are nongeneric since we require both the blowups to occur on a single representative $f^i_C$ of the homology class of $f^i_C$, which cannot be arranged if the blowups were acting at two generic points.

The points on $C$ through which other components of discriminant locus pass can be associated to hypermultiplets in the $6d$ theory charged under some representation of $\fg_C$. Such points and corresponding representations were systematically studied in \cite{Grassi:2011hq}. We have seen above that such points on $C$ can also be associated to some blowups on the collection of surfaces $S_C=\cup_i S^i_C$. Thus, the blowups on $S_C$ can be classified into various sets such that the blowups in a certain set correspond to a certain hyper that is charged under a certain representation of $\fg_C$.

\subsection{Consequences of the structure of elliptic fibration: Collision of curves} \label{multiple}
Let's see how the above correspondence between blowups and $6d$ matter turns out to be useful. Consider two compact holomorphic curves $C$ and $D$ which intersect each other at some point $p$ and both of which are components of the discriminant locus. The collections of surfaces $S_C$ and $S_D$ intersect with each other over $p$. So, $p$ is associated to some gluings between the collections of surfaces $S_C$ and $S_D$. We saw above that, from the point of view of $C$, $p$ is associated to some blowups $x_\alpha^p$ in $S_C$ and some hypers in a representation $R^p_C$ of $\fg_C$. Similarly, from the point of view of $D$, $p$ is associated to some blowups $y_\mu^p$ in $S_D$ and some hypers in a representation $R^p_D$ of $\fg_D$. The fact that $C$ and $D$ intersect at $p$ means that the hypers in $R^p_C$ and hypers in $R^p_D$ are identified with each other to give rise to hypers transforming in the representation $R^p_C\otimes R^p_D$ of $\fg_C\oplus\fg_D$. This suggests that only the corresponding blowups $x_\alpha^p$ and $y_\mu^p$ should participate in the gluing of $S_C$ and $S_D$ associated to $p$. We will see later in this paper that this expectation is correct.

We now describe an important consistency condition that a gluing between $S_C$ and $S_D$ has to satisfy. First recall that the elliptic fiber $f_C$ over $C$ can be written in terms of the components $f^i_C$ as in (\ref{fiber}). Similarly, $f_D$ can be written in terms of its components $f^a_D$ as
\be
f_D=\sum_a m_a f^a_D
\ee
Now, the gluing between $S_C$ and $S_D$ involves curves $f^i_C$ and $x^p_\alpha$ in $S_C$, and curves $f^a_D$ and $y^p_\mu$ in $S_D$. Say the gluing is such that $f^i_C$ is glued to some curve $D[f^i_C]$ in $S_D$ and $f^a_D$ is glued to some curve $C[f^a_D]$ in $S_C$. Since the elliptic fiber $f_C$ should be glued to the elliptic fiber $f_D$ at $p$ we must have that the condition
\be
\sum_i n_i D[f_C^i]=f_D=\sum_a m_a f^a_D \label{glue1}
\ee
holds in $S_D$ and the condition
\be
\sum_a m_a C[f^a_D]=f_C=\sum_i n_i f^i_C \label{glue2}
\ee
holds in $S_C$.

Above we focused our attention only to those compact holomorphic curves in $B$ that are components of discriminant locus. There can also be compact holomorphic curves that are not in the discriminant locus. Consider one such curve $C$. $C$ carries a smooth elliptic fiber which is assigned a Kodaira type I$_0$ even though it is smooth.  Since elliptic fiber $f_C$ over $C$ does not split into multiple pieces, as $f_C$ moves over $C$ it sweeps out a single compact holomorphic surface $S_C$ rather than a collection of surfaces. If $(C\cdot C)|_B\le-3$, then $C$ must carry a non-trivial $\fg_C$ which implies that it must be a component of discriminant locus \cite{Morrison:2012np}. Since in the context of $6d$ SCFTs the self-intersection of $C$ in $B$ has to be negative, it follows that if $C$ carries I$_0$ fiber then $C^2=-1,-2$. If $C^2=-2$, then the fibration of $f_C$ over $C$ is trivial, as is evident from the fact that such a $C$ preserves $16$ supercharges locally. In this case, $S_C=\P^1\times \mbb{T}^2$. If $C^2=-1$, the fibration of $f_C$ over $C$ is non-trivial and it is known that $S_C$ is $\P^2$ blown up at 9 \emph{non-generic} points. We will denote $\P^2$ blown up at 9 points by $(\P^2)^9$.

The first surface $\P^1\times \mbb{T}^2$ is certainly not Hirzebruch but the second surface $(\P^2)^9$ can, in some situations, be represented as the Hirzebruch surface $\F_1$ blown up at 8 points, which we denote as $\F_1^8$. One such situation arises when there is only one compact holomorphic curve $C$ in $B$ with $C^2=-1$ and Kodaira type of $f_C$ is I$_0$. This constructs the well-known E-string theory in $6d$ and in this case $S_C=dP_9$ which is $\P^2$ blown up at 9 \emph{generic} points. Since $\P^2$ blown up at one point equals $\F_1$, $dP_9$ can also be written as $\F_1^8$ with all 8 blowups being generic. In situations where $-1$ curve $C$ intersects some other compact holomorphic curve $D$ carrying a non-trivial $\fg_D$, we will show in Section \ref{sp0} that the blowups are not generic and it is not always possible to provide a description of $S_C$ as $\F_1^8$ even if we make the 8 blowups non-generic. But in all these situations, we can always give a description of $S_C$ as $(\P^2)^9$ with 9 non-generic blowups. So, in this sense, $(\P^2)^9$ should not be regarded as a Hirzebruch surface.

To recap, for each compact holomorphic curve $C$ in the base $B$, we can find a collection of surfaces $S_C=\cup_i S^i_C$ in $\tilde X_\fT$ which capture the behavior of elliptic fiber over $C$.  If $C$ is a component of discriminant locus, then $S^i_C$ is a Hirzebruch surface blown up at some number of points. Each blowup is associated to a hyper transforming in some representation of the $6d$ gauge algebra $\fg_C$ living over $C$, if $\fg_C$ is non-trivial. If $C$ is not a component of discriminant locus, then $S^i_C$  is either $\P^2$ blown up at 9 points, or $\P^1\times T^2$. The blowups on $\P^2$ are not associated to $6d$ hypers since $\fg_C$ is trivial. The surfaces $S^i_C$ are glued to each other and the gluings have to satisfy the Calabi-Yau condition (\ref{CY}). If $D$ is some other compact holomorphic curve that intersects $C$, then the corresponding collections of surfaces $S_C$ and $S_D$ are glued to each other. These gluings not only have to satisfy (\ref{CY}), but also the conditions (\ref{glue1}) and (\ref{glue2}) which state that the elliptic fibers over $C$ and $D$ have to be glued to each other. If both $\fg_C$ and $\fg_D$ are non-trivial, then the intersection of $D$ with $C$ gauges hypers transforming in some representation $R^p_C$ of $\fg_C$, which manifests as the fact that the blowups corresponding to these hypers in $S_C$ are involved in the gluing. Similarly, the intersection also gauges hypers transforming in some representation $R^p_D$ of $\fg_D$, which manifests as the fact that the blowups corresponding to these hypers in $S_D$ are involved in the gluing.

\subsection{Unpacking the Calabi Yau: Mori cone} \label{mori}
In this way, we claim that we can package the relevant data for the compactification of M-theory on $\tilde X_\fT$ into a collection of compact surfaces $S_\fT=\cup_C(\cup_i S^i_C)$ along with gluing rules describing how two surfaces $S^i_C$ and $S^j_D$ are glued to each other. To justify this claim, we will now show how one can extract from this the data listed at the end of Section \ref{M} which is the set of compact holomorphic surfaces in $\tilde X_\fT$, the set of compact holomorphic curves in $\tilde X_\fT$, and the intersection numbers between these curves and surfaces.

The set of compact holomorphic surfaces is simply given by $S^i_C$ for all values of $i$ and $C$. The set of compact holomorphic curves can be specified by positive \emph{integer} linear combinations of a set of linearly independent compact holomorphic curves, which we call generators. These generators are also the generators of the \emph{Mori cone}, which is obtained by tensoring the set of holomorphic curves by $\R^+$. By a slight abuse of terminology, in this paper we will refer to the set of holomorphic curves as the Mori cone and it should always be remembered that we are actually referring to the underlying set of holomorphic curves. 


The set of compact complex curves on the other hand can be described as arbitrary (positive or negative) integer linear combinations of a set of linearly independent compact complex curves. Thus the set of complex curves forms a lattice. The Mori cone can then be thought of as embedded in this lattice, which turns out to be a useful point of view as we will see soon.

We have seen that $S^i_C$ is either $\P^1\times \mbb{T}^2$, or a Hirzebruch surface $\F_n^b$ of degree $n$ and $b$ number of blowups, or $(\P^2)^9$. In the rest of the discussion, we can ignore the surface $\P^1\times \mbb{T}^2$. See Section \ref{rhate} for more details. Let's call the underlying surface without the blowups as $(S^i_C)_u$. $(S^i_C)_u$ is then either a Hirzebruch surface $\F_n$ of degree $n$, or the projective space $\P^2$. Let's review the lattice of complex curves $(\Lambda^i_C)_u$ and the Mori cone $(\cM^i_C)_u$ of $(S^i_C)_u$:
\ben
\item $\F_n$ is a $\P^1$ fibration over $\P^1$. $(\Lambda^i_C)_u$ is generated by $e$ which is a zero section of the fibration, and by $f$ which is the fiber $\P^1$. The intersection numbers are $e^2=-n$, $f^2=0$ and $e\cdot f=0$. Another notable curve in $\F_n$ is a second zero section $h=e+nf$ whose self-intersection is $h^2=+n$. $(\cM^i_C)_u$ is obtained by restricting to positive integer linear combinations of $e$ and $f$.
\item $(\Lambda^i_C)_u$ for $\P^2$ is simply generated by a curve $l$ of self-intersection $l^2=+1$. As above, $(\cM^i_C)_u$ is generated by positive multiples of $l$.
\een
For future use, we have also collected the intersection numbers of curves inside these surfaces as well.

One general property one requires of Mori cone is that its generators must have a well-defined genus. The genus $g$ of a curve $\ell$ in $S$ can be determined by the adjunction formula which states that
\be
(K_S+\ell)\cdot\ell=2g-2 \label{g}
\ee
where $K_S$ is the canonical divisor class of $S$. Thus a curve $\ell$ has a well-defined genus only if $(K_S+\ell)\cdot\ell$ is even and is greater than or equal to $-2$. An example of a curve that does not have a well-defined genus is the curve $2e$ inside $\F_n$. 

Another property one requires captures the geometric intuition that two distinct irreducible curves should intersect in a non-negative number of points. The intuition is that, if two curves $\ell_1$ and $\ell_2$ intersect in a negative number of points, then it must be that the intersection locus of $\ell_1$ and $\ell_2$ involves a curve $\ell_3$ of negative self-intersection, and consequently the curves can be written as $\ell_1=\ell_3+\ell'_1$ and $\ell_2=\ell_3+\ell'_2$ with the mutual intersections between $\ell'_1$, $\ell'_2$ and $\ell_3$ being non-negative. This would imply that either $\ell_1$ or $\ell_2$ is not irreducible depending on whether or not $\ell'_1$ and $\ell'_2$ are non-zero. Thus to capture irreducibility, one demands that if $\ell_a$ is the set of generators of the Mori cone $\cM_S$ of a surface $S$, then it must be that
\be
\ell_a\cdot\ell_b\ge0 \label{i}
\ee
for all $a\neq b$. For a curve $\ell$ that is not a generator, we say that $\ell$ is irreducible if
\be
\ell\cdot\ell_a\ge0
\ee
for all $a$. An example of an irreducible curve that is not a generator is the curve $h$ in $\F_n$. An example of a curve that is not irreducible is $e+f$ in $\F_n$ for $n\ge2$.

We now describe the construction of Mori cone $\cM^i_C$ of the surface $S^i_C$ with the blowups included. Let $x_\alpha$ with $\alpha=1,\cdots,b$ denote the blowups on $S^i_C$. If $(S^i_C)_u$ is a Hirzebruch surface, we let $\Lambda^i_C$ be the lattice generated by $e$, $f$ and $x_\alpha$. If $(S^i_C)_u$ is $\P^2$, we let $\Lambda^i_C$ be the lattice generated by $l$ and $x_\alpha$ where $\alpha=1,\cdots,9$. The intersection product on $\Lambda^i_C$ follows from the intersection product on $(\Lambda^i_C)_u$ and the intersection products $x_\alpha\cdot x_\beta=-\delta_{\alpha\beta}$, $x_\alpha\cdot e=x_\alpha\cdot f=x_\alpha\cdot l=0$. The canonical divisor $K^i_C$ for $S^i_C$ is
\be
K^i_C=\left(K^i_C\right)_u+\sum_\alpha x_\alpha
\ee
where $\left(K^i_C\right)_u$ is the canonical divisor for $(S^i_C)_u$ whose intersection products with $x_\alpha$ are zero by definition. 

Let $\cX^i_C$ denote the positive linear integer span of $x_\alpha$. Then $(\cM^i_C)_b=(\cM^i_C)_u\oplus\cX^i_C$ is a cone inside $\Lambda^i_C$ satisfying (\ref{g}) and (\ref{i}). By definition of $\F_n^b$ and $(\P^2)^9$, the curves contained in $(\cM^i_C)_b$ must be holomorphic, and hence $(\cM^i_C)_b$ must be contained inside $\cM^i_C$. So our task is to determine the elements in $\cM^i_C$ that are not in $(\cM^i_C)_b$. 

First, consider performing one blowup $x$ on $(S^i_C)_u$ at some point $p$. Since the curves of non-negative self-intersection can be deformed, we can always a find a holomorphic representative passing through $p$. Thus the blowup will definitely act on any curve $C$ such that $C^2\ge0$ and split it into $C-x$ and $x$, where $(C-x)^2\ge-1$. On the other hand, since the holomorphic curves of negative self-intersection are rigid, they may or may not pass through $p$. If $p$ is generic, then they will not pass through $p$. If $p$ is sufficiently non-generic, then they will. We can generalize to $b$ blowups as follows. If $(S^i_C)_u$ is a Hirzebruch surface, let $W^i_C$ be the set of curves of the form
\be
n_e e+n_f f-\sum_\alpha n_\alpha x_\alpha \label{-11}
\ee
with $n_e,n_f,n_\alpha\ge0$, $n_e+n_f\ge1$ which have self-intersection greater than or equal to $-1$. Similarly, if $(S^i_C)_u=\P^2$, let $W^i_C$ be the set of curves of the form
\be
n_l l-\sum_\alpha n_\alpha x_\alpha \label{-12}
\ee
with $n_l\ge1$, $n_\alpha\ge0$ which have self-intersection greater than or equal to $-1$. The generalization is that the curves in $W^i_C$ must be holomorphic and hence must be included in $\cM^i_C$. Clearly, these curves do not lie in $(\cM^i_C)_b$ whenever any one of the $n_\alpha$ is non-zero. Similarly, the elements of the set $G^i_C$ of gluing curves in $S^i_C$ are elements of $\Lambda^i_C$ but may not be elements of the cone $(\cM^i_C)_b$. So, we define $\cM^i_C$ to be the \emph{minimal} cone inside $\Lambda^i_C$ that contains all the elements of $(\cM^i_C)_b$, $W^i_C$ and $G^i_C$, and whose generators satisfy (\ref{g}), (\ref{i}). There is an intersection product on $\cM^i_C$ descending from the intersection product on $\Lambda^i_C$ described above. We will explain the reason for demanding the cone to be minimal in Section \ref{hate}.

Finally, we can write the Mori cone $\cM_\fT$ of the threefold $\tilde X_\fT$ as
\be
\cM_\fT=\left[\oplus_C\left(\oplus_i\cM^i_C\right)\right]/\sim
\ee
where $\sim$ means that we identify the curves in different surfaces that are glued to each other, and this also includes self-gluings if any. Note that unlike the Mori cones for surfaces, there is no notion of intersection product on the Mori cone $\cM_\fT$ for the threefold.

\subsection{Unpacking the Calabi Yau: Intersection numbers}\label{num}

However, there are other intersection products that are well defined for threefold $\tilde X_\fT$. The first one is the triple intersection product $S^i_C\cdot S^j_D\cdot S^k_E$ of three distinct surfaces of $S^i_C$, $S^j_D$ and $S^k_D$. This number can be written as an intersection product of gluing curves inside any of the three surfaces. Let $\ell^{ij}$ in $S^i_C$ and $\ell^{ji}$ in $S^j_D$ be the sum of gluing curves participating in the gluing of $S^i_C$ and $S^j_D$; $\ell^{jk}$ in $S^j_D$ and $\ell^{kj}$ $S^k_E$ be the sum of all the gluing curves participating in the gluing of $S^j_D$ and $S^k_E$, and $\ell^{ki}$ in $S^k_E$ and $\ell^{ik}$ in $S^i_C$ be the sum of all the gluing curves participating in the gluin of $S^k_E$ and $S^i_C$. Then,
\be
S^i_C\cdot S^j_D\cdot S^k_E=(\ell^{ij}\cdot\ell^{ik})|_{S^i_C}=(\ell^{ji}\cdot\ell^{jk})|_{S^j_D}=(\ell^{ki}\cdot\ell^{kj})|_{S^k_E} \label{any}
\ee
If only two of the surfaces in a triple intersection product are distinct, that is if we are computing $S^i_C\cdot S^i_C\cdot S^j_D$, then it can be written as
\be
S^i_C\cdot S^i_C\cdot S^j_D=(\ell^{ji}\cdot\ell^{ji})|_{S^j_D}=(K'^i_C\cdot\ell^{ij})|_{S^i_C}
\ee
where $K'^i_C=K^i_C$ denotes the canonical divisor of $S^i_C$ if there are no self-gluings. If there are self-gluings, it means that we have some curves $g_\mu$ and $h_\mu$ for $\mu=1,\cdots,N$ in $S^i_C$ such that $g_\mu$ is glued to $h_\mu$ for all $\mu$. In such a situation, these curves involved in the self-gluing contribute to $K'^i_C$ and we have
\be
K'^i_C=K^i_C+\sum_{\mu=1}^N\left(g_\mu+h_\mu\right)\label{Ksh}
\ee
If all the three surfaces are same, then
\be
S^i_C\cdot S^i_C\cdot S^i_C=(K'^i_C\cdot K'^i_C)|_{S^i_C} \label{K'}
\ee
Thus, all the triple intersections of $S^i_C$ can be recovered from the intersection products on $\cM^i_C$.

Another intersection product that is defined for $\tilde X_\fT$ is the intersection of a surface $S^i_C$ with that of a curve $\ell$ in $\cM_\fT$. If $\ell$ lives inside $\cM^i_C$, then
\be
\ell\cdot S^i_C=(\ell\cdot K'^i_C)|_{S^i_C}
\ee
If $\ell$ lives inside some other surface $S^j_D$ then
\be
\ell\cdot S^i_C=(\ell\cdot \ell^{ji})|_{S^j_D}
\ee 
where $\ell^{ji}$ in $S^j_D$ is the sum of gluing curves participating in the gluing between $S^i_C$ and $S^j_D$. Thus, all the intersections between compact holomorphic surfaces and curves can also be recovered from the intersection products on $\cM^i_C$. This completes our argument that the data about $S_\fT$ plus the gluing rules provided in this paper fully capture the data (listed at the end of Section \ref{M}) of the Coulomb branch of the $5d$ KK theory originating from the compactification of M-theory on $\tilde X_\fT$.

\subsection{RG flows via flops and blowdowns} \label{RG}

As discussed in Section \ref{intro}, to classify $5d$ SCFTs we have to determine all the RG flows of all the $5d$ KK theories which preserve the rank of the Coulomb branch. We now show how these RG flows can be implemented in our formalism in terms of blowdowns of rational curves of self-intersection $-1$ in $S^i_C$. We will refer to rational curves of self-intersection $-1$ as ``$-1$ curves'' in what follows. 

Let $\ell$ be a $-1$ curve in $S^i_C$. Blowing down $\ell$ induces a map $\Lambda^i_C\to\Lambda'^i_C$ under which any other curve $\ell'$ transforms as
\be
\ell'\to\ell'+(\ell'\cdot\ell)\ell \label{bd}
\ee
Consistency of this map with complex structure requires that holomorphic curves be sent to holomorphic curves and non-holomorphic curves be sent to non-holomorphic curves. This requirement implies that $\ell$ must be a generator of $\cM^i_C$. Because, if it is not, then it can be written as
\be
\ell=(\ell-\ell')+\ell'
\ee
where $\ell'$ and $\ell-\ell'$ are holomorphic. We see that if we blow down $\ell$ then
\begin{align}
\ell'&\to\ell'+(\ell'\cdot\ell)\ell\\
\ell'-\ell&\to\ell'-\ell+(\ell'\cdot\ell)\ell-(\ell\cdot\ell)\ell=\ell'+(\ell'\cdot\ell)\ell
\end{align}
that is, both $\ell'$ and $\ell'-\ell$ are transformed to the same curve. Since $\ell'$ is holomorphic and $\ell'-\ell$ is non-holomorphic, this implies that the resulting curve $\ell'+(\ell'\cdot\ell)\ell$ is both holomorphic and non-holomorphic, which is a contradiction.

So, the relevant objects for the study of RG flows are $-1$ curves that are generators of $\cM^i_C$ for some $i,C$. As first explained in Section 5 of \cite{Bhardwaj:2018yhy}, we can divide the subsequent analysis into three cases which we expound:
\bit
\item First assume that the curve $\ell$ in $S^i_C$ being blown down has zero intersection with all the gluing curves in $S^i_C$. Then we can simply do the blowdown. This replaces $S^i_C$ with the blowndown surface $S'^i_C$ while all the other surfaces remain the same. The Mori cone $\cM^i_C$ transforms to the Mori cone $\cM'^i_C$ according to (\ref{bd}). This gives us a new Calabi-Yau threefold $\tilde X'_\fT$.
\item Next assume that $\ell$ only has non-positive intersections with the gluing curves in $S^i_C$. In other words, the gluing curve for some of the gluings is $\ell$ itself, and $\ell$ does not intersect any other gluing curve. Say $\ell$ is glued to curves $\ell_p$ in surfaces $S^{i_p}_{C_p}$ for $p=1,\cdots k$. By (\ref{CY}), $\ell_p$ is then a $-1$ curve in $S^{i_p}_{C_p}$. As we blow down $\ell$ in $S^i_C$, we must simultaneously blow down $\ell_p$ in $S^{i_p}_{C_p}$ leading to a simultaneous blowdown $S^i_C\to S'^i_C$, $S^{i_p}_{C_p}\to S'^{i_p}_{C_p}$. The gluings $\ell\sim\ell_p$ are simply thrown out. The removal of these gluings might cause the set of holomorphic compact surfaces to split into multiple subsets such that only the surfaces in a single subset are glued to each other, and none of the surfaces in one subset is glued to any of the surfaces in another subset. This means that at the endpoint of RG flow we obtain multiple decoupled lower rank SCFTs. So such RG flows can be discarded. The other possibility is that there are enough gluings among $S^i_C$ that even after the removal of gluings $\ell\sim\ell_p$, all the surfaces remain glued to each other. In this situation, the RG flow is rank preserving and hence we have to keep track of such flows.
\item Finally, let's assume that $\ell$ has positive intersections with some of the gluing curves $\ell_\alpha$ in $S^i_C$ gluing $S^i_C$ to $S^{i_\alpha}_{C_\alpha}$ for $\alpha=1,\cdots,\mu$, and that $\ell$ is glued to $\ell_p$ in $S^{i_p}_{C_p}$ for $p=1,\cdots k$. Let us denote the curve in $S^{i_\alpha}_{C_\alpha}$ that is glued to $\ell_\alpha$ as $\ell^\alpha$. Since $\ell$ intersects $\ell_\alpha$, we have a non-trivial triple intersection of $S^i_C$, $S^{i_\alpha}_{C_\alpha}$ and $S^{i_p}_{C_p}$ for all values of $\alpha,p$. The triple intersection, in particular, requires that $S^{i_\alpha}_{C_\alpha}$ must be glued to $S^{i_p}_{C_p}$. Let's call the corresponding gluing curve as $\ell_{\alpha p}$ in $S^{i_\alpha}_{C_\alpha}$ and $\ell_{p\alpha}$ in $S^{i_p}_{C_p}$. The consistency of triple intersection requires that 
\be
\ell\cdot\ell_\alpha=\ell_p\cdot\ell^\alpha=S^{i_\alpha}_{C_\alpha}\cdot S^i_C\cdot S^{i_p}_{C_p}=n_{\alpha} \label{hit}
\ee 
which does not depend on $p$. 

If we now do a simultaneous blowdown $\ell\sim\ell_p$, we change the self-intersections of $\ell_\alpha$ and $\ell_{p\alpha}$ to $\ell^2_\alpha+n^2_\alpha$ and $\ell^2_{p\alpha}+n^2_\alpha$ respectively, thus violating the Calabi-Yau condition (\ref{CY}) for the gluing between $S^i_C$ and $S^{i_\alpha}_{C_\alpha}$, and the gluing between $S^{i_p}_{C_p}$ and $S^{i_\alpha}_{C_\alpha}$. To rectify this, we can do a blow-up inside each $S^{i_\alpha}_{C_\alpha}$ which hits $\ell^\alpha$ and $\ell_{\alpha p}$ at $n_{\alpha}$ number of points. This changes the self-intersections of $\ell^\alpha$ and $\ell_{\alpha p}$ to $(\ell^\alpha)^2-n^2_\alpha$ and $\ell^2_{\alpha p}-n^2_\alpha$ respectively, thus restoring back the Calabi-Yau condition. 

Under the combined process of blowdown and blowup, $S^i_C$, $S^{i_p}_{C_p}$ are sent to $S'^i_C$, $S'^{i_p}_{C_p}$, and $S^{i_\alpha}_{C_\alpha}$ are sent to $(S^{i_\alpha}_{C_\alpha})^1$ which is $S^{i_\alpha}_{C_\alpha}$ blown up at one point such that $\ell^\alpha-n_\alpha x$ and $\ell_{\alpha p}-n_\alpha x$ (where $x$ is the exceptional divisor created by the blowup) must exist but otherwise the blowup is generic. We have already discussed above how to figure out the Mori cone of the surface in the presence of these kind of non-generic blowups. Using those ideas, we can compute the Mori cone $\cM'^{i_\alpha}_{C_\alpha}$ for $(S^{i_\alpha}_{C_\alpha})^1$.

This combined process of blowdown and subsequent blowup is referred to as a \emph{flop}. Two Calabi-Yaus related by a flop describe different chambers of the Coulomb branch of the same $5d$ theory. Thus, the above flop is not an example of RG flow. Instead, the flop transforms $\tilde X_\fT$ to a different Calabi-Yau which we denote as $\tilde X^\ell_\fT$. $\tilde X^\ell_\fT$ describes a different chamber of the Coulomb branch of the same $5d$ KK theory $\fT_{KK}$ associated to the $6d$ SCFT $\fT$. However, $\tilde X^\ell_\fT$ will have its own set of $-1$ curves that could be blown down leading to $5d$ SCFTs that, in general, cannot be obtained via a blowdown of $\tilde X_\fT$.
\eit
Note that above we studied a single blowdown or a single flop starting from $\tilde X_\fT$. But we can continue this process in exactly the same way as above and perform a sequence of flops and blowdowns. In this way, we generate the full list of $5d$ SCFTs that arise from rank preserving RG flows of $\fT_{KK}$.

\subsection{Decoupled states}\label{hate}
In this section, we fix a loose end from Section \ref{mori}. There we demanded that the Mori cone $\cM^i_C$ of each surface $S^i_C$ be the minimal cone satisfying certain requirements that we discussed there. One of these requirements was that curves displayed in (\ref{-11}) and (\ref{-12}) must be a part of the Mori cone. As a consequence all the \emph{possible} curves that have self-intersection greater than or equal to $-1$ must be in $\cM^i_C$. Of course, there still are other curves of self-intersection $\ge -1$ in $\Lambda ^i_C$ but not in $\cM^i_C$, that is those that are not of the form displayed in (\ref{-11}) and (\ref{-12}). However, the blow-downs of such curves are non-holomorphic, and hence they cannot be a part of $\cM^i_C$.

Now consider a non-minimal cone $\tilde \cM^i_C$ satisfying all the consistency requirements detailed in Section \ref{mori}. It is going to have some extra curves compared to the minimal cone $\cM^i_C$. As we have seen above, these extra curves must have self-intersection $\le-2$. The addition of these extra curves can only reduce the number of generators that are $-1$ curves, and hence we must choose the minimal cone $\cM^i_C$ if we want to be able to access the full spectrum of RG flows.

We can complement the above argument by noticing that by definition $\cM^i_C$ embeds inside $\tilde\cM^i_C$. Let us call the KK theory obtained by choosing $\tilde \cM^i_C$ to be the Mori cone as $\tilde \fT_{KK}$. The set of BPS strings is same for both $\fT_{KK}$ and $\tilde\fT_{KK}$. But the set of BPS particles for $\fT_{KK}$ embeds into the set of BPS particles of $\tilde \fT_{KK}$. This suggests that the whole theory $\fT_{KK}$ embeds into the theory $\tilde\fT_{KK}$. We interpret this to mean that $\tilde\fT_{KK}$ is the same as $\fT_{KK}$ but carries some extra states that can be decoupled at those loci in the Coulomb branch where these states become massless. We identify these ``decoupled states'' as the extra elements in $\tilde \cM^i_C$ as compared to $\cM^i_C$. These extra states are certainly not decoupled from the point of view of a generic point on the Coulomb branch, as can be seen for instance by noticing that they have a nontrivial Dirac pairing with the BPS string arising from $S^i_C$. We are only claiming that it is possible to decouple these extra states when they are massless.

A well-known example of decoupled states is given by the threefold with a single surface $\F_2$. This is claimed to construct the same $5d$ SCFT as constructed by the threefold containing the single surface $\F_0$. It has been demonstrated by computations of Nekrasov partition functions that $\F_2$ contains extra decoupled states as compared to $\F_0$. See, for instance, \cite{Taki:2014pba} and references therein. In this case, it can be easily seen that the Mori cone of $\F_0$ embeds into the Mori cone of $\F_2$ via $e\to e+f$ and $f\to f$. Note that this map preserves intersection numbers in the surface.

Thus we propose that the correct Mori cone describing Coulomb branch of the $5d$ SCFT without any decoupled states is actually the minimal cone $\cM^i_C$. Notice that this explanation fits neatly with the above mentioned fact that $\tilde \cM^i_C$ allows for fewer RG flows as compared to $\cM^i_C$. The reason is that the extra decoupled states might not have a consistent coupling to the Coulomb branch of the theory obtained after the RG flow. This manifests itself in the fact that the $-1$ curve responsible for the RG flow becomes a non-generator element in $\tilde \cM^i_C$.

\subsection{Formal gauge algebra}\label{rhate}
For $6d$ SCFTs whose gauge algebras have low enough rank, it is often possible to find multiple F-theory configurations leading to the same $6d$ SCFT in the infrared. For example, fibers of types II, III and IV typically give rise to such extra constructions where they appear as the more singular versions of fibers of types I$_1$, I$_2$ and I$_3$. We conjecture that, in these cases, $\tilde X_\fT$ arising from constructions involving more singular fibers have extra decoupled states\footnote{The exact sense in which these states are decoupled is explained in Section \ref{hate}.} compared to $\tilde X_\fT$ arising from the configuration involving the less singular fibers. So in such cases we do not need to understand the collection of surfaces $S_C$ corresponding to fibers of types II, III and IV. We can replace $S_C$ by the collection of surfaces corresponding to fibers of types I$_1$, I$_2$ and I$_3$.


Now, there do exist a handful of $6d$ SCFTs that can only be constructed if we use fibers of types II, III and IV, and don't admit a construction involving the less singular fibers of types I$_1$, I$_2$ and I$_3$. However, if we play the same game as above and assign the collection of surfaces corresponding to the less singular fiber types, then we find that these collections of surfaces can be glued together without any inconsistency! Let us provide an example to make things clearer. Consider the following F-theory configuration
\begin{align}
\label{na}
\begin{array}{c}
\begin{tikzpicture}[scale=1.5]
\begin{scope}[]
\draw  (-3.5,4) ellipse (0.5 and 0.5);
\node at (-3.5253,3.8046) {$-2$};
\node at (-3.4865,4.219) {I$_2$};
\end{scope}
\begin{scope}[shift={(1,0)}]
\draw  (-3.5,4) ellipse (0.5 and 0.5);
\node at (-3.5253,3.8046) {$-2$};
\node at (-3.4865,4.219) {I$^s_3$};
\end{scope}
\begin{scope}[shift={(0,1)}]
\draw  (-3.5,4) ellipse (0.5 and 0.5);
\node at (-3.5253,3.8046) {$-2$};
\node at (-3.4865,4.219) {I$_1$};
\end{scope}
\begin{scope}[shift={(-1,0)}]
\draw  (-3.5,4) ellipse (0.5 and 0.5);
\node at (-3.5253,3.8046) {$-2$};
\node at (-3.4865,4.219) {I$_1$};
\end{scope}
\end{tikzpicture}
\end{array}
\end{align}
where each circle represents a $\P^1$ in $B$ and the different $\P^1$ intersect transversely in the pattern shown above. We have also shown the self-intersection of each $\P^1$ in $B$ and the Kodaira-Tate fiber type over it. The above F-theory configuration is not consistent. However, the following more singular version
\begin{align}
\label{a}
\begin{array}{c}
\begin{tikzpicture}[scale=1.5]
\begin{scope}[]
\draw  (-3.5,4) ellipse (0.5 and 0.5);
\node at (-3.5253,3.8046) {$-2$};
\node at (-3.4865,4.219) {IV$^{ns}$};
\end{scope}
\begin{scope}[shift={(1,0)}]
\draw  (-3.5,4) ellipse (0.5 and 0.5);
\node at (-3.5253,3.8046) {$-2$};
\node at (-3.4865,4.219) {IV$^s$};
\end{scope}
\begin{scope}[shift={(0,1)}]
\draw  (-3.5,4) ellipse (0.5 and 0.5);
\node at (-3.5253,3.8046) {$-2$};
\node at (-3.4865,4.219) {II};
\end{scope}
\begin{scope}[shift={(-1,0)}]
\draw  (-3.5,4) ellipse (0.5 and 0.5);
\node at (-3.5253,3.8046) {$-2$};
\node at (-3.4865,4.219) {II};
\end{scope}
\end{tikzpicture}
\end{array}
\end{align}
of the theory is consistent. We find that if were to ignore the fact that the configuration (\ref{na}) is not allowed and try to glue the collections of surfaces corresponding to each curve in (\ref{na}), then we can do so consistently. We interpret the existence of consistent gluings to mean that even though the collections of surfaces corresponding to configuration (\ref{na}) form a Calabi-Yau threefold, for some subtle reason the threefold doesn't admit an elliptic fibration structure until we add some decoupled states to reach the Calabi-Yau formed by collections of surfaces corresponding to (\ref{a}) We would like to note that this phenomenon is specific to low rank gauge algebras only and for slightly higher rank versions of (\ref{na}), the corresponding collections of surfaces do not admit consistent gluings. For example, the following F-theory configuration
\begin{align}
\label{nae}
\begin{array}{c}
\begin{tikzpicture}[scale=1.5]
\begin{scope}[]
\draw  (-3.5,4) ellipse (0.5 and 0.5);
\node at (-3.5253,3.8046) {$-2$};
\node at (-3.4865,4.219) {I$_4^s$};
\end{scope}
\begin{scope}[shift={(1,0)}]
\draw  (-3.5,4) ellipse (0.5 and 0.5);
\node at (-3.5253,3.8046) {$-2$};
\node at (-3.4865,4.219) {I$^s_5$};
\end{scope}
\begin{scope}[shift={(0,1)}]
\draw  (-3.5,4) ellipse (0.5 and 0.5);
\node at (-3.5253,3.8046) {$-2$};
\node at (-3.4865,4.219) {I$_2$};
\end{scope}
\begin{scope}[shift={(-1,0)}]
\draw  (-3.5,4) ellipse (0.5 and 0.5);
\node at (-3.5253,3.8046) {$-2$};
\node at (-3.4865,4.219) {I$_2$};
\end{scope}
\end{tikzpicture}
\end{array}
\end{align}
is not consistent for a reason that is similar to the reason due to which the configuration (\ref{na}) is inconsistent. However, unlike (\ref{na}), the theory corresponding to (\ref{nae}) does not admit a more singular construction. Correspondingly, we find that there are no consistent gluings for the collections of surfaces in (\ref{nae}).

\begin{table}[h]
\begin{center}
  \begin{tabular}{ | l | c | c | l | }
    \hline
    Fiber type & $C^2=-k$ & Formal gauge algebra & Matter content \\ \hline
    I$^s_n,n\geq3$ & $k=1,2$ & $\su(n)$ & \parbox[t]{6.5cm}{$kn+16-8k$ hypers in fundamental\\$2-k$ hypers in antisymmetric} \\ \hline
    I$^s_6$, tuned & $k=1$ & $\su(\tilde 6)$ & \parbox[t]{6.5cm}{15 hypers in fundamental\\ $\frac{1}{2}$-hyper in 3-index antisymmetric} \\ \hline
    IV$^s$ & $1\le k\le3$ & $\su(3)$ & \parbox[t]{6.5cm}{$6(3-k)$ hypers in fundamental} \\ \hline
    I$_2$, III, IV$^{ns}$ & $k=2$ & $\su(2)$ & \parbox[t]{6.5cm}{$4$ hypers in fundamental} \\ \hline
    I$_0$, I$_1$, II & $k=2$ & $\su(1)$ & \parbox[t]{6.5cm}{No matter} \\ \hline
    \parbox[t]{2.1cm}{I$^{ns}_{2n},n\geq2$\\I$^{ns}_{2n+1},n\geq1$} & $k=1$ & $\sp(n)$ & \parbox[t]{6.5cm}{$2n+8$ hypers in fundamental} \\ \hline
    I$_2$, III, IV$^{ns}$ & $k=1$ & $\sp(1)=\su(2)$ & \parbox[t]{6.5cm}{$10$ hypers in fundamental} \\ \hline
    I$_0$, I$_1$, II & $k=1$ & $\sp(0)$ & \parbox[t]{6.5cm}{No matter} \\ \hline
    I$^{*s}_{r-4},r\geq4$ & $k=4$ & $\so(2r)$ & \parbox[t]{6.5cm}{$2r-8$ hypers in vector} \\ \hline
    I$^{*ns}_{r-3},r\geq4$ & $k=4$ & $\so(2r+1)$ & \parbox[t]{6.5cm}{$2r-7$ hypers in vector} \\ \hline
    I$^{*s}_{2}$ & $1\le k\le 3$ & $\so(12)$ & \parbox[t]{6.5cm}{$8-k$ hypers in vector\\$4-k$ half-hypers in Weyl spinor} \\ \hline
    I$^{*s}_{1}$ & $1\le k\le 3$ & $\so(10)$ & \parbox[t]{6.5cm}{$6-k$ hypers in vector\\$4-k$ hypers in Weyl spinor} \\ \hline
    I$^{*ns}_{2}$ & $1\le k\le 3$ & $\so(11)$ & \parbox[t]{6.5cm}{$7-k$ hypers in vector\\$4-k$ half-hypers in spinor} \\ \hline
    I$^{*s}_{0}$ & $1\le k\le 3$ & $\so(8)$ & \parbox[t]{6.5cm}{$4-k$ hypers in vector\\$4-k$ half-hypers in spinor\\$4-k$ hypers in cospinor} \\ \hline
    I$^{*ns}_{1}$ & $1\le k\le 3$ & $\so(9)$ & \parbox[t]{6.5cm}{$5-k$ hypers in vector\\$4-k$ hypers in spinor} \\ \hline
    I$^{*ss}_{0}$ & $1\le k\le3$ & $\so(7)$ & \parbox[t]{6.5cm}{$3-k$ hypers in vector\\$2(4-k)$ hypers in spinor} \\ \hline
    I$^{*ns}_{0}$ & $1\le k\le3$ & $\fg_2$ & \parbox[t]{6.5cm}{$10-3k$ hypers in 7-dimensional irrep} \\ \hline
    IV$^{*s}$ & $1\le k\le6$ & $\fe_6$ & \parbox[t]{6.5cm}{$6-k$ hypers in 27-dimensional irrep} \\ \hline
    IV$^{*ns}$ & $1\le k\le5$ & $\ff_4$ & \parbox[t]{6.5cm}{$5-k$ hypers in 26-dimensional irrep} \\ \hline
    III$^{*}$ & $1\le k\le8$ & $\fe_7$ & \parbox[t]{6.5cm}{$8-k$ hypers in 56-dimensional irrep} \\ \hline
    II$^{*}$ & $k=12$ & $\fe_8$ & \parbox[t]{6.5cm}{No matter} \\
    \hline
  \end{tabular}
\end{center}
\caption{We assign a formal gauge algebra to each Kodaira-Tate fiber type. A tuned version of $I^s_6$ on $C^2=-1$ is shown separately and is assigned a formal gauge algebra of its own as it leads to different matter content from a generic I$^s_6$ on $C^2=-1$.}
\label{table}
\end{table}

In general, given a consistent F-theory configuration $X_\fT$ constructing a $6d$ SCFT $\fT$, we propose to perform the following replacements:
\bit
\item IV$^s$ on $C^2=-1,-2$ $\lra$ I$^s_3$ on $C^2=-1,-2$ respectively.
\item IV$^{ns}$ on $C^2=-1,-2$ $\lra$ I$_2$ on $C^2=-1,-2$ respectively.
\item III on $C^2=-1,-2$ $\lra$ I$_2$ on $C^2=-1,-2$ respectively.
\item II on $C^2=-2$ $\lra$ I$_1$ on $C^2=-2$.
\item I$_1$, II on $C^2=-1$ $\lra$ I$_0$ on $C^2=-1$.
\item I$_{2n+1}^{ns}$ on $C^2=-1$ $\lra$ I$_{2n}^{ns}$ on $C^2=-1$ for $n\ge2$.
\item I$_{3}^{ns}$ on $C^2=-1$ $\lra$ I$_{2}$ on $C^2=-1$.
\item I$_0$ on $C^2=-2$ $\lra$ I$_1$ on $C^2=-2$.
\eit
We claim that all of these replacements except the last one remove decoupled states, and hence maximize the available RG flows. See Section \ref{hate}. The last replacement does not remove any states but we claim that it maximizes RG flows based on the following observation. 

Consider KK theories corresponding to $(2,0)$ SCFTs in $6d$. These theories admit an $ADE$ classification. The KK theory corresponding to type $\fg$ $(2,0)$ SCFT is $5d$ $\cN=2$ SYM with a simply laced simple gauge algebra $\fg$, which can be viewed as an $\cN=1$ gauge theory with gauge algebra $\fg$ and an adjoint hyper. Turning on the mass of the adjoint hyper, we obtain an RG flow to pure $\cN=1$ gauge theory with gauge algebra $\fg$. 

The F-theory construction of $(2,0)$ SCFT of type $\fg$ involves rational curves of self-intersection $-2$ in $B$ carrying I$_0$ fibers intersecting in the pattern of the finite Dynkin diagram of type $\fg$. Each I$_0$ on $-2$ leads to a $\P^1\times T^2$ which does not have any $-1$ curves and this would incorrectly suggest that there are no RG flows. To see the above mentioned RG flow triggered by adjoint mass, we should use the surface associated to I$_1$ fiber over each $-2$ instead of $\P^1\times T^2$. We will see in Section \ref{gluing} that the surfaces corresponding I$_1$ can be glued together consistently for each choice of $\fg$. However, if $\fg$ is of $DE$ type, then the resulting collection of surfaces does not admit an elliptic fibration and hence cannot be used as a consistent F-theory background. So, to see the RG flow, we must use a collection of surfaces which seems inconsistent from the point of view of F-theory.

To account for the above replacements, we associate a \emph{formal gauge algebra} to each Kodaira-Tate fiber type. If the fiber type leads to a $6d$ gauge algebra, then the formal gauge algebra coincides with the $6d$ gauge algebra, except for $\su(6)$ which has two formal versions $\su(6)$ and $\su(\tilde 6)$. If the fiber type does not lead to any gauge algebra in $6d$, we associate to it a purely formal gauge algebra which is either $\su(1)$ or $\sp(0)$, corresponding respectively to tensor branches of $A_1$ $(2,0)$ SCFT and E-string theory. For the fiber types I$_0$, I$_1$, I$_2$, III and IV$^{ns}$, we also have to specify the self-intersection of the curve in the base to completely specify the formal gauge algebra. We collect our assignment of formal gauge algebras in Table \ref{table}. We take this opportunity to also list, in the context of $6d$ SCFTs, all the possible self-intersections of the curve $C$ in the base over which a specific fiber type can be realized, and the matter content appearing for a given choice of fiber type and self-intersection.

\subsection{Algorithm for building $\tilde X_\fT$} \label{close}
We close this section by describing the algorithm which, using the results in this paper, produces the data listed at the end of Section \ref{M} for $\tilde X_\fT$ starting from the input an F-theory configuration $X_\fT$ constructing the $6d$ SCFT $\fT$:
\ben
\item The input provides intersection pattern of rational curves in the base $B$, the self-intersections of all rational curves in $B$, and the Kodaira-Tate type of singularity over each rational curve.
\item To each Kodaira-Tate type of singularity, we associate a formal gauge algebra using Table \ref{table}.
\item To the data of each formal gauge algebra along with the self-intersection of the corresponding rational curve, we associate a graph specified in Section \ref{flop-eq}
\item To each transverse intersection of two rational curves, we glue the corresponding graphs using the gluing rules described in Section \ref{gluing}. We note that the gluing rules only depend on the formal gauge algebra and do not depend on the self-intersection.
\item Now using the proposals of Sections \ref{mori} and \ref{num}, we can convert this graph into the desired output.
\een

\section{Computational techniques}\label{over}
Our main computations can be categorized into two categories: computation of the triple intersection numbers between surfaces $S^i_C$ and computation of the degrees of the surfaces $S^i_C$ whenever they are Hirzebruch surfaces. We review our methods of computation and some mathematical background in this section.

\subsection{Tate form of the Weierstrass model}
We work with the Tate form \cite{Bershadsky:1996nh,Katz:2011qp} of the Weierstrass equation which can be written as
\begin{align}
W_0=y^2z+a_1 x yz+a_3 yz^2-(x^3 +a_2 x^2z+a_4 xz^2+a_6z^3 ) =0\label{W01}
\end{align} 
where again $[x:y:z]$ are homogeneous coordinates defining a $\P^2$. $z$ is a function on $B$ and $x,y,a_n$ are sections of $-2K_B,-3K_B,-nK_B$ respectively where $K_B$ is the canonical line bundle on $B$. $f$ and $g$ can be written as
\begin{align}
f&=-\frac{1}{3}\left(\frac{a_1^2}{4}+a_2\right)^2+\frac{a_1a_3}{2}+a_4\\
g&=\frac{2}{27}\left(\frac{a_1^2}{4}+a_2\right)^3+\frac{a_3^2}{4}-\frac{1}{3}\left(\frac{a_1^2}{4}+a_2\right)\left(\frac{a_1a_3}{2}+a_4\right)+a_6
\end{align}
and $\Delta$ can be computed using (\ref{dis}). The minimal orders of vanishing of $a_n$ for each Kodaira-Tate singularity type can be found in Table 2 of \cite{Katz:2011qp}.

We can think of $W_0=0$ as cutting out a Calabi-yau threefold $X_0$ inside a rank two projective bundle $Y_0$ defined by $[x:y:z]$ over $B$. In the context of this paper, the reader should think of $X_0=X_\fT$. Consider a compact holomorphic curve $C$ defined by a local coordinate $e_0=0$ in $B$ over which the orders of vanishing of $a_n$ are $q_n$. Then we can write $W_0$ as
\begin{align}
W_0=y^2z+a_{1,q_1} e_0^{q_1} x yz+a_{3,q_3} e_0^{q_3} yz^2-(x^3 +a_{2,q_2} e_0^{q_2} x^2z+a_{4,q_4} e_0^{q_4} xz^2+a_{6,q_6} e_0^{q_6} z^3 ) =0 \label{W02}
\end{align}
where $a_n=e_0^{q_n} a_{n,q_n}$.

$X_0$ is singular at $e_0=0$ if the elliptic fiber over $C$ is not of type I$_0$, I$_1$ or II. We can verify this by noticing that the system of equations
\begin{align}
W_0|_{e_0=0} = (\partial_x W_0)|_{e_0=0} =(\partial_y W_0)|_{e_0=0}=(\partial_{e_0} W_0)|_{e_0=0}=0 \label{sing}
\end{align}
which locates a singularity along $e_0=0$ admits $x=y=0$ as a solution whenever $q_3\ge1$, $q_4\ge1$ and $q_6\ge2$, which precisely happens whenever the Kodaira-Tate fiber type is not equal to I$_0$, I$_1$ or II. To resolve the singularity, we need to perform blowups.

\subsection{Blowups and resolution}
describe some general facts about blowups. Suppose the equation $W(y_i) =0$ describes a singular projective variety $X \subset Y$ realized as a hypersurface of an ambient projective space $Y$ with homogeneous coordinates $y_i$. Furthermore, suppose that the singular locus of $X$ is a subset of the complete intersection $g_1(y_i) = g_2(y_i) = \cdots =g_n=0$, where $g_i(y_j)$ are homogeneous polynomials in $y_j$. We would like to (partially) resolve the singular locus of $X$ by blowing up the locus $g_1 = g_2 = \cdots=g_n =0$; we call this locus the \emph{center} of the blowup and $n$ as the \emph{length} of the blowup center. A blowup can thus be described in terms of its center $(g_1,g_2,\dots,g_n)$, and a section $e$ whose zero locus $e=0$ is the exceptional divisor $E$ of the blowup. We adopt the following succinct notation:
	\begin{align}
		(g_1,g_2,\dots | e),
	\end{align}
which means we make the substitution
	\begin{align}
		g_1 = e g_1',~~g_2 = e g_2',~~\cdots 	\label{subblow}
	\end{align}
and introduce a new ambient projective space $Y'\rightarrow Y$ in which the locus $g_1=g_2=\cdots =0$ of the original projective bundle $Y$ has been replaced by a projective space $[g_1':g_2':\cdots]$ located at $e=0$ in $Y'$. Note that unlike $g_i$, $g'_i$ are merely variables defining homogeneous coordinates and not polynomials in the coordinates $y_j$ of the ambient space $Y$. In practice we abuse notation and simply use the symbol $g_i$ to refer to $g_i'$. $g_i'$ can be thought of as proper transforms of $g_i$ after the blowup.

Substituting (\ref{subblow}) into $W$, we can write it as
\be
W=e^pW'(y_i)
\ee
where we have dropped primes on the coordinates and $p$ is some number such that $e$ does not divide $W'$. Then we replace $X$ by the hypersurface $X'$ defined by $W'=0$ in $Y'$. $X'$ can be thought of as the proper transform of $X$ under the blowup. This procedure defines a map of hypersurfaces $X' \rightarrow X$ where $X' \subset Y'$ is said to be the blowup of $X$ along the center $g_1=g_2=\cdots =W=0$ (note that center must intersect $X$).

In order to obtain a smooth elliptic fibration starting from $X_0$, we identify a sequence of blowups which do not change the canonical class of the threefold, 
	\begin{align}
		X_r \overset{f_r}{\rightarrow} X_{r-1} \overset{f_{r-1}}{\rightarrow} \cdots \overset{f_2}{\rightarrow} X_{1} \overset{f_1}{\rightarrow} X_{0},
	\end{align}
such that for some choice of positive $r <\infty $ the elliptic fibration $X_r \rightarrow B$ is smooth. Then $X_r=\tilde X_\fT$. Let $n_i$ be the length of blowup center and $e_i$ be the resulting exceptional divisor at the step $X_i\to X_{i-1}$. The proper transform $W_{i-1} =0 $ is defined by
\be
W_{i-1}=e_i^{p_i}W_i
\ee
where $e_i$ does not divide $W_i$. To satisfy the \emph{Calabi-Yau condition} we must require that
\be
p_i=n_i-1
\ee

\noindent \ubf{Example: $\text{I}_2$ model}. 
Let us illustrate the above discussion with a simple example, namely a singular elliptically fibered threefold characterized by a type $\text{I}_2$ Kodaira fiber over a curve $C \subset B$ with $(C^2)_B =-k$. This geometry $X_0$, which we call the `$\text{I}_2$ model', can be realized explicitly by the following Weierstrass equation:
	\begin{align}
		W_0 =y^2 z +a_1 x y z +a_{3,1}e_0 y z^2 - (x^3 +a_{2,1}e_0 x^2 z +a_{4,1}e_0 x z^2  +a_{6,2}e_0^2 z^3 )=0. 
	\end{align} 
The above equation cuts out a hypersurface in the ambient 4-fold $Y_0 \rightarrow B$ whose fibers have homogeneous coordinates $[x:y:z]$. By looking for solutions to the equations $W_0 = \partial_x W_0 = \partial_y W_0 = \partial_{e_0} W_0 = 0$, one can easily show that the singular locus of $X_0$ is
	\begin{align}
		 e_0 = x = y=0. 
	\end{align}
We resolve $X_0$ by blowing up $Y_0$ along the above singular locus, which implies we make the substitution
	\begin{align}
		(x,y,e_0|e_1)~:~ x= e_1 x_1 ,~~ y= e_1 y_1,~~ e_0 =e_{1} e_{0,1}. 
	\end{align}
The new ambient space $Y_1 \rightarrow B$ has fibers with homogeneous coordinates $[e_1x_1:e_1 y_1:z][x_1:y_1:e_{0,1}]$. The total transform of $X_0$ under the blowup is described by the following equation:
	\begin{align}
		e_1^2 W_1 = 0	
	\end{align}
where the zero locus of the section
	\begin{align}
	\begin{split}
		&W_1 = 	y_1^2 z +a_1 x_1 y_1 z +a_{3,1}e_{0,1} y_1 z^2 \\
		&- (e_1 x_1^3 +a_{2,1}e_{0,1} e_1 x_1^2 z +a_{4,1}e_{0,1} x_1 z^2  +a_{6,2}e_{0,1}^2 z^3 )
	\end{split}
	\end{align}
describes the proper transform $X_1$. By showing that the equations $W_1 = \partial_{x_1} W_1 = \partial_{y_1} W_1 = \partial_{e_{0,1}} W_1 = \partial_{e_1} W_1 =0$ have no solution, one can easily verify that $X_1$ is smooth, and hence $X_1 \rightarrow X_0$ is a resolution. \\

\subsection{Components of elliptic fiber}
After the resolution, the elliptic fiber $f_C$ over any curve $C$ in $B$ splits into irreducible components $f^i_C$ as
\be
f_C=\sum_i n_i f^i_C
\ee
The equations describing $f^i_C$ can be obtained by studying the \emph{arithmetically irreducible} components of $e_i=W_r=0$ where $e_i$ are the proper transforms after all the blowups have been done. Arithmetically irreducible means that while factoring the equation, we remember that all the coefficients are sections over $C$. In particular, it may happen that the equation factorizes if we treat the coefficients as numbers but does not if we treat them as sections. The components obtained by treating them as numbers are called \emph{geometrically irreducible} but only a collection of the geometrically irreducible components are arithmetically irreducible. Stated differently, over a point on $C$, the fiber degenerates into geometrically irreducible components; but as we move over $C$, there can be monodromies sending a geometrically irreducible component to another, and then the collection of geometrically irreducible components invariant under the mondromies is known as an arithmetically irreducible component. It is the arithmetically irreducible components that become fibers $f^i_C$ of the irreducible divisors $S^i_C$.

We illustrate for the I$_2$ model below in which case $e_i=W_r=0$ turn out to be arithmetically irreducible, but in a general case $e_i=W_r=0$ split into some number of arithmetically irreducible components and each component defines a fiber $f^i_C$. The singular fiber $f$ of the resolved $\text{I}_2$ model $X_1 \rightarrow B$ splits into the following irreducible components:
	\begin{align}
		f = f^0 + f^1,
	\end{align}
where we have
	\begin{align}
	\begin{split}
		f^0 ~&:~ e_{0,1} = W_1=0\implies a_1 x_1 y_1 z-e_1 x_1^3+y_1^2 z=0\\
		f^1 ~&:~ e_1 =W_1=0\implies -e_{0,1}^2 z^2 a_{6,2}-e_{0,1} x_1 z a_{4,1}+e_{0,1} y_1 z a_{3,1}+a_1 x_1 y_1+y_1^2 = 0.
	\end{split}
	\end{align}
The two irreducible components $f^0,f^1$ intersect in two distinct points:
	\begin{align}
		f^0 \cap f^1 ~:~ e_{0,1} = e_1 = y_1 (y_1 + x_1 a_1) =0 ~~\subset ~~ [0:0:z][x_1:y_1:0],
	\end{align}
as is illustrated by the following affine Dynkin diagram: 
	\begin{align}
		\begin{array}{c}
			\begin{tikzpicture}
				\node[draw,circle,scale=.4,fill=black,label={above:$f^0$}] (F0) at (0,0) {$f^0$};
				\node[draw,circle,scale=.4,fill=black,label={above:$f^1$}] (F1) at (2,0) {$f^1$};
				\draw[transform canvas={yshift=-.3em}]  (F0) -- (F1);
				\draw[transform canvas={yshift=+.3em}]  (F0) -- (F1);
			\end{tikzpicture}	
		\end{array}
	\end{align}
Observe that the above Dynkin diagram is the affine $A_1$ Dynkin diagram associated to the Lie algebra $\mathfrak{su}(2)$.

As the irreducible components $f^i_C$ move over $C$, they sweep out complex surfaces in the threefold $X_r$. Thus the fibral divisors define a natural basis of divisors in $X_r$ which have the structure of $\mathbb P^1$ bundles $S^i_C \rightarrow C$. In terms of the projection $\varphi : X_r \rightarrow B$, these divisors are the irreducible components of the pullback
	\begin{align}
		\varphi^* C = \sum_{i=0}^r m_i S^i_C.	
	\end{align}
Once we have explicitly computed a resolution $X_r \rightarrow X_0$ of the singular threefold $X_0$, we can describe $X_r$ as local neighborhood of a collection of transversely intersecting surfaces $\cup_{i,C} S^i_C$ by computing the triple intersection numbers $(S^i_C \cdot S^j_D \cdot S^k_E)_{X_r}$ and the degrees $n^i_C$ of $S^i_C$.

\subsection{Computation of triple intersection numbers}
$(S^i_C \cdot S^j_D \cdot S^k_E)_{X_r}$ can be computed by pushing it forward to intersection ring of $B$ as outlined in \cite{Esole:2017kyr}. The first type of pushforward map whose properties we need to understand is the pushforward $f_*$ associated to a blowup $f : Y' \rightarrow Y$. Suppose we blow up $Y$ along the center $g_1 = g_2 = \cdots = g_k = 0$, which we assume to be a complete intersection of hypersurfaces $g_i =0$ meeting transversally in $Y$. Denote by $G_i$ the classes of the divisors $g_i=0$, and let $E$ be the class of the exceptional divisor of $f$. Note that the divisor classes $G_i$ do not depend on $E$. Then, following from a collection of useful results in intersection theory, we have
	\begin{align}
		f_* E^p = (-1)^{k+1} h_{p-k}(G_1,\dots, G_k) \prod_{i=1}^k G_i
	\end{align}
where $h_i(x_1,\dots, x_k)$ is the complete homogeneous symmetric polynomial of degree $i$ with the convention that $h_i$ is identically zero for $i<0$ and $h_0 =1$. An equivalent and possibly more practical way of expressing the above fact is as follows:
	\begin{align}
	\label{eqn:pushe}
		f_*E^p =  \sum_{j=1}^k M_j G_j^p,~~M_j \equiv \prod_{i \ne j} \frac{G_i}{G_i -G_j}.
	\end{align}
Using this formula, computing the pushforward of any formal analytic function $Q(E) = \sum Q_a E^a$ in the intersection ring of $Y'$ is now straightforward:
	\begin{align}
		f_* Q(E) = \sum_{j=1}^k M_j Q(G_j). 
	\end{align}
Observe that the pushforward eliminates the dependence on the exceptional divisor class $E$.

The second type of pushforward map we need to understand is the pushforward $\pi_*$ associated to the projection $\pi : Y_0 \rightarrow B$. Recall that the ambient projective space $Y_0$ is a rank 2 projective bundle $\mathbb P(V) \rightarrow B$ where $V = \mathcal O \oplus \mathcal K_B^{-2} \oplus \mathcal K_B^{-3}$, where $\mathcal K_B^{-1}$ is the anticanonical bundle over the base $B$. Let $-K_B$ be the divisor class associated to the anticanonical bundle. Furthermore, assuming that the $\mathbb P^2$ fibers of $Y_0$ have homogeneous coordinates $[x:y:z]$, we denote by $H$ the class of a hyperplane $\P^1$ defined by $a x + b y + c z =0$ in $Y_0$. 

The pushforward of any formal analytic function of $H$ can be expressed as an analytic function of the class $-K_B$. This fact is a consequence of the properties of a particular type of characteristic class called the Segre class $s(V) \equiv 1/c(V)$ where $c(V) = (1-2K_B)(1-3K_B)$, where $V$ is the tangent bundle of the rank 2 vector bundle $V$. By a theorem, 
	\begin{align}
		\pi_* \left( \frac{1}{1-H} \right) = s(V) = \frac{1}{c(V)} ,~~ c(V) = (1-2K_B)(1-3K_B).
	\end{align}
Performing a formal power series expansion on the both sides of the above expression and matching terms, one finds
	\begin{align}
		\pi_* 1 =0,~~ \pi_* H = 0,~~	\pi_*H^{p+2} = (-2 (-2)^p + 3(-3)^p)(-K_B)^p
	\end{align}
Given a formal power series $Q(H) = \sum  Q_a H^a$, by a straightforward computation one can use the above formulas to show that 
	\begin{align}
	\label{eqn:pushpi}
		\pi_* Q(H) = - 2 \left. \frac{Q(H)}{H^2} \right|_{H=2K_B}  + 3 \left. \frac{Q(H)}{H^2} \right|_{H= 3K_B} + \frac{Q(0)}{6K_B^2}.
	\end{align}
We now possess the necessary tools to compute the pushforward of a triple intersection product in the intersection ring of a resolved elliptically fibered Calabi-Yau threefold $X_r \rightarrow B$ to the intersection ring of $B$. Since $X_r$ is a hypersurface of the ambient projective bundle $Y_r$, we work in the intersection ring of $ Y_r$. Let $[W_r]$ denote the class of the divisor $W_r = 0$ in the intersection ring of $Y_r$. Then, we have\footnote{For notational convenience we suppress the $\cdot$ notation indicating the intersection product.}
	\begin{align}
		(S_i \cdot S_j \cdot S_k)|_{X_r} = (S_i  S_j  S_k  [W_r])|_{Y_r} = Q_{ijk}(E_l,H,-K_B)
	\end{align}
where by definition
	\begin{align}
		E_0 = f^* \circ \pi^* C. 
	\end{align}
In words, the above triple intersection product can be expressed as an analytic function $Q$ of the (appropriate pullbacks of the) divisor classes $E_i, H, -K_B$. Using the fact that 
	\begin{align}
		f = f_1 \cdots \circ f_{r-1} \circ f_r,
	\end{align}
we can use the properties of the pushforward maps $\pi_*, f_{i*}$ described above to explicitly  evaluate the triple intersection:
	\begin{align}
		\pi_* \circ f_* Q_{ijk}(E_l,H,-K_B)  = \pi_* \circ f_{1*} \circ \cdots \circ f_{r*}  Q_{ijk}(E_l,H,-K_B) = Q_{ijk}(C,-K_B)|_{B}.
	\end{align}

\noindent \ubf{Example: $\text{I}_2$ model}. We again return to the $\text{I}_2$ model to illustrate the computation of triple intersection numbers. In this case, there are two fibral divisors $S_0, S_1$ whose divisor classes in $Y_1$ are 
	\begin{align}
		S_0 = E_0 - E_1,~~ S_1 = E_1,
	\end{align}
where we bear in mind $E_0 = f^*_1 \circ \pi^* C$. Suppose we would like to compute $(S_0 \cdot S_0 \cdot S_0)_{X_1}$. We write
	\begin{align}
	\begin{split}
		(S_0^3  [W_1])|_{Y_1} &= (E_0-E_1)^3  (3H - 6K_B - E_1)
	\end{split}\\
	\begin{split}
		&=2 E_1^4+ \left(-6 E_0-3 H+6 K_B\right)E_1^3+ \left(6 E_0^2+9 E_0 H-18 E_0 K_B\right) E_1^2\\
		&+ \left(-2 E_0^3-9 E_0^2 H+18 E_0^2K_B\right)E_1+(-6 E_0^3 K_B+3 E_0^3 H).
	\end{split}
	\end{align}	
First, we use the fact that the divisor classes of the generators $g_i=0$ of the blowup center are
	\begin{align}
		G_{1} = H -2K_B ,~~ G_2 = H-3K_B ,~~ G_3 = E_0
	\end{align}
to compute the pushforward of the above expression to the intersection ring of $Y_0$ by making the substitution (\ref{eqn:pushe}), leading to:
	\begin{align}
		\begin{split}
			f_{1*}(S_0^3  [W_1])|_{Y_1} &= E_0 H^3+ \left(-9 E_0 K_B-4 E_0^2\right)H^2\\
			&+ \left(3 E_0^3+26 E_0 K_B^2+20 E_0^2 K_B\right)H\\
			& +(-24 E_0 K_B^3-24 E_0^2 K_B^2-6 E_0^3 K_B).
		\end{split}
	\end{align}
Next, we use (\ref{eqn:pushpi}) to compute
	\begin{align}
		\pi_* \circ f_{1*} (S_0^3 [W_1])|_{Y_1}=- 4 (C \cdot K_B+ C^2)_B. 
	\end{align}
Using adjunction, namely $(K\cdot C+C^2)_B=2g(C) - 2 = -2$, and the fact that $(C^2)_B = -k$, the above expression evaluates to
	\begin{align}
		(S_0 \cdot S_0 \cdot S_0)|_{X_1}=\pi_* \circ f_{1*} (S_0^3 [W_1])|_{Y_1}=-4 (C \cdot K_B+ C^2)|_B = 4 (2-k+k) = 8. 
	\end{align}


\subsection{Computation of degrees}
\subsubsection{Split case}
Let us assume for now that the Kodaira fiber type is split. Since $S^i_C$ is a $\P^1$ bundle over $C$, we can write it in terms of a line bundle $\cL^i_C$ over $C$ as 
	\begin{align}
		S^i_C = \mathbb P[ \mathcal O \oplus \mathcal L^i_C ] \rightarrow C,
	\end{align}
The degree $n^i_C$ of $S^i_C$ can be computed as 
	\begin{align}
		n^i_C = (L'^i_C \cdot C)|_{S^i_C} 	
	\end{align} 
where $L'^i_C$ is the divisor class dual to the pull back of line bundle $\mathcal L_i$ to $S_i$. We can use the fact that the intersection product is invariant under the pushforward to the base to write 
	\begin{align}
		n^i_C = (L^i_C \cdot C)|_B
	\end{align}
where 
	\begin{align}
		L^i_C = a^i_C K_{B} + b^i_C C	
	\end{align}
for some integers $a^i_C,b^i_C$ is the pushforward of $L'^i_C$. Using the fact that $C$ is a genus zero curve of self-intersection $-k$ in $B$, we can write the degree of $S^i_C$ as
	\begin{align}
		n^i_C =  (C \cdot (a^i_C K_B + b^i_C C))|_B = a^i_C (k-2) - b^i_C k.
	\end{align}

\noindent \ubf{Example: $\text{I}_2$ model}. The resolved $\text{I}_2$ model consists of two divisors $S^0,S^1$ associated to the irreducible components $f^0,f^1$ of the resolved elliptic fiber $f=f^0+f^1$. We now illustrate the computation of degree of $S^0$. $f^0$ is defined by 
	\begin{align}
	e_{0,1} = a_1 x_1 y_1 z-e_1 x_1^3+y_1^2 z=0 ~~\subset ~~ [e_1 x_1:e_1 y_1 :z] [ x_1:y_1].
	\end{align}
We work in an affine open set $x \ne 0$ of the ambient space $Y_1$. Using the fact that the homogeneous coordinates are invariant under the following scaling,
	\begin{align}
		[e_1 x_1:e_1 y_1 :z] [ x_1:y_1] \cong [\lambda_0 e_1 x_1: \lambda_0 e_1 y_1 : \lambda_0 z] [ \lambda_1 x_1: \lambda_1 y_1]
	\end{align}
for $\lambda_i \in \mathbb C^{\times}$ (where above we have made implicit use of the scaling behavior $e_1 \cong \lambda_1^{-1} e_1$), we set $\lambda_0 = 1/z$. Defining $x' = x_1/z, y'= y_1/z$, we solve for $e_1$:
	\begin{align}
		e_1 = \frac{y'^2}{x'^3} + \frac{a_1 y'}{x'^2}
	\end{align}
and set $\lambda_1 = 1/x'$. Defining $\psi = y'/x'$, we find that $F_0$ is locally parametrized by 
	\begin{align}
		[\psi(\psi + a_1) : \psi^2 ( \psi + a_1) :1][1: \psi].
	\end{align}
There is an obvious regular map (i.e. a projection) of the above algebraic variety onto the following affine variety:
	\begin{align}
		(\psi(\psi + a_1) , \psi^2 ( \psi + a_1) ) \subset \mathbb C^2.
	\end{align}
The above variety is a rational parametrization of a nodal curve by the parameter $\psi$. In order to determine the divisor class $L_0$ associated to the line bundle $\mathcal L_0$ of which $\psi$ is a section, we simply compute the divisor class $[\psi]$ associated to the hyperplane $\psi =0$, namely
	\begin{align}
		L_0 = [\psi] = [x'/y'] = H+ 2L -E_1 - (H + 3L - E_1) = -L = K_B. 
	\end{align}	  
Thus in this case we did not have to use the pushforward formulas and find the degree of $S^0$ to be
	\begin{align}
		n = k-2
	\end{align}	

\subsubsection{Non-split case}
In the non-split case, the divisor $S^i_C$ can be regarded as an $s^i_C$-cover of a $\mathbb P^1$ bundle over $C$ ramified over a finite number of points where a subset of the distinct $\mathbb P^1$ fibers collapse into a higher multiplicity $\mathbb P^1$. In such cases, we would still like to be able to interpret $S^i_C$ as a ruled surface over a higher genus curve. 

Suppose $f: S \to C$ is a ramified $s$-cover a ruled surface $p : S' \to C$. That is, suppose 
	\begin{align}
		S \overset{\pi}{\rightarrow} S' \overset{p}{\rightarrow} C.
	\end{align}
The above equation implies that we can find a factorization $f = p \circ \pi$. In practice, when computing degrees, the actual object we study is the $\mathbb P^1$ bundle $p : S' \rightarrow C$. So, we need to understand how to use our understanding of the bundle $S' \rightarrow C$ to say something about the bundle $S \rightarrow C$. 

The key to understanding $S$ as a ruled surface is a particular interpretation of the projection $f$ of the fibration as being a different composition of maps than the one described above. This composition is called the \emph{Stein factorization} \cite{Esole:2017rgz}:
	\begin{align}
		S \overset{f'}{\rightarrow} C' \overset{\pi'}{\rightarrow} C,
	\end{align}
where $\pi' : C' \rightarrow C$ is a degree $d$ map, and we assume that $\pi'$ is a $s$-cover of $C$. There is a theorem which guarantees the existence of a factorization $f = \pi' \circ f'$ under mild assumptions about the nature of the morphism $f: S \rightarrow C$. Thus it ends up being the case that one can use the Stein factorization of the projection $f$ to interpret $S$ as a ruled surface according to the following commutative diagram:
	\begin{align}
		\begin{array}{c}
			\begin{tikzpicture}[scale=2]
				\node(D) at (0,1) {$S$};
				\node(D') at (1,1) {$S'$};
				\node(S') at (0,0) {$C'$};
				\node(S) at (1,0) {$C$};
				\draw[big arrow] (D) -- node[above,midway]{$\pi$} (D');
				\draw[big arrow] (D) -- node[left,midway]{$f'$} (S');
				\draw[big arrow] (S') -- node[above,midway]{$\pi'$} (S);
				\draw[big arrow] (D') -- node[right,midway]{$p$} (S); 
				\draw[big arrow] (D) -- node[above,midway]{$f$} (S);
			\end{tikzpicture}	
		\end{array}
	\end{align}
In other words, we are exploiting the fact that 
	\begin{align}
		f = p \circ \pi = \pi' \circ f'. 
	\end{align}

We now describe how to compute the degree of such a surface along with the genus of the curve defining the base of the fibration. Let $\pi: S \rightarrow S'$ be a (possibly branched) $s$-cover of a ruled surface $p: S' \rightarrow C'$ over a smooth base curve $C'$ of genus $g'$. We assume that the degree of $S'$ is $-C'^2 = n$. Now let $C = \pi^* C'$ be a (possibly branched) $s$-cover of $C'$. Using the properties of the pushforward, we find the self-intersection of $C$ to be
	\begin{align}
		C^2|_{S} =\pi_* (C^2)|_{S'} = s C'^2|_{S'} = -sn. 
	\end{align}

Next, we compute the genus of $C$. Suppose the $s$-cover $ \pi : S \rightarrow S'$ is branched along $2b$ fibers. Denoting by $F'$ the class of a generic fiber of $S'$ and defining $F \subset S$ such that $sF = \pi^* F'$, we find that the canonical class of $S$ is given by
	\begin{align}
		K_S = \pi^* K_{S'} + b \pi^* F'.
	\end{align}
Using the fact that that $K_{S'} = -2 C' + ( 2g' -2 - n ) F'$, we find 
	\begin{align}
		K_S = -2 \pi^* C'+ (2g' -2 -n)\pi^* F'+ b \pi^* F' =- 2 \pi^* C' + s (2g' - 2 -n +  b) F,
	\end{align}
where we assume $(\pi^*C' \cdot F)|_{S} = 1$. Using adjunction and the fact that $C^2|_S = -sn$ we find that $C$ has genus 
	\begin{align}
		g(C) = \frac{s}{2} ( 2g' +b -2 +\frac{2}{s}).
	\end{align}
The specific case of a double cover (i.e. $s=2$) covers nearly every example in which we are interested. Setting $s=2$ and assuming $g'=0$ we find 
	\begin{align}
		g(C) = b-1.
	\end{align}
Hence $S$ is a ruled surface over a smooth curve of genus $b-1$.

\subsection{Computation of the number of blowups}
So, in general, $f^i_C$ can be thought of as living in a ruled surface $\tilde S^i_C$ over a curve of genus $g^i_C$ and carrying some number $b'^i_C$ of blowups on top of it. In practice, in all of the examples, $\tilde S^i_C$ arises via self-gluings of a Hirzebruch surface $S^i_C$, and for the purposes of computing the Mori cone, it is better to work in terms of $S^i_C$ rather than $\tilde S^i_C$. To go from $S^i_C$ to $\tilde S^i_C$, we first perform $2g^i_C$ number of blowups on $S^i_C$. Let's pair them up as $x_\mu,y_\mu$ where $\mu=1,\cdots,g^i_C$. Then we self-glue $x_\mu\sim y_\mu$ for each $\mu$. So the number of blowups $b^i_C$ carried by $S^i_C$ equals
\be
b^i_C=b'^i_C+2g^i_C\label{bu}
\ee

The number of blowups can be determined via the triple intersection number $S^i_C\cdot S^i_C\cdot S^i_C$ which according to (\ref{K'}) is equal to $(K'^i_C\cdot K'^i_C)|_{S^i_C}$ where $K'^i_C$ is the canonical class of $S^i_C$ modified by the curves used for self-gluing in accordance with (\ref{Ksh}). In our case, 
\be
K'^i_C=(K^i_C)_u+2\sum_\mu(x_\mu+y_\mu)+\sum_\alpha x_\alpha
\ee
where $x_\alpha,\alpha=1,\cdots,b'^i_C$ describes the blowups not used for self-gluing. Using the fact that $(K^i_C)_u\cdot(K^i_C)_u=8$ for all Hirzebruch surfaces, we can compute that
\be
b^i_C=8-6g^i_C-(S^i_C)^3
\ee

\section{Single curve}\label{flop-eq}
We start with the analysis of elliptic fibration over a single curve $C$ in the base $B$ of the threefold $\tilde X_\fT$. We will associate a graph to each possible $C$ which will capture the degrees of Hirzebruch surfaces $S^i_C$ and the gluings between them. We refer the reader to Section 3.2 of \cite{Bhardwaj:2018yhy} for full information on the graphical notation that we use. The only minor change in the notation is that the labels $i$ for the Hirzebruch surfaces $S^i_C$ are chosen to be consistent with Figure \ref{hell}. This allows for an easy identification of the elliptic fiber in each collection of surfaces. Another point to note is that the graph that we associate here will in general be equal to the graph associated in \cite{Bhardwaj:2018yhy} only up to flops. This is done to give a uniform description of gluing rules in Section \ref{gluing}. We also note that some of the results appearing in this section had their first appearance in \cite{DelZotto:2017pti}.

\subsection{$\su(n),n\ge1$ on $-2$ curve}
For even $n=2m\ge2$, we associate the graph
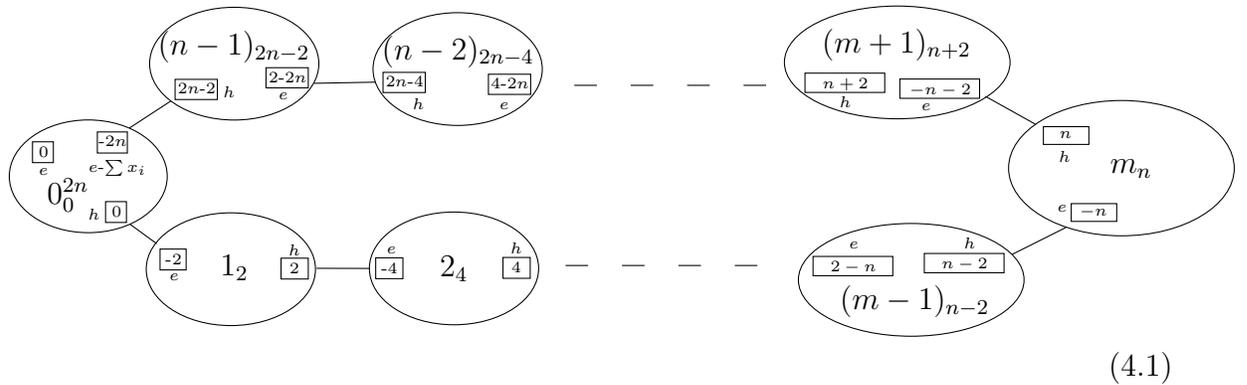
\begin{align}
\begin{array}{c}
\begin{tikzpicture}[scale=1.5]
\draw  (-3.5372,3.9161) ellipse (0.7 and 0.5);
\node (v1) at (-3.7154,3.7502) {$0_{0}^{2n}$};
\draw  (-2.2435,4.9153) ellipse (0.75 and 0.5);
\node (v2) at (-2.2479,5.0646) {$(n-1)_{2n-2}$};
\draw  (-0.2634,4.8673) ellipse (0.75 and 0.5);
\node (v4) at (-0.2607,5.0258) {$(n-2)_{2n-4}$};
\draw  (-2.2735,3.0921) ellipse (0.75 and 0.5);
\node (v3) at (-2.2418,3.1032) {$1_2$};
\draw  (-0.2957,3.0991) ellipse (0.75 and 0.5);
\node (v5) at (-0.2957,3.0991) {$2_4$};
\draw  (3.6325,4.9262) ellipse (1 and 0.5);
\node at (3.6265,5.0868) {$(m+1)_{n+2}$};
\draw  (5.6194,3.9883) ellipse (1 and 0.6);
\node at (5.7066,3.9861) {$m_{n}$};
\draw (-3.1727,4.344) -- (-2.8035,4.5852);
\draw (-3.1658,3.49) -- (-2.9309,3.3184);
\draw (-1.534,4.741) -- (-0.9899,4.7512);
\draw (-1.5132,3.1001) -- (-1.0394,3.0989);
\draw (4.4136,4.6256) -- (4.8564,4.3786);
\draw (5.1271,3.4619) -- (4.6488,3.2223);
\draw (0.7704,4.7231) -- (0.9704,4.7231) (1.2704,4.7231) -- (1.4704,4.7231);
\draw (1.7704,4.7231) -- (1.9704,4.7231) (2.2704,4.7231) -- (2.4704,4.7231);
\begin{scope}[shift={(0.6928,-1.875)}]
\draw (0,5) -- (0.2,5) (0.5,5) -- (0.7,5);
\draw (1,5) -- (1.2,5) (1.5,5) -- (1.7,5);
\end{scope}
\draw  (-3.4568,4.3097) rectangle (-3.191,4.1225);
\node at (-3.3192,4.2301) {\tiny{-2$n$}};
\begin{scope}[shift={(3.4937,0.4636)}]
\draw  (-3.4845,4.3579) rectangle (-3.1101,4.179);
\node at (-3.2917,4.2752) {\tiny{4-$2n$}};
\end{scope}
\begin{scope}[shift={(0.0074,-0.6727)}]
\draw  (-3.3913,4.3642) rectangle (-3.2064,4.1794);
\node at (-3.2917,4.2752) {\tiny{0}};
\end{scope}
\begin{scope}[shift={(-0.6401,-0.142)}]
\draw  (-3.3913,4.3642) rectangle (-3.2064,4.1794);
\node at (-3.2917,4.2752) {\tiny{0}};
\end{scope}
\begin{scope}[shift={(0.5102,-1.0935)}]
\draw  (-3.4113,4.3642) rectangle (-3.18,4.1794);
\node at (-3.2917,4.2752) {\tiny{-2}};
\end{scope}
\begin{scope}[shift={(1.5826,-1.1743)}]
\draw  (-3.4113,4.3642) rectangle (-3.17,4.1794);
\node at (-3.2917,4.2752) {\tiny{2}};
\end{scope}
\begin{scope}[shift={(2.4216,-1.1712)}]
\draw  (-3.4113,4.3642) rectangle (-3.17,4.1794);
\node at (-3.2917,4.2752) {\tiny{-4}};
\end{scope}
\begin{scope}[shift={(3.5496,-1.168)}]
\draw  (-3.4113,4.3642) rectangle (-3.17,4.1794);
\node at (-3.2917,4.2752) {\tiny{4}};
\end{scope}
\begin{scope}[shift={(7.3066,0.4078)}]
\draw  (-3.6509,4.3711) rectangle (-2.9441,4.2);
\node at (-3.2917,4.2752) {\tiny{$-n-2$}};
\end{scope}
\begin{scope}[shift={(8.5787,-0.0147)}]
\draw  (-3.6509,4.3711) rectangle (-3.2477,4.2136);
\node at (-3.4547,4.3027) {\tiny{$n$}};
\end{scope}
\begin{scope}[shift={(8.8255,-0.6974)}]
\draw  (-3.6509,4.3711) rectangle (-3.2477,4.2136);
\node at (-3.4547,4.3027) {\tiny{$-n$}};
\end{scope}
\begin{scope}[shift={(6.466,0.46)}]
\draw  (-3.6509,4.3711) rectangle (-2.9441,4.2);
\node at (-3.2917,4.2752) {\tiny{$n+2$}};
\end{scope}
\begin{scope}[shift={(0.7414,-2.0111)}]
\draw  (3.0151,5.0088) ellipse (1 and 0.5);
\node at (3.0457,4.8218) {$(m-1)_{n-2}$};
\begin{scope}[shift={(6.7819,0.8716)}]
\draw  (-3.6509,4.3711) rectangle (-2.9441,4.2);
\node at (-3.2917,4.2752) {\tiny{$n-2$}};
\end{scope}
\begin{scope}[shift={(5.796,0.8453)}]
\draw  (-3.6509,4.3711) rectangle (-2.9441,4.2);
\node at (-3.2917,4.2752) {\tiny{$2-n$}};
\end{scope}
\end{scope}
\node at (-3.9344,3.9561) {\tiny{$e$}};
\node at (-2.2899,4.6699) {\tiny{$h$}};
\node at (-3.4859,3.5787) {\tiny{$h$}};
\node at (-1.7089,3.26) {\tiny{$h$}};
\node at (0.1498,4.544) {\tiny{$e$}};
\node at (3.8879,4.5326) {\tiny{$e$}};
\node at (4.2762,3.3293) {\tiny{$h$}};
\node at (5.1162,4.0884) {\tiny{$h$}};
\node at (5.093,3.6313) {\tiny{$e$}};
\node at (-0.8508,3.2665) {\tiny{$e$}};
\node at (-2.7762,3.0228) {\tiny{$e$}};
\node at (-1.8095,4.6266) {\tiny{$e$}};
\node at (0.2576,3.2786) {\tiny{$h$}};
\node at (-0.6101,4.5623) {\tiny{$h$}};
\node at (3.1714,4.5736) {\tiny{$h$}};
\node at (3.2478,3.3035) {\tiny{$e$}};
\node at (-3.2769,3.9913) {\tiny{$e$-$\sum x_i$}};
\begin{scope}[shift={(2.5598,0.4801)}]
\draw  (-3.4845,4.3579) rectangle (-3.1101,4.179);
\node at (-3.2917,4.2752) {\tiny{$2n$-4}};
\end{scope}
\begin{scope}[shift={(0.7169,0.4181)}]
\draw  (-3.4845,4.3579) rectangle (-3.1101,4.179);
\node at (-3.2917,4.2752) {\tiny{$2n$-2}};
\end{scope}
\begin{scope}[shift={(1.5227,0.5215)}]
\draw  (-3.4845,4.3579) rectangle (-3.1101,4.179);
\node at (-3.2917,4.2752) {\tiny{2-$2n$}};
\end{scope}
\end{tikzpicture}
\end{array}
\end{align}
where $x_i$, $i=1,\cdots,2n$ denote the exceptional curves generated due to $2n$ blowups on $S^0_C$, and $-\sum x_i$ denotes negative of the sum over all the exceptional curves\footnote{In what follows, we will adopt these conventions for representing exceptional curves, unless otherwise specified}. Here the $2n$ blowups correspond to $2n$ hypers in the fundamental representation of the gauge algebra $\mf{su}(n)$. This graph is flop equivalent to the one presented in \cite{Bhardwaj:2018yhy}.

Notice that the curve labeled $e$ in $S^0_C$ above does not have any edge associated to it. We remind the reader that in our notation this is supposed to mean that this curve acts as the gluing curve that glues the collection of surfaces $S_C$ to the base $B$ of the elliptic fibration. In this example, the gluing curve in the base happens to be a rational curve of self-intersection $-2$, and hence it is consistent with the Calabi-Yau condition \ref{CY} to glue it along a rational curve of self-intersection $0$ inside $S_C$.

We reproduce the graph for $n=2$ below because it is a limiting case of the above family of graphs
\begin{align}\label{su2I}
\begin{array}{c}
\begin{tikzpicture}[scale=1.5]
\draw  (-2.2735,3.0921) ellipse (0.75 and 0.5);
\node (v3) at (-2.2427,3.0001) {$0_0^4$};
\draw  (0.4646,3.0996) ellipse (0.75 and 0.5);
\node (v5) at (0.4646,3.0996) {$1_2$};
\begin{scope}[shift={(0.4647,-1.0852)}]
\draw  (-3.4113,4.3642) rectangle (-3.18,4.1794);
\node at (-3.2917,4.2752) {\tiny{0}};
\end{scope}
\begin{scope}[shift={(1.5826,-1.1743)}]
\draw  (-3.4113,4.3642) rectangle (-3.17,4.1794);
\node at (-3.2917,4.2752) {\tiny{-4}};
\end{scope}
\begin{scope}[shift={(3.1819,-1.1707)}]
\draw  (-3.4113,4.3642) rectangle (-3.17,4.1794);
\node at (-3.2917,4.2752) {\tiny{0}};
\end{scope}
\node at (-2.0194,3.2975) {\tiny{$h$,$e$-$\sum x_i$}};
\node at (-0.0269,3.2997) {\tiny{$e$,$h$}};
\node at (-2.8163,3.0004) {\tiny{$e$}};
\begin{scope}[shift={(2.3886,-1.166)}]
\draw  (-3.4113,4.3642) rectangle (-3.17,4.1794);
\node at (-3.2917,4.2752) {\tiny{2}};
\end{scope}
\draw (-1.5223,3.0935) -- (-1.0209,3.0935) (-0.7818,3.0935) -- (-0.2892,3.0961);
\end{tikzpicture}
\end{array}
\end{align}
Here there are two gluing curves between $S^0_C$ and $S^1_C$. Our convention is such that the first gluing curves on each side are glued to each other, and the same is true for the second gluing curves\footnote{We will use this convention in what follows, unless otherwise stated.}. That is, $h$ in $S^0_C$ is glued to $e$ in $S^1_C$ and $e-\sum x_i$ is $S^0_C$ is glued to $h$ in $S^1_C$.

We will use another graph for $n=2$ which is flop equivalent to (\ref{su2I}) when $\su(2)$ does not have a neighboring $\fg_2$. But if such a neighbor is present, then the two graphs cannot be flopped into each other. The second graph is
\begin{align}\label{su2II}
\begin{array}{c}
\begin{tikzpicture}[scale=1.5]
\draw  (-2.2735,3.0921) ellipse (0.75 and 0.5);
\node (v3) at (-2.2427,3.0001) {$0_0^3$};
\draw  (0.4646,3.0996) ellipse (0.75 and 0.5);
\node (v5) at (0.4646,3.0996) {$1^1_2$};
\begin{scope}[shift={(0.4647,-1.0852)}]
\draw  (-3.4113,4.3642) rectangle (-3.18,4.1794);
\node at (-3.2917,4.2752) {\tiny{0}};
\end{scope}
\begin{scope}[shift={(1.5826,-1.1743)}]
\draw  (-3.4113,4.3642) rectangle (-3.17,4.1794);
\node at (-3.2917,4.2752) {\tiny{-3}};
\end{scope}
\begin{scope}[shift={(3.1819,-1.1707)}]
\draw  (-3.4113,4.3642) rectangle (-3.17,4.1794);
\node at (-3.2917,4.2752) {\tiny{-1}};
\end{scope}
\node at (-2.0194,3.2975) {\tiny{$h$,$e$-$\sum x_i$}};
\node at (0.0561,3.2889) {\tiny{$e$-$x$,$h$}};
\node at (-2.8163,3.0004) {\tiny{$e$}};
\begin{scope}[shift={(2.3886,-1.166)}]
\draw  (-3.4113,4.3642) rectangle (-3.17,4.1794);
\node at (-3.2917,4.2752) {\tiny{2}};
\end{scope}
\draw (-1.5223,3.0935) -- (-1.0209,3.0935) (-0.7818,3.0935) -- (-0.2892,3.0961);
\end{tikzpicture}
\end{array}
\end{align}


For odd $n=2m+1\ge3$, we associate the graph
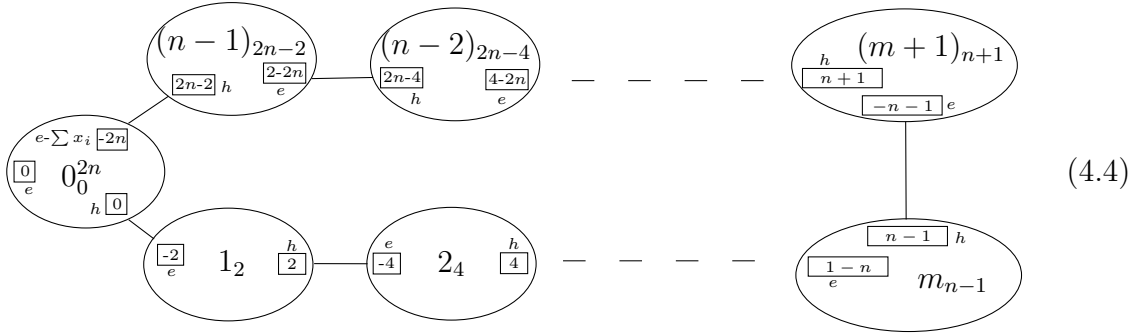
\begin{align}
\begin{array}{c}
\begin{tikzpicture}[scale=1.5]
\draw  (-3.5372,3.9161) ellipse (0.7 and 0.5);
\node (v1) at (-3.5709,3.8658) {$0_{0}^{2n}$};
\draw  (-2.2435,4.9153) ellipse (0.75 and 0.5);
\node (v2) at (-2.2479,5.0646) {$(n-1)_{2n-2}$};
\draw  (-0.2634,4.8673) ellipse (0.75 and 0.5);
\node (v4) at (-0.2607,5.0258) {$(n-2)_{2n-4}$};
\draw  (-2.2735,3.0921) ellipse (0.75 and 0.5);
\node (v3) at (-2.2418,3.1032) {$1_2$};
\draw  (-0.2957,3.0991) ellipse (0.75 and 0.5);
\node (v5) at (-0.2957,3.0991) {$2_4$};
\draw  (3.7193,4.8594) ellipse (1 and 0.5);
\node at (3.9529,5.0124) {$(m+1)_{n+1}$};
\draw (-3.1588,4.3443) -- (-2.8035,4.5852);
\draw (-3.1567,3.4893) -- (-2.9309,3.3184);
\draw (-1.534,4.741) -- (-0.9899,4.7512);
\draw (-1.5132,3.1001) -- (-1.0394,3.0989);
\draw (0.7704,4.7231) -- (0.9704,4.7231) (1.2704,4.7231) -- (1.4704,4.7231);
\draw (1.7704,4.7231) -- (1.9704,4.7231) (2.2704,4.7231) -- (2.4704,4.7231);
\begin{scope}[shift={(0.6928,-1.875)}]
\draw (0,5) -- (0.2,5) (0.5,5) -- (0.7,5);
\draw (1,5) -- (1.2,5) (1.5,5) -- (1.7,5);
\end{scope}
\draw  (-3.4353,4.2943) rectangle (-3.1695,4.1071);
\node at (-3.2977,4.2147) {\tiny{-2$n$}};
\begin{scope}[shift={(3.4937,0.4636)}]
\draw  (-3.4845,4.3579) rectangle (-3.1101,4.179);
\node at (-3.2917,4.2752) {\tiny{4-$2n$}};
\end{scope}
\begin{scope}[shift={(0.0334,-0.643)}]
\draw  (-3.3913,4.3642) rectangle (-3.2064,4.1794);
\node at (-3.2917,4.2752) {\tiny{0}};
\end{scope}
\begin{scope}[shift={(-0.7814,-0.3555)}]
\draw  (-3.3913,4.3642) rectangle (-3.2064,4.1794);
\node at (-3.2917,4.2752) {\tiny{0}};
\end{scope}
\begin{scope}[shift={(0.5102,-1.0935)}]
\draw  (-3.4113,4.3642) rectangle (-3.18,4.1794);
\node at (-3.2917,4.2752) {\tiny{-2}};
\end{scope}
\begin{scope}[shift={(1.5826,-1.1743)}]
\draw  (-3.4113,4.3642) rectangle (-3.17,4.1794);
\node at (-3.2917,4.2752) {\tiny{2}};
\end{scope}
\begin{scope}[shift={(2.4216,-1.1712)}]
\draw  (-3.4113,4.3642) rectangle (-3.17,4.1794);
\node at (-3.2917,4.2752) {\tiny{-4}};
\end{scope}
\begin{scope}[shift={(3.5496,-1.168)}]
\draw  (-3.4113,4.3642) rectangle (-3.17,4.1794);
\node at (-3.2917,4.2752) {\tiny{4}};
\end{scope}
\begin{scope}[shift={(6.9967,0.2178)}]
\draw  (-3.6509,4.3711) rectangle (-2.9441,4.2);
\node at (-3.2917,4.2752) {\tiny{$-n-1$}};
\end{scope}
\begin{scope}[shift={(6.466,0.46)}]
\draw  (-3.6509,4.3711) rectangle (-2.9441,4.2);
\node at (-3.2917,4.2752) {\tiny{$n+1$}};
\end{scope}
\begin{scope}[shift={(0.7414,-2.0111)}]
\draw  (3.0151,5.0088) ellipse (1 and 0.5);
\node at (3.4059,4.9283) {$m_{n-1}$};
\begin{scope}[shift={(6.3026,1.0699)}]
\draw  (-3.6509,4.3711) rectangle (-2.9441,4.2);
\node at (-3.2917,4.2752) {\tiny{$n-1$}};
\end{scope}
\begin{scope}[shift={(5.7754,0.7998)}]
\draw  (-3.6509,4.3711) rectangle (-2.9441,4.2);
\node at (-3.2917,4.2752) {\tiny{$1-n$}};
\end{scope}
\end{scope}
\node at (-3.7335,4.2162) {\tiny{$e$-$\sum x_i$}};
\node at (-2.2899,4.6699) {\tiny{$h$}};
\node at (-3.4496,3.5928) {\tiny{$h$}};
\node at (-1.7089,3.26) {\tiny{$h$}};
\node at (0.1498,4.544) {\tiny{$e$}};
\node at (4.1333,4.4999) {\tiny{$e$}};
\node at (4.2161,3.3324) {\tiny{$h$}};
\node at (-0.8508,3.2665) {\tiny{$e$}};
\node at (-2.7762,3.0228) {\tiny{$e$}};
\node at (-1.8095,4.6266) {\tiny{$e$}};
\node at (0.2576,3.2786) {\tiny{$h$}};
\node at (-0.6101,4.5623) {\tiny{$h$}};
\node at (3.0151,4.9134) {\tiny{$h$}};
\node at (3.0867,2.9192) {\tiny{$e$}};
\node at (-4.0406,3.7456) {\tiny{$e$}};
\begin{scope}[shift={(2.5598,0.4801)}]
\draw  (-3.4845,4.3579) rectangle (-3.1101,4.179);
\node at (-3.2917,4.2752) {\tiny{$2n$-4}};
\end{scope}
\begin{scope}[shift={(0.7169,0.4181)}]
\draw  (-3.4845,4.3579) rectangle (-3.1101,4.179);
\node at (-3.2917,4.2752) {\tiny{$2n$-2}};
\end{scope}
\begin{scope}[shift={(1.5227,0.5215)}]
\draw  (-3.4845,4.3579) rectangle (-3.1101,4.179);
\node at (-3.2917,4.2752) {\tiny{2-$2n$}};
\end{scope}
\draw (3.7296,4.3592) -- (3.7338,3.4915);
\end{tikzpicture}
\end{array}
\end{align}
which is flop equivalent to the one presented in \cite{Bhardwaj:2018yhy}. Again the $2n$ blowups correspond to $2n$ hypers in fundamental of $\su(n)$.

The degenerate case $n=1$ is a limiting case of the above series of graphs
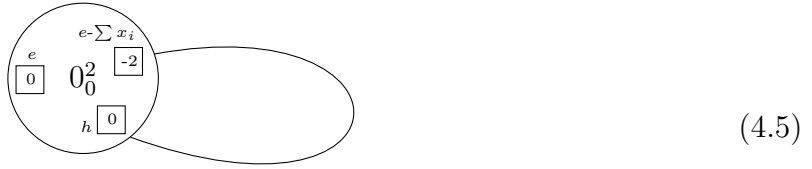
\begin{align}
\begin{array}{c}
\begin{tikzpicture}[scale=2]
\draw  (-3.5,4) ellipse (0.5 and 0.5);
\node (v1) at (-3.5,4) {$0^2_0$};
\draw  (-3.2902,4.2051) rectangle (-3.1053,4.0203);
\node at (-3.1906,4.1161) {\tiny{-2}};
\begin{scope}[shift={(-0.0137,-0.5476)}]
\draw  (-3.3913,4.3642) rectangle (-3.2064,4.1794);
\node at (-3.2917,4.2752) {\tiny{0}};
\end{scope}
\begin{scope}[shift={(-0.5545,-0.2807)}]
\draw  (-3.3913,4.3642) rectangle (-3.2064,4.1794);
\node at (-3.2917,4.2752) {\tiny{0}};
\end{scope}
\node at (-3.3418,4.3033) {\tiny{$e$-$\sum x_i$}};
\node at (-3.4769,3.681) {\tiny{$h$}};
\node at (-3.8392,4.1532) {\tiny{$e$}};
\draw (-3.1893,3.61) .. controls (-1.2581,2.9132) and (-1.2077,4.4972) .. (-3.0281,4.1613);
\end{tikzpicture}
\end{array}
\end{align}
which is a self-glued $\F_0^2$. Notice that due to self-gluing $f$ in $\F_0^2$ becomes the genus one elliptic fiber having a nodal singularity. The genus of $f$ turns out to be one if one uses adjunction (\ref{g}) with the shifted canonical class defined in (\ref{Ksh}).

Let us study the case of I$^s_n$ fiber in more detail. If the fibration is completely generic except for the fact that we have I$_n^s$ on $C$, then there are $2n$ special points $p_i,i=1,\cdots,2n$ on $C$ over which the fibration degenerates further to I$_{n+1}$. The point $p_i$ is associated with $i$-th hyper in the fundamental representation, and we find that over $p_i$ $f^0_C$ decomposes as a sum of $f^0_C-x_i$ and $x_i$. This justifies our claim that each blowup is associated to a hyper in the fundamental.


\subsection{$\su(n),n\ge3,n\neq\tilde6$ on $-1$ curve}
For even $n=2m\ge 4$, we associate the graph
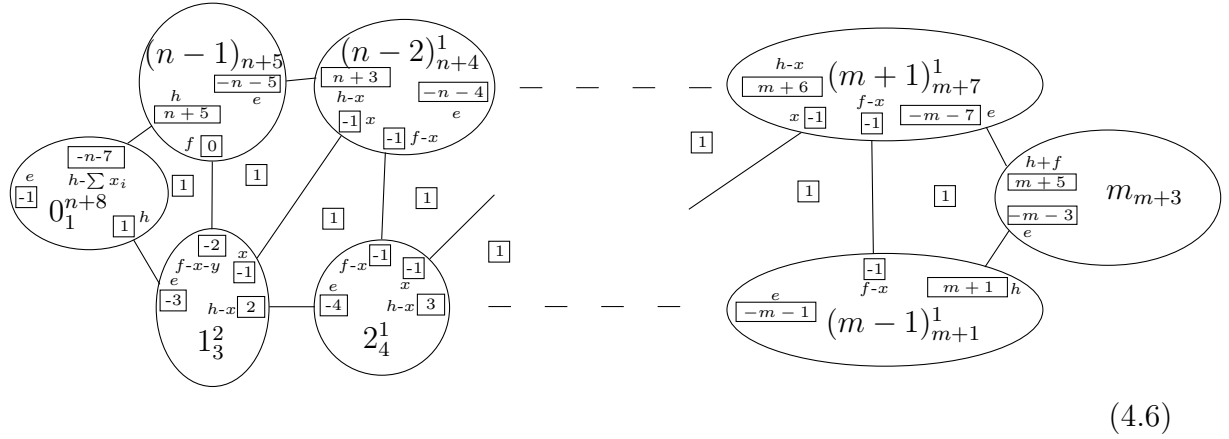
\begin{align}
\label{asyme}
\begin{array}{c}
\begin{tikzpicture}[scale=1.5]
\draw  (-3.5933,4.0135) ellipse (0.7 and 0.5);
\node (v1) at (-3.6884,3.8485) {$0_{1}^{n+8}$};
\draw  (-2.5,5) ellipse (0.65 and 0.7);
\node (v2) at (-2.4848,5.2395) {$(n-1)_{n+5}$};
\draw  (-0.799,4.956) ellipse (0.8 and 0.6);
\node (v4) at (-0.7526,5.2579) {$(n-2)_{n+4}^1$};
\draw  (-2.5,3) ellipse (0.5 and 0.7);
\node (v3) at (-2.5156,2.6911) {$1_3^2$};
\draw  (-1,3) ellipse (0.6 and 0.6);
\node (v5) at (-1.043,2.7213) {$2_4^1$};
\draw  (3.4586,4.9793) ellipse (1.4 and 0.5);
\node at (3.6307,5.0449) {$(m+1)_{m+7}^1$};
\draw  (5.4345,3.9749) ellipse (1 and 0.6);
\node at (5.7621,3.9947) {$m_{m+3}$};
\draw (-3.2535,4.4565) -- (-3.0386,4.6116);
\draw (-3.2,3.6) -- (-2.9679,3.2062);
\draw (-1.8481,5.0102) -- (-1.5807,5.0366);
\draw (-1.9977,3.0075) -- (-1.5933,3.0074);
\draw (4.559,3.6825) -- (4.3448,3.3593);
\draw (0.2222,4.9504) -- (0.4222,4.9504) (0.7222,4.9504) -- (0.9222,4.9504);
\draw (1.2222,4.9504) -- (1.4222,4.9504) (1.7222,4.9504) -- (1.9222,4.9504);
\begin{scope}[shift={(-0.0558,-1.9909)}]
\draw (0,5) -- (0.2,5) (0.5,5) -- (0.7,5);
\draw (1,5) -- (1.2,5) (1.5,5) -- (1.7,5);
\end{scope}
\draw  (-3.7805,4.4276) rectangle (-3.282,4.2384);
\node at (-3.5445,4.3427) {\tiny{-$n$-7}};
\begin{scope}[shift={(2.0572,0.759)}]
\draw  (-3.5934,4.3703) rectangle (-2.9834,4.1997);
\node at (-3.2917,4.2752) {\tiny{$n+3$}};
\end{scope}
\begin{scope}[shift={(0.0115,-0.5518)}]
\draw  (-3.3913,4.3642) rectangle (-3.2064,4.1794);
\node at (-3.2917,4.2752) {\tiny{1}};
\end{scope}
\begin{scope}[shift={(0.4458,-1.2087)}]
\draw  (-3.4113,4.3642) rectangle (-3.18,4.1794);
\node at (-3.2917,4.2752) {\tiny{-3}};
\end{scope}
\begin{scope}[shift={(1.1318,-1.2659)}]
\draw  (-3.4113,4.3642) rectangle (-3.17,4.1794);
\node at (-3.2917,4.2752) {\tiny{2}};
\end{scope}
\begin{scope}[shift={(1.8647,-1.2523)}]
\draw  (-3.4113,4.3642) rectangle (-3.17,4.1794);
\node at (-3.2917,4.2752) {\tiny{-4}};
\end{scope}
\begin{scope}[shift={(2.726,-1.2454)}]
\draw  (-3.4113,4.3642) rectangle (-3.17,4.1794);
\node at (-3.2917,4.2752) {\tiny{3}};
\end{scope}
\begin{scope}[shift={(7.2486,0.4223)}]
\draw  (-3.6509,4.3711) rectangle (-2.9441,4.2);
\node at (-3.2917,4.2752) {\tiny{$-m-7$}};
\end{scope}
\begin{scope}[shift={(8.2011,-0.186)}]
\draw  (-3.6509,4.3711) rectangle (-3.0476,4.2282);
\node at (-3.366,4.2929) {\tiny{$m+5$}};
\end{scope}
\begin{scope}[shift={(8.2858,-0.4761)}]
\draw  (-3.736,4.3764) rectangle (-3.1252,4.2142);
\node at (-3.4526,4.2849) {\tiny{$-m-3$}};
\end{scope}
\begin{scope}[shift={(5.8467,0.6735)}]
\draw  (-3.6509,4.3711) rectangle (-2.9441,4.2);
\node at (-3.2917,4.2752) {\tiny{$m+6$}};
\end{scope}
\begin{scope}[shift={(0.4434,-2.0286)}]
\draw  (3.0151,5.0088) ellipse (1.4 and 0.5);
\node at (3.1809,4.8747) {$(m-1)_{m+1}^1$};
\begin{scope}[shift={(7.0448,0.9242)}]
\draw  (-3.6509,4.3711) rectangle (-2.9441,4.2);
\node at (-3.2917,4.2752) {\tiny{$m+1$}};
\end{scope}
\begin{scope}[shift={(5.3469,0.7052)}]
\draw  (-3.6509,4.3711) rectangle (-2.9441,4.2);
\node at (-3.2917,4.2752) {\tiny{$-m-1$}};
\end{scope}
\end{scope}
\node at (-3.5166,4.1241) {\tiny{$h$-$\sum x_i$}};
\node at (-2.8258,4.8793) {\tiny{$h$}};
\node at (-3.1026,3.8095) {\tiny{$h$}};
\node at (-2.4179,2.9941) {\tiny{$h$-$x$}};
\node at (4.399,4.7223) {\tiny{$e$}};
\node at (4.6205,3.1718) {\tiny{$h$}};
\node at (4.7232,3.6563) {\tiny{$e$}};
\node at (-1.432,3.1734) {\tiny{$e$}};
\node at (-2.1078,4.8351) {\tiny{$e$}};
\node at (-0.3379,4.7125) {\tiny{$e$}};
\node at (2.4878,3.1019) {\tiny{$e$}};
\begin{scope}[shift={(0.7884,0.1602)}]
\draw  (-3.3913,4.3642) rectangle (-3.2064,4.1794);
\node at (-3.2917,4.2752) {\tiny{0}};
\end{scope}
\node at (-2.705,4.4473) {\tiny{$f$}};
\begin{scope}[shift={(2.4083,0.2285)}]
\draw  (-3.3913,4.3642) rectangle (-3.2064,4.1794);
\node at (-3.2917,4.2752) {\tiny{-1}};
\end{scope}
\begin{scope}[shift={(2.0145,0.3788)}]
\draw  (-3.3913,4.3642) rectangle (-3.2064,4.1794);
\node at (-3.2917,4.2752) {\tiny{-1}};
\end{scope}
\node at (-1.0992,4.6566) {\tiny{$x$}};
\begin{scope}[shift={(0.7862,-0.7233)}]
\draw  (-3.4113,4.3642) rectangle (-3.17,4.1794);
\node at (-3.2917,4.2752) {\tiny{-2}};
\end{scope}
\begin{scope}[shift={(1.0795,-0.9591)}]
\draw  (-3.3913,4.3642) rectangle (-3.2064,4.1794);
\node at (-3.2917,4.2752) {\tiny{-1}};
\end{scope}
\node at (-2.8398,3.2311) {\tiny{$e$}};
\node at (-2.6343,3.3722) {\tiny{$f$-$x$-$y$}};
\node at (-2.229,3.4713) {\tiny{$x$}};
\begin{scope}[shift={(2.2832,-0.8078)}]
\draw  (-3.3913,4.3642) rectangle (-3.2064,4.1794);
\node at (-3.2917,4.2752) {\tiny{-1}};
\end{scope}
\begin{scope}[shift={(2.5808,-0.9223)}]
\draw  (-3.3913,4.3642) rectangle (-3.2064,4.1794);
\node at (-3.2917,4.2752) {\tiny{-1}};
\end{scope}
\node at (-1.255,3.3728) {\tiny{$f$-$x$}};
\node at (-0.791,3.1924) {\tiny{$x$}};
\node at (-0.833,3.0177) {\tiny{$h$-$x$}};
\begin{scope}[shift={(6.6421,0.3617)}]
\draw  (-3.3913,4.3642) rectangle (-3.2064,4.1794);
\node at (-3.2917,4.2752) {\tiny{-1}};
\end{scope}
\begin{scope}[shift={(6.6649,-0.9309)}]
\draw  (-3.3913,4.3642) rectangle (-3.2064,4.1794);
\node at (-3.2917,4.2752) {\tiny{-1}};
\end{scope}
\begin{scope}[shift={(6.1397,0.4119)}]
\draw  (-3.3913,4.3642) rectangle (-3.2064,4.1794);
\node at (-3.2917,4.2752) {\tiny{-1}};
\end{scope}
\node at (3.3302,4.824) {\tiny{$f$-$x$}};
\node at (3.3631,3.1684) {\tiny{$f$-$x$}};
\node at (2.6596,4.6838) {\tiny{$x$}};
\node at (-1.2764,4.8738) {\tiny{$h$-$x$}};
\node at (2.6165,5.1607) {\tiny{$h$-$x$}};
\node at (-0.6253,4.5078) {\tiny{$f$-$x$}};
\node at (4.8674,4.2852) {\tiny{$h$+$f$}};
\begin{scope}[shift={(0.5344,-0.1913)}]
\draw  (-3.3913,4.3642) rectangle (-3.2064,4.1794);
\node at (-3.2917,4.2752) {\tiny{1}};
\end{scope}
\begin{scope}[shift={(1.1875,-0.0954)}]
\draw  (-3.3913,4.3642) rectangle (-3.2064,4.1794);
\node at (-3.2917,4.2752) {\tiny{1}};
\end{scope}
\begin{scope}[shift={(1.868,-0.4836)}]
\draw  (-3.3913,4.3642) rectangle (-3.2064,4.1794);
\node at (-3.2917,4.2752) {\tiny{1}};
\end{scope}
\begin{scope}[shift={(2.6947,-0.3009)}]
\draw  (-3.3913,4.3642) rectangle (-3.2064,4.1794);
\node at (-3.2917,4.2752) {\tiny{1}};
\end{scope}
\begin{scope}[shift={(5.1337,0.1924)}]
\draw  (-3.3913,4.3642) rectangle (-3.2064,4.1794);
\node at (-3.2917,4.2752) {\tiny{1}};
\end{scope}
\begin{scope}[shift={(6.0792,-0.2415)}]
\draw  (-3.3913,4.3642) rectangle (-3.2064,4.1794);
\node at (-3.2917,4.2752) {\tiny{1}};
\end{scope}
\begin{scope}[shift={(3.3387,-0.7759)}]
\draw  (-3.3913,4.3642) rectangle (-3.2064,4.1794);
\node at (-3.2917,4.2752) {\tiny{1}};
\end{scope}
\begin{scope}[shift={(7.2896,-0.2826)}]
\draw  (-3.3913,4.3642) rectangle (-3.2064,4.1794);
\node at (-3.2917,4.2752) {\tiny{1}};
\end{scope}
\draw (-2.5028,4.2956) -- (-2.5063,3.6954);
\draw (-2.1064,3.4386) -- (-1.3607,4.523) (-0.9692,4.3753) -- (-1.0014,3.6065) (-0.5753,3.4325) -- (-0.0023,4.0006);
\draw (2.7165,4.5523) -- (1.7208,3.8717) (3.3423,4.4792) -- (3.356,3.4789);
\draw (4.356,4.6016) -- (4.5359,4.2555);
\begin{scope}[shift={(-0.8527,-0.2924)}]
\draw  (-3.3913,4.3642) rectangle (-3.2064,4.1794);
\node at (-3.2917,4.2752) {\tiny{-1}};
\end{scope}
\node at (-4.1214,4.1541) {\tiny{$e$}};
\begin{scope}[shift={(2.9216,0.6283)}]
\draw  (-3.5934,4.3703) rectangle (-2.9834,4.1997);
\node at (-3.2917,4.2752) {\tiny{$-n-4$}};
\end{scope}
\begin{scope}[shift={(1.0369,0.7139)}]
\draw  (-3.5199,4.3472) rectangle (-2.9381,4.196);
\node at (-3.2388,4.2752) {\tiny{$-n-5$}};
\end{scope}
\begin{scope}[shift={(0.5051,0.4414)}]
\draw  (-3.5199,4.3472) rectangle (-2.9381,4.196);
\node at (-3.2388,4.2752) {\tiny{$n+5$}};
\end{scope}
\end{tikzpicture}
\end{array}
\end{align}
where the $n+8$ blowups on $S^0_C$ correspond to $n+8$ hypers in the fundamental representation of $\mf{su}(n)$ and all the other blowups correspond to a hyper in the two-index antisymmetric representation of $\mf{su}(n)$. This graph can be checked to be flop equivalent to the one presented in \cite{Bhardwaj:2018yhy}.

For odd $n=2m+1\ge 5$, the associated graph is
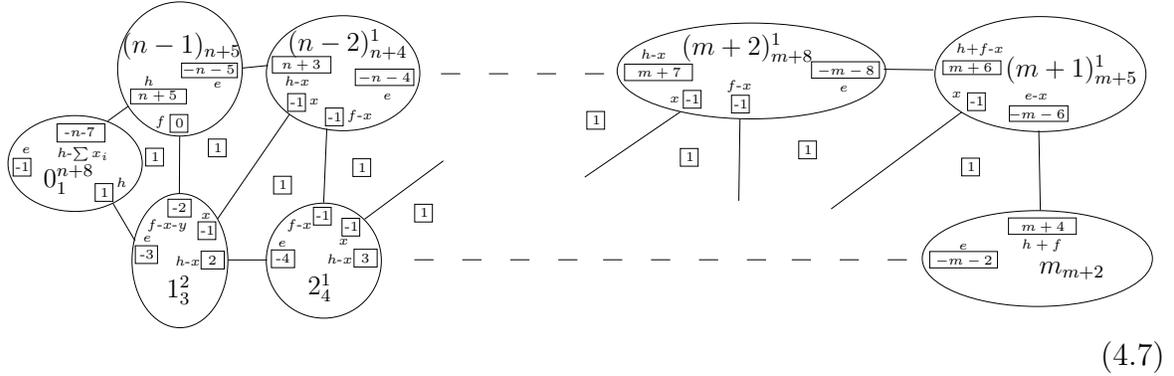
\begin{align}
\label{asymo}
 \scalebox{0.85}{$
\begin{array}{c}
\begin{tikzpicture}[scale=1.5]
\draw  (-3.5933,4.0135) ellipse (0.7 and 0.5);
\node (v1) at (-3.66,3.854) {$0_{1}^{n+8}$};
\draw  (-2.5,5) ellipse (0.65 and 0.7);
\node (v2) at (-2.4848,5.2395) {$(n-1)_{n+5}$};
\draw  (-0.799,4.956) ellipse (0.8 and 0.6);
\node (v4) at (-0.7526,5.2579) {$(n-2)_{n+4}^1$};
\draw  (-2.5,3) ellipse (0.5 and 0.7);
\node (v3) at (-2.5156,2.6911) {$1_3^2$};
\draw  (-1,3) ellipse (0.6 and 0.6);
\node (v5) at (-1.043,2.7213) {$2_4^1$};
\draw  (3.4586,4.9793) ellipse (1.4 and 0.5);
\node at (3.4169,5.2156) {$(m+2)_{m+8}^1$};
\draw  (6.4721,4.9467) ellipse (1.1 and 0.6);
\node at (6.7936,4.9903) {$(m+1)^1_{m+5}$};
\draw (-3.2535,4.4565) -- (-3.0386,4.6116);
\draw (-3.2,3.6) -- (-2.9679,3.2062);
\draw (-1.8481,5.0102) -- (-1.5807,5.0366);
\draw (-1.9977,3.0075) -- (-1.5933,3.0074);
\draw (6.4551,4.3562) -- (6.471,3.5265);
\draw (0.2222,4.9504) -- (0.4222,4.9504) (0.7222,4.9504) -- (0.9222,4.9504);
\draw (1.2222,4.9504) -- (1.4222,4.9504) (1.7222,4.9504) -- (1.9222,4.9504);
\begin{scope}[shift={(-0.0558,-1.9909)}]
\draw (0,5) -- (0.2,5) (0.5,5) -- (0.7,5);
\draw (1,5) -- (1.2,5) (1.5,5) -- (1.7,5);
\end{scope}
\draw  (-3.7805,4.4276) rectangle (-3.282,4.2384);
\node at (-3.5445,4.3427) {\tiny{-$n$-7}};
\begin{scope}[shift={(2.0572,0.759)}]
\draw  (-3.5934,4.3703) rectangle (-2.9834,4.1997);
\node at (-3.2917,4.2752) {\tiny{$n+3$}};
\end{scope}
\begin{scope}[shift={(0.0073,-0.5602)}]
\draw  (-3.3913,4.3642) rectangle (-3.2064,4.1794);
\node at (-3.2917,4.2752) {\tiny{1}};
\end{scope}
\begin{scope}[shift={(0.4458,-1.2087)}]
\draw  (-3.4113,4.3642) rectangle (-3.18,4.1794);
\node at (-3.2917,4.2752) {\tiny{-3}};
\end{scope}
\begin{scope}[shift={(1.1318,-1.2659)}]
\draw  (-3.4113,4.3642) rectangle (-3.17,4.1794);
\node at (-3.2917,4.2752) {\tiny{2}};
\end{scope}
\begin{scope}[shift={(1.8647,-1.2523)}]
\draw  (-3.4113,4.3642) rectangle (-3.17,4.1794);
\node at (-3.2917,4.2752) {\tiny{-4}};
\end{scope}
\begin{scope}[shift={(2.726,-1.2454)}]
\draw  (-3.4113,4.3642) rectangle (-3.17,4.1794);
\node at (-3.2917,4.2752) {\tiny{3}};
\end{scope}
\begin{scope}[shift={(7.7367,0.7044)}]
\draw  (-3.6509,4.3711) rectangle (-2.9441,4.2);
\node at (-3.2917,4.2752) {\tiny{$-m-8$}};
\end{scope}
\begin{scope}[shift={(9.1074,0.7143)}]
\draw  (-3.6509,4.3711) rectangle (-3.0476,4.2282);
\node at (-3.366,4.2929) {\tiny{$m+6$}};
\end{scope}
\begin{scope}[shift={(9.8838,0.2393)}]
\draw  (-3.736,4.3764) rectangle (-3.1252,4.2142);
\node at (-3.4526,4.2849) {\tiny{$-m-6$}};
\end{scope}
\begin{scope}[shift={(5.7871,0.6795)}]
\draw  (-3.6509,4.3711) rectangle (-2.9441,4.2);
\node at (-3.2917,4.2752) {\tiny{$m+7$}};
\end{scope}
\begin{scope}[shift={(3.6214,-1.9809)}]
\draw  (2.8184,5.0028) ellipse (1.2 and 0.5);
\node at (3.1809,4.8747) {$m_{m+2}$};
\begin{scope}[shift={(6.1684,1.0554)}]
\draw  (-3.6509,4.3711) rectangle (-2.9441,4.2);
\node at (-3.2917,4.2752) {\tiny{$m+4$}};
\end{scope}
\begin{scope}[shift={(5.3469,0.7052)}]
\draw  (-3.6509,4.3711) rectangle (-2.9441,4.2);
\node at (-3.2917,4.2752) {\tiny{$-m-2$}};
\end{scope}
\end{scope}
\node at (-3.5133,4.1161) {\tiny{$h$-$\sum x_i$}};
\node at (-2.8258,4.8793) {\tiny{$h$}};
\node at (-3.1072,3.8222) {\tiny{$h$}};
\node at (-2.4179,2.9941) {\tiny{$h$-$x$}};
\node at (4.4228,4.7938) {\tiny{$e$}};
\node at (6.4867,3.1599) {\tiny{$h+f$}};
\node at (6.4265,4.7082) {\tiny{$e$-$x$}};
\node at (-1.432,3.1734) {\tiny{$e$}};
\node at (-2.1078,4.8351) {\tiny{$e$}};
\node at (-0.3379,4.7125) {\tiny{$e$}};
\node at (5.6658,3.1496) {\tiny{$e$}};
\begin{scope}[shift={(0.7884,0.1602)}]
\draw  (-3.3913,4.3642) rectangle (-3.2064,4.1794);
\node at (-3.2917,4.2752) {\tiny{0}};
\end{scope}
\node at (-2.705,4.4473) {\tiny{$f$}};
\begin{scope}[shift={(2.4083,0.2285)}]
\draw  (-3.3913,4.3642) rectangle (-3.2064,4.1794);
\node at (-3.2917,4.2752) {\tiny{-1}};
\end{scope}
\begin{scope}[shift={(2.0145,0.3788)}]
\draw  (-3.3913,4.3642) rectangle (-3.2064,4.1794);
\node at (-3.2917,4.2752) {\tiny{-1}};
\end{scope}
\node at (-1.0992,4.6566) {\tiny{$x$}};
\begin{scope}[shift={(0.7862,-0.7233)}]
\draw  (-3.4113,4.3642) rectangle (-3.17,4.1794);
\node at (-3.2917,4.2752) {\tiny{-2}};
\end{scope}
\begin{scope}[shift={(1.0795,-0.9591)}]
\draw  (-3.3913,4.3642) rectangle (-3.2064,4.1794);
\node at (-3.2917,4.2752) {\tiny{-1}};
\end{scope}
\node at (-2.8398,3.2311) {\tiny{$e$}};
\node at (-2.6343,3.3722) {\tiny{$f$-$x$-$y$}};
\node at (-2.229,3.4713) {\tiny{$x$}};
\begin{scope}[shift={(2.2832,-0.8078)}]
\draw  (-3.3913,4.3642) rectangle (-3.2064,4.1794);
\node at (-3.2917,4.2752) {\tiny{-1}};
\end{scope}
\begin{scope}[shift={(2.5808,-0.9223)}]
\draw  (-3.3913,4.3642) rectangle (-3.2064,4.1794);
\node at (-3.2917,4.2752) {\tiny{-1}};
\end{scope}
\node at (-1.255,3.3728) {\tiny{$f$-$x$}};
\node at (-0.791,3.1924) {\tiny{$x$}};
\node at (-0.833,3.0177) {\tiny{$h$-$x$}};
\begin{scope}[shift={(6.6421,0.3617)}]
\draw  (-3.3913,4.3642) rectangle (-3.2064,4.1794);
\node at (-3.2917,4.2752) {\tiny{-1}};
\end{scope}
\begin{scope}[shift={(9.1036,0.3868)}]
\draw  (-3.3913,4.3642) rectangle (-3.2064,4.1794);
\node at (-3.2917,4.2752) {\tiny{-1}};
\end{scope}
\begin{scope}[shift={(6.1397,0.4119)}]
\draw  (-3.3913,4.3642) rectangle (-3.2064,4.1794);
\node at (-3.2917,4.2752) {\tiny{-1}};
\end{scope}
\node at (3.3302,4.824) {\tiny{$f$-$x$}};
\node at (5.5991,4.7305) {\tiny{$x$}};
\node at (2.6596,4.6838) {\tiny{$x$}};
\node at (-1.2764,4.8738) {\tiny{$h$-$x$}};
\node at (2.4397,5.1594) {\tiny{$h$-$x$}};
\node at (-0.6253,4.5078) {\tiny{$f$-$x$}};
\node at (5.841,5.2002) {\tiny{$h$+$f$-$x$}};
\begin{scope}[shift={(0.5344,-0.1913)}]
\draw  (-3.3913,4.3642) rectangle (-3.2064,4.1794);
\node at (-3.2917,4.2752) {\tiny{1}};
\end{scope}
\begin{scope}[shift={(1.1875,-0.0954)}]
\draw  (-3.3913,4.3642) rectangle (-3.2064,4.1794);
\node at (-3.2917,4.2752) {\tiny{1}};
\end{scope}
\begin{scope}[shift={(1.868,-0.4836)}]
\draw  (-3.3913,4.3642) rectangle (-3.2064,4.1794);
\node at (-3.2917,4.2752) {\tiny{1}};
\end{scope}
\begin{scope}[shift={(2.6947,-0.3009)}]
\draw  (-3.3913,4.3642) rectangle (-3.2064,4.1794);
\node at (-3.2917,4.2752) {\tiny{1}};
\end{scope}
\begin{scope}[shift={(5.1337,0.1924)}]
\draw  (-3.3913,4.3642) rectangle (-3.2064,4.1794);
\node at (-3.2917,4.2752) {\tiny{1}};
\end{scope}
\begin{scope}[shift={(7.3432,-0.1163)}]
\draw  (-3.3913,4.3642) rectangle (-3.2064,4.1794);
\node at (-3.2917,4.2752) {\tiny{1}};
\end{scope}
\begin{scope}[shift={(3.3387,-0.7759)}]
\draw  (-3.3913,4.3642) rectangle (-3.2064,4.1794);
\node at (-3.2917,4.2752) {\tiny{1}};
\end{scope}
\begin{scope}[shift={(9.0425,-0.2945)}]
\draw  (-3.3913,4.3642) rectangle (-3.2064,4.1794);
\node at (-3.2917,4.2752) {\tiny{1}};
\end{scope}
\draw (-2.5028,4.2956) -- (-2.5063,3.6954);
\draw (-2.1064,3.4386) -- (-1.3607,4.523) (-0.9692,4.3753) -- (-1.0014,3.6065) (-0.5753,3.4325) -- (0.2469,4.0419);
\draw (2.7165,4.5523) -- (1.7208,3.8717) (4.2897,3.5394) -- (5.6532,4.5516);
\draw (4.8381,4.997) -- (5.3557,4.9913);
\begin{scope}[shift={(-0.8486,-0.2966)}]
\draw  (-3.3913,4.3642) rectangle (-3.2064,4.1794);
\node at (-3.2917,4.2752) {\tiny{-1}};
\end{scope}
\node at (-4.1021,4.1535) {\tiny{$e$}};
\begin{scope}[shift={(2.9216,0.6283)}]
\draw  (-3.5934,4.3703) rectangle (-2.9834,4.1997);
\node at (-3.2917,4.2752) {\tiny{$-n-4$}};
\end{scope}
\begin{scope}[shift={(1.0369,0.7139)}]
\draw  (-3.5199,4.3472) rectangle (-2.9381,4.196);
\node at (-3.2388,4.2752) {\tiny{$-n-5$}};
\end{scope}
\begin{scope}[shift={(0.5051,0.4414)}]
\draw  (-3.5199,4.3472) rectangle (-2.9381,4.196);
\node at (-3.2388,4.2752) {\tiny{$n+5$}};
\end{scope}
\begin{scope}[shift={(1.9118,-1.9909)}]
\draw (0,5) -- (0.2,5) (0.5,5) -- (0.7,5);
\draw (1,5) -- (1.2,5) (1.5,5) -- (1.7,5);
\end{scope}
\begin{scope}[shift={(3.8793,-1.9909)}]
\draw (0,5) -- (0.2,5) (0.5,5) -- (0.7,5);
\draw (1,5) -- (1.2,5);
\end{scope}
\begin{scope}[shift={(6.109,-0.1938)}]
\draw  (-3.3913,4.3642) rectangle (-3.2064,4.1794);
\node at (-3.2917,4.2752) {\tiny{1}};
\end{scope}
\draw (3.339,4.4773) -- (3.3271,3.6247);
\end{tikzpicture}
\end{array}
$}
\end{align}
where again the $n+8$ blowups on $S^0_C$ correspond to fundamental representation and all the other blowups correspond to two-index antisymmetric representation. The graph is flop equivalent to the one presented in \cite{Bhardwaj:2018yhy}.

The graph that we associate to the special case of $n=3$ is
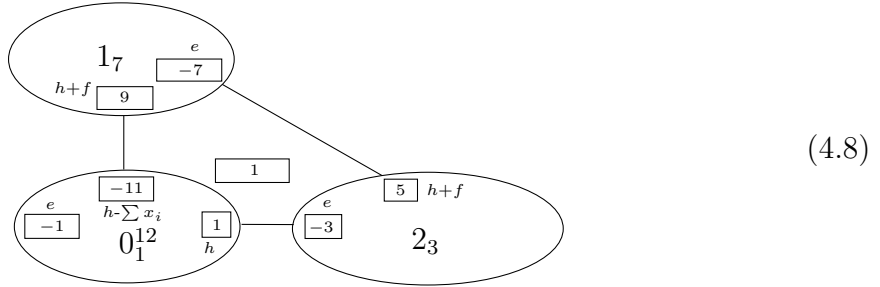
\begin{align}
\begin{array}{c}
\begin{tikzpicture}[scale=1.5]
\begin{scope}[shift={(2.4713,0.0223)}]
\draw  (-4.4603,4.9901) ellipse (1 and 0.5);
\node (v1) at (-4.3588,4.8276) {$0^{12}_{1}$};
\draw (-3.4464,5.0091) -- (-2.9847,5.0073);
\begin{scope}[shift={(-1.8354,0.7134)}]
\draw  (-3.5321,4.3845) rectangle (-3.0504,4.1716);
\node at (-3.2917,4.2752) {\tiny{$-1$}};
\end{scope}
\begin{scope}[shift={(-0.3754,0.7057)}]
\draw  (-3.4209,4.4137) rectangle (-3.1665,4.189);
\node at (-3.2901,4.2995) {\tiny{$1$}};
\end{scope}
\begin{scope}[shift={(-1.1756,1.0455)}]
\draw  (-3.5321,4.3845) rectangle (-3.0504,4.1716);
\node at (-3.2917,4.2752) {\tiny{$-11$}};
\end{scope}
\node at (-4.4107,5.1089) {\tiny{$h$-$\sum x_i$}};
\end{scope}
\begin{scope}[shift={(5.1361,0.0036)}]
\draw  (-4.4603,4.9901) ellipse (1.2 and 0.5);
\node (v1) at (-4.4811,4.879) {$2_{3}$};
\node at (-5.3519,5.2155) {\tiny{$e$}};
\begin{scope}[shift={(-2.1142,0.7232)}]
\draw  (-3.4342,4.3914) rectangle (-3.1098,4.1794);
\node at (-3.2917,4.2752) {\tiny{$-3$}};
\end{scope}
\begin{scope}[shift={(-1.3924,1.0617)}]
\draw  (-3.4542,4.367) rectangle (-3.1597,4.1724);
\node at (-3.308,4.2726) {\tiny{$5$}};
\end{scope}
\node at (-4.3034,5.3236) {\tiny{$h$+$f$}};
\end{scope}
\node at (-2.6675,5.1964) {\tiny{$e$}};
\node at (-1.2595,4.8245) {\tiny{$h$}};
\begin{scope}[shift={(2.4199,1.5062)}]
\draw  (-4.4603,4.9901) ellipse (1 and 0.5);
\node (v1) at (-4.5578,4.9971) {$1_{7}$};
\node at (-3.8099,5.0963) {\tiny{$e$}};
\node at (-4.887,4.7214) {\tiny{$h$+$f$}};
\begin{scope}[shift={(-0.5576,0.611)}]
\draw  (-3.5896,4.3807) rectangle (-3.0131,4.1849);
\node at (-3.292,4.2758) {\tiny{$-7$}};
\end{scope}
\begin{scope}[shift={(-1.1411,0.3576)}]
\draw  (-3.5352,4.3891) rectangle (-3.0397,4.1868);
\node at (-3.2901,4.2995) {\tiny{$9$}};
\end{scope}
\end{scope}
\begin{scope}[shift={(2.4325,1.2108)}]
\draw  (-3.6392,4.4011) rectangle (-2.9855,4.1922);
\node at (-3.2901,4.2995) {\tiny{$1$}};
\end{scope}
\draw (-2.019,5.9932) -- (-2.019,5.5182);
\draw (-1.1458,6.2729) -- (0.2707,5.464);
\end{tikzpicture}
\end{array}
\end{align}
which is not a limit of the above series of graphs. The $12$ blowups correspond to $12$ hypers in the fundamental of $\su(3)$.

%

Let us study the case of I$^s_n$ fiber in more detail. For a generic fibration keeping I$_n^s$ on $C$, we have two kinds of special points. $n+8$ of the special points correspond to a further degeneration to I$_{n+1}$ which is again implemented by the splittings $f^0_C-x_i,x_i$ where $x_i$ are the blowups corresponding to hypers in the fundamental of $\su(n)$. There is another special point $p$ at which the fibers $f^i_C$ split according to the blowups corresponding to the antisymmetric representation, and moreover the resulting components are combined according to the gluing rules presented in (\ref{asyme}) and (\ref{asymo}). For example, for $n=4$ the fiber at $p$ can be depicted as
\begin{align}
\begin{array}{c}
\begin{tikzpicture}
\node (v1) at (3.4322,2.0211) {$f^0_C$};
\node (v2) at (5.4322,2.0211) {$f^3_C$};
\node (v3) at (7.4322,2.0211) {$f^2_C$};
\node (v4) at (5.4322,3.5211) {$x$};
\node (v5) at (5.4322,2.0211-1.5) {$y$};
\draw  (v1) -- (v2);
\draw  (v2) -- (v3);
\draw  (v2) edge (v4);
\draw  (v2) edge (v5);
\end{tikzpicture}
\end{array}
\end{align}
where $f^3_C= f^1_C-x-y$ over $p$ and the edges depict transverse intersections between different components. Hence, $f_C$ over $p$ can be written in terms of its components as $f^0_C+2f^3_C+f^2_C+x+y$. In other words, we obtain an I$_0^*$ fiber over $p$. Turning it backwards, since components of the elliptic fiber go on top of each other over $p$, this manifests as extra gluings between $S^i_C$ that are not visible from the intersections of $f^i_C$.

\subsection{$\su(\tilde6)$ on $-1$ curve}
We associate the graph
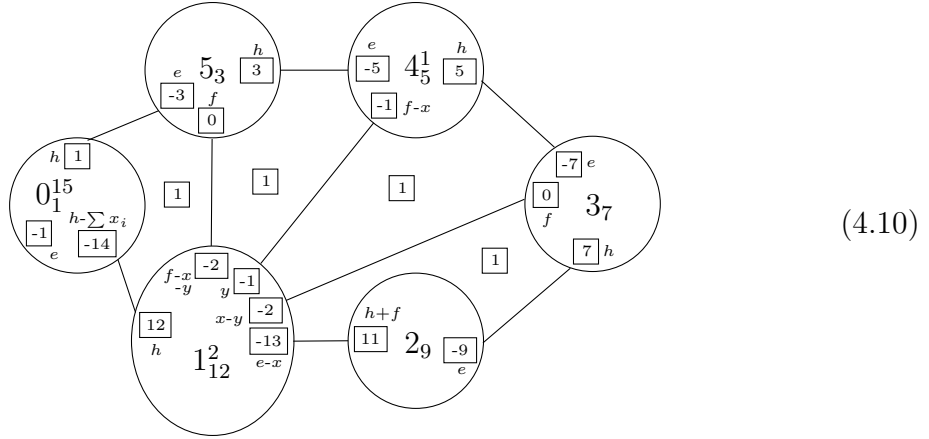
\begin{align}
\begin{array}{c}
\begin{tikzpicture}[scale=1.8]
\draw  (-3.5,4) ellipse (0.5 and 0.5);
\node (v1) at (-3.6636,4.0736) {$0^{15}_1$};
\draw  (-2.5,5) ellipse (0.5 and 0.5);
\node (v2) at (-2.5,5) {$5_3$};
\draw  (-1,5) ellipse (0.5 and 0.5);
\node (v4) at (-0.9827,5.0196) {$4_5^1$};
\draw  (-2.5,3) ellipse (0.6 and 0.7);
\node (v3) at (-2.5056,2.8479) {$1_{12}^2$};
\draw  (-1,3) ellipse (0.5 and 0.5);
\node (v5) at (-0.9791,2.966) {$2_9$};
\draw  (0.3058,4.0117) ellipse (0.5 and 0.5);
\node at (0.3576,3.9889) {$3_7$};
\draw (-3.4246,4.4813) -- (-2.9,4.7);
\draw (-3.2,3.6) -- (-3.0699,3.2091);
\draw (-2,5) -- (-1.5,5);
\draw (-1.9057,2.997) -- (-1.5,3);
\begin{scope}[shift={(0.7818,-0.7181)}]
\draw  (-3.4113,4.3642) rectangle (-3.17,4.1794);
\node at (-3.2917,4.2752) {\tiny{-2}};
\end{scope}
\draw  (-3.5929,4.4561) rectangle (-3.408,4.2713);
\node at (-3.4933,4.3671) {\tiny{1}};
\begin{scope}[shift={(0.5286,0.5394)}]
\draw  (-3.4113,4.3642) rectangle (-3.17,4.1794);
\node at (-3.2917,4.2752) {\tiny{-3}};
\end{scope}
\begin{scope}[shift={(1.1159,0.7257)}]
\draw  (-3.4113,4.3642) rectangle (-3.17,4.1794);
\node at (-3.2917,4.2752) {\tiny{3}};
\end{scope}
\begin{scope}[shift={(1.9717,0.7394)}]
\draw  (-3.4113,4.3642) rectangle (-3.17,4.1794);
\node at (-3.2917,4.2752) {\tiny{-5}};
\end{scope}
\begin{scope}[shift={(2.6153,0.7188)}]
\draw  (-3.4113,4.3642) rectangle (-3.17,4.1794);
\node at (-3.2917,4.2752) {\tiny{5}};
\end{scope}
\begin{scope}[shift={(1.2053,-1.2764)}]
\draw  (-3.4302,4.3706) rectangle (-3.1529,4.1767);
\node at (-3.2917,4.2752) {\tiny{-13}};
\end{scope}
\begin{scope}[shift={(-0.4847,-0.4782)}]
\draw  (-3.3913,4.3642) rectangle (-3.2064,4.1794);
\node at (-3.2917,4.2752) {\tiny{-1}};
\end{scope}
\begin{scope}[shift={(0.3718,-1.1552)}]
\draw  (-3.4113,4.3642) rectangle (-3.18,4.1794);
\node at (-3.2917,4.2752) {\tiny{12}};
\end{scope}
\begin{scope}[shift={(-0.061,-0.556)}]
\draw  (-3.4302,4.3706) rectangle (-3.1529,4.1767);
\node at (-3.2917,4.2752) {\tiny{-14}};
\end{scope}
\begin{scope}[shift={(1.958,-1.2597)}]
\draw  (-3.4113,4.3642) rectangle (-3.17,4.1794);
\node at (-3.2917,4.2752) {\tiny{11}};
\end{scope}
\begin{scope}[shift={(2.6188,-1.342)}]
\draw  (-3.4113,4.3642) rectangle (-3.17,4.1794);
\node at (-3.2917,4.2752) {\tiny{-9}};
\end{scope}
\node at (-3.6659,4.3616) {\tiny{$h$}};
\node at (-2.1625,5.1528) {\tiny{$h$}};
\node at (-3.3439,3.8972) {\tiny{$h$-$\sum x_i$}};
\node at (-2.0843,2.828) {\tiny{$e$-$x$}};
\node at (-0.0276,3.8905) {\tiny{$f$}};
\node at (0.4264,3.6659) {\tiny{$h$}};
\node at (0.2942,4.3073) {\tiny{$e$}};
\node at (-2.7601,4.9723) {\tiny{$e$}};
\node at (-0.6607,5.1664) {\tiny{$h$}};
\node at (-3.672,3.6265) {\tiny{$e$}};
\begin{scope}[shift={(0.7867,0.3557)}]
\draw  (-3.3913,4.3642) rectangle (-3.2064,4.1794);
\node at (-3.2917,4.2752) {\tiny{0}};
\end{scope}
\node at (-2.4975,4.7835) {\tiny{$f$}};
\begin{scope}[shift={(2.064,0.4663)}]
\draw  (-3.3913,4.3642) rectangle (-3.2064,4.1794);
\node at (-3.2917,4.2752) {\tiny{-1}};
\end{scope}
\begin{scope}[shift={(1.1834,-1.0477)}]
\draw  (-3.4113,4.3642) rectangle (-3.17,4.1794);
\node at (-3.2917,4.2752) {\tiny{-2}};
\end{scope}
\begin{scope}[shift={(1.0492,-0.8312)}]
\draw  (-3.3913,4.3642) rectangle (-3.2064,4.1794);
\node at (-3.2917,4.2752) {\tiny{-1}};
\end{scope}
\node at (-2.9161,2.9374) {\tiny{$h$}};
\node at (-2.7588,3.4811) {\tiny{$f$-$x$}};
\node at (-2.715,3.3757) {\tiny{-$y$}};
\node at (-2.4079,3.3499) {\tiny{$y$}};
\node at (-2.3781,3.1478) {\tiny{$x$-$y$}};
\begin{scope}[shift={(3.43,0.0318)}]
\draw  (-3.3913,4.3642) rectangle (-3.2064,4.1794);
\node at (-3.2917,4.2752) {\tiny{-7}};
\end{scope}
\begin{scope}[shift={(3.5558,-0.6124)}]
\draw  (-3.3913,4.3642) rectangle (-3.2064,4.1794);
\node at (-3.2917,4.2752) {\tiny{7}};
\end{scope}
\node at (-0.6601,2.7787) {\tiny{$e$}};
\node at (-1.2427,3.1989) {\tiny{$h$+$f$}};
\node at (-1.3237,5.1789) {\tiny{$e$}};
\node at (-0.9942,4.7211) {\tiny{$f$-$x$}};
\begin{scope}[shift={(0.5344,-0.1913)}]
\draw  (-3.3913,4.3642) rectangle (-3.2064,4.1794);
\node at (-3.2917,4.2752) {\tiny{1}};
\end{scope}
\begin{scope}[shift={(1.1875,-0.0954)}]
\draw  (-3.3913,4.3642) rectangle (-3.2064,4.1794);
\node at (-3.2917,4.2752) {\tiny{1}};
\end{scope}
\begin{scope}[shift={(2.1945,-0.1439)}]
\draw  (-3.3913,4.3642) rectangle (-3.2064,4.1794);
\node at (-3.2917,4.2752) {\tiny{1}};
\end{scope}
\begin{scope}[shift={(3.2583,-0.1941)}]
\draw  (-3.3913,4.3642) rectangle (-3.2064,4.1794);
\node at (-3.2917,4.2752) {\tiny{0}};
\end{scope}
\begin{scope}[shift={(2.882,-0.68)}]
\draw  (-3.3913,4.3642) rectangle (-3.2064,4.1794);
\node at (-3.2917,4.2752) {\tiny{1}};
\end{scope}
\draw (-2.5045,4.4911) -- (-2.5107,3.7006);
\draw (-2.1452,3.571) -- (-1.3112,4.6105);
\draw (-0.5158,4.9219) -- (0.0237,4.4304) (-0.198,4.0458) -- (-1.9581,3.298) (-0.5042,2.9842) -- (0.1412,3.5357);
\end{tikzpicture}
\end{array}
\end{align}
which is flop equivalent to the one presented in \cite{Bhardwaj:2018yhy}. Here the $15$ blowups correspond to $15$ hypers in fundamental representation of $\su(6)$, and all the other blowups correspond to a half-hyper in three-index antisymmetric representation of $\su(6)$.

For this formal gauge algebra, the only possible fiber is a slightly tuned non-generic version of I$_6^s$. The generic version gives rise to $n=6$ version of (\ref{asyme}) and has a hyper in two-index antisymmetric along with 14 hypers in fundamental of $\su(6)$. Under the tuning, the hyper in two index antisymmetric is traded for a hyper in fundamental and a half-hyper in three-index antisymmetric. If we let the fibration be generic except for keeping $\su(\tilde 6)$ on $C$, then there are two kinds of special points. Over 15 of the special points $p_i,i=1,\cdots,15$, $f^0_C$ splits into $f^0_C-x_i,x_i$. Over another special point $p$, the fibers $f^i_C$ split according to blowups corresponding to the three-index antisymmetric representation, and furthermore the resulting components are combined according to the gluing rules shown in the graph above.

\subsection{$\sp(n),n\ge0$ on $-1$ curve}\label{sp}
We will associate two graphs to this case which are flop equivalent when $\sp(n)$ does not have a neighboring $\so(2m+1)$ or $\fg_2$, but are not flop equivalent when such a neighbor is present. The first associated graph is
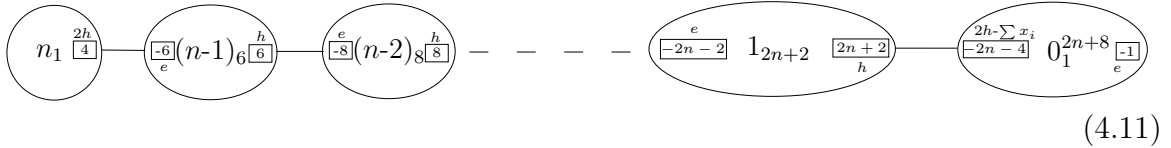
\begin{align} \label{sp1}
\scalebox{.9}{$
\begin{array}{c}
\begin{tikzpicture}[scale=1.4]
\draw  (-3.9189,4.9308) ellipse (0.5 and 0.5);
\node (v1) at (-3.9645,4.9308) {$n_1$};
\draw  (-2.2693,4.9351) ellipse (0.7 and 0.5);
\node (v2) at (-2.2693,4.9351) {$(n$-$1)_6$};
\draw  (-0.3896,4.946) ellipse (0.7 and 0.5);
\node (v4) at (-0.3896,4.946) {$(n$-$2)_8$};
\draw  (3.6555,4.97) ellipse (1.3 and 0.5);
\node at (3.7229,4.9682) {$1_{2n+2}$};
\draw (-3.4164,4.9595) -- (-2.961,4.9554);
\draw (-1.5775,4.9449) -- (-1.0775,4.9449);
\draw (0.4606,4.9383) -- (0.6606,4.9383) (0.9606,4.9383) -- (1.1606,4.9383);
\draw (1.4606,4.9383) -- (1.6606,4.9383) (1.9606,4.9383) -- (2.1606,4.9383);
\begin{scope}[shift={(1.5471,0.6515)}]
\draw  (-3.4113,4.3642) rectangle (-3.17,4.1794);
\node at (-3.2917,4.2752) {\tiny{6}};
\end{scope}
\begin{scope}[shift={(2.3942,0.6843)}]
\draw  (-3.4113,4.3642) rectangle (-3.17,4.1794);
\node at (-3.2917,4.2752) {\tiny{-8}};
\end{scope}
\begin{scope}[shift={(3.4077,0.655)}]
\draw  (-3.4113,4.3642) rectangle (-3.17,4.1794);
\node at (-3.2917,4.2752) {\tiny{8}};
\end{scope}
\begin{scope}[shift={(7.9172,0.6889)}]
\draw  (-3.6209,4.3874) rectangle (-3.0318,4.196);
\node at (-3.3153,4.289) {\tiny{$2n+2$}};
\end{scope}
\begin{scope}[shift={(6.1095,0.6935)}]
\draw  (-3.6329,4.3783) rectangle (-2.9362,4.1728);
\node at (-3.2917,4.2752) {\tiny{$-2n-2$}};
\end{scope}
\node at (-1.7361,5.1076) {\tiny{$h$}};
\node at (7.2985,4.7637) {\tiny{$e$}};
\node at (-0.887,5.1182) {\tiny{$e$}};
\node at (-2.7619,4.775) {\tiny{$e$}};
\node at (0.1065,5.0953) {\tiny{$h$}};
\node at (2.8359,5.1764) {\tiny{$e$}};
\node at (-3.6211,5.14) {\tiny{$2h$}};
\begin{scope}[shift={(9.4713,0.9424)}]
\draw  (-2.8576,4.0223) ellipse (1 and 0.5);
\node (v1) at (-2.5786,4.0047) {$0^{2n+8}_{1}$};
\draw (-3.8646,4.0367) -- (-4.5163,4.0368);
\node at (-3.3554,4.2093) {\tiny{$2h$-$\sum x_i$}};
\end{scope}
\begin{scope}[shift={(10.6964,0.6749)}]
\draw  (-3.4113,4.3642) rectangle (-3.17,4.1794);
\node at (-3.2917,4.2752) {\tiny{-1}};
\end{scope}
\begin{scope}[shift={(-0.3067,0.6902)}]
\draw  (-3.4113,4.3642) rectangle (-3.17,4.1794);
\node at (-3.2917,4.2752) {\tiny{4}};
\end{scope}
\begin{scope}[shift={(0.517,0.6804)}]
\draw  (-3.4113,4.3642) rectangle (-3.17,4.1794);
\node at (-3.2917,4.2752) {\tiny{-6}};
\end{scope}
\node at (4.6078,4.7834) {\tiny{$h$}};
\begin{scope}[shift={(9.34,0.6792)}]
\draw  (-3.6675,4.3771) rectangle (-2.9831,4.1983);
\node at (-3.3431,4.2881) {\tiny{$-2n-4$}};
\end{scope}
\end{tikzpicture}
\end{array}
$}
\end{align}
which is flop equivalent to the one presented in \cite{Bhardwaj:2018yhy}. Here the $2n+8$ blowups on $S^0_C$ correspond to $2n+8$ hypers in the fundamental representation of $\mf{sp}(n)$. If we have a generic fibration with I$_{2n}^{ns}$ on $C$, then we obtain $2n+8$ special points corresponding to hypers in fundamental where (in the resolution corresponding to the above geometry) $f^0_C$ splits into $f^0_C-x_i,x_i$ for $i=1,\cdots,2n+8$.

\noindent The second associated graph is
\begin{align} \label{sp2}
\scalebox{.9}{$
\begin{array}{c}
\begin{tikzpicture}[scale=1.4]
\draw  (-3.9189,4.9308) ellipse (0.5 and 0.5);
\node (v1) at (-4.0052,4.8627) {$n^1_1$};
\draw  (-2.2693,4.9351) ellipse (0.7 and 0.5);
\node (v2) at (-2.2693,4.9351) {$(n$-$1)_5$};
\draw  (-0.3896,4.946) ellipse (0.7 and 0.5);
\node (v4) at (-0.3896,4.946) {$(n$-$2)_7$};
\draw  (3.6555,4.97) ellipse (1.3 and 0.5);
\node at (3.7229,4.9682) {$1_{2n+1}$};
\draw (-3.4164,4.9595) -- (-2.961,4.9554);
\draw (-1.5775,4.9449) -- (-1.0775,4.9449);
\draw (0.4606,4.9383) -- (0.6606,4.9383) (0.9606,4.9383) -- (1.1606,4.9383);
\draw (1.4606,4.9383) -- (1.6606,4.9383) (1.9606,4.9383) -- (2.1606,4.9383);
\begin{scope}[shift={(1.5471,0.6515)}]
\draw  (-3.4113,4.3642) rectangle (-3.17,4.1794);
\node at (-3.2917,4.2752) {\tiny{5}};
\end{scope}
\begin{scope}[shift={(2.3942,0.6843)}]
\draw  (-3.4113,4.3642) rectangle (-3.17,4.1794);
\node at (-3.2917,4.2752) {\tiny{-7}};
\end{scope}
\begin{scope}[shift={(3.4077,0.655)}]
\draw  (-3.4113,4.3642) rectangle (-3.17,4.1794);
\node at (-3.2917,4.2752) {\tiny{7}};
\end{scope}
\begin{scope}[shift={(7.9172,0.6889)}]
\draw  (-3.6209,4.3874) rectangle (-3.0318,4.196);
\node at (-3.3153,4.289) {\tiny{$2n+1$}};
\end{scope}
\begin{scope}[shift={(6.1095,0.6935)}]
\draw  (-3.6329,4.3783) rectangle (-2.9362,4.1728);
\node at (-3.2917,4.2752) {\tiny{$-2n-1$}};
\end{scope}
\node at (-1.7361,5.1076) {\tiny{$h$}};
\node at (7.2985,4.7637) {\tiny{$e$}};
\node at (-0.887,5.1182) {\tiny{$e$}};
\node at (-2.7619,4.775) {\tiny{$e$}};
\node at (0.1065,5.0953) {\tiny{$h$}};
\node at (2.8359,5.1764) {\tiny{$e$}};
\node at (-3.7624,5.1758) {\tiny{$2h$-$x$}};
\begin{scope}[shift={(9.4713,0.9424)}]
\draw  (-2.8576,4.0223) ellipse (1 and 0.5);
\node (v1) at (-2.5786,4.0047) {$0^{2n+7}_{1}$};
\draw (-3.8646,4.0367) -- (-4.5163,4.0368);
\node at (-3.3554,4.2093) {\tiny{$2h$-$\sum x_i$}};
\end{scope}
\begin{scope}[shift={(10.6964,0.6749)}]
\draw  (-3.4113,4.3642) rectangle (-3.17,4.1794);
\node at (-3.2917,4.2752) {\tiny{-1}};
\end{scope}
\begin{scope}[shift={(-0.3067,0.6902)}]
\draw  (-3.4113,4.3642) rectangle (-3.17,4.1794);
\node at (-3.2917,4.2752) {\tiny{3}};
\end{scope}
\begin{scope}[shift={(0.517,0.6804)}]
\draw  (-3.4113,4.3642) rectangle (-3.17,4.1794);
\node at (-3.2917,4.2752) {\tiny{-5}};
\end{scope}
\node at (4.6078,4.7834) {\tiny{$h$}};
\begin{scope}[shift={(9.34,0.6792)}]
\draw  (-3.6675,4.3771) rectangle (-2.9831,4.1983);
\node at (-3.3431,4.2881) {\tiny{$-2n-3$}};
\end{scope}
\end{tikzpicture}
\end{array}
$}
\end{align}
which is flop equivalent to the one presented above. Here the $2n+7$ blowups on $S^0_C$ correspond to $2n+7$ hypers in the fundamental representation of $\mf{sp}(n)$. Similarly, the blowup on $S^n_C$ corresponds to a hyper in the fundamental of $\mf{sp}(n)$. If we have a generic fibration with I$_{2n}^{ns}$ on $C$, then we obtain $2n+8$ special points corresponding to hypers in fundamental where (in the resolution corresponding to the above geometry), at $2n+7$ of those points, $f^0_C$ splits into $f^0_C-x_i,x_i$ for $i=1,\cdots,2n+7$, and at the $(2n+8)^{\text{th}}$ point $f^n_C$ splits into $f^n_C-x,x$.

\noindent In both the cases, the elliptic fiber $f_C$ can be recognized as $f^0_C+2(f^1_C+\cdots+f_C^{n-1})+f^n_C$ due to the monodromy.

The case $n=0$ can be recognized as a limit of the above series of geometries which yields $S_C=\F_1^8$ where $S_C$ is glued to the base $B$ along the $e$ curve. However, this is only a trick and the correct answer is the closely related surface $S_C=(\P^2)^9$ where $S_C$ is glued to $B$ along one of the blowups, say $x_9$. See Section \ref{multiple} and Section \ref{sp0} for more discussion on this point.

\subsection{$\so(2r),r\ge4$ on $-4$ curve}
The associated graph is
\begin{align}
\begin{array}{c}
\begin{tikzpicture}[scale=1.4]
\draw  (-3.5,4) ellipse (0.5 and 0.5);
\node (v1) at (-3.5,4) {$0_2$};
\draw  (-2.5,5) ellipse (0.5 and 0.5);
\node (v2) at (-2.5,5) {$2_0$};
\draw  (-1,5) ellipse (0.5 and 0.5);
\node (v4) at (-1,5) {$3_2$};
\draw  (3.4586,4.9793) ellipse (1.4 and 0.5);
\node at (3.389,4.9889) {$(r-2)_{2r-8}$};
\draw (-3.2,4.4) -- (-2.9,4.7);
\draw (-2,5) -- (-1.5,5);
\draw (-0.1,5) -- (0.1,5) (0.4,5) -- (0.6,5);
\draw (0.9,5) -- (1.1,5) (1.4,5) -- (1.6,5);
\begin{scope}[shift={(-0.0636,1.4373)}]
\draw  (-3.4113,4.3642) rectangle (-3.17,4.1794);
\node at (-3.2917,4.2752) {\tiny{-2}};
\end{scope}
\begin{scope}[shift={(1.1159,0.7257)}]
\draw  (-3.4113,4.3642) rectangle (-3.17,4.1794);
\node at (-3.2917,4.2752) {\tiny{0}};
\end{scope}
\begin{scope}[shift={(1.9717,0.7394)}]
\draw  (-3.4113,4.3642) rectangle (-3.17,4.1794);
\node at (-3.2917,4.2752) {\tiny{-2}};
\end{scope}
\begin{scope}[shift={(2.6153,0.7188)}]
\draw  (-3.4113,4.3642) rectangle (-3.17,4.1794);
\node at (-3.2917,4.2752) {\tiny{2}};
\end{scope}
\begin{scope}[shift={(7.3822,0.4349)}]
\draw  (-3.5851,4.3727) rectangle (-3.0318,4.196);
\node at (-3.2917,4.2752) {\tiny{$2r-8$}};
\end{scope}
\begin{scope}[shift={(5.6849,0.6908)}]
\draw  (-3.5608,4.3783) rectangle (-3.0453,4.2148);
\node at (-3.2917,4.2752) {\tiny{$8-2r$}};
\end{scope}
\node at (-2.1625,5.1528) {\tiny{$h$}};
\node at (-3.5652,3.5724) {\tiny{$h$}};
\node at (-1.3095,5.1733) {\tiny{$e$}};
\node at (-2.7275,4.6352) {\tiny{$h$}};
\node at (-0.6859,5.1591) {\tiny{$h$}};
\node at (2.3865,5.1337) {\tiny{$e$}};
\node at (-3.3582,4.4131) {\tiny{$e$}};
\begin{scope}[shift={(0.0001,1.9902)}]
\draw  (-3.5,4) ellipse (0.5 and 0.5);
\node (v1) at (-3.5,4) {$1_2$};
\draw (-3.1971,3.6066) -- (-2.8839,3.3294);
\begin{scope}[shift={(-0.0137,-0.5476)}]
\end{scope}
\node at (-3.1558,3.7385) {\tiny{$e$}};
\end{scope}
\begin{scope}[shift={(8.9112,2.3311)}]
\draw  (-3.3027,4.011) ellipse (0.9 and 0.6);
\node (v1) at (-3.2831,4.122) {$(r-1)_{2r-6}$};
\draw (-3.9808,3.6346) -- (-4.6003,3.0354);
\begin{scope}[shift={(-0.0181,-0.7126)}]
\draw  (-3.3913,4.3642) rectangle (-3.2064,4.1794);
\node at (-3.2917,4.2752) {\tiny{0}};
\end{scope}
\node at (-3.1436,3.5684) {\tiny{$f$}};
\node at (-4.0415,3.8069) {\tiny{$e$}};
\end{scope}
\begin{scope}[shift={(8.917,-0.1552)}]
\draw  (-3.1597,3.9773) ellipse (0.9 and 0.6);
\node (v1) at (-3.0415,3.8049) {$r^{4r-16}_{2r-6}$};
\draw (-3.9884,4.2124) -- (-4.6002,4.7342);
\node at (-3.0506,4.1836) {\tiny{$f-x_i$}};
\node at (-3.7387,3.9269) {\tiny{$e$}};
\end{scope}
\begin{scope}[shift={(-0.3948,-0.5205)}]
\draw  (-3.4113,4.3642) rectangle (-3.17,4.1794);
\node at (-3.2917,4.2752) {\tiny{2}};
\end{scope}
\begin{scope}[shift={(-0.0224,-0.0146)}]
\draw  (-3.4113,4.3642) rectangle (-3.17,4.1794);
\node at (-3.2917,4.2752) {\tiny{-2}};
\end{scope}
\begin{scope}[shift={(0.5457,0.5375)}]
\draw  (-3.4113,4.3642) rectangle (-3.17,4.1794);
\node at (-3.2917,4.2752) {\tiny{0}};
\end{scope}
\begin{scope}[shift={(0.5516,0.949)}]
\draw  (-3.4113,4.3642) rectangle (-3.17,4.1794);
\node at (-3.2917,4.2752) {\tiny{0}};
\end{scope}
\node at (-2.7003,5.3738) {\tiny{$h$}};
\begin{scope}[shift={(7.3866,0.9712)}]
\draw  (-3.5928,4.3712) rectangle (-3.0377,4.2089);
\node at (-3.2917,4.2752) {\tiny{$2r-8$}};
\end{scope}
\node at (4.4531,5.2234) {\tiny{$h$}};
\node at (4.4466,4.727) {\tiny{$h$}};
\begin{scope}[shift={(8.5285,-0.3324)}]
\draw  (-3.5483,4.3533) rectangle (-3.0379,4.185);
\node at (-3.2917,4.2752) {\tiny{$6-2r$}};
\end{scope}
\begin{scope}[shift={(8.5214,1.7981)}]
\draw  (-3.5742,4.3532) rectangle (-3.0511,4.1911);
\node at (-3.2917,4.2752) {\tiny{$6-2r$}};
\end{scope}
\begin{scope}[shift={(8.916,-0.0201)}]
\draw  (-3.5388,4.3703) rectangle (-3.0562,4.1503);
\node at (-3.3049,4.2549) {\tiny{16-4$r$}};
\end{scope}
\node at (6.2703,4.0317) {\tiny{$-y_i$}};
\begin{scope}[shift={(8.3714,1.1427)}]
\draw  (-3.5739,4.3711) rectangle (-3.0318,4.196);
\node at (-3.2917,4.2752) {\tiny{$2r-8$}};
\end{scope}
\begin{scope}[shift={(8.8915,0.7249)}]
\draw  (-3.5833,4.3591) rectangle (-3.0318,4.196);
\node at (-3.2917,4.2752) {\tiny{$2r-8$}};
\end{scope}
\draw (5.5873,4.9147) -- (5.5937,4.4173);
\draw (5.5873,5.0892) -- (5.6002,5.7353);
\end{tikzpicture}
\end{array}\label{pattern}
\end{align} 
where we have paired up the $4r-16$ blowups on $S^r_C$ into $2r-8$ pairs, and labeled them by $x_i$ and $y_i$ where $i=1,\cdots,2r-8$. We remind the reader that in our notation the box in the middle of an edge labels the number of curves participating in the gluing between the two surfaces joined by the edge. Above we have an edge labeled by $2r-8$ between $S^r_C$ and $S^{r-1}_C$ which denotes that there are $2r-8$ gluing curves in between them. Each of these gluing curves has self intersection $-2$ in $S^r_C$ and self-intersection $0$ in $S^{r-1}_C$. The $2r-8$ gluing curves in $S^{r-1}_C$ are simply $2r-8$ copies of $f$ and the $2r-8$ gluing curves in $S^r_C$ are $f-x_i-y_i$ where $i$ ranges from $1$ to $2r-8$. We will use this notation in what follows whenever there are multiple similar-looking gluings between two surfaces.

Each of the $2r-8$ pairs of blowups corresponds to a hyper in the vector representation of $\mf{so}(2r)$, thus totalling to $2r-8$ such hypers. This is the same graph as the one presented in \cite{Bhardwaj:2018yhy}.

For I$^{*s}_{r-4}$ fiber realizing the above graph, there are $2r-8$ special points $p_i,i=1,\cdots,2r-8$ such that at $p_i$ $f^r_C$ splits into $f^r_C-x_i-y_i,x_i,y_i$ and $f^r_C-x_i-y_i\sim f^{r-1}_C$. Thus, the fiber at each $p_i$ enhances to I$^*_{r-3}$. This also explains the $2r-8$ intersections between $S^{r-1}_C$ and $S^r_C$ shown in the above graph. This explains the gluings between $S^{r-1}_C$ and $S^r_C$ which one would not expect naively since $f^{r-1}_C$ and $f^r_C$ do not intersect with each other. Similar comments about special points hold for all the cases discussed below, but we won't discuss them since we hope the pattern is clear to the reader. Moreover, our main objective with the discussion of special points was to justify the appearance of extra gluings not visible from the intersection pattern of the Kodaira fiber. We hope that this objective has been fulfilled with the discussion of special points in this subsection and previous subsections.

\subsection{$\so(2r+1),r\ge4$ on $-4$ curve}
The associated graph is
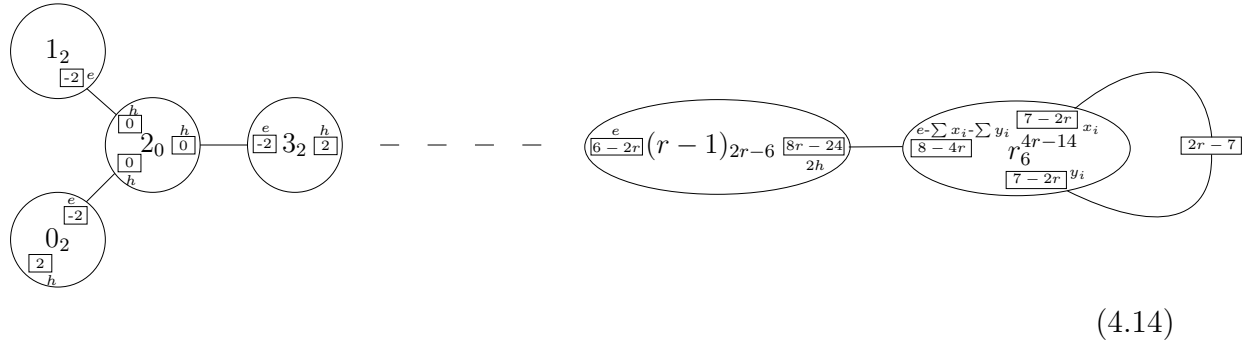
\begin{align}
\scalebox{.9}{$
\begin{array}{c}
\noindent\begin{tikzpicture}[scale=1.4]
\draw  (-3.5,4) ellipse (0.5 and 0.5);
\node (v1) at (-3.5,4) {$0_2$};
\draw  (-2.5,5) ellipse (0.5 and 0.5);
\node (v2) at (-2.5,5) {$2_0$};
\draw  (-1,5) ellipse (0.5 and 0.5);
\node (v4) at (-1,5) {$3_2$};
\draw  (3.4586,4.9793) ellipse (1.4 and 0.5);
\node at (3.389,4.9889) {$(r-1)_{2r-6}$};
\draw (-3.2,4.4) -- (-2.9,4.7);
\draw (-2,5) -- (-1.5,5);
\draw (-0.1,5) -- (0.1,5) (0.4,5) -- (0.6,5);
\draw (0.9,5) -- (1.1,5) (1.4,5) -- (1.6,5);
\begin{scope}[shift={(-0.0636,1.4373)}]
\draw  (-3.4113,4.3642) rectangle (-3.17,4.1794);
\node at (-3.2917,4.2752) {\tiny{-2}};
\end{scope}
\begin{scope}[shift={(1.1159,0.7257)}]
\draw  (-3.4113,4.3642) rectangle (-3.17,4.1794);
\node at (-3.2917,4.2752) {\tiny{0}};
\end{scope}
\begin{scope}[shift={(1.9717,0.7394)}]
\draw  (-3.4113,4.3642) rectangle (-3.17,4.1794);
\node at (-3.2917,4.2752) {\tiny{-2}};
\end{scope}
\begin{scope}[shift={(2.6153,0.7188)}]
\draw  (-3.4113,4.3642) rectangle (-3.17,4.1794);
\node at (-3.2917,4.2752) {\tiny{2}};
\end{scope}
\begin{scope}[shift={(7.8258,0.6985)}]
\draw  (-3.6675,4.3771) rectangle (-3.0318,4.196);
\node at (-3.3431,4.2881) {\tiny{$8r-24$}};
\end{scope}
\begin{scope}[shift={(5.6849,0.6908)}]
\draw  (-3.5608,4.3783) rectangle (-3.0178,4.2033);
\node at (-3.2917,4.2752) {\tiny{$6-2r$}};
\end{scope}
\node at (-2.1625,5.1528) {\tiny{$h$}};
\node at (-3.5652,3.5724) {\tiny{$h$}};
\node at (-1.3095,5.1733) {\tiny{$e$}};
\node at (-2.7275,4.6352) {\tiny{$h$}};
\node at (-0.6859,5.1591) {\tiny{$h$}};
\node at (2.3865,5.1337) {\tiny{$e$}};
\node at (-3.3582,4.4131) {\tiny{$e$}};
\begin{scope}[shift={(0.0001,1.9902)}]
\draw  (-3.5,4) ellipse (0.5 and 0.5);
\node (v1) at (-3.5,4) {$1_2$};
\draw (-3.1971,3.6066) -- (-2.8839,3.3294);
\begin{scope}[shift={(-0.0137,-0.5476)}]
\end{scope}
\node at (-3.1558,3.7385) {\tiny{$e$}};
\end{scope}
\begin{scope}[shift={(9.4713,0.9424)}]
\draw  (-2.8576,4.0223) ellipse (1.2 and 0.5);
\node (v1) at (-2.5786,4.0047) {$r^{4r-14}_{6}$};
\draw (-4.0584,4.0405) -- (-4.6269,4.0368);
\node at (-3.4162,4.214) {\tiny{$e$-$\sum x_i$-$\sum y_i$}};
\end{scope}
\begin{scope}[shift={(-0.3948,-0.5205)}]
\draw  (-3.4113,4.3642) rectangle (-3.17,4.1794);
\node at (-3.2917,4.2752) {\tiny{2}};
\end{scope}
\begin{scope}[shift={(-0.0224,-0.0146)}]
\draw  (-3.4113,4.3642) rectangle (-3.17,4.1794);
\node at (-3.2917,4.2752) {\tiny{-2}};
\end{scope}
\begin{scope}[shift={(0.5457,0.5375)}]
\draw  (-3.4113,4.3642) rectangle (-3.17,4.1794);
\node at (-3.2917,4.2752) {\tiny{0}};
\end{scope}
\begin{scope}[shift={(0.5516,0.949)}]
\draw  (-3.4113,4.3642) rectangle (-3.17,4.1794);
\node at (-3.2917,4.2752) {\tiny{0}};
\end{scope}
\node at (-2.7003,5.3738) {\tiny{$h$}};
\node at (4.4924,4.7834) {\tiny{$2h$}};
\begin{scope}[shift={(9.1694,0.6792)}]
\draw  (-3.6675,4.3771) rectangle (-3.0318,4.196);
\node at (-3.3431,4.2881) {\tiny{$8-4r$}};
\end{scope}
\begin{scope}[shift={(10.2873,0.9805)}]
\draw  (-3.6675,4.3771) rectangle (-3.0318,4.196);
\node at (-3.3431,4.2881) {\tiny{$7-2r$}};
\end{scope}
\begin{scope}[shift={(10.172,0.3438)}]
\draw  (-3.6675,4.3771) rectangle (-3.0318,4.196);
\node at (-3.3431,4.2881) {\tiny{$7-2r$}};
\end{scope}
\node at (7.4027,5.1681) {\tiny{$x_i$}};
\node at (7.2614,4.6873) {\tiny{$y_i$}};
\begin{scope}[shift={(12.018,0.707)}]
\draw  (-3.6675,4.3771) rectangle (-3.0318,4.196);
\node at (-3.3431,4.2881) {\tiny{$2r-7$}};
\end{scope}
\draw (7.2273,5.3828) .. controls (8.035,6.0453) and (8.6035,5.7974) .. (8.6739,5.0808);
\draw (7.1504,4.5153) .. controls (8.082,4.0538) and (8.6889,4.1691) .. (8.6718,4.8999);
\end{tikzpicture}
\end{array}
$}
\end{align}
where the $4r-14$ blowups are again paired up into $2r-7$ pairs of blowups and the pairs correspond to $2r-7$ hypers in the vector representation of $\mf{so}(2r+1)$. The two blowups in each pair are glued to each other leading to self-gluings of the surface $S^r_C$. This is the same graph as the one presented in \cite{Bhardwaj:2018yhy}, however the notation is slightly different for convenience in later sections. There $S^r_C$ was instead represented as a genus $2r-7$ ruled surface of degree $4r-8$. This is because gluing two $-1$ curves $c$ and $d$ inside a surface $S$ change the canonical divisor of the surface as $K_S\to K_S+c+d$ which changes the genus of curves via adjunction. This change in the canonical class can be used to check that the genus of the curve $e-\sum x_i-\sum y_i$ in $S^r_C$ is $2r-7$ and hence $S^r_C$ can also be represented as a ruled surface\footnote{More precisely, a smoothing of self-glued $\F^{4r-16}_6$ equals a ruled surface of genus $2r-7$ and degree $4r-8$.} of genus $2r-7$. Notice that the curve $2h$ in $S^{r-1}_C=\F_{2r-6}$ also has genus $2r-7$ which is an important consistency check as the gluing curves must have same genus on both sides. Moreover, the gluing between $S^{r-1}_C$ and $S^r_C$ satisfies Calabi-Yau condition because $(8r-24)+(8-4r)=2g-2$ where $g=2r-7$.

Due to the monodromy, the elliptic fiber can be recognized as $f_C=f^0_C+f^1_C+2(f^2_C+\cdots+f^r_C)$.


\subsection{$\so(n),8\le n\le12$ and $\fg_2$ on $-k$ curve, $1\le k\le 3$}
For $\so(8)$, the associated graph is
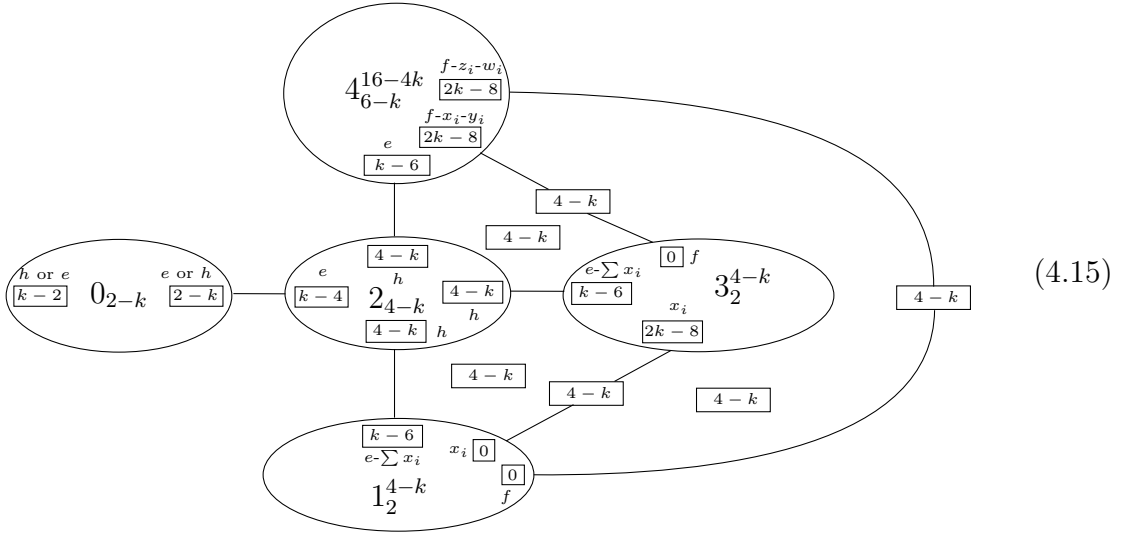
\begin{align}
\begin{array}{c}
\begin{tikzpicture}[scale=1.5]
\draw  (-4.4603,4.9901) ellipse (1 and 0.5);
\node (v1) at (-4.4603,4.9901) {$0_{2-k}$};
\draw (-3.4464,5.0091) -- (-2.9847,5.0073);
\node at (-5.1218,5.1869) {\tiny{$h$ or $e$}};
\begin{scope}[shift={(-1.8741,0.7147)}]
\draw  (-3.5178,4.3618) rectangle (-3.0475,4.1821);
\node at (-3.2917,4.2752) {\tiny{$k-2$}};
\end{scope}
\begin{scope}[shift={(2.4713,0.0223)}]
\draw  (-4.4603,4.9901) ellipse (1 and 0.5);
\node (v1) at (-4.4631,4.916) {$2_{4-k}$};
\draw (-3.464,5.0089) -- (-2.9847,5.0073);
\begin{scope}[shift={(-0.4057,-0.6827)}]
\draw  (-3.3869,4.3749) rectangle (-3.1958,4.1846);
\node at (-3.2917,4.2752) {\tiny{$0$}};
\end{scope}
\begin{scope}[shift={(-1.1756,1.0455)}]
\draw  (-3.5508,4.3685) rectangle (-3.0201,4.1746);
\node at (-3.2917,4.2752) {\tiny{$4-k$}};
\end{scope}
\node at (-4.4576,5.1232) {\tiny{$h$}};
\node at (-5.1248,5.1706) {\tiny{$e$}};
\begin{scope}[shift={(-1.1904,0.3801)}]
\draw  (-3.5508,4.3685) rectangle (-3.0201,4.1746);
\node at (-3.2917,4.2752) {\tiny{$4-k$}};
\end{scope}
\begin{scope}[shift={(-0.5101,0.7276)}]
\draw  (-3.5508,4.3685) rectangle (-3.0201,4.1746);
\node at (-3.2917,4.2752) {\tiny{$4-k$}};
\end{scope}
\end{scope}
\begin{scope}[shift={(5.1361,0.0036)}]
\draw  (-4.4603,4.9901) ellipse (1.2 and 0.5);
\node (v1) at (-4.0643,5.0706) {$3_{2}^{4-k}$};
\node at (-5.2126,5.2166) {\tiny{$e$-$\sum x_i$}};
\begin{scope}[shift={(-1.3924,1.0617)}]
\draw  (-3.4027,4.3566) rectangle (-3.2043,4.1827);
\node at (-3.308,4.2726) {\tiny{$0$}};
\end{scope}
\node at (-4.4923,5.3311) {\tiny{$f$}};
\begin{scope}[shift={(-2.0334,0.735)}]
\draw  (-3.5508,4.3685) rectangle (-3.0201,4.1746);
\node at (-3.2917,4.2752) {\tiny{$k-6$}};
\end{scope}
\begin{scope}[shift={(-1.4048,0.3948)}]
\draw  (-3.5508,4.3685) rectangle (-3.0201,4.1746);
\node at (-3.2917,4.2752) {\tiny{$2k-8$}};
\end{scope}
\node at (-4.6328,4.8727) {\tiny{$x_i$}};
\end{scope}
\node at (-1.5987,4.6745) {\tiny{$h$}};
\node at (-1.3116,4.8264) {\tiny{$h$}};
\begin{scope}[shift={(2.4199,1.5062)}]
\draw  (-4.4204,5.2786) ellipse (1 and 0.8);
\node (v1) at (-4.5217,5.3005) {$4_{6-k}^{16-4k}$};
\node at (-3.9029,5.0735) {\tiny{$f$-$x_i$-$y_i$}};
\node at (-4.4988,4.8246) {\tiny{$e$}};
\begin{scope}[shift={(-0.6486,0.611)}]
\draw  (-3.5666,4.3692) rectangle (-3.0246,4.1887);
\node at (-3.292,4.2758) {\tiny{$2k-8$}};
\end{scope}
\begin{scope}[shift={(-1.1411,0.3576)}]
\draw  (-3.5655,4.3723) rectangle (-2.9835,4.1949);
\node at (-3.277,4.2798) {\tiny{$k-6$}};
\end{scope}
\node at (-3.7509,5.5169) {\tiny{$f$-$z_i$-$w_i$}};
\begin{scope}[shift={(-0.4709,1.0329)}]
\draw  (-3.5666,4.3692) rectangle (-3.0246,4.1887);
\node at (-3.292,4.2758) {\tiny{$2k-8$}};
\end{scope}
\end{scope}
\begin{scope}[shift={(2.8763,1.5294)}]
\draw  (-3.6392,4.4011) rectangle (-2.9855,4.1922);
\node at (-3.2901,4.2995) {\tiny{$4-k$}};
\end{scope}
\begin{scope}[shift={(2.4325,1.2108)}]
\draw  (-3.6392,4.4011) rectangle (-2.9855,4.1922);
\node at (-3.2901,4.2995) {\tiny{$4-k$}};
\end{scope}
\draw (-2.019,5.9932) -- (-2.019,5.5182);
\draw (-0.3102,5.7204) -- (0.2707,5.464);
\draw (-0.6531,5.9337) -- (-1.2595,6.2581);
\begin{scope}[shift={(2.9943,-0.1738)}]
\draw  (-3.6392,4.4011) rectangle (-2.9855,4.1922);
\node at (-3.2901,4.2995) {\tiny{$4-k$}};
\end{scope}
\begin{scope}[shift={(2.1269,-0.018)}]
\draw  (-3.6392,4.4011) rectangle (-2.9855,4.1922);
\node at (-3.2901,4.2995) {\tiny{$4-k$}};
\end{scope}
\begin{scope}[shift={(2.4773,-1.5882)}]
\draw  (-4.4603,4.9901) ellipse (1.2 and 0.5);
\node (v1) at (-4.4536,4.8042) {$1_2^{4-k}$};
\node at (-4.5078,5.1433) {\tiny{$e$-$\sum x_i$}};
\begin{scope}[shift={(-0.1406,0.72)}]
\draw  (-3.4027,4.3566) rectangle (-3.2043,4.1827);
\node at (-3.308,4.2726) {\tiny{$0$}};
\end{scope}
\node at (-3.5048,4.7952) {\tiny{$f$}};
\begin{scope}[shift={(-1.2274,1.0603)}]
\draw  (-3.5508,4.3685) rectangle (-3.0201,4.1746);
\node at (-3.2917,4.2752) {\tiny{$k-6$}};
\end{scope}
\node at (-3.9242,5.1963) {\tiny{$x_i$}};
\end{scope}
\draw (-2.0184,4.5162) -- (-2.0184,3.9095);
\begin{scope}[shift={(6.0714,0.6765)}]
\draw  (-3.6392,4.4011) rectangle (-2.9855,4.1922);
\node at (-3.2901,4.2995) {\tiny{$4-k$}};
\end{scope}
\begin{scope}[shift={(4.2974,-0.2308)}]
\draw  (-3.6392,4.4011) rectangle (-2.9855,4.1922);
\node at (-3.2901,4.2995) {\tiny{$4-k$}};
\end{scope}
\draw (-0.7849,3.4023) .. controls (0.9706,3.3868) and (2.752,3.6851) .. (2.7681,4.8619);
\draw (-0.9961,6.7964) .. controls (1.1963,6.7561) and (2.7681,6.3692) .. (2.7601,5.0715);
\begin{scope}[shift={(-0.4913,0.7147)}]
\draw  (-3.5178,4.3618) rectangle (-3.0475,4.1821);
\node at (-3.2917,4.2752) {\tiny{$2-k$}};
\end{scope}
\node at (-3.8642,5.1943) {\tiny{$e$ or $h$}};
\begin{scope}[shift={(0.6178,0.7073)}]
\draw  (-3.5178,4.3618) rectangle (-3.0475,4.1821);
\node at (-3.2917,4.2752) {\tiny{$k-4$}};
\end{scope}
\draw (-1.0292,3.7022) -- (-0.4318,4.0241) (-0.0681,4.2311) -- (0.4346,4.5061);
\end{tikzpicture}
\end{array}
\end{align}
where the label ``$h$ or $e$'' indicates that the curve will be labeled either as $h$ or as $e$ depending on the value of $k$. The $4-k$ blowups on $S^3_C$ and $S^1_C$ correspond to $4-k$ hypers in the vector representation. $8-2k$ blowups out of $16-4k$ blowups on $S^4_C$ have been paired into $4-k$ pairs denoted by $z_i,w_i$ where $i=1,\cdots,4-k$, and correspond to $4-k$ hypers in the cospinor representation of $\so(8)$. The rest of the $8-2k$ blowups on $S^4_C$ have been paired into $4-k$ pairs denoted by $x_i,y_i$ where $i=1,\cdots,4-k$, and correspond to $4-k$ hypers in the spinor representation of $\so(8)$. This graph is flop equivalent to the one presented in \cite{Bhardwaj:2018yhy}.


For $\so(7)$, the associated graph is
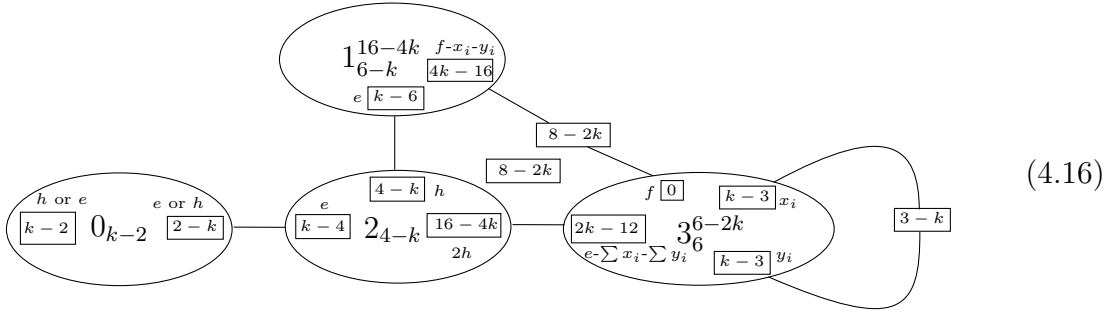
\begin{align}
\begin{array}{c}
\begin{tikzpicture}[scale=1.5]
\draw  (-4.4603,4.9901) ellipse (1 and 0.5);
\node (v1) at (-4.4603,4.9901) {$0_{k-2}$};
\draw (-3.4464,5.0091) -- (-2.9847,5.0073);
\node at (-3.937,5.2302) {\tiny{$e$ or $h$}};
\node at (-4.9686,5.2627) {\tiny{$h$ or $e$}};
\begin{scope}[shift={(-1.8167,0.7227)}]
\draw  (-3.5199,4.4183) rectangle (-3.0317,4.1342);
\node at (-3.2917,4.2752) {\tiny{$k-2$}};
\end{scope}
\begin{scope}[shift={(-0.4988,0.715)}]
\draw  (-3.5352,4.3891) rectangle (-3.0397,4.1868);
\node at (-3.2901,4.2995) {\tiny{$2-k$}};
\end{scope}
\begin{scope}[shift={(2.4713,0.0223)}]
\draw  (-4.4603,4.9901) ellipse (1 and 0.5);
\node (v1) at (-4.507,4.9528) {$2_{4-k}$};
\draw (-3.4464,5.0091) -- (-2.9847,5.0073);
\begin{scope}[shift={(-1.8354,0.7134)}]
\draw  (-3.5321,4.3845) rectangle (-3.0504,4.1716);
\node at (-3.2917,4.2752) {\tiny{$k-4$}};
\end{scope}
\begin{scope}[shift={(-0.5642,0.7057)}]
\draw  (-3.6392,4.4011) rectangle (-2.9855,4.1922);
\node at (-3.2901,4.2995) {\tiny{$16-4k$}};
\end{scope}
\begin{scope}[shift={(-1.1756,1.0455)}]
\draw  (-3.5321,4.3845) rectangle (-3.0504,4.1716);
\node at (-3.2917,4.2752) {\tiny{$4-k$}};
\end{scope}
\node at (-4.0918,5.3074) {\tiny{$h$}};
\end{scope}
\begin{scope}[shift={(5.1361,0.0036)}]
\draw  (-4.4603,4.9901) ellipse (1.2 and 0.5);
\node (v1) at (-4.3518,4.9493) {$3_{6}^{6-2k}$};
\node at (-5.0126,4.7657) {\tiny{$e$-$\sum x_i$-$\sum y_i$}};
\begin{scope}[shift={(-1.9849,0.7134)}]
\draw  (-3.6041,4.3993) rectangle (-2.9476,4.1525);
\node at (-3.2917,4.2752) {\tiny{$2k-12$}};
\end{scope}
\begin{scope}[shift={(-1.3924,1.0617)}]
\draw  (-3.4027,4.3566) rectangle (-3.2043,4.1827);
\node at (-3.308,4.2726) {\tiny{$0$}};
\end{scope}
\node at (-4.8955,5.3048) {\tiny{$f$}};
\node at (-3.6661,5.22) {\tiny{$x_i$}};
\node at (-3.6962,4.7424) {\tiny{$y_i$}};
\end{scope}
\node at (-2.6424,5.2031) {\tiny{$e$}};
\node at (-1.423,4.7873) {\tiny{$2h$}};
\begin{scope}[shift={(2.4199,1.5062)}]
\draw  (-4.4603,4.9901) ellipse (1 and 0.5);
\node (v1) at (-4.5578,4.9971) {$1_{6-k}^{16-4k}$};
\node at (-3.8099,5.0963) {\tiny{$f$-$x_i$-$y_i$}};
\node at (-4.7699,4.6439) {\tiny{$e$}};
\begin{scope}[shift={(-0.5576,0.611)}]
\draw  (-3.5896,4.3807) rectangle (-3.0131,4.1849);
\node at (-3.292,4.2758) {\tiny{$4k-16$}};
\end{scope}
\begin{scope}[shift={(-1.1411,0.3576)}]
\draw  (-3.5352,4.3891) rectangle (-3.0397,4.1868);
\node at (-3.2901,4.2995) {\tiny{$k-6$}};
\end{scope}
\end{scope}
\begin{scope}[shift={(2.8763,1.5294)}]
\draw  (-3.6392,4.4011) rectangle (-2.9855,4.1922);
\node at (-3.2901,4.2995) {\tiny{$8-2k$}};
\end{scope}
\begin{scope}[shift={(2.4325,1.2108)}]
\draw  (-3.6392,4.4011) rectangle (-2.9855,4.1922);
\node at (-3.2901,4.2995) {\tiny{$8-2k$}};
\end{scope}
\draw (-2.019,5.9932) -- (-2.019,5.5182);
\draw (-0.3102,5.7204) -- (0.2707,5.464);
\draw (-0.6531,5.9337) -- (-1.1906,6.2395);
\begin{scope}[shift={(4.3953,0.9835)}]
\draw  (-3.5352,4.3891) rectangle (-3.0397,4.1868);
\node at (-3.2901,4.2995) {\tiny{$k-3$}};
\end{scope}
\begin{scope}[shift={(4.3483,0.4041)}]
\draw  (-3.5352,4.3891) rectangle (-3.0397,4.1868);
\node at (-3.2901,4.2995) {\tiny{$k-3$}};
\end{scope}
\begin{scope}[shift={(5.9451,0.7857)}]
\draw  (-3.5352,4.3891) rectangle (-3.0397,4.1868);
\node at (-3.2901,4.2995) {\tiny{$3-k$}};
\end{scope}
\draw (1.3395,5.4064) .. controls (2.3853,5.9763) and (2.6255,5.7125) .. (2.6302,5.1661);
\draw (1.2924,4.5679) .. controls (2.4653,4.0593) and (2.6396,4.2665) .. (2.6349,4.9683);
\end{tikzpicture}
\end{array}
\end{align}
where we pair up $16-4k$ blowups on $S^1_C$ into $8-2k$ pairs denoted by $x_i,y_i$ where $i=1,\cdots,8-2k$, and pair up the $6-2k$ blowups on $S^3_C$ into $3-k$ pairs denoted by $x_i,y_i$ where $i=1,\cdots,3-k$. The first kind of pairs correspond to $8-2k$ hypers in the spinor representation of $\mf{so}(7)$, and the second kind of pairs correspond to $3-k$ hypers in the vector representation of $\mf{so}(7)$. This graph is the same as the one presented in \cite{Bhardwaj:2018yhy}. Due to monodromy, the elliptic fiber is represented as $f_C=f^0_C+f^1_C+2(f^2_C+f^3_C)$. 


For $\so(10)$, the associated graph is
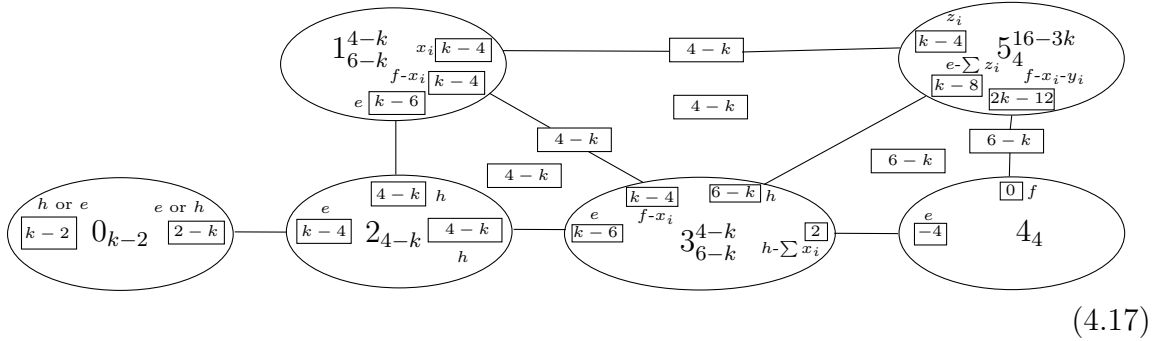
\begin{align}
\label{eqn:SO10samp}
\begin{array}{c}
\begin{tikzpicture}[scale=1.5]
\draw  (-4.4603,4.9901) ellipse (1 and 0.5);
\node (v1) at (-4.4603,4.9901) {$0_{k-2}$};
\draw (-3.4464,5.0091) -- (-2.9847,5.0073);
\node at (-3.937,5.2302) {\tiny{$e$ or $h$}};
\node at (-4.9686,5.2627) {\tiny{$h$ or $e$}};
\begin{scope}[shift={(-1.8167,0.7227)}]
\draw  (-3.5199,4.4183) rectangle (-3.0317,4.1342);
\node at (-3.2917,4.2752) {\tiny{$k-2$}};
\end{scope}
\begin{scope}[shift={(-0.4988,0.715)}]
\draw  (-3.5352,4.3891) rectangle (-3.0397,4.1868);
\node at (-3.2901,4.2995) {\tiny{$2-k$}};
\end{scope}
\begin{scope}[shift={(2.4713,0.0223)}]
\draw  (-4.4603,4.9901) ellipse (1 and 0.5);
\node (v1) at (-4.507,4.9528) {$2_{4-k}$};
\draw (-3.4464,5.0091) -- (-2.9847,5.0073);
\begin{scope}[shift={(-1.8354,0.7134)}]
\draw  (-3.5321,4.3845) rectangle (-3.0504,4.1716);
\node at (-3.2917,4.2752) {\tiny{$k-4$}};
\end{scope}
\begin{scope}[shift={(-0.5642,0.7057)}]
\draw  (-3.6392,4.4011) rectangle (-2.9855,4.1922);
\node at (-3.2901,4.2995) {\tiny{$4-k$}};
\end{scope}
\begin{scope}[shift={(-1.1756,1.0455)}]
\draw  (-3.5321,4.3845) rectangle (-3.0504,4.1716);
\node at (-3.2917,4.2752) {\tiny{$4-k$}};
\end{scope}
\node at (-4.0918,5.3074) {\tiny{$h$}};
\end{scope}
\begin{scope}[shift={(5.1361,0.0036)}]
\draw  (-4.4603,4.9901) ellipse (1.2 and 0.5);
\node (v1) at (-4.3814,4.9083) {$3_{6-k}^{4-k}$};
\node at (-5.3928,5.1471) {\tiny{$e$}};
\begin{scope}[shift={(-2.0408,0.7209)}]
\draw  (-3.5524,4.3484) rectangle (-3.1035,4.1956);
\node at (-3.3438,4.2761) {\tiny{$k-6$}};
\end{scope}
\begin{scope}[shift={(-1.595,1.033)}]
\draw  (-3.5156,4.3674) rectangle (-3.0604,4.1989);
\node at (-3.2917,4.2752) {\tiny{$k-4$}};
\end{scope}
\node at (-4.8506,5.1441) {\tiny{$f$-$x_i$}};
\begin{scope}[shift={(-0.1234,0.722)}]
\draw  (-3.4041,4.3635) rectangle (-3.206,4.2062);
\node at (-3.3103,4.2939) {\tiny{$2$}};
\end{scope}
\node at (-3.651,4.8569) {\tiny{$h$-$\sum x_i$}};
\begin{scope}[shift={(-0.8189,1.0748)}]
\draw  (-3.5524,4.3484) rectangle (-3.1035,4.1956);
\node at (-3.3438,4.2761) {\tiny{$6-k$}};
\end{scope}
\node at (-3.8317,5.3301) {\tiny{$h$}};
\end{scope}
\node at (-2.6424,5.2031) {\tiny{$e$}};
\node at (-1.423,4.7873) {\tiny{$h$}};
\begin{scope}[shift={(2.4199,1.5062)}]
\draw  (-4.4603,4.9901) ellipse (1 and 0.5);
\node (v1) at (-4.7515,5.1201) {$1_{6-k}^{4-k}$};
\node at (-4.1752,5.115) {\tiny{$x_i$}};
\node at (-4.7699,4.6439) {\tiny{$e$}};
\begin{scope}[shift={(-1.1411,0.3576)}]
\draw  (-3.5352,4.3891) rectangle (-3.0397,4.1868);
\node at (-3.2901,4.2995) {\tiny{$k-6$}};
\end{scope}
\begin{scope}[shift={(-0.6121,0.5364)}]
\draw  (-3.5352,4.3891) rectangle (-3.0397,4.1868);
\node at (-3.2901,4.2995) {\tiny{$k-4$}};
\end{scope}
\begin{scope}[shift={(-0.5525,0.8269)}]
\draw  (-3.5352,4.3891) rectangle (-3.0397,4.1868);
\node at (-3.2901,4.2995) {\tiny{$k-4$}};
\end{scope}
\node at (-4.3291,4.8865) {\tiny{$f$-$x_i$}};
\end{scope}
\begin{scope}[shift={(2.8763,1.5294)}]
\draw  (-3.6392,4.4011) rectangle (-2.9855,4.1922);
\node at (-3.2901,4.2995) {\tiny{$4-k$}};
\end{scope}
\begin{scope}[shift={(2.4325,1.2108)}]
\draw  (-3.6392,4.4011) rectangle (-2.9855,4.1922);
\node at (-3.2901,4.2995) {\tiny{$4-k$}};
\end{scope}
\draw (-2.019,5.9932) -- (-2.019,5.5182);
\draw (-0.3102,5.7204) -- (0.1701,5.4491);
\draw (-0.6531,5.9337) -- (-1.1906,6.2395);
\begin{scope}[shift={(7.901,0.0069)}]
\draw  (-4.4603,4.9901) ellipse (1 and 0.5);
\node (v1) at (-4.2995,4.9925) {$4_4$};
\node at (-5.1985,5.1414) {\tiny{$e$}};
\begin{scope}[shift={(-1.8953,0.7227)}]
\draw  (-3.4282,4.3553) rectangle (-3.1432,4.1635);
\node at (-3.2917,4.2752) {\tiny{$-4$}};
\end{scope}
\begin{scope}[shift={(-1.159,1.0796)}]
\draw  (-3.4041,4.3635) rectangle (-3.206,4.2062);
\node at (-3.3103,4.2939) {\tiny{$0$}};
\end{scope}
\node at (-4.2784,5.3538) {\tiny{$f$}};
\end{scope}
\draw (1.8669,4.9968) -- (2.432,4.9939);
\begin{scope}[shift={(7.8974,1.5551)}]
\draw  (-4.4603,4.9901) ellipse (1 and 0.5);
\node (v1) at (-4.2364,5.0884) {$5_4^{16-3k}$};
\node at (-4.9626,5.3376) {\tiny{$z_i$}};
\begin{scope}[shift={(-1.7979,0.8707)}]
\draw  (-3.5112,4.3662) rectangle (-3.0665,4.1777);
\node at (-3.2917,4.2752) {\tiny{$k-4$}};
\end{scope}
\begin{scope}[shift={(-1.0659,0.3588)}]
\draw  (-3.5906,4.3632) rectangle (-3.0156,4.1777);
\node at (-3.3001,4.2753) {\tiny{$2k-12$}};
\end{scope}
\begin{scope}[shift={(-1.654,0.4786)}]
\draw  (-3.5112,4.3662) rectangle (-3.0665,4.1777);
\node at (-3.2917,4.2752) {\tiny{$k-8$}};
\end{scope}
\node at (-4.7957,4.9353) {\tiny{$e$-$\sum z_i$}};
\node at (-4.0809,4.8324) {\tiny{$f$-$x_i$-$y_i$}};
\end{scope}
\begin{scope}[shift={(4.0467,2.3257)}]
\draw  (-3.6392,4.4011) rectangle (-2.9855,4.1922);
\node at (-3.2901,4.2995) {\tiny{$4-k$}};
\end{scope}
\begin{scope}[shift={(6.7072,1.5173)}]
\draw  (-3.6392,4.4011) rectangle (-2.9855,4.1922);
\node at (-3.2901,4.2995) {\tiny{$6-k$}};
\end{scope}
\begin{scope}[shift={(4.0834,1.8194)}]
\draw  (-3.6392,4.4011) rectangle (-2.9855,4.1922);
\node at (-3.2901,4.2995) {\tiny{$4-k$}};
\end{scope}
\begin{scope}[shift={(5.834,1.3412)}]
\draw  (-3.6392,4.4011) rectangle (-2.9855,4.1922);
\node at (-3.2901,4.2995) {\tiny{$6-k$}};
\end{scope}
\draw (-1.0684,6.6169) -- (0.4036,6.6109) (1.0491,6.6048) -- (2.4486,6.641);
\draw (1.2421,5.4285) -- (2.6779,6.2127) (3.4396,6.0414) -- (3.4263,5.9185) (3.4237,5.7112) -- (3.4186,5.5015);
\end{tikzpicture}
\end{array}
\end{align}
where we pair up the $12-2k$ blowups out of $16-3k$ blowups on $S^5_C$ into $6-k$ pairs labeled by $x_i$, $y_i$ for $i=1,\cdots,6-k$. These pairs represent $6-k$ hypers in the vector representation of $\mf{so}(10)$. The rest of the $4-k$ blowups on $S^5_C$ denoted by $z_i$ for $i=1,\cdots,4-k$, along with blowups on the other surfaces represent $\left(4-k\right)$ hypers in the Weyl spinor representation of $\mf{so}(10)$. This graph is same as the one presented in \cite{Bhardwaj:2018yhy}.


For $\so(9)$, the associated graph is
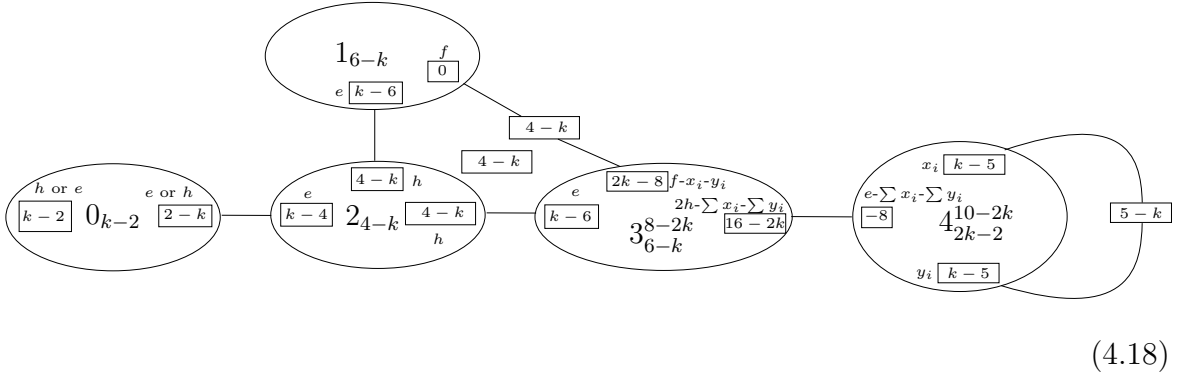
\begin{align}
\scalebox{.95}{$
\begin{array}{c}
\begin{tikzpicture}[scale=1.5]
\draw  (-4.4603,4.9901) ellipse (1 and 0.5);
\node (v1) at (-4.4603,4.9901) {$0_{k-2}$};
\draw (-3.4464,5.0091) -- (-2.9847,5.0073);
\node at (-3.937,5.2302) {\tiny{$e$ or $h$}};
\node at (-4.9686,5.2627) {\tiny{$h$ or $e$}};
\begin{scope}[shift={(-1.8167,0.7227)}]
\draw  (-3.5199,4.4183) rectangle (-3.0317,4.1342);
\node at (-3.2917,4.2752) {\tiny{$k-2$}};
\end{scope}
\begin{scope}[shift={(-0.4988,0.715)}]
\draw  (-3.5352,4.3891) rectangle (-3.0397,4.1868);
\node at (-3.2901,4.2995) {\tiny{$2-k$}};
\end{scope}
\begin{scope}[shift={(2.4713,0.0223)}]
\draw  (-4.4603,4.9901) ellipse (1 and 0.5);
\node (v1) at (-4.507,4.9528) {$2_{4-k}$};
\draw (-3.4464,5.0091) -- (-2.9847,5.0073);
\begin{scope}[shift={(-1.8354,0.7134)}]
\draw  (-3.5321,4.3845) rectangle (-3.0504,4.1716);
\node at (-3.2917,4.2752) {\tiny{$k-4$}};
\end{scope}
\begin{scope}[shift={(-0.5642,0.7057)}]
\draw  (-3.6392,4.4011) rectangle (-2.9855,4.1922);
\node at (-3.2901,4.2995) {\tiny{$4-k$}};
\end{scope}
\begin{scope}[shift={(-1.1756,1.0455)}]
\draw  (-3.5321,4.3845) rectangle (-3.0504,4.1716);
\node at (-3.2917,4.2752) {\tiny{$4-k$}};
\end{scope}
\node at (-4.0918,5.3074) {\tiny{$h$}};
\end{scope}
\begin{scope}[shift={(5.1361,0.0036)}]
\draw  (-4.4603,4.9901) ellipse (1.2 and 0.5);
\node (v1) at (-4.4744,4.8202) {$3_{6-k}^{8-2k}$};
\node at (-5.2811,5.2179) {\tiny{$e$}};
\begin{scope}[shift={(-1.9849,0.7134)}]
\draw  (-3.5835,4.3891) rectangle (-3.0825,4.1694);
\node at (-3.3438,4.2761) {\tiny{$k-6$}};
\end{scope}
\begin{scope}[shift={(-1.4125,1.0405)}]
\draw  (-3.5769,4.3717) rectangle (-3.0077,4.19);
\node at (-3.2917,4.2752) {\tiny{$2k-8$}};
\end{scope}
\node at (-4.1392,5.3182) {\tiny{$f$-$x_i$-$y_i$}};
\begin{scope}[shift={(-0.3134,0.6475)}]
\draw  (-3.5769,4.3717) rectangle (-3.0077,4.19);
\node at (-3.2917,4.2752) {\tiny{$16-2k$}};
\end{scope}
\node at (-3.8406,5.0988) {\tiny{$2h$-$\sum x_i$-$\sum y_i$}};
\end{scope}
\node at (-2.6424,5.2031) {\tiny{$e$}};
\node at (-1.423,4.7873) {\tiny{$h$}};
\begin{scope}[shift={(2.4199,1.5062)}]
\draw  (-4.4603,4.9901) ellipse (1 and 0.5);
\node (v1) at (-4.5578,4.9971) {$1_{6-k}$};
\node at (-3.7953,5.0293) {\tiny{$f$}};
\node at (-4.7699,4.6439) {\tiny{$e$}};
\begin{scope}[shift={(-1.1411,0.3576)}]
\draw  (-3.5352,4.3891) rectangle (-3.0397,4.1868);
\node at (-3.2901,4.2995) {\tiny{$k-6$}};
\end{scope}
\begin{scope}[shift={(-0.5147,0.5858)}]
\draw  (-3.4282,4.3553) rectangle (-3.1432,4.1635);
\node at (-3.2917,4.2752) {\tiny{$0$}};
\end{scope}
\end{scope}
\begin{scope}[shift={(2.8763,1.5294)}]
\draw  (-3.6392,4.4011) rectangle (-2.9855,4.1922);
\node at (-3.2901,4.2995) {\tiny{$4-k$}};
\end{scope}
\begin{scope}[shift={(2.4325,1.2108)}]
\draw  (-3.6392,4.4011) rectangle (-2.9855,4.1922);
\node at (-3.2901,4.2995) {\tiny{$4-k$}};
\end{scope}
\draw (-2.019,5.9932) -- (-2.019,5.5182);
\draw (-0.3102,5.7204) -- (0.2707,5.464);
\draw (-0.6531,5.9337) -- (-1.1906,6.2395);
\begin{scope}[shift={(7.901,0.0069)}]
\draw  (-4.4603,4.9901) ellipse (1 and 0.7);
\node (v1) at (-4.3022,4.9249) {$4_{2k-2}^{10-2k}$};
\node at (-4.8911,5.1947) {\tiny{$e$-$\sum x_i$-$\sum y_i$}};
\begin{scope}[shift={(-1.9495,0.7227)}]
\draw  (-3.4282,4.3553) rectangle (-3.1432,4.1635);
\node at (-3.2917,4.2752) {\tiny{$-8$}};
\end{scope}
\begin{scope}[shift={(-1.0338,1.1946)}]
\draw  (-3.5769,4.3717) rectangle (-3.0077,4.19);
\node at (-3.2917,4.2752) {\tiny{$k-5$}};
\end{scope}
\begin{scope}[shift={(-1.088,0.1816)}]
\draw  (-3.5769,4.3717) rectangle (-3.0077,4.19);
\node at (-3.2917,4.2752) {\tiny{$k-5$}};
\end{scope}
\node at (-4.7336,5.4541) {\tiny{$x_i$}};
\node at (-4.7882,4.4542) {\tiny{$y_i$}};
\begin{scope}[shift={(0.5207,0.745)}]
\draw  (-3.5769,4.3717) rectangle (-3.0077,4.19);
\node at (-3.2917,4.2752) {\tiny{$5-k$}};
\end{scope}
\end{scope}
\draw (1.8669,4.9968) -- (2.432,4.9939);
\draw (3.8578,5.6192) .. controls (4.952,5.9821) and (5.1308,5.6084) .. (5.1412,5.1269);
\draw (3.8253,4.3517) .. controls (5.0414,3.9875) and (5.1317,4.5371) .. (5.1406,4.9356);
\end{tikzpicture}
\end{array}
$}
\end{align}
where we pair up the $10-2k$ blowups on $S^4_C$ into $5-k$ pairs representing $5-k$ hypers in the vector representation of $\mf{so}(9)$. The $8-2k$ blowups on $S^3_C$ have been paired and represent $4-k$ hypers in the spinor representation of $\mf{so}(9)$. This graph is same as the one presented in \cite{Bhardwaj:2018yhy}. Due to monodromy, the elliptic fiber is represented as $f_C=f^0_C+f^1_C+2(f^2_C+f^3_C+f^4_C)$. 


For $\so(12)$ and $k=3$, the associated graph is
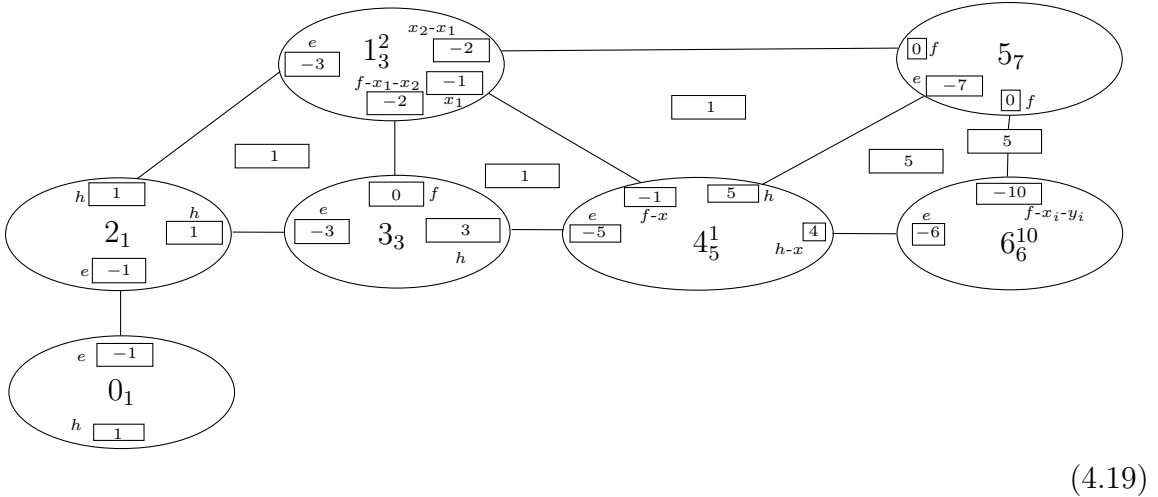
\begin{align}
\begin{array}{c}
\begin{tikzpicture}[scale=1.5]
\draw  (-4.4603,4.9901) ellipse (1 and 0.5);
\node (v1) at (-4.4603,4.9901) {$2_{1}$};
\draw (-3.4464,5.0091) -- (-2.9847,5.0073);
\node at (-3.785,5.1805) {\tiny{$h$}};
\node at (-4.7982,5.3284) {\tiny{$h$}};
\node at (-4.7578,4.6477) {\tiny{$e$}};
\begin{scope}[shift={(-1.1644,0.3871)}]
\draw  (-3.5262,4.3886) rectangle (-3.0539,4.1721);
\node at (-3.2917,4.2752) {\tiny{$-1$}};
\end{scope}
\begin{scope}[shift={(-0.4988,0.715)}]
\draw  (-3.5352,4.3891) rectangle (-3.0397,4.1868);
\node at (-3.2901,4.2995) {\tiny{$1$}};
\end{scope}
\begin{scope}[shift={(-1.186,1.057)}]
\draw  (-3.5352,4.3891) rectangle (-3.0397,4.1868);
\node at (-3.2901,4.2995) {\tiny{$1$}};
\end{scope}
\begin{scope}[shift={(2.4713,0.0223)}]
\draw  (-4.4603,4.9901) ellipse (1 and 0.5);
\node (v1) at (-4.507,4.9528) {$3_{3}$};
\draw (-3.4464,5.0091) -- (-2.9847,5.0073);
\begin{scope}[shift={(-1.8354,0.7134)}]
\draw  (-3.5321,4.3845) rectangle (-3.0504,4.1716);
\node at (-3.2917,4.2752) {\tiny{$-3$}};
\end{scope}
\begin{scope}[shift={(-0.5642,0.7057)}]
\draw  (-3.6392,4.4011) rectangle (-2.9855,4.1922);
\node at (-3.2901,4.2995) {\tiny{$3$}};
\end{scope}
\begin{scope}[shift={(-1.1756,1.0455)}]
\draw  (-3.5321,4.3845) rectangle (-3.0504,4.1716);
\node at (-3.2917,4.2752) {\tiny{$0$}};
\end{scope}
\node at (-4.133,5.3201) {\tiny{$f$}};
\end{scope}
\begin{scope}[shift={(5.1361,0.0036)}]
\draw  (-4.4603,4.9901) ellipse (1.2 and 0.5);
\node (v1) at (-4.3814,4.9083) {$4_{5}^{1}$};
\node at (-5.3928,5.1471) {\tiny{$e$}};
\begin{scope}[shift={(-2.0408,0.7209)}]
\draw  (-3.5524,4.3484) rectangle (-3.1035,4.1956);
\node at (-3.3438,4.2761) {\tiny{$-5$}};
\end{scope}
\begin{scope}[shift={(-1.595,1.033)}]
\draw  (-3.5156,4.3674) rectangle (-3.0604,4.1989);
\node at (-3.2917,4.2752) {\tiny{$-1$}};
\end{scope}
\node at (-4.8506,5.1441) {\tiny{$f$-$x$}};
\begin{scope}[shift={(-0.1234,0.722)}]
\draw  (-3.4041,4.3635) rectangle (-3.206,4.2062);
\node at (-3.3103,4.2939) {\tiny{$4$}};
\end{scope}
\node at (-3.651,4.8569) {\tiny{$h$-$x$}};
\begin{scope}[shift={(-0.8189,1.0748)}]
\draw  (-3.5524,4.3484) rectangle (-3.1035,4.1956);
\node at (-3.3438,4.2761) {\tiny{$5$}};
\end{scope}
\node at (-3.8317,5.3301) {\tiny{$h$}};
\end{scope}
\node at (-2.6424,5.2031) {\tiny{$e$}};
\node at (-1.423,4.7873) {\tiny{$h$}};
\begin{scope}[shift={(2.4199,1.5062)}]
\draw  (-4.4603,4.9901) ellipse (1 and 0.5);
\node (v1) at (-4.593,5.1091) {$1_{3}^{2}$};
\node at (-3.9028,4.6583) {\tiny{$x_1$}};
\node at (-4.0954,5.2962) {\tiny{$x_2$-$x_1$}};
\begin{scope}[shift={(-1.1411,0.3576)}]
\draw  (-3.5352,4.3891) rectangle (-3.0397,4.1868);
\node at (-3.2901,4.2995) {\tiny{$-2$}};
\end{scope}
\begin{scope}[shift={(-0.6121,0.5364)}]
\draw  (-3.5352,4.3891) rectangle (-3.0397,4.1868);
\node at (-3.2901,4.2995) {\tiny{$-1$}};
\end{scope}
\begin{scope}[shift={(-0.5525,0.8269)}]
\draw  (-3.5352,4.3891) rectangle (-3.0397,4.1868);
\node at (-3.2901,4.2995) {\tiny{$-2$}};
\end{scope}
\node at (-5.1598,5.1746) {\tiny{$e$}};
\node at (-4.496,4.823) {\tiny{$f$-$x_1$-$x_2$}};
\begin{scope}[shift={(-1.8702,0.7134)}]
\draw  (-3.5321,4.3845) rectangle (-3.0504,4.1716);
\node at (-3.2917,4.2752) {\tiny{$-3$}};
\end{scope}
\end{scope}
\begin{scope}[shift={(2.4325,1.2108)}]
\draw  (-3.6392,4.4011) rectangle (-2.9855,4.1922);
\node at (-3.2901,4.2995) {\tiny{$1$}};
\end{scope}
\draw (-1.1799,6.2403) -- (0.1701,5.4491);
\begin{scope}[shift={(7.901,0.0069)}]
\draw  (-4.4603,4.9901) ellipse (1 and 0.5);
\node (v1) at (-4.3723,4.8912) {$6_6^{10}$};
\node at (-5.1985,5.1414) {\tiny{$e$}};
\begin{scope}[shift={(-1.8953,0.7227)}]
\draw  (-3.4282,4.3553) rectangle (-3.1432,4.1635);
\node at (-3.2917,4.2752) {\tiny{$-6$}};
\end{scope}
\begin{scope}[shift={(-1.1641,1.0713)}]
\draw  (-3.5906,4.3632) rectangle (-3.0156,4.1777);
\node at (-3.3001,4.2753) {\tiny{$-10$}};
\end{scope}
\node at (-4.0633,5.1687) {\tiny{$f$-$x_i$-$y_i$}};
\end{scope}
\draw (1.8669,4.9968) -- (2.432,4.9939);
\begin{scope}[shift={(7.8974,1.5551)}]
\draw  (-4.4603,4.9901) ellipse (1 and 0.5);
\node (v1) at (-4.4453,5.0113) {$5_{7}$};
\begin{scope}[shift={(-1.1387,0.3525)}]
\draw  (-3.3931,4.3603) rectangle (-3.2271,4.1785);
\node at (-3.3131,4.2709) {\tiny{$0$}};
\end{scope}
\begin{scope}[shift={(-1.654,0.4786)}]
\draw  (-3.5428,4.3599) rectangle (-3.045,4.1827);
\node at (-3.2917,4.2752) {\tiny{$-7$}};
\end{scope}
\node at (-5.2824,4.8066) {\tiny{$e$}};
\node at (-4.2741,4.6108) {\tiny{$f$}};
\begin{scope}[shift={(-1.9683,0.8116)}]
\draw  (-3.3931,4.3603) rectangle (-3.2271,4.1785);
\node at (-3.3131,4.2709) {\tiny{$0$}};
\end{scope}
\node at (-5.1264,5.0805) {\tiny{$f$}};
\end{scope}
\begin{scope}[shift={(6.7072,1.5173)}]
\draw  (-3.6392,4.4011) rectangle (-2.9855,4.1922);
\node at (-3.2901,4.2995) {\tiny{$5$}};
\end{scope}
\begin{scope}[shift={(4.0834,1.8194)}]
\draw  (-3.6392,4.4011) rectangle (-2.9855,4.1922);
\node at (-3.2901,4.2995) {\tiny{$1$}};
\end{scope}
\begin{scope}[shift={(5.834,1.3412)}]
\draw  (-3.6392,4.4011) rectangle (-2.9855,4.1922);
\node at (-3.2901,4.2995) {\tiny{$5$}};
\end{scope}
\draw (-1.0633,6.6163) -- (2.4486,6.641);
\draw (1.2421,5.4285) -- (2.6779,6.2127) (3.4396,6.0414) -- (3.4263,5.9185) (3.4237,5.7112) -- (3.4186,5.5015);
\begin{scope}[shift={(0.0298,-1.4012)}]
\draw  (-4.4603,4.9901) ellipse (1 and 0.5);
\node (v1) at (-4.4603,4.9901) {$0_{1}$};
\node at (-4.8154,5.2984) {\tiny{$e$}};
\node at (-4.8626,4.7003) {\tiny{$h$}};
\begin{scope}[shift={(-1.2024,0.349)}]
\draw  (-3.5067,4.3572) rectangle (-3.0602,4.2134);
\node at (-3.2917,4.2752) {\tiny{$1$}};
\end{scope}
\begin{scope}[shift={(-1.1448,1.0332)}]
\draw  (-3.5352,4.3891) rectangle (-3.0397,4.1868);
\node at (-3.2901,4.2995) {\tiny{$-1$}};
\end{scope}
\end{scope}
\draw (-4.2916,5.484) -- (-3.025,6.434) (-4.4404,4.4897) -- (-4.4404,4.0939);
\begin{scope}[shift={(0.2145,1.3862)}]
\draw  (-3.6392,4.4011) rectangle (-2.9855,4.1922);
\node at (-3.2901,4.2995) {\tiny{$1$}};
\end{scope}
\draw (-2.0101,5.511) -- (-2.0104,5.9952);
\end{tikzpicture}
\end{array}
\end{align}
For $k=2$, we have
\begin{align}
\begin{array}{c}
\begin{tikzpicture}[scale=1.5]
\draw  (-4.4603,4.9901) ellipse (1 and 0.5);
\node (v1) at (-4.4603,4.9901) {$2_{2}$};
\draw (-3.4464,5.0091) -- (-2.9847,5.0073);
\node at (-3.785,5.1805) {\tiny{$h$}};
\node at (-4.7982,5.3284) {\tiny{$h$}};
\node at (-4.7578,4.6477) {\tiny{$e$}};
\begin{scope}[shift={(-1.1644,0.3871)}]
\draw  (-3.5262,4.3886) rectangle (-3.0539,4.1721);
\node at (-3.2917,4.2752) {\tiny{$-2$}};
\end{scope}
\begin{scope}[shift={(-0.4988,0.715)}]
\draw  (-3.5352,4.3891) rectangle (-3.0397,4.1868);
\node at (-3.2901,4.2995) {\tiny{$2$}};
\end{scope}
\begin{scope}[shift={(-1.186,1.057)}]
\draw  (-3.5352,4.3891) rectangle (-3.0397,4.1868);
\node at (-3.2901,4.2995) {\tiny{$2$}};
\end{scope}
\begin{scope}[shift={(2.4713,0.0223)}]
\draw  (-4.4603,4.9901) ellipse (1 and 0.5);
\node (v1) at (-4.507,4.9528) {$3_{4}$};
\draw (-3.4464,5.0091) -- (-2.9847,5.0073);
\begin{scope}[shift={(-1.8354,0.7134)}]
\draw  (-3.5321,4.3845) rectangle (-3.0504,4.1716);
\node at (-3.2917,4.2752) {\tiny{$-4$}};
\end{scope}
\begin{scope}[shift={(-0.5642,0.7057)}]
\draw  (-3.6392,4.4011) rectangle (-2.9855,4.1922);
\node at (-3.2901,4.2995) {\tiny{$4$}};
\end{scope}
\begin{scope}[shift={(-1.1756,1.0455)}]
\draw  (-3.5321,4.3845) rectangle (-3.0504,4.1716);
\node at (-3.2917,4.2752) {\tiny{$0$}};
\end{scope}
\node at (-4.0675,5.2942) {\tiny{$f$}};
\end{scope}
\begin{scope}[shift={(5.1361,0.0036)}]
\draw  (-4.4603,4.9901) ellipse (1.2 and 0.5);
\node (v1) at (-4.3814,4.9083) {$4_{6}^{2}$};
\node at (-5.3928,5.1471) {\tiny{$e$}};
\begin{scope}[shift={(-2.0408,0.7209)}]
\draw  (-3.5524,4.3484) rectangle (-3.1035,4.1956);
\node at (-3.3438,4.2761) {\tiny{$-6$}};
\end{scope}
\begin{scope}[shift={(-1.595,1.033)}]
\draw  (-3.5156,4.3674) rectangle (-3.0604,4.1989);
\node at (-3.2917,4.2752) {\tiny{$-2$}};
\end{scope}
\node at (-4.8506,5.1441) {\tiny{$f$-$x_i$}};
\begin{scope}[shift={(-0.1234,0.722)}]
\draw  (-3.4041,4.3635) rectangle (-3.206,4.2062);
\node at (-3.3103,4.2939) {\tiny{$4$}};
\end{scope}
\node at (-3.651,4.8569) {\tiny{$h$-$\sum x_i$}};
\begin{scope}[shift={(-0.8189,1.0748)}]
\draw  (-3.5524,4.3484) rectangle (-3.1035,4.1956);
\node at (-3.3438,4.2761) {\tiny{$6$}};
\end{scope}
\node at (-3.8317,5.3301) {\tiny{$h$}};
\end{scope}
\node at (-2.6424,5.2031) {\tiny{$e$}};
\node at (-1.423,4.7873) {\tiny{$h$}};
\begin{scope}[shift={(2.4199,1.5062)}]
\draw  (-4.4627,5.3025) ellipse (1 and 0.8);
\node (v1) at (-4.4951,5.3728) {$1_{4}^{4}$};
\node at (-4.1603,5.0698) {\tiny{$x_1$,$x_3$}};
\node at (-4.3589,5.879) {\tiny{$x_2$-$x_1$,}};
\begin{scope}[shift={(-1.1411,0.3576)}]
\draw  (-3.4304,4.3869) rectangle (-3.1527,4.1998);
\node at (-3.2901,4.2995) {\tiny{$-4$}};
\end{scope}
\begin{scope}[shift={(-0.4805,0.7681)}]
\draw  (-3.446,4.3787) rectangle (-3.1287,4.1848);
\node at (-3.2901,4.2823) {\tiny{$-2$}};
\end{scope}
\begin{scope}[shift={(-0.6037,1.3302)}]
\draw  (-3.5352,4.3891) rectangle (-3.0397,4.1868);
\node at (-3.2901,4.2995) {\tiny{$-4$}};
\end{scope}
\node at (-5.1977,5.3046) {\tiny{$e$}};
\node at (-4.7418,4.9691) {\tiny{$f$-$x_1$-$x_3$,}};
\begin{scope}[shift={(-1.7942,1.22)}]
\draw  (-3.5321,4.3845) rectangle (-3.2451,4.1773);
\node at (-3.3955,4.2703) {\tiny{$-4$}};
\end{scope}
\end{scope}
\begin{scope}[shift={(2.888,1.6316)}]
\draw  (-3.6392,4.4011) rectangle (-2.9855,4.1922);
\node at (-3.2901,4.2995) {\tiny{$2$}};
\end{scope}
\begin{scope}[shift={(2.4325,1.2108)}]
\draw  (-3.6392,4.4011) rectangle (-2.9855,4.1922);
\node at (-3.2901,4.2995) {\tiny{$2$}};
\end{scope}
\draw (-0.2985,5.8226) -- (0.1701,5.4491);
\draw (-0.6414,6.0359) -- (-1.1459,6.4617);
\begin{scope}[shift={(7.901,0.0069)}]
\draw  (-4.4603,4.9901) ellipse (1 and 0.5);
\node (v1) at (-4.3723,4.8912) {$6_6^{12}$};
\node at (-5.1985,5.1414) {\tiny{$e$}};
\begin{scope}[shift={(-1.8953,0.7227)}]
\draw  (-3.4282,4.3553) rectangle (-3.1432,4.1635);
\node at (-3.2917,4.2752) {\tiny{$-6$}};
\end{scope}
\begin{scope}[shift={(-1.1641,1.0713)}]
\draw  (-3.5906,4.3632) rectangle (-3.0156,4.1777);
\node at (-3.3001,4.2753) {\tiny{$-12$}};
\end{scope}
\node at (-4.0633,5.1687) {\tiny{$f$-$x_i$-$y_i$}};
\end{scope}
\draw (1.8669,4.9968) -- (2.432,4.9939);
\begin{scope}[shift={(7.8974,1.5551)}]
\draw  (-4.4603,4.9901) ellipse (1 and 0.5);
\node (v1) at (-4.4453,5.0113) {$5_{8}$};
\begin{scope}[shift={(-1.1387,0.3525)}]
\draw  (-3.3931,4.3603) rectangle (-3.2271,4.1785);
\node at (-3.3131,4.2709) {\tiny{$0$}};
\end{scope}
\begin{scope}[shift={(-1.654,0.4786)}]
\draw  (-3.5428,4.3599) rectangle (-3.045,4.1827);
\node at (-3.2917,4.2752) {\tiny{$-8$}};
\end{scope}
\node at (-5.2824,4.8066) {\tiny{$e$}};
\node at (-4.2741,4.6108) {\tiny{$f$}};
\begin{scope}[shift={(-1.9683,0.8116)}]
\draw  (-3.3931,4.3603) rectangle (-3.2271,4.1785);
\node at (-3.3131,4.2709) {\tiny{$0$}};
\end{scope}
\node at (-5.1264,5.0805) {\tiny{$f$}};
\end{scope}
\begin{scope}[shift={(4.0664,2.5302)}]
\draw  (-3.6392,4.4011) rectangle (-2.9855,4.1922);
\node at (-3.2901,4.2995) {\tiny{$2$}};
\end{scope}
\begin{scope}[shift={(6.7072,1.5173)}]
\draw  (-3.6392,4.4011) rectangle (-2.9855,4.1922);
\node at (-3.2901,4.2995) {\tiny{$6$}};
\end{scope}
\begin{scope}[shift={(3.9733,1.9806)}]
\draw  (-3.6392,4.4011) rectangle (-2.9855,4.1922);
\node at (-3.2901,4.2995) {\tiny{$2$}};
\end{scope}
\begin{scope}[shift={(5.834,1.3412)}]
\draw  (-3.6392,4.4011) rectangle (-2.9855,4.1922);
\node at (-3.2901,4.2995) {\tiny{$6$}};
\end{scope}
\draw (-1.1078,7.0966) -- (0.4253,6.8846) (1.0814,6.8121) -- (2.4486,6.641);
\draw (1.2421,5.4285) -- (2.6779,6.2127) (3.4396,6.0414) -- (3.4263,5.9185) (3.4237,5.7112) -- (3.4186,5.5015);
\begin{scope}[shift={(0.0298,-1.4012)}]
\draw  (-4.4603,4.9901) ellipse (1 and 0.5);
\node (v1) at (-4.4603,4.9901) {$0_{0}$};
\node at (-4.7787,5.3106) {\tiny{$e$}};
\node at (-4.7984,4.6725) {\tiny{$e$}};
\begin{scope}[shift={(-1.2024,0.349)}]
\draw  (-3.5067,4.3572) rectangle (-3.0602,4.2134);
\node at (-3.2917,4.2752) {\tiny{$0$}};
\end{scope}
\begin{scope}[shift={(-1.1448,1.0332)}]
\draw  (-3.5352,4.3891) rectangle (-3.0397,4.1868);
\node at (-3.2901,4.2995) {\tiny{$0$}};
\end{scope}
\end{scope}
\begin{scope}[shift={(1.3147,1.4736)}]
\draw  (-3.6392,4.4011) rectangle (-2.9855,4.1922);
\node at (-3.2901,4.2995) {\tiny{$2$}};
\end{scope}
\draw (-2.0116,6.009) -- (-2.0086,5.8672) (-2.0117,5.6582) -- (-2.0149,5.5125);
\draw (-4.2916,5.484) -- (-3.0273,6.9303) (-4.4404,4.4897) -- (-4.4404,4.0939);
\begin{scope}[shift={(0.2145,1.3862)}]
\draw  (-3.6392,4.4011) rectangle (-2.9855,4.1922);
\node at (-3.2901,4.2995) {\tiny{$2$}};
\end{scope}
\node at (-2.3468,6.3534) {\tiny{$f$-$x_2$-$x_4$}};
\node at (-1.9704,7.2727) {\tiny{$x_4$-$x_3$}};
\end{tikzpicture}
\end{array}
\end{align}
For $k=1$, we have
\begin{align}
\begin{array}{c}
\begin{tikzpicture}[scale=1.5]
\draw  (-4.4603,4.9901) ellipse (1 and 0.5);
\node (v1) at (-4.4603,4.9901) {$2_{3}$};
\draw (-3.4464,5.0091) -- (-2.9847,5.0073);
\node at (-3.785,5.1805) {\tiny{$h$}};
\node at (-4.7982,5.3284) {\tiny{$h$}};
\node at (-4.7578,4.6477) {\tiny{$e$}};
\begin{scope}[shift={(-1.1644,0.3871)}]
\draw  (-3.5262,4.3886) rectangle (-3.0539,4.1721);
\node at (-3.2917,4.2752) {\tiny{$-3$}};
\end{scope}
\begin{scope}[shift={(-0.4988,0.715)}]
\draw  (-3.5352,4.3891) rectangle (-3.0397,4.1868);
\node at (-3.2901,4.2995) {\tiny{$3$}};
\end{scope}
\begin{scope}[shift={(-1.186,1.057)}]
\draw  (-3.5352,4.3891) rectangle (-3.0397,4.1868);
\node at (-3.2901,4.2995) {\tiny{$3$}};
\end{scope}
\begin{scope}[shift={(2.4713,0.0223)}]
\draw  (-4.4603,4.9901) ellipse (1 and 0.5);
\node (v1) at (-4.507,4.9528) {$3_{5}$};
\draw (-3.4464,5.0091) -- (-2.9847,5.0073);
\begin{scope}[shift={(-1.8354,0.7134)}]
\draw  (-3.5321,4.3845) rectangle (-3.0504,4.1716);
\node at (-3.2917,4.2752) {\tiny{$-5$}};
\end{scope}
\begin{scope}[shift={(-0.5642,0.7057)}]
\draw  (-3.6392,4.4011) rectangle (-2.9855,4.1922);
\node at (-3.2901,4.2995) {\tiny{$5$}};
\end{scope}
\begin{scope}[shift={(-1.1756,1.0455)}]
\draw  (-3.5321,4.3845) rectangle (-3.0504,4.1716);
\node at (-3.2917,4.2752) {\tiny{$0$}};
\end{scope}
\node at (-4.0675,5.2942) {\tiny{$f$}};
\end{scope}
\begin{scope}[shift={(5.1361,0.0036)}]
\draw  (-4.4603,4.9901) ellipse (1.2 and 0.5);
\node (v1) at (-4.3814,4.9083) {$4_{7}^{3}$};
\node at (-5.3928,5.1471) {\tiny{$e$}};
\begin{scope}[shift={(-2.0408,0.7209)}]
\draw  (-3.5524,4.3484) rectangle (-3.1035,4.1956);
\node at (-3.3438,4.2761) {\tiny{$-7$}};
\end{scope}
\begin{scope}[shift={(-1.595,1.033)}]
\draw  (-3.5156,4.3674) rectangle (-3.0604,4.1989);
\node at (-3.2917,4.2752) {\tiny{$-3$}};
\end{scope}
\node at (-4.8506,5.1441) {\tiny{$f$-$x_i$}};
\begin{scope}[shift={(-0.1234,0.722)}]
\draw  (-3.4041,4.3635) rectangle (-3.206,4.2062);
\node at (-3.3103,4.2939) {\tiny{$4$}};
\end{scope}
\node at (-3.651,4.8569) {\tiny{$h$-$\sum x_i$}};
\begin{scope}[shift={(-0.8189,1.0748)}]
\draw  (-3.5524,4.3484) rectangle (-3.1035,4.1956);
\node at (-3.3438,4.2761) {\tiny{$7$}};
\end{scope}
\node at (-3.8317,5.3301) {\tiny{$h$}};
\end{scope}
\node at (-2.6424,5.2031) {\tiny{$e$}};
\node at (-1.423,4.7873) {\tiny{$h$}};
\begin{scope}[shift={(2.4199,1.5062)}]
\draw  (-4.4627,5.3025) ellipse (1 and 0.8);
\node (v1) at (-4.654,5.4692) {$1_{5}^{6}$};
\node at (-4.12,5.2187) {\tiny{$x_1$,$x_3$,$x_5$}};
\node at (-4.3559,5.9815) {\tiny{$x_2$-$x_1$,}};
\begin{scope}[shift={(-1.1411,0.3576)}]
\draw  (-3.4304,4.3869) rectangle (-3.1527,4.1998);
\node at (-3.2901,4.2995) {\tiny{$-6$}};
\end{scope}
\begin{scope}[shift={(-0.4805,0.7681)}]
\draw  (-3.446,4.3787) rectangle (-3.1287,4.1848);
\node at (-3.2901,4.2823) {\tiny{$-3$}};
\end{scope}
\begin{scope}[shift={(-0.6037,1.3302)}]
\draw  (-3.5352,4.3891) rectangle (-3.0397,4.1868);
\node at (-3.2901,4.2995) {\tiny{$-6$}};
\end{scope}
\node at (-5.1977,5.3046) {\tiny{$e$}};
\node at (-4.8139,5.0848) {\tiny{$f$-$x_1$-$x_3$,}};
\begin{scope}[shift={(-1.7942,1.22)}]
\draw  (-3.5321,4.3845) rectangle (-3.2451,4.1773);
\node at (-3.3955,4.2703) {\tiny{$-5$}};
\end{scope}
\end{scope}
\begin{scope}[shift={(2.888,1.6316)}]
\draw  (-3.6392,4.4011) rectangle (-2.9855,4.1922);
\node at (-3.2901,4.2995) {\tiny{$3$}};
\end{scope}
\begin{scope}[shift={(2.4325,1.2108)}]
\draw  (-3.6392,4.4011) rectangle (-2.9855,4.1922);
\node at (-3.2901,4.2995) {\tiny{$3$}};
\end{scope}
\draw (-0.2985,5.8226) -- (0.1701,5.4491);
\draw (-0.6414,6.0359) -- (-1.1459,6.4617);
\begin{scope}[shift={(7.901,0.0069)}]
\draw  (-4.4603,4.9901) ellipse (1 and 0.5);
\node (v1) at (-4.3723,4.8912) {$6_6^{14}$};
\node at (-5.1985,5.1414) {\tiny{$e$}};
\begin{scope}[shift={(-1.8953,0.7227)}]
\draw  (-3.4282,4.3553) rectangle (-3.1432,4.1635);
\node at (-3.2917,4.2752) {\tiny{$-6$}};
\end{scope}
\begin{scope}[shift={(-1.1641,1.0713)}]
\draw  (-3.5906,4.3632) rectangle (-3.0156,4.1777);
\node at (-3.3001,4.2753) {\tiny{$-14$}};
\end{scope}
\node at (-4.0633,5.1687) {\tiny{$f$-$x_i$-$y_i$}};
\end{scope}
\draw (1.8669,4.9968) -- (2.432,4.9939);
\begin{scope}[shift={(7.8974,1.5551)}]
\draw  (-4.4603,4.9901) ellipse (1 and 0.5);
\node (v1) at (-4.4453,5.0113) {$5_{9}$};
\begin{scope}[shift={(-1.1387,0.3525)}]
\draw  (-3.3931,4.3603) rectangle (-3.2271,4.1785);
\node at (-3.3131,4.2709) {\tiny{$0$}};
\end{scope}
\begin{scope}[shift={(-1.654,0.4786)}]
\draw  (-3.5428,4.3599) rectangle (-3.045,4.1827);
\node at (-3.2917,4.2752) {\tiny{$-9$}};
\end{scope}
\node at (-5.2824,4.8066) {\tiny{$e$}};
\node at (-4.2741,4.6108) {\tiny{$f$}};
\begin{scope}[shift={(-1.9683,0.8116)}]
\draw  (-3.3931,4.3603) rectangle (-3.2271,4.1785);
\node at (-3.3131,4.2709) {\tiny{$0$}};
\end{scope}
\node at (-5.1264,5.0805) {\tiny{$f$}};
\end{scope}
\begin{scope}[shift={(4.0664,2.5302)}]
\draw  (-3.6392,4.4011) rectangle (-2.9855,4.1922);
\node at (-3.2901,4.2995) {\tiny{$3$}};
\end{scope}
\begin{scope}[shift={(6.7072,1.5173)}]
\draw  (-3.6392,4.4011) rectangle (-2.9855,4.1922);
\node at (-3.2901,4.2995) {\tiny{$7$}};
\end{scope}
\begin{scope}[shift={(3.9733,1.9806)}]
\draw  (-3.6392,4.4011) rectangle (-2.9855,4.1922);
\node at (-3.2901,4.2995) {\tiny{$3$}};
\end{scope}
\begin{scope}[shift={(5.834,1.3412)}]
\draw  (-3.6392,4.4011) rectangle (-2.9855,4.1922);
\node at (-3.2901,4.2995) {\tiny{$7$}};
\end{scope}
\draw (-1.1078,7.0966) -- (0.4253,6.8846) (1.0814,6.8121) -- (2.4486,6.641);
\draw (1.2421,5.4285) -- (2.6779,6.2127) (3.4396,6.0414) -- (3.4263,5.9185) (3.4237,5.7112) -- (3.4186,5.5015);
\begin{scope}[shift={(0.0298,-1.4012)}]
\draw  (-4.4603,4.9901) ellipse (1 and 0.5);
\node (v1) at (-4.4603,4.9901) {$0_{1}$};
\node at (-4.7787,5.3106) {\tiny{$h$}};
\node at (-4.7984,4.6725) {\tiny{$e$}};
\begin{scope}[shift={(-1.2024,0.349)}]
\draw  (-3.5067,4.3572) rectangle (-3.0602,4.2134);
\node at (-3.2917,4.2752) {\tiny{$-1$}};
\end{scope}
\begin{scope}[shift={(-1.1448,1.0332)}]
\draw  (-3.5352,4.3891) rectangle (-3.0397,4.1868);
\node at (-3.2901,4.2995) {\tiny{$1$}};
\end{scope}
\end{scope}
\begin{scope}[shift={(1.3147,1.4736)}]
\draw  (-3.6392,4.4011) rectangle (-2.9855,4.1922);
\node at (-3.2901,4.2995) {\tiny{$3$}};
\end{scope}
\draw (-2.0116,6.009) -- (-2.0086,5.8672) (-2.0117,5.6582) -- (-2.0149,5.5125);
\draw (-4.2916,5.484) -- (-3.0273,6.9303) (-4.4404,4.4897) -- (-4.4404,4.0939);
\begin{scope}[shift={(0.2145,1.3862)}]
\draw  (-3.6392,4.4011) rectangle (-2.9855,4.1922);
\node at (-3.2901,4.2995) {\tiny{$3$}};
\end{scope}
\node at (-2.3916,6.4632) {\tiny{$f$-$x_2$-$x_4$,}};
\node at (-1.929,7.3854) {\tiny{$x_4$-$x_3$,}};
\node at (-2.4253,6.3325) {\tiny{$f$-$x_5$-$x_6$}};
\node at (-1.9629,7.2654) {\tiny{$x_6$-$x_5$}};
\end{tikzpicture}
\end{array}
\end{align}
In each of the cases above, we have paired up the $16-2k$ blowups on $S^6_C$ into $8-k$ pairs representing $8-k$ hypers in the vector representation of $\mf{so}(12)$. The rest of the blowups represent $\frac{1}{2}\left(4-k\right)$ hypers in the Weyl spinor representation of $\mf{so}(12)$. This graph is same as the one presented in \cite{Bhardwaj:2018yhy} for $k=3$, but different for $k=1,2$. We believe that the proposal made in \cite{Bhardwaj:2018yhy} for $k=1,2$ is incorrect and the one that we have presented is the correct one. One can verify that the gluing curves are as we claim them to be by studying the behavior of fiber $f_C$ over the special points on $C$.


For $\so(11)$ and $k=3$, the associated graph is
\begin{align}
\begin{array}{c}
\begin{tikzpicture}[scale=1.5]
\draw  (-4.4603,4.9901) ellipse (1 and 0.5);
\node (v1) at (-4.4603,4.9901) {$2_{1}$};
\draw (-3.4464,5.0091) -- (-2.9847,5.0073);
\node at (-3.785,5.1805) {\tiny{$h$}};
\node at (-4.7982,5.3284) {\tiny{$h$}};
\node at (-4.7578,4.6477) {\tiny{$e$}};
\begin{scope}[shift={(-1.1644,0.3871)}]
\draw  (-3.5262,4.3886) rectangle (-3.0539,4.1721);
\node at (-3.2917,4.2752) {\tiny{$-1$}};
\end{scope}
\begin{scope}[shift={(-0.4988,0.715)}]
\draw  (-3.5352,4.3891) rectangle (-3.0397,4.1868);
\node at (-3.2901,4.2995) {\tiny{$1$}};
\end{scope}
\begin{scope}[shift={(-1.186,1.057)}]
\draw  (-3.5352,4.3891) rectangle (-3.0397,4.1868);
\node at (-3.2901,4.2995) {\tiny{$1$}};
\end{scope}
\begin{scope}[shift={(2.4713,0.0223)}]
\draw  (-4.4603,4.9901) ellipse (1 and 0.5);
\node (v1) at (-4.507,4.9528) {$3_{3}$};
\draw (-3.4464,5.0091) -- (-2.9847,5.0073);
\begin{scope}[shift={(-1.8354,0.7134)}]
\draw  (-3.5321,4.3845) rectangle (-3.0504,4.1716);
\node at (-3.2917,4.2752) {\tiny{$-3$}};
\end{scope}
\begin{scope}[shift={(-0.5642,0.7057)}]
\draw  (-3.6392,4.4011) rectangle (-2.9855,4.1922);
\node at (-3.2901,4.2995) {\tiny{$3$}};
\end{scope}
\begin{scope}[shift={(-1.1756,1.0455)}]
\draw  (-3.5321,4.3845) rectangle (-3.0504,4.1716);
\node at (-3.2917,4.2752) {\tiny{$0$}};
\end{scope}
\node at (-4.133,5.3201) {\tiny{$f$}};
\end{scope}
\begin{scope}[shift={(5.1361,0.0036)}]
\draw  (-4.4603,4.9901) ellipse (1.2 and 0.5);
\node (v1) at (-4.3814,4.9083) {$4_{5}^{1}$};
\node at (-5.3928,5.1471) {\tiny{$e$}};
\begin{scope}[shift={(-2.0408,0.7209)}]
\draw  (-3.5524,4.3484) rectangle (-3.1035,4.1956);
\node at (-3.3438,4.2761) {\tiny{$-5$}};
\end{scope}
\begin{scope}[shift={(-1.595,1.033)}]
\draw  (-3.5156,4.3674) rectangle (-3.0604,4.1989);
\node at (-3.2917,4.2752) {\tiny{$-1$}};
\end{scope}
\node at (-4.8506,5.1441) {\tiny{$f$-$x$}};
\begin{scope}[shift={(-0.1234,0.722)}]
\draw  (-3.4041,4.3635) rectangle (-3.206,4.2062);
\node at (-3.3103,4.2939) {\tiny{$19$}};
\end{scope}
\node at (-3.651,4.8569) {\tiny{$2h$-$x$}};
\end{scope}
\node at (-2.6424,5.2031) {\tiny{$e$}};
\node at (-1.423,4.7873) {\tiny{$h$}};
\begin{scope}[shift={(2.4199,1.5062)}]
\draw  (-4.4603,4.9901) ellipse (1 and 0.5);
\node (v1) at (-4.593,5.1091) {$1_{3}^{2}$};
\node at (-3.9028,4.6583) {\tiny{$x_1$}};
\node at (-4.0954,5.2962) {\tiny{$x_2$-$x_1$}};
\begin{scope}[shift={(-1.1411,0.3576)}]
\draw  (-3.5352,4.3891) rectangle (-3.0397,4.1868);
\node at (-3.2901,4.2995) {\tiny{$-2$}};
\end{scope}
\begin{scope}[shift={(-0.6121,0.5364)}]
\draw  (-3.5352,4.3891) rectangle (-3.0397,4.1868);
\node at (-3.2901,4.2995) {\tiny{$-1$}};
\end{scope}
\begin{scope}[shift={(-0.5525,0.8269)}]
\draw  (-3.5352,4.3891) rectangle (-3.0397,4.1868);
\node at (-3.2901,4.2995) {\tiny{$-2$}};
\end{scope}
\node at (-5.1598,5.1746) {\tiny{$e$}};
\node at (-4.496,4.823) {\tiny{$f$-$x_1$-$x_2$}};
\begin{scope}[shift={(-1.8702,0.7134)}]
\draw  (-3.5321,4.3845) rectangle (-3.0504,4.1716);
\node at (-3.2917,4.2752) {\tiny{$-3$}};
\end{scope}
\end{scope}
\begin{scope}[shift={(2.4325,1.2108)}]
\draw  (-3.6392,4.4011) rectangle (-2.9855,4.1922);
\node at (-3.2901,4.2995) {\tiny{$1$}};
\end{scope}
\draw (-1.1799,6.2403) -- (0.1701,5.4491);
\begin{scope}[shift={(7.901,0.0069)}]
\draw  (-4.4603,4.9901) ellipse (1 and 0.5);
\node (v1) at (-4.3416,4.996) {$5_5^{8}$};
\node at (-4.8567,4.7577) {\tiny{$e$-$\sum x_i$-$\sum y_i$}};
\begin{scope}[shift={(-1.8953,0.7227)}]
\draw  (-3.4282,4.3553) rectangle (-3.1432,4.1635);
\node at (-3.2917,4.2752) {\tiny{-$13$}};
\end{scope}
\begin{scope}[shift={(-1.1641,1.0713)}]
\draw  (-3.4372,4.3534) rectangle (-3.1711,4.1839);
\node at (-3.3001,4.2753) {\tiny{$-4$}};
\end{scope}
\node at (-4.2119,5.3273) {\tiny{$x_i$}};
\begin{scope}[shift={(-0.7158,0.4543)}]
\draw  (-3.4372,4.3534) rectangle (-3.1711,4.1839);
\node at (-3.3001,4.2753) {\tiny{$-4$}};
\end{scope}
\node at (-3.9832,4.9074) {\tiny{$y_i$}};
\begin{scope}[shift={(-1.6711,0.9528)}]
\draw  (-3.4372,4.3534) rectangle (-3.1711,4.1839);
\node at (-3.3001,4.2753) {\tiny{$0$}};
\end{scope}
\node at (-4.7399,5.1645) {\tiny{$f$}};
\end{scope}
\draw (1.8669,4.9968) -- (2.432,4.9939);
\begin{scope}[shift={(8.5743,0.7394)}]
\draw  (-3.6392,4.4011) rectangle (-2.9855,4.1922);
\node at (-3.2901,4.2995) {\tiny{$4$}};
\end{scope}
\begin{scope}[shift={(3.9496,1.4813)}]
\draw  (-3.6392,4.4011) rectangle (-2.9855,4.1922);
\node at (-3.2901,4.2995) {\tiny{$1$}};
\end{scope}
\draw (-1.0633,6.6163) -- (2.6892,5.3246);
\begin{scope}[shift={(0.0298,-1.4012)}]
\draw  (-4.4603,4.9901) ellipse (1 and 0.5);
\node (v1) at (-4.4603,4.9901) {$0_{1}$};
\node at (-4.8154,5.2984) {\tiny{$e$}};
\node at (-4.8626,4.7003) {\tiny{$h$}};
\begin{scope}[shift={(-1.2024,0.349)}]
\draw  (-3.5067,4.3572) rectangle (-3.0602,4.2134);
\node at (-3.2917,4.2752) {\tiny{$1$}};
\end{scope}
\begin{scope}[shift={(-1.1448,1.0332)}]
\draw  (-3.5352,4.3891) rectangle (-3.0397,4.1868);
\node at (-3.2901,4.2995) {\tiny{$-1$}};
\end{scope}
\end{scope}
\draw (-4.2916,5.484) -- (-3.025,6.434) (-4.4404,4.4897) -- (-4.4404,4.0939);
\begin{scope}[shift={(0.2145,1.3862)}]
\draw  (-3.6392,4.4011) rectangle (-2.9855,4.1922);
\node at (-3.2901,4.2995) {\tiny{$1$}};
\end{scope}
\draw (-2.0101,5.511) -- (-2.0104,5.9952);
\draw (3.4629,5.4849) .. controls (4.0723,6.4439) and (5.2015,6.2557) .. (5.2557,5.1371);
\draw (5.265,4.9319) .. controls (5.2549,3.8797) and (4.314,3.8797) .. (3.9649,4.5677);
\end{tikzpicture}
\end{array}
\end{align}
For $k=2$, we have
\begin{align}
\begin{array}{c}
\begin{tikzpicture}[scale=1.5]
\draw  (-4.4603,4.9901) ellipse (1 and 0.5);
\node (v1) at (-4.4603,4.9901) {$2_{2}$};
\draw (-3.4464,5.0091) -- (-2.9847,5.0073);
\node at (-3.785,5.1805) {\tiny{$h$}};
\node at (-4.7982,5.3284) {\tiny{$h$}};
\node at (-4.7578,4.6477) {\tiny{$e$}};
\begin{scope}[shift={(-1.1644,0.3871)}]
\draw  (-3.5262,4.3886) rectangle (-3.0539,4.1721);
\node at (-3.2917,4.2752) {\tiny{$-2$}};
\end{scope}
\begin{scope}[shift={(-0.4988,0.715)}]
\draw  (-3.5352,4.3891) rectangle (-3.0397,4.1868);
\node at (-3.2901,4.2995) {\tiny{$2$}};
\end{scope}
\begin{scope}[shift={(-1.186,1.057)}]
\draw  (-3.5352,4.3891) rectangle (-3.0397,4.1868);
\node at (-3.2901,4.2995) {\tiny{$2$}};
\end{scope}
\begin{scope}[shift={(2.4713,0.0223)}]
\draw  (-4.4603,4.9901) ellipse (1 and 0.5);
\node (v1) at (-4.507,4.9528) {$3_{4}$};
\draw (-3.4464,5.0091) -- (-2.9847,5.0073);
\begin{scope}[shift={(-1.8354,0.7134)}]
\draw  (-3.5321,4.3845) rectangle (-3.0504,4.1716);
\node at (-3.2917,4.2752) {\tiny{$-4$}};
\end{scope}
\begin{scope}[shift={(-0.5642,0.7057)}]
\draw  (-3.6392,4.4011) rectangle (-2.9855,4.1922);
\node at (-3.2901,4.2995) {\tiny{$4$}};
\end{scope}
\begin{scope}[shift={(-1.1756,1.0455)}]
\draw  (-3.5321,4.3845) rectangle (-3.0504,4.1716);
\node at (-3.2917,4.2752) {\tiny{$0$}};
\end{scope}
\node at (-4.0675,5.2942) {\tiny{$f$}};
\end{scope}
\begin{scope}[shift={(5.1361,0.0036)}]
\draw  (-4.4603,4.9901) ellipse (1.2 and 0.5);
\node (v1) at (-4.3814,4.9083) {$4_{6}^{2}$};
\node at (-5.3928,5.1471) {\tiny{$e$}};
\begin{scope}[shift={(-2.0408,0.7209)}]
\draw  (-3.5524,4.3484) rectangle (-3.1035,4.1956);
\node at (-3.3438,4.2761) {\tiny{$-6$}};
\end{scope}
\begin{scope}[shift={(-1.595,1.033)}]
\draw  (-3.5156,4.3674) rectangle (-3.0604,4.1989);
\node at (-3.2917,4.2752) {\tiny{$-2$}};
\end{scope}
\node at (-4.8506,5.1441) {\tiny{$f$-$x_i$}};
\begin{scope}[shift={(-0.1234,0.722)}]
\draw  (-3.4825,4.3967) rectangle (-3.206,4.2062);
\node at (-3.3499,4.3031) {\tiny{$22$}};
\end{scope}
\node at (-3.7036,4.847) {\tiny{$2h$-$\sum x_i$}};
\end{scope}
\node at (-2.6424,5.2031) {\tiny{$e$}};
\node at (-1.423,4.7873) {\tiny{$h$}};
\begin{scope}[shift={(2.4199,1.5062)}]
\draw  (-4.4627,5.3025) ellipse (1 and 0.8);
\node (v1) at (-4.4951,5.3728) {$1_{4}^{4}$};
\node at (-4.1603,5.0698) {\tiny{$x_1$,$x_3$}};
\node at (-4.3589,5.879) {\tiny{$x_2$-$x_1$,}};
\begin{scope}[shift={(-1.1411,0.3576)}]
\draw  (-3.4304,4.3869) rectangle (-3.1527,4.1998);
\node at (-3.2901,4.2995) {\tiny{$-4$}};
\end{scope}
\begin{scope}[shift={(-0.4805,0.7681)}]
\draw  (-3.446,4.3787) rectangle (-3.1287,4.1848);
\node at (-3.2901,4.2823) {\tiny{$-2$}};
\end{scope}
\begin{scope}[shift={(-0.6037,1.3302)}]
\draw  (-3.5352,4.3891) rectangle (-3.0397,4.1868);
\node at (-3.2901,4.2995) {\tiny{$-4$}};
\end{scope}
\node at (-5.1977,5.3046) {\tiny{$e$}};
\node at (-4.7418,4.9691) {\tiny{$f$-$x_1$-$x_3$,}};
\begin{scope}[shift={(-1.7942,1.22)}]
\draw  (-3.5321,4.3845) rectangle (-3.2451,4.1773);
\node at (-3.3955,4.2703) {\tiny{$-4$}};
\end{scope}
\end{scope}
\begin{scope}[shift={(2.888,1.6316)}]
\draw  (-3.6392,4.4011) rectangle (-2.9855,4.1922);
\node at (-3.2901,4.2995) {\tiny{$2$}};
\end{scope}
\begin{scope}[shift={(2.4325,1.2108)}]
\draw  (-3.6392,4.4011) rectangle (-2.9855,4.1922);
\node at (-3.2901,4.2995) {\tiny{$2$}};
\end{scope}
\draw (-0.2985,5.8226) -- (0.1701,5.4491);
\draw (-0.6414,6.0359) -- (-1.1459,6.4617);
\begin{scope}[shift={(7.901,0.0069)}]
\draw  (-4.4603,4.9901) ellipse (1 and 0.5);
\node (v1) at (-4.4062,4.9931) {$5_4^{10}$};
\node at (-4.8164,4.7848) {\tiny{$e$-$\sum x_i$-$\sum y_i$}};
\begin{scope}[shift={(-1.8953,0.7227)}]
\draw  (-3.4282,4.3553) rectangle (-3.1432,4.1635);
\node at (-3.2917,4.2752) {\tiny{-$14$}};
\end{scope}
\begin{scope}[shift={(-0.946,1.0647)}]
\draw  (-3.4712,4.3565) rectangle (-3.1365,4.1763);
\node at (-3.3001,4.2753) {\tiny{$-1$}};
\end{scope}
\node at (-3.9283,5.2363) {\tiny{$x_i$}};
\begin{scope}[shift={(-0.7056,0.4685)}]
\draw  (-3.4712,4.3565) rectangle (-3.1365,4.1763);
\node at (-3.3001,4.2753) {\tiny{$-1$}};
\end{scope}
\node at (-3.9615,4.914) {\tiny{$y_i$}};
\begin{scope}[shift={(-1.62,0.9755)}]
\draw  (-3.4712,4.3565) rectangle (-3.1365,4.1763);
\node at (-3.3001,4.2753) {\tiny{$0$}};
\end{scope}
\node at (-4.6733,5.288) {\tiny{$f$}};
\end{scope}
\draw (1.8669,4.9968) -- (2.432,4.9939);
\begin{scope}[shift={(4.3205,2.1099)}]
\draw  (-3.6392,4.4011) rectangle (-2.9855,4.1922);
\node at (-3.2901,4.2995) {\tiny{$2$}};
\end{scope}
\begin{scope}[shift={(3.9606,1.6198)}]
\draw  (-3.6392,4.4011) rectangle (-2.9855,4.1922);
\node at (-3.2901,4.2995) {\tiny{$2$}};
\end{scope}
\draw (-1.1078,7.0966) -- (0.6794,6.4643) (1.2211,6.3069) -- (2.8284,5.3862);
\begin{scope}[shift={(0.0298,-1.4012)}]
\draw  (-4.4603,4.9901) ellipse (1 and 0.5);
\node (v1) at (-4.4603,4.9901) {$0_{0}$};
\node at (-4.7787,5.3106) {\tiny{$e$}};
\node at (-4.7984,4.6725) {\tiny{$e$}};
\begin{scope}[shift={(-1.2024,0.349)}]
\draw  (-3.5067,4.3572) rectangle (-3.0602,4.2134);
\node at (-3.2917,4.2752) {\tiny{$0$}};
\end{scope}
\begin{scope}[shift={(-1.1448,1.0332)}]
\draw  (-3.5352,4.3891) rectangle (-3.0397,4.1868);
\node at (-3.2901,4.2995) {\tiny{$0$}};
\end{scope}
\end{scope}
\begin{scope}[shift={(1.3147,1.4736)}]
\draw  (-3.6392,4.4011) rectangle (-2.9855,4.1922);
\node at (-3.2901,4.2995) {\tiny{$2$}};
\end{scope}
\draw (-2.0116,6.009) -- (-2.0086,5.8672) (-2.0117,5.6582) -- (-2.0149,5.5125);
\draw (-4.2916,5.484) -- (-3.0273,6.9303) (-4.4404,4.4897) -- (-4.4404,4.0939);
\begin{scope}[shift={(0.2145,1.3862)}]
\draw  (-3.6392,4.4011) rectangle (-2.9855,4.1922);
\node at (-3.2901,4.2995) {\tiny{$2$}};
\end{scope}
\node at (-2.3468,6.3534) {\tiny{$f$-$x_2$-$x_4$}};
\node at (-1.9704,7.2727) {\tiny{$x_4$-$x_3$}};
\begin{scope}[shift={(8.5878,0.775)}]
\draw  (-3.6392,4.4011) rectangle (-2.9855,4.1922);
\node at (-3.2901,4.2995) {\tiny{$5$}};
\end{scope}
\draw (3.7419,5.4759) .. controls (4.2827,6.1723) and (5.2029,6.4956) .. (5.2776,5.1774);
\draw (5.2776,4.9661) .. controls (5.1781,3.7164) and (4.3636,3.8842) .. (4.0527,4.6117);
\end{tikzpicture}
\end{array}
\end{align}
For $k=1$, we have
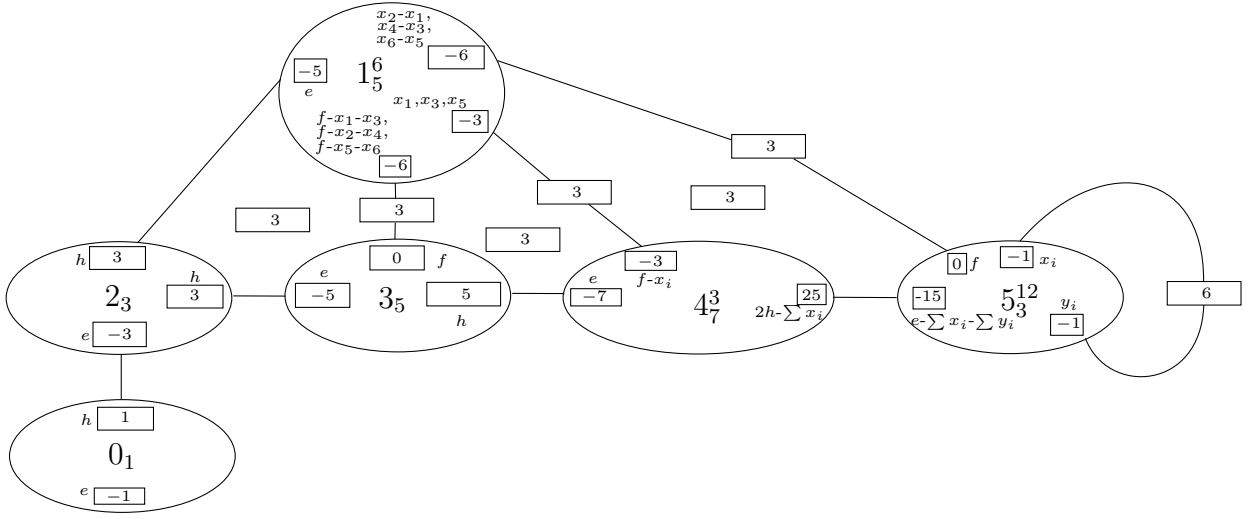
\begin{align}
\begin{array}{c}
\begin{tikzpicture}[scale=1.5]
\draw  (-4.4603,4.9901) ellipse (1 and 0.5);
\node (v1) at (-4.4603,4.9901) {$2_{3}$};
\draw (-3.4464,5.0091) -- (-2.9847,5.0073);
\node at (-3.785,5.1805) {\tiny{$h$}};
\node at (-4.7982,5.3284) {\tiny{$h$}};
\node at (-4.7578,4.6477) {\tiny{$e$}};
\begin{scope}[shift={(-1.1644,0.3871)}]
\draw  (-3.5262,4.3886) rectangle (-3.0539,4.1721);
\node at (-3.2917,4.2752) {\tiny{$-3$}};
\end{scope}
\begin{scope}[shift={(-0.4988,0.715)}]
\draw  (-3.5352,4.3891) rectangle (-3.0397,4.1868);
\node at (-3.2901,4.2995) {\tiny{$3$}};
\end{scope}
\begin{scope}[shift={(-1.186,1.057)}]
\draw  (-3.5352,4.3891) rectangle (-3.0397,4.1868);
\node at (-3.2901,4.2995) {\tiny{$3$}};
\end{scope}
\begin{scope}[shift={(2.4713,0.0223)}]
\draw  (-4.4603,4.9901) ellipse (1 and 0.5);
\node (v1) at (-4.507,4.9528) {$3_{5}$};
\draw (-3.4464,5.0091) -- (-2.9847,5.0073);
\begin{scope}[shift={(-1.8354,0.7134)}]
\draw  (-3.5321,4.3845) rectangle (-3.0504,4.1716);
\node at (-3.2917,4.2752) {\tiny{$-5$}};
\end{scope}
\begin{scope}[shift={(-0.5642,0.7057)}]
\draw  (-3.6392,4.4011) rectangle (-2.9855,4.1922);
\node at (-3.2901,4.2995) {\tiny{$5$}};
\end{scope}
\begin{scope}[shift={(-1.1756,1.0455)}]
\draw  (-3.5321,4.3845) rectangle (-3.0504,4.1716);
\node at (-3.2917,4.2752) {\tiny{$0$}};
\end{scope}
\node at (-4.0675,5.2942) {\tiny{$f$}};
\end{scope}
\begin{scope}[shift={(5.1361,0.0036)}]
\draw  (-4.4603,4.9901) ellipse (1.2 and 0.5);
\node (v1) at (-4.3814,4.9083) {$4_{7}^{3}$};
\node at (-5.3928,5.1471) {\tiny{$e$}};
\begin{scope}[shift={(-2.0408,0.7209)}]
\draw  (-3.5524,4.3484) rectangle (-3.1035,4.1956);
\node at (-3.3438,4.2761) {\tiny{$-7$}};
\end{scope}
\begin{scope}[shift={(-1.595,1.033)}]
\draw  (-3.5156,4.3674) rectangle (-3.0604,4.1989);
\node at (-3.2917,4.2752) {\tiny{$-3$}};
\end{scope}
\node at (-4.8506,5.1441) {\tiny{$f$-$x_i$}};
\begin{scope}[shift={(-0.1234,0.722)}]
\draw  (-3.4613,4.3873) rectangle (-3.206,4.2062);
\node at (-3.3354,4.292) {\tiny{$25$}};
\end{scope}
\node at (-3.651,4.8569) {\tiny{$2h$-$\sum x_i$}};
\end{scope}
\node at (-2.6424,5.2031) {\tiny{$e$}};
\node at (-1.423,4.7873) {\tiny{$h$}};
\begin{scope}[shift={(2.4199,1.5062)}]
\draw  (-4.4627,5.3025) ellipse (1 and 0.8);
\node (v1) at (-4.654,5.4692) {$1_{5}^{6}$};
\node at (-4.12,5.2187) {\tiny{$x_1$,$x_3$,$x_5$}};
\node at (-4.3559,5.9815) {\tiny{$x_2$-$x_1$,}};
\begin{scope}[shift={(-1.1411,0.3576)}]
\draw  (-3.4304,4.3869) rectangle (-3.1527,4.1998);
\node at (-3.2901,4.2995) {\tiny{$-6$}};
\end{scope}
\begin{scope}[shift={(-0.4805,0.7681)}]
\draw  (-3.446,4.3787) rectangle (-3.1287,4.1848);
\node at (-3.2901,4.2823) {\tiny{$-3$}};
\end{scope}
\begin{scope}[shift={(-0.6037,1.3302)}]
\draw  (-3.5352,4.3891) rectangle (-3.0397,4.1868);
\node at (-3.2901,4.2995) {\tiny{$-6$}};
\end{scope}
\node at (-5.1977,5.3046) {\tiny{$e$}};
\node at (-4.8139,5.0848) {\tiny{$f$-$x_1$-$x_3$,}};
\begin{scope}[shift={(-1.7942,1.22)}]
\draw  (-3.5321,4.3845) rectangle (-3.2451,4.1773);
\node at (-3.3955,4.2703) {\tiny{$-5$}};
\end{scope}
\end{scope}
\begin{scope}[shift={(2.888,1.6316)}]
\draw  (-3.6392,4.4011) rectangle (-2.9855,4.1922);
\node at (-3.2901,4.2995) {\tiny{$3$}};
\end{scope}
\begin{scope}[shift={(2.4325,1.2108)}]
\draw  (-3.6392,4.4011) rectangle (-2.9855,4.1922);
\node at (-3.2901,4.2995) {\tiny{$3$}};
\end{scope}
\draw (-0.2985,5.8226) -- (0.1701,5.4491);
\draw (-0.6414,6.0359) -- (-1.1459,6.4617);
\begin{scope}[shift={(7.901,0.0069)}]
\draw  (-4.4603,4.9901) ellipse (1 and 0.5);
\node (v1) at (-4.3755,4.9579) {$5_3^{12}$};
\node at (-4.8804,4.7761) {\tiny{$e$-$\sum x_i$-$\sum y_i$}};
\begin{scope}[shift={(-1.8953,0.7227)}]
\draw  (-3.4282,4.3553) rectangle (-3.1432,4.1635);
\node at (-3.2917,4.2752) {\tiny{-$15$}};
\end{scope}
\begin{scope}[shift={(-1.0879,1.0745)}]
\draw  (-3.4616,4.3641) rectangle (-3.173,4.1782);
\node at (-3.3001,4.2753) {\tiny{$-1$}};
\end{scope}
\node at (-4.1201,5.3044) {\tiny{$x_i$}};
\begin{scope}[shift={(-1.6222,1.018)}]
\draw  (-3.3931,4.3603) rectangle (-3.2271,4.1785);
\node at (-3.3131,4.2709) {\tiny{$0$}};
\end{scope}
\node at (-4.7803,5.2869) {\tiny{$f$}};
\begin{scope}[shift={(-0.6433,0.4744)}]
\draw  (-3.4616,4.3641) rectangle (-3.173,4.1782);
\node at (-3.3001,4.2753) {\tiny{$-1$}};
\end{scope}
\node at (-3.9395,4.9242) {\tiny{$y_i$}};
\end{scope}
\draw (1.8669,4.9968) -- (2.432,4.9939);
\begin{scope}[shift={(4.6093,2.039)}]
\draw  (-3.6392,4.4011) rectangle (-2.9855,4.1922);
\node at (-3.2901,4.2995) {\tiny{$3$}};
\end{scope}
\begin{scope}[shift={(8.4694,0.7355)}]
\draw  (-3.6392,4.4011) rectangle (-2.9855,4.1922);
\node at (-3.2901,4.2995) {\tiny{$6$}};
\end{scope}
\begin{scope}[shift={(4.2491,1.5842)}]
\draw  (-3.6392,4.4011) rectangle (-2.9855,4.1922);
\node at (-3.2901,4.2995) {\tiny{$3$}};
\end{scope}
\draw (-1.1078,7.0966) -- (0.9682,6.3934) (1.519,6.2255) -- (2.8762,5.4111);
\begin{scope}[shift={(0.0298,-1.4012)}]
\draw  (-4.4603,4.9901) ellipse (1 and 0.5);
\node (v1) at (-4.4603,4.9901) {$0_{1}$};
\node at (-4.7787,5.3106) {\tiny{$h$}};
\node at (-4.7984,4.6725) {\tiny{$e$}};
\begin{scope}[shift={(-1.2024,0.349)}]
\draw  (-3.5067,4.3572) rectangle (-3.0602,4.2134);
\node at (-3.2917,4.2752) {\tiny{$-1$}};
\end{scope}
\begin{scope}[shift={(-1.1448,1.0332)}]
\draw  (-3.5352,4.3891) rectangle (-3.0397,4.1868);
\node at (-3.2901,4.2995) {\tiny{$1$}};
\end{scope}
\end{scope}
\begin{scope}[shift={(1.3147,1.4736)}]
\draw  (-3.6392,4.4011) rectangle (-2.9855,4.1922);
\node at (-3.2901,4.2995) {\tiny{$3$}};
\end{scope}
\draw (-2.0116,6.009) -- (-2.0086,5.8672) (-2.0117,5.6582) -- (-2.0149,5.5125);
\draw (-4.2916,5.484) -- (-3.0273,6.9303) (-4.4404,4.4897) -- (-4.4404,4.0939);
\begin{scope}[shift={(0.2145,1.3862)}]
\draw  (-3.6392,4.4011) rectangle (-2.9855,4.1922);
\node at (-3.2901,4.2995) {\tiny{$3$}};
\end{scope}
\node at (-2.3916,6.4632) {\tiny{$f$-$x_2$-$x_4$,}};
\node at (-1.929,7.3854) {\tiny{$x_4$-$x_3$,}};
\node at (-2.4253,6.3325) {\tiny{$f$-$x_5$-$x_6$}};
\node at (-1.9629,7.2654) {\tiny{$x_6$-$x_5$}};
\draw (3.5235,5.4927) .. controls (4.1935,6.2769) and (5.1589,6.1834) .. (5.1523,5.1402);
\draw (5.1538,4.9279) .. controls (5.1292,4.1657) and (4.3015,4.1165) .. (4.1048,4.6328);
\end{tikzpicture}
\end{array}
\end{align}
In each of cases above, we have paired up the $14-2k$ blowups on $S^5_C$ into $7-k$ pairs representing $7-k$ hypers in the vector representation of $\mf{so}(11)$. The rest of the blowups represent $\frac{1}{2}\left(4-k\right)$ hypers in the spinor representation of $\mf{so}(11)$. We also pair up the $8-2k$ blowups on $S^1_C$ into $4-k$ pairs. This graph is same as the one presented in \cite{Bhardwaj:2018yhy} for $k=3$. We believe that the proposal made in \cite{Bhardwaj:2018yhy} for $k=1,2$ is incorrect and the one that we have presented is the correct one. One can verify that the gluing curves are as we claim them to be by studying the behavior of fiber $f_C$ over the special points on $C$. \\
Due to monodromy, the elliptic fiber is represented as $f_C=f^0_C+f^1_C+2(f^2_C+f^3_C+f^4_C+f^5_C)$. 


For $\fg_2$, the associated graph is
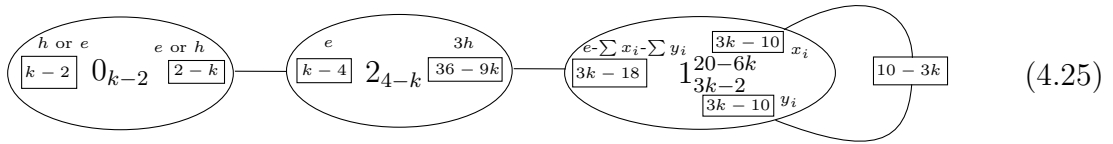
\begin{align}
\label{eqn:G2}
\begin{array}{c}
\noindent\begin{tikzpicture}[scale=1.5]
\draw  (-4.4603,4.9901) ellipse (1 and 0.5);
\node (v1) at (-4.4603,4.9901) {$0_{k-2}$};
\draw (-3.4464,5.0091) -- (-2.9847,5.0073);
\node at (-3.937,5.2302) {\tiny{$e$ or $h$}};
\node at (-4.9686,5.2627) {\tiny{$h$ or $e$}};
\begin{scope}[shift={(-1.8167,0.7227)}]
\draw  (-3.5199,4.4183) rectangle (-3.0317,4.1342);
\node at (-3.2917,4.2752) {\tiny{$k-2$}};
\end{scope}
\begin{scope}[shift={(-0.4988,0.715)}]
\draw  (-3.5352,4.3891) rectangle (-3.0397,4.1868);
\node at (-3.2901,4.2995) {\tiny{$2-k$}};
\end{scope}
\begin{scope}[shift={(2.4713,0.0223)}]
\draw  (-4.4603,4.9901) ellipse (1 and 0.5);
\node (v1) at (-4.507,4.9528) {$2_{4-k}$};
\draw (-3.4464,5.0091) -- (-2.9847,5.0073);
\begin{scope}[shift={(-1.8354,0.7134)}]
\draw  (-3.5321,4.3845) rectangle (-3.0504,4.1716);
\node at (-3.2917,4.2752) {\tiny{$k-4$}};
\end{scope}
\begin{scope}[shift={(-0.5642,0.7057)}]
\draw  (-3.6392,4.4011) rectangle (-2.9855,4.1922);
\node at (-3.2901,4.2995) {\tiny{$36-9k$}};
\end{scope}
\end{scope}
\begin{scope}[shift={(5.1361,0.0036)}]
\draw  (-4.4603,4.9901) ellipse (1.2 and 0.5);
\node (v1) at (-4.3,4.9771) {$1_{3k-2}^{20-6k}$};
\node at (-5.0346,5.215) {\tiny{$e$-$\sum x_i$-$\sum y_i$}};
\begin{scope}[shift={(-1.9849,0.7134)}]
\draw  (-3.6041,4.3993) rectangle (-2.9476,4.1525);
\node at (-3.2917,4.2752) {\tiny{$3k-18$}};
\end{scope}
\begin{scope}[shift={(-0.7461,0.9867)}]
\draw  (-3.6037,4.3711) rectangle (-2.9759,4.1855);
\node at (-3.2917,4.2752) {\tiny{$3k-10$}};
\end{scope}
\begin{scope}[shift={(-0.8309,0.4215)}]
\draw  (-3.6037,4.3711) rectangle (-2.9759,4.1855);
\node at (-3.2917,4.2752) {\tiny{$3k-10$}};
\end{scope}
\node at (-3.5698,5.1726) {\tiny{$x_i$}};
\node at (-3.6701,4.7374) {\tiny{$y_i$}};
\begin{scope}[shift={(0.6811,0.7323)}]
\draw  (-3.6041,4.3993) rectangle (-2.9476,4.1525);
\node at (-3.2917,4.2752) {\tiny{$10-3k$}};
\end{scope}
\end{scope}
\node at (-2.6169,5.2536) {\tiny{$e$}};
\node at (-1.4157,5.2653) {\tiny{$3h$}};
\draw (1.4196,5.3778) .. controls (2.1498,5.7966) and (2.5501,5.538) .. (2.5583,5.1281);
\draw (1.3443,4.5745) .. controls (2.2063,4.2401) and (2.5124,4.3625) .. (2.552,4.8922);
\end{tikzpicture}
\end{array}
\end{align}
where the $20-6k$ blowups on $S^1_C$ are paired up into $10-3k$ pairs of blowups corresponding to $10-3k$ hypers in the $7$-dimensional irreducible representation of $\mf{g}_2$. This graph is same as the one presented in \cite{Bhardwaj:2018yhy}. The elliptic fiber is $f_C=f^0_C+3f^1_C+2f^2_C$. 

\subsection{$\fe_6$, $\fe_7$, $\fe_8$ and $\ff_4$}
For $\fe_6$ on $-k$ curve, $1\le k\le 6$, the associated graph is
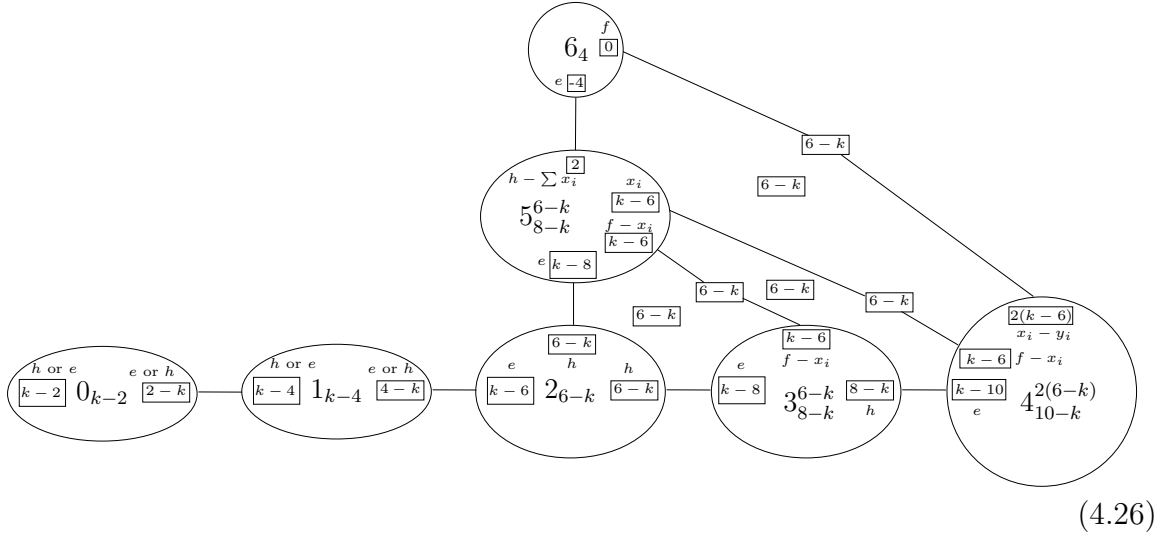
\begin{align}\label{e6g}
\scalebox{.9}{$
\begin{array}{c}
\noindent\begin{tikzpicture}[scale=1.4]
\draw  (-4.4603,4.9901) ellipse (1 and 0.5);
\node (v1) at (-4.4603,4.9901) {$0_{k-2}$};
\draw (-3.4464,5.0091) -- (-2.9847,5.0073);
\node at (-3.937,5.2302) {\tiny{$e$ or $h$}};
\node at (-4.9686,5.2627) {\tiny{$h$ or $e$}};
\begin{scope}[shift={(-1.8167,0.7227)}]
\draw  (-3.5199,4.4183) rectangle (-3.0317,4.1342);
\node at (-3.2917,4.2752) {\tiny{$k-2$}};
\end{scope}
\begin{scope}[shift={(-0.4988,0.715)}]
\draw  (-3.5352,4.3891) rectangle (-3.0397,4.1868);
\node at (-3.2901,4.2995) {\tiny{$2-k$}};
\end{scope}
\begin{scope}[shift={(2.4713,0.0223)}]
\draw  (-4.4603,4.9901) ellipse (1 and 0.5);
\node (v1) at (-4.4603,4.9901) {$1_{k-4}$};
\draw (-3.4464,5.0091) -- (-2.9847,5.0073);
\begin{scope}[shift={(-1.8167,0.7227)}]
\draw  (-3.5199,4.4183) rectangle (-3.0317,4.1342);
\node at (-3.2917,4.2752) {\tiny{$k-4$}};
\end{scope}
\begin{scope}[shift={(-0.4988,0.715)}]
\draw  (-3.5352,4.3891) rectangle (-3.0397,4.1868);
\node at (-3.2901,4.2995) {\tiny{$4-k$}};
\end{scope}
\end{scope}
\begin{scope}[shift={(4.9399,0.0223)}]
\draw  (-4.4603,4.9901) ellipse (1 and 0.7);
\node (v1) at (-4.4603,4.9901) {$2_{6-k}$};
\draw (-3.4464,5.0091) -- (-2.9847,5.0073);
\node at (-3.8462,5.2361) {\tiny{$h$}};
\node at (-5.0811,5.2563) {\tiny{$e$}};
\begin{scope}[shift={(-1.8167,0.7227)}]
\draw  (-3.5199,4.4183) rectangle (-3.0317,4.1342);
\node at (-3.2917,4.2752) {\tiny{$k-6$}};
\end{scope}
\begin{scope}[shift={(-0.4988,0.715)}]
\draw  (-3.5352,4.3891) rectangle (-3.0397,4.1868);
\node at (-3.2901,4.2995) {\tiny{$6-k$}};
\end{scope}
\end{scope}
\begin{scope}[shift={(7.426,0.0223)}]
\draw  (-4.4603,4.9901) ellipse (1 and 0.7);
\node (v1) at (-4.4259,4.8705) {$3^{6-k}_{8-k}$};
\draw (-3.4464,5.0091) -- (-2.9847,5.0073);
\node at (-3.7797,4.7913) {\tiny{$h$}};
\node at (-5.1466,5.2663) {\tiny{$e$}};
\begin{scope}[shift={(-1.8625,0.7278)}]
\draw  (-3.5199,4.4183) rectangle (-3.0317,4.1342);
\node at (-3.2917,4.2752) {\tiny{$k-8$}};
\end{scope}
\begin{scope}[shift={(-0.4988,0.715)}]
\draw  (-3.5352,4.3891) rectangle (-3.0397,4.1868);
\node at (-3.2901,4.2995) {\tiny{$8-k$}};
\end{scope}
\begin{scope}[shift={(-1.1701,1.2542)}]
\draw  (-3.5352,4.3891) rectangle (-3.0397,4.1868);
\node at (-3.2901,4.2995) {\tiny{$k-6$}};
\end{scope}
\node at (-4.4576,5.3148) {\tiny{$f-x_i$}};
\end{scope}
\begin{scope}[shift={(9.9121,0.0223)}]
\draw  (-4.4603,4.9901) ellipse (1 and 1);
\node (v1) at (-4.2758,4.8851) {$4^{2(6-k)}_{10-k}$};
\node at (-5.1443,4.7784) {\tiny{$e$}};
\begin{scope}[shift={(-1.8241,0.7315)}]
\draw  (-3.5803,4.3865) rectangle (-3.0157,4.1716);
\node at (-3.2917,4.2752) {\tiny{$k-10$}};
\end{scope}
\begin{scope}[shift={(-1.7545,1.0524)}]
\draw  (-3.5696,4.3865) rectangle (-3.0341,4.1871);
\node at (-3.2917,4.2752) {\tiny{$k-6$}};
\end{scope}
\begin{scope}[shift={(-1.1716,1.507)}]
\draw  (-3.6354,4.3787) rectangle (-2.9623,4.193);
\node at (-3.2917,4.2752) {\tiny{$2(k-6)$}};
\end{scope}
\node at (-4.483,5.3393) {\tiny{$f-x_i$}};
\node at (-4.4334,5.599) {\tiny{$x_i-y_i$}};
\end{scope}
\node at (-2.4487,5.3096) {\tiny{$h$ or $e$}};
\node at (-1.4157,5.2653) {\tiny{$e$ or $h$}};
\begin{scope}[shift={(3.7837,1.2228)}]
\draw  (-3.5352,4.3891) rectangle (-3.0397,4.1868);
\node at (-3.2901,4.2995) {\tiny{$6-k$}};
\end{scope}
\node at (0.496,5.3158) {\tiny{$h$}};
\begin{scope}[shift={(4.9906,1.8718)}]
\draw  (-4.4603,4.9901) ellipse (1 and 0.7);
\node (v1) at (-4.7772,4.9799) {$5^{6-k}_{8-k}$};
\node at (-4.7985,5.3804) {\tiny{$h-\sum x_i$}};
\node at (-4.8141,4.5057) {\tiny{$e$}};
\begin{scope}[shift={(-1.2082,0.2019)}]
\draw  (-3.5199,4.4183) rectangle (-3.0317,4.1342);
\node at (-3.2917,4.2752) {\tiny{$k-8$}};
\end{scope}
\begin{scope}[shift={(-0.6096,0.4237)}]
\draw  (-3.5352,4.3891) rectangle (-3.0397,4.1868);
\node at (-3.2901,4.2995) {\tiny{$k-6$}};
\end{scope}
\begin{scope}[shift={(-1.1678,1.2435)}]
\draw  (-3.3836,4.3758) rectangle (-3.1988,4.2014);
\node at (-3.2901,4.2995) {\tiny{2}};
\end{scope}
\begin{scope}[shift={(-0.5382,0.8499)}]
\draw  (-3.5352,4.3891) rectangle (-3.0397,4.1868);
\node at (-3.2901,4.2995) {\tiny{$k-6$}};
\end{scope}
\node at (-3.8901,4.8966) {\tiny{$f-x_i$}};
\node at (-3.8342,5.3391) {\tiny{$x_i$}};
\end{scope}
\begin{scope}[shift={(4.9957,3.6315)}]
\draw  (-4.4603,4.9901) ellipse (0.5 and 0.5);
\node (v1) at (-4.4505,4.9971) {$6_4$};
\node at (-4.1444,5.2221) {\tiny{$f$}};
\node at (-4.6425,4.6524) {\tiny{$e$}};
\begin{scope}[shift={(-1.1627,0.3484)}]
\draw  (-3.3836,4.3758) rectangle (-3.1988,4.2014);
\node at (-3.2901,4.2995) {\tiny{-4}};
\end{scope}
\begin{scope}[shift={(-0.822,0.7248)}]
\draw  (-3.3836,4.3758) rectangle (-3.1988,4.2014);
\node at (-3.2901,4.2995) {\tiny{0}};
\end{scope}
\end{scope}
\begin{scope}[shift={(5.3412,1.7739)}]
\draw  (-3.5352,4.3891) rectangle (-3.0397,4.1868) node (v2) {};
\node at (-3.2901,4.2995) {\tiny{$6-k$}};
\end{scope}
\begin{scope}[shift={(7.1327,1.667)}]
\draw  (-3.5352,4.3891) rectangle (-3.0397,4.1868);
\node at (-3.2901,4.2995) {\tiny{$6-k$}};
\end{scope}
\begin{scope}[shift={(6.4588,3.3302)}]
\draw  (-3.5352,4.3891) rectangle (-3.0397,4.1868);
\node at (-3.2901,4.2995) {\tiny{$6-k$}};
\end{scope}
\draw (0.5114,6.1638) -- (0.5114,5.7178);
\draw (0.5314,7.5621) -- (0.5355,8.1165);
\draw (1.3953,6.5134) -- (1.8855,6.1678) (v2);
\draw (2.2954,5.9991) -- (2.8981,5.7178) (4.5695,5.5048) -- (3.9709,5.8585) (1.5239,6.9272) -- (3.5851,6.0232);
\draw (1.0326,8.6029) -- (2.9927,7.7162) (3.3381,7.5108) -- (5.3636,6.0174);
\begin{scope}[shift={(5.9921,2.8915)}]
\draw  (-3.5352,4.3891) rectangle (-3.0397,4.1868);
\node at (-3.2901,4.2995) {\tiny{$6-k$}};
\end{scope}
\begin{scope}[shift={(6.0855,1.7994)}]
\draw  (-3.5352,4.3891) rectangle (-3.0397,4.1868);
\node at (-3.2901,4.2995) {\tiny{$6-k$}};
\end{scope}
\begin{scope}[shift={(4.6761,1.5194)}]
\draw  (-3.5352,4.3891) rectangle (-3.0397,4.1868);
\node at (-3.2901,4.2995) {\tiny{$6-k$}};
\end{scope}
\end{tikzpicture}
\end{array}
$}
\end{align}
where the blowups correspond to $6-k$ hypers in the 27-dimensional irreducible representation of $\fe_6$. This graph is same as the one presented in \cite{Bhardwaj:2018yhy}. There are $6-k$ special points $p_i,i=1,\cdots,6-k$. Over $p_i$, $f^3_C$ splits into $f^3_C-x_i,x_i$, $f^4_C$ splits into $f^4_C-x_i,x_i-y_i,y_i$, and $f^5_C$ splits into $f^5_C-x_i,x_i$. Moreover, $f^3_C-x_i$ in $S^3_C$ is identified with $f^5_C-x_i$ in $S^5_C$; $f^4_C-x_i$ in $S^4_C$ is identified with $x_i$ in $S^5_C$; and $x_i-y_i$ in $S^4_C$ is identified with $f^6_C$ in $S^6_C$.

For $\ff_4$ on $-k$ curve, $1\le k\le 5$, the associated graph is
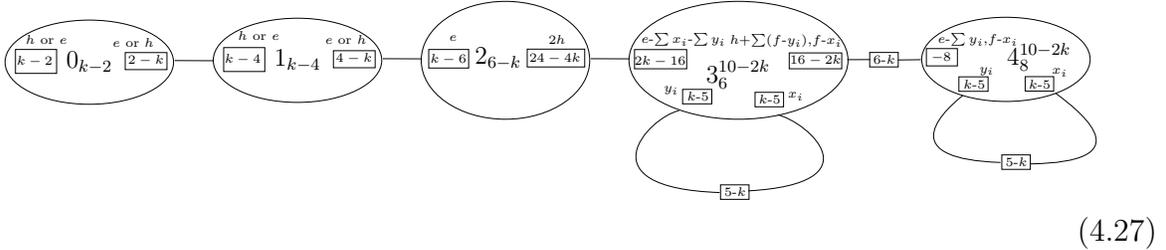
\begin{align}
\scalebox{.8}{$
\begin{array}{c}
\noindent\begin{tikzpicture}[scale=1.4]
\draw  (-4.4603,4.9901) ellipse (1 and 0.5);
\node (v1) at (-4.4603,4.9901) {$0_{k-2}$};
\draw (-3.4464,5.0091) -- (-2.9847,5.0073);
\node at (-3.937,5.2302) {\tiny{$e$ or $h$}};
\node at (-4.9686,5.2627) {\tiny{$h$ or $e$}};
\begin{scope}[shift={(-1.8167,0.7227)}]
\draw  (-3.5199,4.4183) rectangle (-3.0317,4.1342);
\node at (-3.2917,4.2752) {\tiny{$k-2$}};
\end{scope}
\begin{scope}[shift={(-0.4988,0.715)}]
\draw  (-3.5352,4.3891) rectangle (-3.0397,4.1868);
\node at (-3.2901,4.2995) {\tiny{$2-k$}};
\end{scope}
\begin{scope}[shift={(2.4713,0.0223)}]
\draw  (-4.4603,4.9901) ellipse (1 and 0.5);
\node (v1) at (-4.4603,4.9901) {$1_{k-4}$};
\draw (-3.4464,5.0091) -- (-2.9847,5.0073);
\begin{scope}[shift={(-1.8167,0.7227)}]
\draw  (-3.5199,4.4183) rectangle (-3.0317,4.1342);
\node at (-3.2917,4.2752) {\tiny{$k-4$}};
\end{scope}
\begin{scope}[shift={(-0.4988,0.715)}]
\draw  (-3.5352,4.3891) rectangle (-3.0397,4.1868);
\node at (-3.2901,4.2995) {\tiny{$4-k$}};
\end{scope}
\end{scope}
\begin{scope}[shift={(4.9399,0.0223)}]
\draw  (-4.4603,4.9901) ellipse (1 and 0.7);
\node (v1) at (-4.5471,5.0079) {$2_{6-k}$};
\draw (-3.4464,5.0091) -- (-2.9847,5.0073);
\node at (-3.8462,5.2361) {\tiny{$2h$}};
\node at (-5.0811,5.2563) {\tiny{$e$}};
\begin{scope}[shift={(-1.8447,0.7227)}]
\draw  (-3.5266,4.3959) rectangle (-3.0525,4.1755);
\node at (-3.2917,4.2752) {\tiny{$k-6$}};
\end{scope}
\begin{scope}[shift={(-0.5829,0.7149)}]
\draw  (-3.6193,4.4079) rectangle (-2.9369,4.1961);
\node at (-3.2901,4.2995) {\tiny{$24-4k$}};
\end{scope}
\end{scope}
\begin{scope}[shift={(7.706,0.0223)}]
\draw  (-4.4603,4.9901) ellipse (1.3 and 0.7);
\node (v1) at (-4.4638,4.8214) {$3^{10-2k}_{6}$};
\node at (-3.8973,5.2208) {\tiny{$h$+$\sum(f$-$y_i)$,$f$-$x_i$}};
\node at (-5.1118,5.2382) {\tiny{$e$-$\sum x_i$-$\sum y_i$}};
\begin{scope}[shift={(-2.1745,0.6991)}]
\draw  (-3.5199,4.4183) rectangle (-2.9009,4.1903);
\node at (-3.217,4.2845) {\tiny{$2k-16$}};
\end{scope}
\begin{scope}[shift={(-0.195,0.7027)}]
\draw  (-3.6632,4.3751) rectangle (-3.0397,4.1868);
\node at (-3.3477,4.2872) {\tiny{$16-2k$}};
\end{scope}
\begin{scope}[shift={(-0.6827,0.2622)}]
\draw  (-3.5812,4.3474) rectangle (-3.2427,4.1755);
\node at (-3.4079,4.2629) {\tiny{$k$-$5$}};
\end{scope}
\node at (-3.7806,4.5621) {\tiny{$x_i$}};
\begin{scope}[shift={(-1.5315,0.2993)}]
\draw  (-3.5922,4.3454) rectangle (-3.2458,4.1735);
\node at (-3.4079,4.2629) {\tiny{$k$-$5$}};
\end{scope}
\node at (-5.2577,4.6334) {\tiny{$y_i$}};
\end{scope}
\begin{scope}[shift={(10.5041,0.0383)}]
\draw  (-4.0844,4.9852) ellipse (1 and 0.5);
\node (v1) at (-3.6903,5.013) {$4_{8}^{10-2k}$};
\node at (-4.4161,5.2208) {\tiny{$e$-$\sum y_i$,$f$-$x_i$}};
\begin{scope}[shift={(-1.4482,0.7266)}]
\draw  (-3.5803,4.3865) rectangle (-3.2119,4.181);
\node at (-3.4038,4.2752) {\tiny{$-8$}};
\end{scope}
\begin{scope}[shift={(-0.2745,0.4326)}]
\draw  (-3.5812,4.3474) rectangle (-3.2427,4.1755);
\node at (-3.4079,4.2629) {\tiny{$k$-$5$}};
\end{scope}
\node at (-3.4433,4.8264) {\tiny{$x_i$}};
\begin{scope}[shift={(-1.0502,0.4322)}]
\draw  (-3.5922,4.3454) rectangle (-3.2458,4.1735);
\node at (-3.4079,4.2629) {\tiny{$k$-$5$}};
\end{scope}
\node at (-4.3113,4.8476) {\tiny{$y_i$}};
\begin{scope}[shift={(-0.5518,-0.4969)}]
\draw  (-3.5812,4.3474) rectangle (-3.2427,4.1755);
\node at (-3.4079,4.2629) {\tiny{$5$-$k$}};
\end{scope}
\begin{scope}[shift={(-3.8907,-0.853)}]
\draw  (-3.5812,4.3474) rectangle (-3.2427,4.1755);
\node at (-3.4079,4.2629) {\tiny{$5$-$k$}};
\end{scope}
\begin{scope}[shift={(-2.1198,0.7199)}]
\draw  (-3.5812,4.3474) rectangle (-3.2427,4.1755);
\node at (-3.4079,4.2629) {\tiny{$6$-$k$}};
\end{scope}
\end{scope}
\node at (-2.4487,5.3096) {\tiny{$h$ or $e$}};
\node at (-1.4157,5.2653) {\tiny{$e$ or $h$}};
\draw (7.0086,4.6218) .. controls (7.3918,4.3408) and (7.9985,3.8024) .. (6.709,3.782);
\draw (5.9079,4.5945) .. controls (5.5761,4.3282) and (5.1668,3.8385) .. (6.3546,3.7989);
\draw (3.7107,4.3651) .. controls (4.087,4.2717) and (4.874,3.5475) .. (3.3621,3.4613);
\draw (2.5418,4.4163) .. controls (2.1706,4.3296) and (1.3929,3.4957) .. (3.0228,3.4415);
\draw (4.5367,5.0198) -- (4.809,5.0209) (5.1395,5.0163) -- (5.4091,5.0197);
\end{tikzpicture}
\end{array}
$}
\end{align}
where the blowups correspond to $5-k$ hypers in the 26-dimensional irreducible representation of $\ff_4$.  These hypers cannot be gauged to produce other $6d$ SCFTs. Notice that there was only a single gluing curve between $S^3_C$ and $S^4_C$ in the proposal of \cite{Bhardwaj:2018yhy}. We have resolved this gluing into multiple gluings in our answer presented above. This resolution can be seen by studying in detail the behavior of $f_C$ over $C$. The presence of these extra gluings can also be seen by folding the geometry for $\fe_6$ presented in (\ref{e6g}). Under the folding process $S^5_C$ and $S^3_C$ in $\fe_6$ geometry descend to $S^3_C$ in the $\ff_4$ geometry, and $S^6_C$ and $S^4_C$ in $\fe_6$ geometry descend to $S^4_C$ in $\ff_4$ geometry. The gluing curves between $S^5_C$ and $S^4_C$ in the $\fe_6$ geometry descend to the extra gluing curves between $S^3_C$ and $S^4_C$ in the $\ff_4$ geometry. Hence, our answer is more complete than the answer presented in \cite{Bhardwaj:2018yhy}.

$\fe_7$ on $-k$ curve, $1\le k\le 8$, the associated graph is
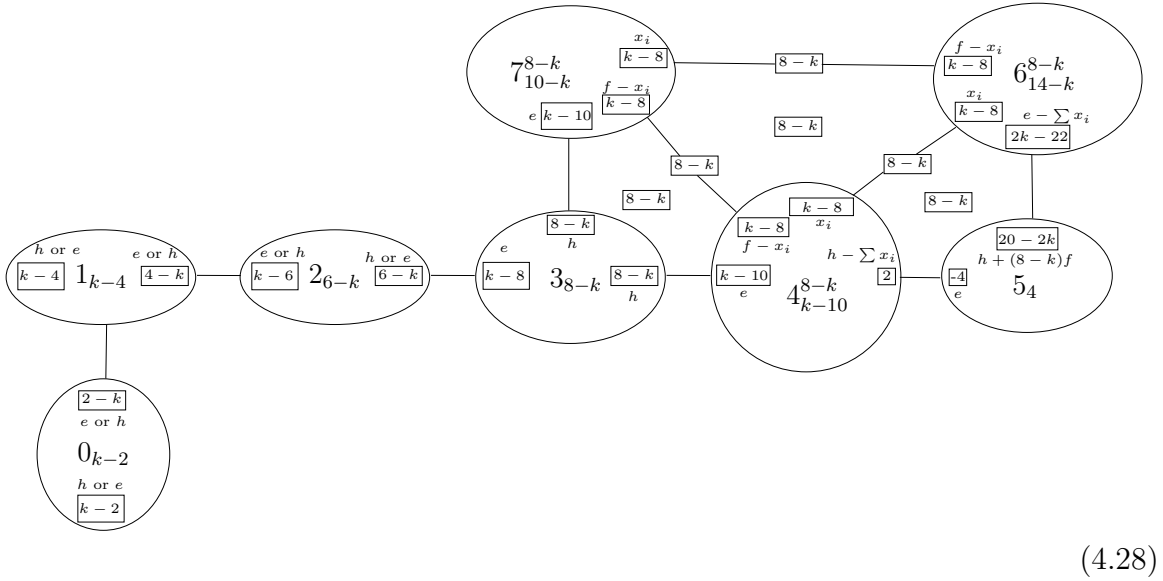
\begin{align}
\scalebox{.9}{$
\begin{array}{c}
\noindent\begin{tikzpicture}[scale=1.4]
\draw  (-1.9624,3.1392) ellipse (0.7 and 0.8);
\node (v1) at (-1.9624,3.1392) {$0_{k-2}$};
\node at (-1.9629,3.4946) {\tiny{$e$ or $h$}};
\node at (-1.996,2.8287) {\tiny{$h$ or $e$}};
\begin{scope}[shift={(1.2877,-1.7077)}]
\draw  (-3.5199,4.4183) rectangle (-3.0317,4.1342);
\node at (-3.2917,4.2752) {\tiny{$k-2$}};
\end{scope}
\begin{scope}[shift={(1.3118,-0.5766)}]
\draw  (-3.5352,4.3891) rectangle (-3.0397,4.1868);
\node at (-3.2901,4.2995) {\tiny{$2-k$}};
\end{scope}
\begin{scope}[shift={(2.4713,0.0223)}]
\draw  (-4.4603,4.9901) ellipse (1 and 0.5);
\node (v1) at (-4.4603,4.9901) {$1_{k-4}$};
\draw (-3.4464,5.0091) -- (-2.9847,5.0073);
\begin{scope}[shift={(-1.8167,0.7227)}]
\draw  (-3.5199,4.4183) rectangle (-3.0317,4.1342);
\node at (-3.2917,4.2752) {\tiny{$k-4$}};
\end{scope}
\begin{scope}[shift={(-0.4988,0.715)}]
\draw  (-3.5352,4.3891) rectangle (-3.0397,4.1868);
\node at (-3.2901,4.2995) {\tiny{$4-k$}};
\end{scope}
\end{scope}
\begin{scope}[shift={(4.9399,0.0223)}]
\draw  (-4.4603,4.9901) ellipse (1 and 0.5);
\node (v1) at (-4.4603,4.9901) {$2_{6-k}$};
\draw (-3.4464,5.0091) -- (-2.9847,5.0073);
\node at (-3.8997,5.2056) {\tiny{$h$ or $e$}};
\node at (-5.0091,5.2503) {\tiny{$e$ or $h$}};
\begin{scope}[shift={(-1.8167,0.7227)}]
\draw  (-3.5199,4.4183) rectangle (-3.0317,4.1342);
\node at (-3.2917,4.2752) {\tiny{$k-6$}};
\end{scope}
\begin{scope}[shift={(-0.4988,0.715)}]
\draw  (-3.5352,4.3891) rectangle (-3.0397,4.1868);
\node at (-3.2901,4.2995) {\tiny{$6-k$}};
\end{scope}
\end{scope}
\begin{scope}[shift={(7.426,0.0223)}]
\draw  (-4.4603,4.9901) ellipse (1 and 0.7);
\node (v1) at (-4.4026,4.949) {$3_{8-k}$};
\draw (-3.4464,5.0091) -- (-2.9847,5.0073);
\node at (-3.7958,4.7913) {\tiny{$h$}};
\node at (-5.1627,5.2842) {\tiny{$e$}};
\begin{scope}[shift={(-1.8625,0.7278)}]
\draw  (-3.5199,4.4183) rectangle (-3.0317,4.1342);
\node at (-3.2917,4.2752) {\tiny{$k-8$}};
\end{scope}
\begin{scope}[shift={(-0.4988,0.715)}]
\draw  (-3.5352,4.3891) rectangle (-3.0397,4.1868);
\node at (-3.2901,4.2995) {\tiny{$8-k$}};
\end{scope}
\begin{scope}[shift={(-1.1701,1.2542)}]
\draw  (-3.5352,4.3891) rectangle (-3.0397,4.1868);
\node at (-3.2901,4.2995) {\tiny{$8-k$}};
\end{scope}
\node at (-4.4422,5.3521) {\tiny{$h$}};
\end{scope}
\begin{scope}[shift={(9.9121,0.0223)}]
\draw  (-4.4603,4.9901) ellipse (1 and 1);
\node (v1) at (-4.3411,4.7916) {$4^{8-k}_{k-10}$};
\node at (-4.2701,5.5496) {\tiny{$x_i$}};
\node at (-5.1195,4.8156) {\tiny{$e$}};
\begin{scope}[shift={(-1.8241,0.7315)}]
\draw  (-3.5803,4.3865) rectangle (-3.0157,4.1716);
\node at (-3.2917,4.2752) {\tiny{$k-10$}};
\end{scope}
\begin{scope}[shift={(-1.6086,1.2309)}]
\draw  (-3.5696,4.3865) rectangle (-3.0341,4.1871);
\node at (-3.2917,4.2752) {\tiny{$k-8$}};
\end{scope}
\begin{scope}[shift={(-0.999,1.4473)}]
\draw  (-3.6354,4.3787) rectangle (-2.9623,4.193);
\node at (-3.2917,4.2752) {\tiny{$k-8$}};
\end{scope}
\node at (-4.8758,5.3108) {\tiny{$f-x_i$}};
\node at (-3.8739,5.2177) {\tiny{$h-\sum x_i$}};
\begin{scope}[shift={(-0.3191,0.7086)}]
\draw  (-3.3836,4.3758) rectangle (-3.1988,4.2014);
\node at (-3.2901,4.2995) {\tiny{2}};
\end{scope}
\end{scope}
\node at (-2.4487,5.3096) {\tiny{$h$ or $e$}};
\node at (-1.4157,5.2653) {\tiny{$e$ or $h$}};
\begin{scope}[shift={(7.4323,2.1881)}]
\draw  (-4.4603,4.9901) ellipse (1.1 and 0.7);
\node (v1) at (-4.7772,4.9799) {$7^{8-k}_{10-k}$};
\node at (-4.8666,4.5116) {\tiny{$e$}};
\begin{scope}[shift={(-1.2082,0.2506)}]
\draw  (-3.566,4.4142) rectangle (-3.0317,4.1342);
\node at (-3.2917,4.2752) {\tiny{$k-10$}};
\end{scope}
\begin{scope}[shift={(-0.5934,0.3588)}]
\draw  (-3.5352,4.3891) rectangle (-3.0397,4.1868);
\node at (-3.2901,4.2995) {\tiny{$k-8$}};
\end{scope}
\begin{scope}[shift={(-0.4084,0.8499)}]
\draw  (-3.5352,4.3891) rectangle (-3.0397,4.1868);
\node at (-3.2901,4.2995) {\tiny{$k-8$}};
\end{scope}
\node at (-3.8739,4.8317) {\tiny{$f-x_i$}};
\node at (-3.7044,5.3391) {\tiny{$x_i$}};
\end{scope}
\begin{scope}[shift={(8.6658,2.9811)}]
\draw  (-3.5352,4.3891) rectangle (-3.0397,4.1868);
\node at (-3.2901,4.2995) {\tiny{$8-k$}};
\end{scope}
\begin{scope}[shift={(12.2424,0.0386)}]
\draw  (-4.4603,4.9901) ellipse (0.9 and 0.6);
\node (v1) at (-4.4815,4.8695) {$5_4$};
\begin{scope}[shift={(-1.1677,1.0942)}]
\draw  (-3.6091,4.4021) rectangle (-2.9668,4.1676);
\node at (-3.2917,4.2752) {\tiny{$20-2k$}};
\end{scope}
\begin{scope}[shift={(-1.8978,0.6838)}]
\draw  (-3.3836,4.3758) rectangle (-3.1988,4.2014);
\node at (-3.2901,4.2995) {\tiny{-4}};
\end{scope}
\node at (-4.475,5.1525) {\tiny{$h+(8-k)f$}};
\node at (-5.1889,4.7875) {\tiny{$e$}};
\end{scope}
\begin{scope}[shift={(12.3561,2.1151)}]
\draw  (-4.4603,4.9901) ellipse (1.1 and 0.8);
\node (v1) at (-4.3879,5.034) {$6^{8-k}_{14-k}$};
\node at (-4.2582,4.6008) {\tiny{$e-\sum x_i$}};
\begin{scope}[shift={(-1.1476,0.0987)}]
\draw  (-3.6492,4.4029) rectangle (-2.9749,4.1557);
\node at (-3.2917,4.2752) {\tiny{$2k-22$}};
\end{scope}
\begin{scope}[shift={(-1.91,0.8402)}]
\draw  (-3.5352,4.3891) rectangle (-3.0397,4.1868);
\node at (-3.2901,4.2995) {\tiny{$k-8$}};
\end{scope}
\begin{scope}[shift={(-1.7955,0.3551)}]
\draw  (-3.5352,4.3891) rectangle (-3.0397,4.1868);
\node at (-3.2901,4.2995) {\tiny{$k-8$}};
\end{scope}
\node at (-5.0824,5.3266) {\tiny{$f-x_i$}};
\node at (-5.124,4.8171) {\tiny{$x_i$}};
\end{scope}
\begin{scope}[shift={(7.5626,1.9022)}]
\draw  (-3.5352,4.3891) rectangle (-3.0397,4.1868);
\node at (-3.2901,4.2995) {\tiny{$8-k$}};
\end{scope}
\begin{scope}[shift={(9.8095,1.9184)}]
\draw  (-3.5352,4.3891) rectangle (-3.0397,4.1868);
\node at (-3.2901,4.2995) {\tiny{$8-k$}};
\end{scope}
\draw (2.9474,6.4759) -- (2.9474,5.7158) (6.441,5.0107) -- (6.8668,5.0061);
\draw (7.833,6.3065) -- (7.8421,5.6334);
\draw (5.6168,7.2589) -- (6.8027,7.2451) (4.0554,7.2772) -- (5.1223,7.2634) (3.7761,6.7003) -- (4.1195,6.2927) (4.3027,6.0867) -- (4.7148,5.7021) (7.0317,6.5949) -- (6.615,6.3111) (6.267,6.1233) -- (5.9465,5.8761);
\begin{scope}[shift={(8.6566,2.3126)}]
\draw  (-3.5352,4.3891) rectangle (-3.0397,4.1868);
\node at (-3.2901,4.2995) {\tiny{$8-k$}};
\end{scope}
\begin{scope}[shift={(7.054,1.5388)}]
\draw  (-3.5352,4.3891) rectangle (-3.0397,4.1868);
\node at (-3.2901,4.2995) {\tiny{$8-k$}};
\end{scope}
\begin{scope}[shift={(10.2409,1.5159)}]
\draw  (-3.5352,4.3891) rectangle (-3.0397,4.1868);
\node at (-3.2901,4.2995) {\tiny{$8-k$}};
\end{scope}
\draw (-1.9287,4.5105) -- (-1.937,3.9415);
\end{tikzpicture}
\end{array}
$}
\end{align}
where the blowups correspond to $8-k$ half-hypers in the 56-dimensional irreducible representation of $\fe_7$. This graph is same as the one presented in \cite{Bhardwaj:2018yhy}. At special points $p_i,i=1,\cdots,8-k$, $f^6_C$ splits into $f^6_C-x_i,x_i$, $f^7_C$ splits into $f^7_C-x_i,x_i$, and $f^4_C$ splits into $f^4_C-x_i,x_i$. Moreover, $f^4_C-x_i\sim f^7_C-x_i$, $f^6_C-x_i$ is identified with $x_i$ in $S^7_C$, and $x_i$ in $S^4_C$ is identified with $x_i$ in $S^6_C$.

For $\fe_8$ on $-12$ curve, the associated graph is
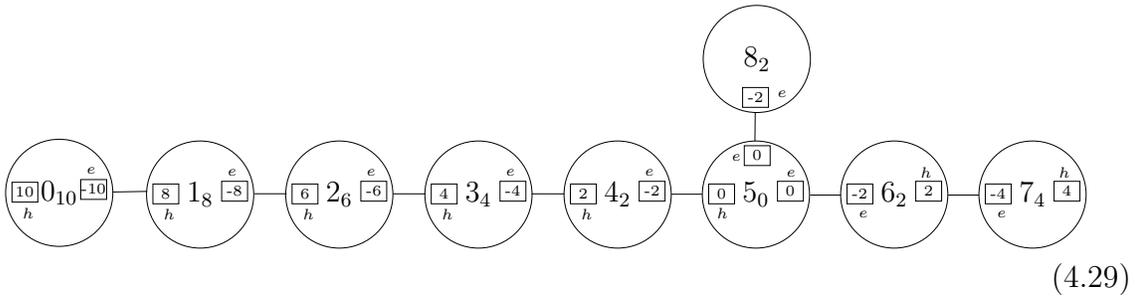
\begin{align}
\scalebox{.95}{$
\begin{array}{c}
\noindent\begin{tikzpicture}[scale=1.5]
\draw  (-3.9479,4.9946) ellipse (0.5 and 0.5);
\node (v1) at (-3.9479,4.9946) {$0_{10}$};
\node at (-4.2364,4.8087) {\tiny{$h$}};
\node at (-3.6501,5.2038) {\tiny{$e$}};
\begin{scope}[shift={(-0.9773,0.7289)}]
\draw  (-3.4113,4.3642) rectangle (-3.17,4.1794);
\node at (-3.2917,4.2752) {\tiny{10}};
\end{scope}
\begin{scope}[shift={(-0.3357,0.754)}]
\draw  (-3.4113,4.3642) rectangle (-3.17,4.1794);
\node at (-3.2917,4.2752) {\tiny{-10}};
\end{scope}
\begin{scope}[shift={(1.3146,-0.0162)}]
\draw  (-3.9479,4.9946) ellipse (0.5 and 0.5);
\node (v1) at (-3.9479,4.9946) {$1_{8}$};
\node at (-4.2364,4.8087) {\tiny{$h$}};
\node at (-3.6501,5.2038) {\tiny{$e$}};
\begin{scope}[shift={(-0.9773,0.7289)}]
\draw  (-3.4113,4.3642) rectangle (-3.17,4.1794);
\node at (-3.2917,4.2752) {\tiny{8}};
\end{scope}
\begin{scope}[shift={(-0.3357,0.754)}]
\draw  (-3.4113,4.3642) rectangle (-3.17,4.1794);
\node at (-3.2917,4.2752) {\tiny{-8}};
\end{scope}
\end{scope}
\begin{scope}[shift={(2.6129,-0.0162)}]
\draw  (-3.9479,4.9946) ellipse (0.5 and 0.5);
\node (v1) at (-3.9479,4.9946) {$2_{6}$};
\node at (-4.2364,4.8087) {\tiny{$h$}};
\node at (-3.6501,5.2038) {\tiny{$e$}};
\begin{scope}[shift={(-0.9773,0.7289)}]
\draw  (-3.4113,4.3642) rectangle (-3.17,4.1794);
\node at (-3.2917,4.2752) {\tiny{6}};
\end{scope}
\begin{scope}[shift={(-0.3357,0.754)}]
\draw  (-3.4113,4.3642) rectangle (-3.17,4.1794);
\node at (-3.2917,4.2752) {\tiny{-6}};
\end{scope}
\end{scope}
\begin{scope}[shift={(3.9112,-0.0162)}]
\draw  (-3.9479,4.9946) ellipse (0.5 and 0.5);
\node (v1) at (-3.9479,4.9946) {$3_{4}$};
\node at (-4.2364,4.8087) {\tiny{$h$}};
\node at (-3.6501,5.2038) {\tiny{$e$}};
\begin{scope}[shift={(-0.9773,0.7289)}]
\draw  (-3.4113,4.3642) rectangle (-3.17,4.1794);
\node at (-3.2917,4.2752) {\tiny{4}};
\end{scope}
\begin{scope}[shift={(-0.3357,0.754)}]
\draw  (-3.4113,4.3642) rectangle (-3.17,4.1794);
\node at (-3.2917,4.2752) {\tiny{-4}};
\end{scope}
\end{scope}
\begin{scope}[shift={(5.2095,-0.0162)}]
\draw  (-3.9479,4.9946) ellipse (0.5 and 0.5);
\node (v1) at (-3.9479,4.9946) {$4_{2}$};
\node at (-4.2364,4.8087) {\tiny{$h$}};
\node at (-3.6501,5.2038) {\tiny{$e$}};
\begin{scope}[shift={(-0.9773,0.7289)}]
\draw  (-3.4113,4.3642) rectangle (-3.17,4.1794);
\node at (-3.2917,4.2752) {\tiny{2}};
\end{scope}
\begin{scope}[shift={(-0.3357,0.754)}]
\draw  (-3.4113,4.3642) rectangle (-3.17,4.1794);
\node at (-3.2917,4.2752) {\tiny{-2}};
\end{scope}
\end{scope}
\begin{scope}[shift={(6.4997,-0.0162)}]
\draw  (-3.9479,4.9946) ellipse (0.5 and 0.5);
\node (v1) at (-3.9479,4.9946) {$5_{0}$};
\node at (-4.2598,4.8234) {\tiny{$h$}};
\node at (-3.6211,5.1939) {\tiny{$e$}};
\begin{scope}[shift={(-0.9773,0.7289)}]
\draw  (-3.4113,4.3642) rectangle (-3.17,4.1794);
\node at (-3.2917,4.2752) {\tiny{0}};
\end{scope}
\begin{scope}[shift={(-0.3357,0.754)}]
\draw  (-3.4113,4.3642) rectangle (-3.17,4.1794);
\node at (-3.2917,4.2752) {\tiny{0}};
\end{scope}
\begin{scope}[shift={(-0.6452,1.0863)}]
\draw  (-3.4113,4.3642) rectangle (-3.17,4.1794);
\node at (-3.2917,4.2752) {\tiny{0}};
\end{scope}
\node at (-4.1293,5.3438) {\tiny{$e$}};
\end{scope}
\begin{scope}[shift={(7.7899,-0.0162)}]
\draw  (-3.9479,4.9946) ellipse (0.5 and 0.5);
\node (v1) at (-3.9479,4.9946) {$6_{2}$};
\node at (-4.2364,4.8087) {\tiny{$e$}};
\node at (-3.6501,5.2038) {\tiny{$h$}};
\begin{scope}[shift={(-0.9773,0.7289)}]
\draw  (-3.4113,4.3642) rectangle (-3.17,4.1794);
\node at (-3.2917,4.2752) {\tiny{-2}};
\end{scope}
\begin{scope}[shift={(-0.3357,0.754)}]
\draw  (-3.4113,4.3642) rectangle (-3.17,4.1794);
\node at (-3.2917,4.2752) {\tiny{2}};
\end{scope}
\end{scope}
\begin{scope}[shift={(9.0801,-0.0162)}]
\draw  (-3.9479,4.9946) ellipse (0.5 and 0.5);
\node (v1) at (-3.9479,4.9946) {$7_{4}$};
\node at (-4.2364,4.8087) {\tiny{$e$}};
\node at (-3.6501,5.2038) {\tiny{$h$}};
\begin{scope}[shift={(-0.9773,0.7289)}]
\draw  (-3.4113,4.3642) rectangle (-3.17,4.1794);
\node at (-3.2917,4.2752) {\tiny{-4}};
\end{scope}
\begin{scope}[shift={(-0.3357,0.754)}]
\draw  (-3.4113,4.3642) rectangle (-3.17,4.1794);
\node at (-3.2917,4.2752) {\tiny{4}};
\end{scope}
\end{scope}
\begin{scope}[shift={(6.5079,1.2497)}]
\draw  (-3.9479,4.9946) ellipse (0.5 and 0.5);
\node (v1) at (-3.9479,4.9946) {$8_{2}$};
\node at (-3.7125,4.6657) {\tiny{$e$}};
\begin{scope}[shift={(-0.669,0.3637)}]
\draw  (-3.4113,4.3642) rectangle (-3.17,4.1794);
\node at (-3.2917,4.2752) {\tiny{-2}};
\end{scope}
\end{scope}
\draw (-3.4455,5.0023) -- (-3.1276,5.0079);
\draw (-0.8335,4.9851) -- (-0.5307,4.985);
\begin{scope}[shift={(1.2939,0)}]
\draw (-0.8335,4.9851) -- (-0.5307,4.985);
\end{scope}
\begin{scope}[shift={(2.5809,-0.0034)}]
\draw (-0.8173,4.9851) -- (-0.5307,4.985);
\end{scope}
\draw (-2.1263,4.9858) -- (-1.8269,4.9858);
\begin{scope}[shift={(3.8744,-0.0116)}]
\draw (-0.8173,4.9851) -- (-0.5307,4.985);
\end{scope}
\begin{scope}[shift={(5.1623,-0.0116)}]
\draw (-0.8173,4.9851) -- (-0.5307,4.985);
\end{scope}
\draw (2.5451,5.7438) -- (2.5423,5.4774);
\end{tikzpicture}
\end{array}
$}
\end{align}
where the corresponding gauge algebra is $\fe_8$ and there are no charged hypers, which goes well with the fact that there are no blowups. This is the same as the graph given in \cite{Bhardwaj:2018yhy}.

\subsection{$\su(3)$ on $-3$}
The associated graph is
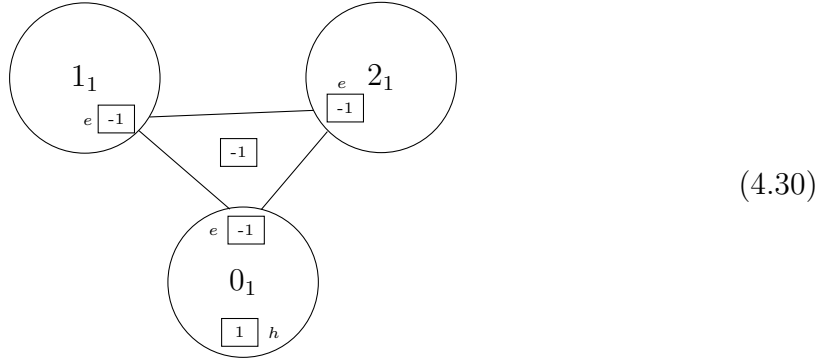
\begin{align}
\begin{array}{c}
\begin{tikzpicture}[scale=2]
\draw  (-3.9479,4.9946) ellipse (0.5 and 0.5);
\node (v1) at (-3.9479,4.9946) {$0_1$};
\node at (-4.1458,5.3303) {\tiny{$e$}};
\begin{scope}[shift={(-0.6371,1.0692)}]
\draw  (-3.4113,4.3642) rectangle (-3.17,4.1794);
\node at (-3.2917,4.2752) {\tiny{-1}};
\end{scope}
\begin{scope}[shift={(-1.0523,1.3581)}]
\draw  (-3.9479,4.9946) ellipse (0.5 and 0.5);
\node (v1) at (-3.9479,4.9946) {$1_1$};
\node at (-3.9282,4.7027) {\tiny{$e$}};
\begin{scope}[shift={(-0.4478,0.4502)}]
\draw  (-3.4113,4.3642) rectangle (-3.17,4.1794);
\node at (-3.2917,4.2752) {\tiny{-1}};
\end{scope}
\end{scope}
\begin{scope}[shift={(0.9176,1.3618)}]
\draw  (-3.9479,4.9946) ellipse (0.5 and 0.5);
\node (v1) at (-3.9479,4.9946) {$2_1$};
\node at (-4.2086,4.9466) {\tiny{$e$}};
\begin{scope}[shift={(-0.8956,0.5194)}]
\draw  (-3.4113,4.3642) rectangle (-3.17,4.1794);
\node at (-3.2917,4.2752) {\tiny{-1}};
\end{scope}
\end{scope}
\begin{scope}[shift={(-0.687,1.5856)}]
\draw  (-3.4113,4.3642) rectangle (-3.17,4.1794);
\node at (-3.2917,4.2752) {\tiny{-1}};
\end{scope}
\draw (-4.6404,6.0026) -- (-4.036,5.4804) (-3.8277,5.4774) -- (-3.3876,5.9967) (-3.4786,6.1376) -- (-4.57,6.0936);
\begin{scope}[shift={(-0.6796,0.3888)}]
\draw  (-3.4113,4.3642) rectangle (-3.17,4.1794);
\node at (-3.2917,4.2752) {\tiny{1}};
\end{scope}
\node at (-3.7425,4.6612) {\tiny{$h$}};
\end{tikzpicture}
\end{array}
\end{align}
The three surfaces intersect each other transversely in a common curve $L$ that can be represented as the curve $e$ in each of the individual surfaces $S^i_C=\F_1$. $L$ is the locus spanned by the point of intersection of the components of type IV elliptic fiber as it moves over $C$. This constructs a pure $\su(3)$ gauge theory.

\section{Gluing rules}\label{gluing}
In this section, we will describe our proposal for gluing the graphs appearing in Section \ref{flop-eq} whenever their corresponding base curves collide. We could draw a combined graph as we drew in the last section but it becomes more and more messy. So, we instead choose to simply tabulate which curve is glued to which curve. From this gluing data and using the graphs in Section \ref{flop-eq}, one can easily construct a combined graph.

In what follows, the first gauge algebra participating in the gluing will be referred to as $\fg_C$ living over a curve $C$, and the second gauge algebra participating in the gluing will be referred to as $\fg_D$ living over a curve $D$. We have $(C\cdot D)|_B =1$ which is the only case we have to worry about in the context of $6d$ SCFTs.

We would like to note that all the gluing rules we propose below are consistent with the triple intersections that can be computed using the techniques described in Section \ref{over} and by using the attached Mathematica notebook.

\subsection{Gluing of $\su(m),m\ge1$ or $m=\tilde6$ and $\su(n),n\ge1$}
This collision forces $n$ number of fundamentals of $\mf{su}(m)$ and $m$ number of fundamentals of $\mf{su}(n)$ to combine and transform in a bifundamental of $\mf{su}(m)\oplus\mf{su}(n)$. So, we separate $n$ blowups on $S^0_C$ to glue to the collection $S_D$, and similarly we separate $m$ blowups on $S^0_D$ to glue to the collection $S_C$. The gluings are:
\bit
\item $f-x_1$ in $S^0_C$ is glued to $f-x_1$ in $S^0_D$. $x_n$ in $S^0_C$ is glued to $x_m$ in $S^0_D$.
\item $x_{i}-x_{i+1}$ in $S^0_C$ is glued to $f$ in $S^i_D$ for $i=1,\cdots,n-1$.
\item $x_{i}-x_{i+1}$ in $S^0_D$ is glued to $f$ in $S^i_C$ for $i=1,\cdots,m-1$.
\eit
In other words, $f$ in $S^0_C$ is glued to the sum of all the $f$ in $S^i_D$ for $i=1,\cdots,n-1$ plus $f-x_1+x_m$ in $S^0_D$; and $f$ in $S^i_C$ for $i=1,\cdots,m-1$ is glued to $x_i-x_{i+1}$ in $S^0_D$. So, in total, $f_C$ is glued to $f_D$, as it should be. This proves that the consistency conditions (\ref{glue1}) and (\ref{glue2}) are satisfied.
 
Another consistency check is as follows. According to (\ref{any}), the triple intersection of three surfaces can be computed as an intersection of the corresponding gluing curves in any of the three surfaces. So the gluing curves must be such that all such intersections lead to the same triple intersection number. One can easily check that the gluing rules we proposed above provide consistent triple intersections of surfaces. In fact, the same holds true for all the gluing rules we will propose in what follows, and we won't be repeating this comment again.

Suppose an $\su(p)$ living on another curve $E$ also glues to $\mf{su}(n)$ living on $D$.  Then the gluing will use a set of $p$ blowups on $S^0_D$. We claim that if $m,p\ge2$, then the set of $m$ blowups used in the gluing to $\mf{su}(m)$ must be distinct from the set of $p$ blowups used in the gluing to $\su(p)$. If the two set of blowups intersect, then it is easy to see that this will create at least one non-zero triple intersection between three surfaces $S^0_C,S^0_D,S^0_E$. For this triple intersection to be consistent, $S^0_C$ must intersect $S^0_E$ which is impossible since $C$ and $E$ do not collide in $B$. If on the other hand, say $m=1$, then the total gluing curve in $S^0_D$ for the gluing between $S_C$ and $S_D$ is $(f^0_D-x_1)+x_1=f^0_D$. Thus $x_1$ can be used in the gluing to $S_E$ without creating any unwanted triple intersections.

As a consequence of the above discussion, we see that we can assign $\su(1)$ to every $-2$ curve in the F-theory configuration constructing a $(2,0)$ SCFT of $DE$ type and still obtain a consistent gluing even though the configuration of I$_1$ fibers intersecting in the pattern of finite Dynkin diagram of $DE$ type does not give rise to a consistent elliptic fibration structure. See the end of Section \ref{rhate} for related discussion.

Similar comments hold for all the other cases below. In general, it should be kept in mind that a curve can be used in multiple gluing rules as long as it does not create any unwanted triple intersections.

Recall that for the case of $\su(2)$ on a $-2$ curve, we assigned two geometries (\ref{su2I}) and (\ref{su2II}). The above gluing rules apply to the case of (\ref{su2I}). However, as we remarked earlier, (\ref{su2I}) cannot be used when $\su(2)$ has a neighboring $\fg_2$. So, consider an $\su(m)=\su(2)$ with a neighboring $\fg_2$. There can be no $\su(n)$ neighbor of such an $\su(2)$ except for $n=1$. So, we need to present separate gluing rules for this case:
\bit
\item $f$ in $S^0_C$ is glued to $f-x_1-x_2$ in $S^0_D$.
\item $f-x_1$ in $S^1_C$ is glued to $x_1$ in $S^0_D$. $x_1$ in $S^1_C$ is glued to $x_2$ in $S^0_D$.
\eit



\subsection{Gluing of $\sp(m),m\ge1$ and $\su(n),n\ge1$}
This collision constructs a hyper in the bifundamental of $\mf{sp}(m)$ and $\mf{su}(n)$. We use $n$ blowups on $S^0_C$ and $2m$ blowups on $S^0_D$ for gluing:
\bit
\item $f-x_1$ in $S^0_C$ is glued to $f-x_1$ in $S^0_D$. $x_n$ in $S^0_C$ is glued to $x_{2m}$ in $S^0_D$.
\item $x_i-x_{i+1}$ in $S^0_C$ is glued to $f$ in $S^i_D$ for $i=1,\cdots,n-1$.
\item $x_i-x_{i+1}$ in $S^0_D$ is glued to $f$ in $S^i_C$ for $i=1,\cdots,m-1$, and $x_i-x_{i+1}$ in $S^0_D$ is glued to another copy of $f$ in $S^{m-i}_C$ for $i=m+1\cdots,2m-1$.
\item $x_m-x_{m+1}$ in $S^0_D$ is glued to $f$ in $S^m_C$.
\eit
In other words, $f$ in $S^0_C$ is glued to the sum of all the $f$ in $S^i_D$ for $i=1,\cdots,n-1$ plus $f-x_1+x_{2m}$ in $S^0_D$; $2f$ in $S^i_C$ for $i=1,\cdots,m-1$ is glued to $x_i-x_{i+1}+x_{2m-i}-x_{2m-i+1}$ in $S^0_D$; and $f$ in $S^m_C$ is glued to $x_m-x_{m+1}$ in $S^0_D$. So, in total, $f_C$ is glued to $f_D$ as required by (\ref{glue1}) and (\ref{glue2}).

The above gluing rules work with either geometry (\ref{sp1}), (\ref{sp2}) for $\sp(m)$ except in the case when $\sp(m)=\sp(1)$ and $\sp(1)$ has another neighboring gauge algebra equal to $\so(19)$. In this case, we have to give separate gluing rules between $\sp(m)=\sp(1)$ and $\su(n)=\su(1)$:
\bit
\item $f$ in $S^0_C$ is glued to $f-x_1-x_2$ in $S^0_D$.
\item $f-x_1$ in $S^1_C$ is glued to $x_1$ in $S^0_D$. $x_1$ in $S^1_C$ is glued to $x_2$ in $S^0_D$.
\eit


\subsection{Gluing of $\sp(m),m\ge1$ and $\so(2r),r\ge4$}
For $r\ge5$, this constructs a half-hyper in the bifundamental of $\mf{sp}(m)$ and $\mf{so}(2r)$. We use $r$ blowups on $S^0_C$ and $m$ pairs of blowups corresponding to fundamental representation on $S^r_D$ for gluing. For $r=4$, this constructs a half-hyper in the fundamental$\otimes$spinor of $\mf{sp}(m)\oplus\mf{so}(8)$. We use $r$ blowups on $S^0_C$ and $m$ pairs of blowups corresponding to spinor representation on $S^r_D=S^4_D$ for gluing. We can give the following gluing rules which work uniformly for both $r=4$ and $r\ge5$ cases:
\bit
\item $f-x_1-x_{2}$ in $S^0_C$ is glued to $f$ in $S^0_D$.
\item $x_{i}-x_{i+1}$ in $S^0_C$ is glued to $f$ in $S^i_D$ for $i=1,\cdots,r-1$.
\item $x_{r-1}$ in $S^0_C$ is glued to $f-x_1$ in $S^r_D$. $x_r$ in $S^0_C$ is glued to $y_1$ in $S^r_D$.
\item $y_{i+1}-y_{i}$ and $x_{i}-x_{i+1}$ in $S^r_D$ are glued to two copies of $f$ in $S^i_C$ for $i=1,\cdots,m-1$.
\item $x_m-y_{m}$ in $S^r_D$ is glued to $f$ in $S^m_C$.
\eit
It can be checked easily that $f_C$ is indeed glued to $f_D$.

\noindent Recall that we actually associated two geometries (\ref{sp1}) and (\ref{sp2}) to $\sp(m)$. Whenever $2m+8>r$, we can use either geometry. When $2m+8=r$, we must use the geometry (\ref{sp1}).


\subsection{Gluing of $\sp(m),m\ge1$ and $\so(2r+1),r\ge4$}\label{half-blowup}
This constructs a half-hyper in the bifundamental of $\mf{sp}(m)$ and $\mf{so}(2r+1)$. In this case, we must use the geometry (\ref{sp2}). We use $r$ blowups on $S^0_C$, the single blowup on $S^m_C$ and $m$ pairs of blowups corresponding to fundamental representation on $S^r_D$ for gluing:
\bit
\item $f-x_1-x_{2}$ in $S^0_C$ is glued to $f$ in $S^0_D$.
\item $x_{i}-x_{i+1}$ in $S^0_C$ is glued to $f$ in $S^i_D$ for $i=1,\cdots,r-1$.
\item $x_{r}$ in $S^0_C$ is glued to $x_1$ in $S^r_D$. $x_{r}$ in $S^0_C$ is glued to $y_1$ in $S^r_D$.
\item $y_{i+1}-y_{i}$ and $x_{i+1}-x_{i}$ in $S^r_D$ are glued to two copies of $f$ in $S^i_C$ for $i=1,\cdots,m-1$.
\item $f-x_m$ in $S^r_D$ is glued to $x$ in $S^m_C$. $f-y_m$ in $S^r_D$ is glued to $f-x$ in $S^m_C$.
\eit
One can check that the corresponding elliptic fibers are identified with each other.


One curious fact that one can notice is that the total gluing curve between $S^m_C$ and $S^r_D$ is identified as $f$ inside $S^m_C$, and does not involve $x$. This means that $x$ does not participate in the intersection numbers related to the above gluing and, hence, can participate in other gluings without spoiling consistency. For example, if we want a configuration $\so(4m-2r+15)-\sp(m)-\so(2r+1)$, then some blowup on $\sp(m)$ must participate in both the gluings $\sp(m)-\so(4m-2r+15)$ and $\sp(m)-\so(2r+1)$, and only the above mentioned blowup $x$ on $S^m$ associated to $\sp(m)$ can do so consistently.


\subsection{Gluing of $\mf{sp}(m),m\ge1$ and $\so(7)$}
This gluing produces a half-hyper in the fundamental$\otimes$spinor of $\mf{sp}(m)\oplus\mf{so}(7)$. We use $4$ blowups on $S^0_C$ and $m$ out of $8-2k$ pairs of blowups associated to the spinor representation on $S^1_D$:
\bit
\item $f-x_1-x_{2}$ in $S^0_C$ is glued to $f$ in $S^0_D$.
\item $x_{2}-x_{3}$ in $S^0_C$ is glued to $f$ in $S^2_D$.
\item $x_1-x_2$ and $x_3-x_4$ in $S^0_C$ are glued to two copies of $f$ in $S^3_D$.
\item $x_{3}$ in $S^0_C$ is glued to $f-x_1$ in $S^1_D$. $x_4$ in $S^0_C$ is glued to $y_1$ in $S^1_D$.
\item $y_{i+1}-y_{i}$ and $x_{i}-x_{i+1}$ in $S^1_D$ are glued to two copies of $f$ in $S^i_C$ for $i=1,\cdots,m-1$.
\item $x_m-y_{m}$ in $S^1_D$ is glued to $f$ in $S^m_C$.
\eit
For $\sp(m)$, we can use either geometry (\ref{sp1}) or (\ref{sp2}).

We should keep in mind that the formal gauge algebra $\sp(1)=\su(2)$. Thus the above gluing rules apply both to $\sp(1)$ on $C^2=-1$ and to $\su(2)$ on $C^2=-2$.

\subsection{Gluing of $\mf{sp}(m),m\ge1$ and $\fg_2$}
This constructs a half-hyper in the fundamental $\otimes\: 7$ of $\mf{sp}(m)\oplus\fg_2$. We must use the geometry (\ref{sp2}) for $\sp(m)$. For gluing, we use $3$ blowups on $S^0_C$, the single blowup on $S^m_C$ and $m$ pairs of blowups on $S^2_D$:
\bit
\item $f-x_1-x_{2}$ in $S^0_C$ is glued to $f$ in $S^0_D$.
\item $x_{2}-x_{3}$ in $S^0_C$ is glued to $f$ in $S^2_D$.
\item $x_1-x_2$ in $S^0_C$ is glued to $f$ in $S^1_D$. $x_{3}$ in $S^0_C$ is glued to $x_1$ in $S^1_D$. $x_3$ in $S^0_C$ is glued to $y_1$ in $S^1_D$.
\item $y_{i+1}-y_{i}$ and $x_{i+1}-x_{i}$ in $S^1_D$ are glued to two copies of $f$ in $S^i_C$ for $i=1,\cdots,m-1$.
\item $f-x_m$ in $S^1_D$ is glued to $x$ in $S^m_C$. $f-y_m$ in $S^1_D$ is glued to $f-x$ in $S^m_C$.
\eit
Similar remarks as towards the end of Section \ref{half-blowup} apply here to $x$ in $S^m_C$. We should keep in mind that the formal gauge algebra $\sp(1)=\su(2)$. Thus the above gluing rules apply both to $\sp(1)$ on $C^2=-1$ and to $\su(2)$ on $C^2=-2$. In the case of $\su(2)$ on $-2$ curve, we must use the geometry (\ref{su2II}).


%

\subsection{Gluings of $\sp(0)=$ E-string}\label{sp0}
In F-theory configurations constructing $6d$ SCFTs, a $-1$ curve $D$ in $B$ carrying $\sp(0)$ can only intersect at most two compact holomorphic curves. The sum of possible gauge algebras carried by compact holomorphic curves intersecting $D$ must be a subalgebra of $\fe_8$ and all the possible values are collected in Table \ref{table2}. Each gauge algebra summand is realized on a compact holomorphic curve intersecting $D$. Thus when we have two summands $\fg_C\oplus\fg_E$, then we have two curves $C$ and $E$ in $B$ carrying $\fg_C$ and $\fg_E$ respectively intersecting $D$. However, when we have a single summand $\fg_C$, then we can either have one curve $C$ carrying $\fg_C$ intersecting $D$, or we can have two curves $C$ and $E$ carrying $\fg_C$ and $\fg_E=\su(1)$ intersecting $D$. See Table \ref{table} for the possible fiber types corresponding to $\su(1)$.

\begin{table}[h]
\begin{center}
  \begin{tabular}{ | c | l | }
    \hline
    Rank & $\fg_C\oplus\fg_E$ \\ \hline
    8 & $\fe_8,\enspace  \so(16),\enspace \su(9),\enspace \fe_7\oplus\su(2),\enspace \fe_6\oplus\su(3),\enspace \so(8)\oplus\so(8),\enspace \so(10)\oplus\su(4)$ \\ \hline
    7 & $\fe_7,\enspace \so(14,15),\enspace \su(8),\enspace \fe_6\oplus\su(2),\enspace \so(8,9)\oplus\so(7),\enspace \so(12,13)\oplus\su(2),$ \\
    {} & $\so(10)\oplus\su(3),\enspace \so(8,9)\oplus\su(4)$ \\ \hline
    6 & $\fe_6,\enspace \so(12,13),\enspace \su(7),\enspace \ff_4\oplus\fg_2,\enspace \ff_4\oplus\su(3),\enspace \so(8,9)\oplus\fg_2,\enspace \so(7)\oplus\so(7),$ \\
    {} & $\so(10,11)\oplus\su(2),\enspace \so(8,9)\oplus\su(3),\enspace \so(7)\oplus\su(4)$ \\ \hline
    5 & $\so(10,11),\enspace \su(6),\enspace \ff_4\oplus\su(2),\enspace \so(7)\oplus\fg_2,\enspace \su(4)\oplus\fg_2,\enspace \so(8,9)\oplus\su(2),$ \\
    {} & $\so(7)\oplus\su(3)$ \\ \hline
    4 & $\so(8,9),\enspace \su(5),\enspace \ff_4,\enspace \fg_2\oplus\fg_2,\enspace \su(3)\oplus\fg_2,\enspace \so(7)\oplus\su(2),\enspace \su(3)\oplus\su(3)$ \\ \hline
    3 & $\so(7),\enspace \su(4),\enspace \su(2)\oplus\fg_2$ \\ \hline
    2 & $\fg_2,\enspace \su(3)$ \\ \hline
    1 & $\su(2)$ \\ \hline
  \end{tabular}
\end{center}
\caption{Possible gauge algebras that can surround a $-1$ curve carrying $\sp(0)$.}
\label{table2}
\end{table}

Recall that the surface corresponding to $\sp(0)$ is $S_D=(\P^2)^9$, that is $\P^2$ blown up at 9 points. $\P^2$ has a single curve $l$ with self-intersection $l^2=1$. Let's denote the nine blowups as $x_1,\cdots,x_9$. We will always choose $x_9$ to be the curve gluing $S_D$ to $B$. The elliptic fiber is the curve $f_D=3l-\sum x_i$, which can be verified to have genus one and self-intersection zero.

Since coupling to $\sp(0)$ does not gauge any matter transforming under $\fg_C$, none of the blowups on the collection of surfaces $S_C$ can be involved in the gluing. This means that the curves gluing $S^i_C$ to $S_D$ must be the fibers $f^i_C$ of the Hirzebruch surface $S^i_C$. Similar comments apply to the gluing of $S_E$ to $S_D$ whenever $D$ has another neighbor $E$.

\subsubsection{Simply laced}

Consider first the case when the formal gauge algebra over $C$ is simply laced. We regard $\sp(1)=\su(2)$ and $\su(1)$ as simply laced. In this case, we claim that the curve gluing $S^i_C$ to $S_D$ must be a \emph{single copy} of the fiber $f^i_C$. Say $f^i_C$ is glued to some curve $D[f^i_C]$ inside $S_D=(\P^2)^9$. By (\ref{CY}), the self-intersection of $D[f^i_C]$ in $S_D$ must be $-2$. Moreover, since $f^i_C$ intersect in the pattern of affine Dynkin diagram for $\fg_C$, the consistency of triple intersections $S_D\cdot S^i_C\cdot S^j_C$ requires that $D[f^i_C]$ must intersect in the pattern of affine Dynkin diagram as well\footnote{This line of thought was independently pursued by Hee-Cheol Kim and conveyed to the authors in a private communication.}. In other words, we have argued that
\be
\left(D[f^i_C]\cdot D[f^j_C]\right)|_{S_D}=M^{ij}_C \label{E}
\ee
where $M^{ij}_C$ is the affine Cartan matrix for $\fg_C$. Since a single copy of $f^i_C$ participates in the gluing, (\ref{glue1}) requires that
\be
\sum_in_iD[f^i_C]=f_D=3l-\sum x_i \label{EE}
\ee
where $n_if^i_C=f_C$ the elliptic fiber over $C$. Recall the property of affine Cartan matrices that if $D[f^i_C]$ satisfy (\ref{E}), then 
\be
\left(\sum_in_iD[f^i_C]\right)\cdot D[f^i_C]=0
\ee
which using (\ref{EE}) becomes
\be
\left(f_D\cdot D[f^i_C]\right)|_{S_D}=0 \label{re}
\ee
for all $i$. Similar comments apply to the gluing of $S_E$ to $S_D$ whenever $D$ has another neighbor $E$. For $6d$ SCFTs, $(C\cdot E)|_B=0$ implying
\be
\left(D[f^i_C]\cdot D[f^k_E]\right)|_{S_D}=0 \label{EEE}
\ee
which means that the curves gluing $S_D$ to $S_C$ and the curves gluing $S_D$ to $S_E$ do not intersect. 

We will now provide a set of gluing curves in $(\P^2)^9$ for each of the possibilities of $\fg_C$ and $\fg_E$ shown in Table \ref{table2} as long as $\fg_C$ and $\fg_E$ are simply laced. Let us first get $\fg_E=\su(1)$ out of the way. In this case, $D[f^0_E]=f_D$ due to (\ref{EE}) since $f_E=f^0_E$. (\ref{EEE}) demands that $(D[f^0_E]\cdot D[f^i_C])|_{S_D}=0$ which follows straightforwardly from (\ref{re}). Thus, coupling to $\fg_E=\su(1)$ introduces no constraints on the coupling to $\fg_C$, so we need to provide gluing curves only for the cases $\fg_C\neq\su(1)$ and $\fg_E\neq\su(1)$.

\begin{center}
\ubf{$\fg_C=\fe_8$}\,:
\end{center}

\begin{align}
\nn
\begin{array}{c}
\begin{tikzpicture}
\node (v1) at (0,0) {\scriptsize{$x_8-x_9$}};
\node (v2) at (2,0) {\scriptsize{$x_7-x_8$}};
\node (v3) at (4,0) {\scriptsize{$x_6-x_7$}};
\node (v4) at (6,0) {\scriptsize{$x_5-x_6$}};
\node (v5) at (8,0) {\scriptsize{$x_4-x_5$}};
\node (v1_1) at (10,0) {\scriptsize{$x_1-x_4$}};
\node (v1_2) at (12,0) {\scriptsize{$x_2-x_1$}};
\node (v1_3) at (14,0) {\scriptsize{$x_3-x_2$}};
\node (v1_4) at (10,1) {\scriptsize{$l-x_1-x_2-x_3$}};
\draw  (v1) edge (v2);
\draw  (v2) edge (v3);
\draw  (v3) edge (v4);
\draw  (v4) edge (v5);
\draw  (v5) edge (v1_1);
\draw  (v1_1) edge (v1_2);
\draw  (v1_2) edge (v1_3);
\draw  (v1_1) edge (v1_4);
\end{tikzpicture}
\end{array}
\end{align}
One can check that all the curves have self-intersection $-2$ and that they intersect in the pattern displayed in the above graph, which is the pattern of affine Dynkin diagram for $\fe_8$. Moreover, one can check that if one adds all the curves with the multiplicities shown in Figure \ref{hell}, then one obtains $3l-\sum x_i$. Similar comments apply to all of the diagrams below and so we won't repeat them.

Now we can also tie up a loose end from Section \ref{multiple}. There we claimed that the surface corresponding to E-string theory should be regarded as $\P^2$ blown up at 9 points rather than $\F_1$ blown up at 8 points, even though the two descriptions match when all the blowups are generic. We claimed that when we have collisions of E-string with other curves, the blowups are not generic and for some cases we are forced to use some non-generic configuration of blowups that can be written as $(\P^2)^9$ but not as $\F_1^8$. 

In fact, the above configuration of gluing curves implies that $S_D$ cannot be written as $\F_1$ blown up at 8 non-generic points. For if it was possible, then $x_9$ would become the curve $e$ since these are the curves that glue to $B$. Moreover, $x_i,i=1,\cdots,8$ would become the 8 blowups on $\F_1$. But then the curve gluing $x_8-x_9$ gluing $S_D$ to $S^0_C$ would become $x_8-e$ which cannot be holomorphic in $\F^8_1$ thus leading to a contradiction.

\begin{center}
\ubf{$\fg_C=\so(16)$}\,:
\end{center}

\begin{align}
\nn
\begin{array}{c}
\begin{tikzpicture}
\node (v1) at (0,0) {\scriptsize{$x_8-x_9$}};
\node (v2) at (2,0) {\scriptsize{$x_7-x_8$}};
\node (v3) at (4,0) {\scriptsize{$x_6-x_7$}};
\node (v4) at (6,0) {\scriptsize{$x_5-x_6$}};
\node (v5) at (8,0) {\scriptsize{$x_4-x_5$}};
\node (v1_1) at (10,0) {\scriptsize{$x_1-x_4$}};
\node (v1_2) at (12,0) {\scriptsize{$x_2-x_1$}};
\node (v1_3) at (2,1) {\scriptsize{$2l-x_1-x_2-x_4-x_5-x_6-x_7$}};
\node (v1_4) at (10,1) {\scriptsize{$l-x_1-x_2-x_3$}};
\draw  (v1) edge (v2);
\draw  (v2) edge (v3);
\draw  (v3) edge (v4);
\draw  (v4) edge (v5);
\draw  (v5) edge (v1_1);
\draw  (v1_1) edge (v1_2);
\draw  (v1_1) edge (v1_4);
\draw  (v1_3) edge (v2);
\end{tikzpicture}
\end{array}
\end{align}\\

\begin{center}
\ubf{$\fg_C=\su(9)$}\,:
\end{center}

\begin{align}
\nn
\begin{array}{c}
\begin{tikzpicture}
\node (v1) at (0,0) {\scriptsize{$x_8-x_9$}};
\node (v2) at (2,0) {\scriptsize{$x_7-x_8$}};
\node (v3) at (4,0) {\scriptsize{$x_6-x_7$}};
\node (v4) at (6,0) {\scriptsize{$x_5-x_6$}};
\node (v5) at (8,0) {\scriptsize{$x_4-x_5$}};
\node (v1_1) at (10,0) {\scriptsize{$x_1-x_4$}};
\node (v1_2) at (12,0) {\scriptsize{$x_2-x_1$}};
\node (v1_3) at (14,0) {\scriptsize{$x_3-x_2$}};
\node (v1_4) at (7,2) {\scriptsize{$3l-x_1-x_2-2x_3-x_4-x_5-x_6-x_7-x_8$}};
\draw  (v1) edge (v2);
\draw  (v2) edge (v3);
\draw  (v3) edge (v4);
\draw  (v4) edge (v5);
\draw  (v5) edge (v1_1);
\draw  (v1_1) edge (v1_2);
\draw  (v1_2) edge (v1_3);
\draw  (v1_4) edge (v1);
\draw  (v1_4) edge (v1_3);
\end{tikzpicture}
\end{array}
\end{align}\\

\begin{center}
\ubf{$\fg_C\oplus\fg_E=\fe_7\oplus\su(2)$}\,:
\end{center}

\begin{align}
\nn
\begin{array}{c}
\begin{tikzpicture}
\node (v1) at (8,-1.5) {\scriptsize{$x_8-x_9$}};
\node (v1_1) at (13,-1.5) {\scriptsize{$3l-x_1-x_2-x_3-x_4-x_5-x_6-x_7-2x_8$}};
\node (v2) at (16.25,0) {\scriptsize{$l-x_3-x_8-x_9$}};
\node (v3) at (4,0) {\scriptsize{$x_6-x_7$}};
\node (v4) at (6,0) {\scriptsize{$x_5-x_6$}};
\node (v5) at (8,0) {\scriptsize{$x_4-x_5$}};
\node (v1_1_1) at (10,0) {\scriptsize{$x_1-x_4$}};
\node (v1_2) at (12,0) {\scriptsize{$x_2-x_1$}};
\node (v1_3) at (14,0) {\scriptsize{$x_3-x_2$}};
\node (v1_4) at (10,1) {\scriptsize{$l-x_1-x_2-x_3$}};
\draw  (v3) edge (v4);
\draw  (v4) edge (v5);
\draw  (v5) edge (v1_1_1);
\draw  (v1_1_1) edge (v1_2);
\draw  (v1_2) edge (v1_3);
\draw  (v1_1_1) edge (v1_4);
\draw  (v1_3) edge (v2);
\draw[double]  (v1_1) -- (v1);
\end{tikzpicture}
\end{array}
\end{align}
where the connected graph towards the top is used to glue to $\fe_7$ and the connected graph towards the bottom is used to glue to $\su(2)$. The double edge denotes the fact that the adjacent curves intersect at two points. The fact that the graphs are disconnected denotes the fact that none of the curves in the top graph intersect any of the curves in the bottom graph.

\begin{center}
\ubf{$\fg_C\oplus\fg_E=\fe_6\oplus\su(3)$}\,:
\end{center}

\begin{align}
\nn
\begin{array}{c}
\begin{tikzpicture}
\node (v1) at (8,-2) {\scriptsize{$x_8-x_9$}};
\node (v1_1) at (12,-2) {\scriptsize{$x_7-x_8$}};
\node (v1_1_1) at (10,-1) {\scriptsize{$3l-x_1-x_2-x_3-x_4-x_5-x_6-2x_7-x_8$}};
\node (v2) at (10,2) {\scriptsize{$l-x_7-x_8-x_9$}};
\node (v4) at (6,0) {\scriptsize{$x_5-x_6$}};
\node (v5) at (8,0) {\scriptsize{$x_4-x_5$}};
\node (v1_1_1_1) at (10,0) {\scriptsize{$x_1-x_4$}};
\node (v1_2) at (12,0) {\scriptsize{$x_2-x_1$}};
\node (v1_3) at (14,0) {\scriptsize{$x_3-x_2$}};
\node (v1_4) at (10,1) {\scriptsize{$l-x_1-x_2-x_3$}};
\draw  (v4) edge (v5);
\draw  (v5) edge (v1_1_1_1);
\draw  (v1_1_1_1) edge (v1_2);
\draw  (v1_2) edge (v1_3);
\draw  (v1_1_1_1) edge (v1_4);
\draw  (v1_1_1) -- (v1_1);
\draw  (v1_4) edge (v2);
\draw  (v1) edge (v1_1_1);
\draw  (v1) edge (v1_1);
\end{tikzpicture}
\end{array}
\end{align}\\

\begin{center}
\ubf{$\fg_C\oplus\fg_E=\so(8)\oplus\so(8)$}\,:
\end{center}

\begin{align}
\nn
\begin{array}{c}
\begin{tikzpicture}
\node (v1) at (8,3.5) {\scriptsize{$x_8-x_9$}};
\node (v1_1) at (11,3.5) {\scriptsize{$x_7-x_8$}};
\node (v1_1_1) at (11,5) {\scriptsize{$2l-x_1-x_2-x_4-x_5-x_6-x_7$}};
\node (v2) at (11,-2) {\scriptsize{$2l-x_1-x_2-x_6-x_7-x_8-x_9$}};
\node (v4) at (14,3.5) {\scriptsize{$x_6-x_7$}};
\node (v5) at (8,-0.5) {\scriptsize{$x_4-x_5$}};
\node (v1_1_1_1) at (11,-0.5) {\scriptsize{$x_1-x_4$}};
\node (v1_2) at (14,-0.5) {\scriptsize{$x_2-x_1$}};
\node (v1_3) at (11,2) {\scriptsize{$l-x_3-x_6-x_7$}};
\node (v1_4) at (11,1) {\scriptsize{$l-x_1-x_2-x_3$}};
\draw  (v5) edge (v1_1_1_1);
\draw  (v1_1_1_1) edge (v1_2);
\draw  (v1_1_1_1) edge (v1_4);
\draw  (v1) edge (v1_1);
\draw  (v1_1) edge (v4);
\draw  (v1_1_1_1) edge (v2);
\draw  (v1_1_1) edge (v1_1);
\draw  (v1_1) edge (v1_3);
\end{tikzpicture}
\end{array}
\end{align}\\

\begin{center}
\ubf{$\fg_C\oplus\fg_E=\so(10)\oplus\su(4)$}\,:
\end{center}

\begin{align}
\nn
\begin{array}{c}
\begin{tikzpicture}
\node (v1) at (10,-2) {\scriptsize{$x_8-x_9$}};
\node (v1_1) at (12,-3) {\scriptsize{$x_7-x_8$}};
\node (v1_1_1) at (12,-1) {\scriptsize{$3l-x_1-x_2-x_3-x_4-x_5-2x_6-x_7-x_8$}};
\node (v2) at (14,1) {\scriptsize{$2l-x_2-x_3-x_6-x_7-x_8-x_9$}};
\node (v4) at (14,-2) {\scriptsize{$x_6-x_7$}};
\node (v5) at (8,0) {\scriptsize{$x_4-x_5$}};
\node (v1_1_1_1) at (10,0) {\scriptsize{$x_1-x_4$}};
\node (v1_2) at (14,0) {\scriptsize{$x_2-x_1$}};
\node (v1_3) at (16,0) {\scriptsize{$x_3-x_2$}};
\node (v1_4) at (10,1) {\scriptsize{$l-x_1-x_2-x_3$}};
\draw  (v5) edge (v1_1_1_1);
\draw  (v1_1_1_1) edge (v1_2);
\draw  (v1_2) edge (v1_3);
\draw  (v1_1_1_1) edge (v1_4);
\draw  (v1) edge (v1_1_1);
\draw  (v1) edge (v1_1);
\draw  (v2) edge (v1_2);
\draw  (v1_1) edge (v4);
\draw  (v4) edge (v1_1_1);
\end{tikzpicture}
\end{array}
\end{align}\\

\begin{center}
\ubf{$\fg_C=\fe_7$}\,:
\end{center}

\begin{align}
\nn
\begin{array}{c}
\begin{tikzpicture}
\node (v2) at (16.25,0) {\scriptsize{$l-x_3-x_8-x_9$}};
\node (v3) at (4,0) {\scriptsize{$x_6-x_7$}};
\node (v4) at (6,0) {\scriptsize{$x_5-x_6$}};
\node (v5) at (8,0) {\scriptsize{$x_4-x_5$}};
\node (v1_1_1) at (10,0) {\scriptsize{$x_1-x_4$}};
\node (v1_2) at (12,0) {\scriptsize{$x_2-x_1$}};
\node (v1_3) at (14,0) {\scriptsize{$x_3-x_2$}};
\node (v1_4) at (10,1) {\scriptsize{$l-x_1-x_2-x_3$}};
\draw  (v3) edge (v4);
\draw  (v4) edge (v5);
\draw  (v5) edge (v1_1_1);
\draw  (v1_1_1) edge (v1_2);
\draw  (v1_2) edge (v1_3);
\draw  (v1_1_1) edge (v1_4);
\draw  (v1_3) edge (v2);
\end{tikzpicture}
\end{array}
\end{align}\\

\begin{center}
\ubf{$\fg_C=\so(14)$}\,:
\end{center}

\begin{align}
\nn
\begin{array}{c}
\begin{tikzpicture}
\node (v1) at (0,0) {\scriptsize{$x_8-x_9$}};
\node (v2) at (2,0) {\scriptsize{$x_7-x_8$}};
\node (v3) at (4,0) {\scriptsize{$x_6-x_7$}};
\node (v4) at (6,0) {\scriptsize{$x_5-x_6$}};
\node (v5) at (8,0) {\scriptsize{$x_4-x_5$}};
\node (v1_1) at (10,0) {\scriptsize{$x_1-x_4$}};
\node (v1_3) at (2,1) {\scriptsize{$2l-x_1-x_2-x_4-x_5-x_6-x_7$}};
\node (v1_4) at (8,1) {\scriptsize{$l-x_1-x_3-x_4$}};
\draw  (v1) edge (v2);
\draw  (v2) edge (v3);
\draw  (v3) edge (v4);
\draw  (v4) edge (v5);
\draw  (v5) edge (v1_1);
\draw  (v5) edge (v1_4);
\draw  (v1_3) edge (v2);
\end{tikzpicture}
\end{array}
\end{align}\\

\begin{center}
\ubf{$\fg_C=\su(8)$}\,:
\end{center}

\begin{align}
\nn
\begin{array}{c}
\begin{tikzpicture}
\node (v1) at (0,0) {\scriptsize{$x_8-x_9$}};
\node (v2) at (2,0) {\scriptsize{$x_7-x_8$}};
\node (v3) at (4,0) {\scriptsize{$x_6-x_7$}};
\node (v4) at (6,0) {\scriptsize{$x_5-x_6$}};
\node (v5) at (8,0) {\scriptsize{$x_4-x_5$}};
\node (v1_1) at (10,0) {\scriptsize{$x_1-x_4$}};
\node (v1_2) at (12,0) {\scriptsize{$x_2-x_1$}};
\node (v1_4) at (6,2) {\scriptsize{$3l-x_1-2x_2-x_3-x_4-x_5-x_6-x_7-x_8$}};
\draw  (v1) edge (v2);
\draw  (v2) edge (v3);
\draw  (v3) edge (v4);
\draw  (v4) edge (v5);
\draw  (v5) edge (v1_1);
\draw  (v1_1) edge (v1_2);
\draw  (v1_4) edge (v1);
\draw  (v1_4) edge (v1_2);
\end{tikzpicture}
\end{array}
\end{align}\\

\begin{center}
\ubf{$\fg_C\oplus\fg_E=\fe_6\oplus\su(2)$}\,:
\end{center}

\begin{align}
\nn
\begin{array}{c}
\begin{tikzpicture}
\node (v1) at (7,-1.5) {\scriptsize{$x_8-x_9$}};
\node (v1_1) at (12,-1.5) {\scriptsize{$3l-x_1-x_2-x_3-x_4-x_5-x_6-x_7-2x_8$}};
\node (v2) at (10,2) {\scriptsize{$l-x_7-x_8-x_9$}};
\node (v4) at (6,0) {\scriptsize{$x_5-x_6$}};
\node (v5) at (8,0) {\scriptsize{$x_4-x_5$}};
\node (v1_1_1_1) at (10,0) {\scriptsize{$x_1-x_4$}};
\node (v1_2) at (12,0) {\scriptsize{$x_2-x_1$}};
\node (v1_3) at (14,0) {\scriptsize{$x_3-x_2$}};
\node (v1_4) at (10,1) {\scriptsize{$l-x_1-x_2-x_3$}};
\draw  (v4) edge (v5);
\draw  (v5) edge (v1_1_1_1);
\draw  (v1_1_1_1) edge (v1_2);
\draw  (v1_2) edge (v1_3);
\draw  (v1_1_1_1) edge (v1_4);
\draw  (v1_4) edge (v2);
\draw[double]  (v1_1) -- (v1);
\end{tikzpicture}
\end{array}
\end{align}\\

\begin{center}
\ubf{$\fg_C\oplus\fg_E=\so(12)\oplus\su(2)$}\,:
\end{center}

\begin{align}
\nn
\begin{array}{c}
\begin{tikzpicture}
\node (v1) at (4.5,-1.5) {\scriptsize{$x_8-x_9$}};
\node (v2) at (9.5,-1.5) {\scriptsize{$3l-x_1-x_2-x_3-x_4-x_5-x_6-x_7-2x_8$}};
\node (v3) at (4,0) {\scriptsize{$x_6-x_7$}};
\node (v4) at (6,0) {\scriptsize{$x_5-x_6$}};
\node (v5) at (8,0) {\scriptsize{$x_4-x_5$}};
\node (v1_1) at (10,0) {\scriptsize{$x_1-x_4$}};
\node (v1_2) at (12,0) {\scriptsize{$x_2-x_1$}};
\node (v1_3) at (6,1) {\scriptsize{$2l-x_1-x_2-x_4-x_5-x_8-x_9$}};
\node (v1_4) at (10,1) {\scriptsize{$l-x_1-x_2-x_3$}};
\draw  (v3) edge (v4);
\draw  (v4) edge (v5);
\draw  (v5) edge (v1_1);
\draw  (v1_1) edge (v1_2);
\draw  (v1_1) edge (v1_4);
\draw  (v1_3) edge (v4);
\draw[double]  (v2) -- (v1);
\end{tikzpicture}
\end{array}
\end{align}\\

\begin{center}
\ubf{$\fg_C\oplus\fg_E=\so(10)\oplus\su(3)$}\,:
\end{center}

\begin{align}
\nn
\begin{array}{c}
\begin{tikzpicture}
\node (v1) at (10,-2) {\scriptsize{$x_8-x_9$}};
\node (v1_1) at (13.5,-2) {\scriptsize{$x_7-x_8$}};
\node (v1_1_1) at (12,-1) {\scriptsize{$3l-x_1-x_2-x_3-x_4-x_5-x_6-2x_7-x_8$}};
\node (v2) at (14,1) {\scriptsize{$2l-x_2-x_3-x_6-x_7-x_8-x_9$}};
\node (v5) at (8,0) {\scriptsize{$x_4-x_5$}};
\node (v1_1_1_1) at (10,0) {\scriptsize{$x_1-x_4$}};
\node (v1_2) at (14,0) {\scriptsize{$x_2-x_1$}};
\node (v1_3) at (16,0) {\scriptsize{$x_3-x_2$}};
\node (v1_4) at (10,1) {\scriptsize{$l-x_1-x_2-x_3$}};
\draw  (v5) edge (v1_1_1_1);
\draw  (v1_1_1_1) edge (v1_2);
\draw  (v1_2) edge (v1_3);
\draw  (v1_1_1_1) edge (v1_4);
\draw  (v1) edge (v1_1_1);
\draw  (v1) edge (v1_1);
\draw  (v2) edge (v1_2);
\draw  (v1_1_1) edge (v1_1);
\end{tikzpicture}
\end{array}
\end{align}\\

\begin{center}
\ubf{$\fg_C\oplus\fg_E=\so(8)\oplus\su(4)$}\,:
\end{center}

\begin{align}
\nn
\begin{array}{c}
\begin{tikzpicture}
\node (v1) at (10,-2) {\scriptsize{$x_8-x_9$}};
\node (v1_1) at (12,-3) {\scriptsize{$x_7-x_8$}};
\node (v1_1_1) at (12,-1) {\scriptsize{$3l-x_1-x_2-x_3-x_4-x_5-2x_6-x_7-x_8$}};
\node (v2) at (12,2) {\scriptsize{$2l-x_2-x_3-x_6-x_7-x_8-x_9$}};
\node (v4) at (14,-2) {\scriptsize{$x_6-x_7$}};
\node (v1_1_1_1) at (10,1) {\scriptsize{$x_1-x_4$}};
\node (v1_2) at (12,1) {\scriptsize{$x_2-x_1$}};
\node (v1_3) at (14,1) {\scriptsize{$x_3-x_2$}};
\node (v1_4) at (12,0) {\scriptsize{$l-x_2-x_3-x_5$}};
\draw  (v1_2) edge (v1_3);
\draw  (v1) edge (v1_1_1);
\draw  (v1) edge (v1_1);
\draw  (v2) edge (v1_2);
\draw  (v1_1) edge (v4);
\draw  (v4) edge (v1_1_1);
\draw  (v1_1_1_1) edge (v1_2);
\draw  (v1_4) edge (v1_2);
\end{tikzpicture}
\end{array}
\end{align}\\

\begin{center}
\ubf{$\fg_C=\fe_6$}\,:
\end{center}

\begin{align}
\nn
\begin{array}{c}
\begin{tikzpicture}
\node (v2) at (10,2) {\scriptsize{$l-x_7-x_8-x_9$}};
\node (v4) at (6,0) {\scriptsize{$x_5-x_6$}};
\node (v5) at (8,0) {\scriptsize{$x_4-x_5$}};
\node (v1_1_1_1) at (10,0) {\scriptsize{$x_1-x_4$}};
\node (v1_2) at (12,0) {\scriptsize{$x_2-x_1$}};
\node (v1_3) at (14,0) {\scriptsize{$x_3-x_2$}};
\node (v1_4) at (10,1) {\scriptsize{$l-x_1-x_2-x_3$}};
\draw  (v4) edge (v5);
\draw  (v5) edge (v1_1_1_1);
\draw  (v1_1_1_1) edge (v1_2);
\draw  (v1_2) edge (v1_3);
\draw  (v1_1_1_1) edge (v1_4);
\draw  (v1_4) edge (v2);
\end{tikzpicture}
\end{array}
\end{align}\\

\begin{center}
\ubf{$\fg_C=\so(12)$}\,:
\end{center}

\begin{align}
\nn
\begin{array}{c}
\begin{tikzpicture}
\node (v3) at (4,0) {\scriptsize{$x_6-x_7$}};
\node (v4) at (6,0) {\scriptsize{$x_5-x_6$}};
\node (v5) at (8,0) {\scriptsize{$x_4-x_5$}};
\node (v1_1) at (10,0) {\scriptsize{$x_1-x_4$}};
\node (v1_2) at (12,0) {\scriptsize{$x_2-x_1$}};
\node (v1_3) at (6,1) {\scriptsize{$2l-x_1-x_2-x_4-x_5-x_8-x_9$}};
\node (v1_4) at (10,1) {\scriptsize{$l-x_1-x_2-x_3$}};
\draw  (v3) edge (v4);
\draw  (v4) edge (v5);
\draw  (v5) edge (v1_1);
\draw  (v1_1) edge (v1_2);
\draw  (v1_1) edge (v1_4);
\draw  (v1_3) edge (v4);
\end{tikzpicture}
\end{array}
\end{align}\\

\begin{center}
\ubf{$\fg_C=\su(7)$}\,:
\end{center}

\begin{align}
\nn
\begin{array}{c}
\begin{tikzpicture}
\node (v1) at (0,0) {\scriptsize{$x_8-x_9$}};
\node (v2) at (2,0) {\scriptsize{$x_7-x_8$}};
\node (v3) at (4,0) {\scriptsize{$x_6-x_7$}};
\node (v4) at (6,0) {\scriptsize{$x_5-x_6$}};
\node (v5) at (8,0) {\scriptsize{$x_4-x_5$}};
\node (v1_1) at (10,0) {\scriptsize{$x_1-x_4$}};
\node (v1_4) at (5,2) {\scriptsize{$3l-2x_1-x_2-x_3-x_4-x_5-x_6-x_7-x_8$}};
\draw  (v1) edge (v2);
\draw  (v2) edge (v3);
\draw  (v3) edge (v4);
\draw  (v4) edge (v5);
\draw  (v5) edge (v1_1);
\draw  (v1_4) edge (v1);
\draw  (v1_4) edge (v1_1);
\end{tikzpicture}
\end{array}
\end{align}\\

\begin{center}
\ubf{$\fg_C\oplus\fg_E=\so(10)\oplus\su(2)$}\,:
\end{center}

\begin{align}
\nn
\begin{array}{c}
\begin{tikzpicture}
\node (v1) at (8.5,-1.5) {\scriptsize{$x_8-x_9$}};
\node (v1_1_1) at (13.5,-1.5) {\scriptsize{$3l-x_1-x_2-x_3-x_4-x_5-x_6-x_7-2x_8$}};
\node (v2) at (14,1) {\scriptsize{$2l-x_2-x_3-x_6-x_7-x_8-x_9$}};
\node (v5) at (8,0) {\scriptsize{$x_4-x_5$}};
\node (v1_1_1_1) at (10,0) {\scriptsize{$x_1-x_4$}};
\node (v1_2) at (14,0) {\scriptsize{$x_2-x_1$}};
\node (v1_3) at (16,0) {\scriptsize{$x_3-x_2$}};
\node (v1_4) at (10,1) {\scriptsize{$l-x_1-x_2-x_3$}};
\draw  (v5) edge (v1_1_1_1);
\draw  (v1_1_1_1) edge (v1_2);
\draw  (v1_2) edge (v1_3);
\draw  (v1_1_1_1) edge (v1_4);
\draw[double]  (v1) -- (v1_1_1);
\draw  (v2) edge (v1_2);
\end{tikzpicture}
\end{array}
\end{align}\\

\begin{center}
\ubf{$\fg_C\oplus\fg_E=\so(8)\oplus\su(3)$}\,:
\end{center}

\begin{align}
\nn
\begin{array}{c}
\begin{tikzpicture}
\node (v1) at (10,-2) {\scriptsize{$x_8-x_9$}};
\node (v1_1) at (14,-2) {\scriptsize{$x_7-x_8$}};
\node (v1_1_1) at (12,-1) {\scriptsize{$3l-x_1-x_2-x_3-x_4-x_5-x_6-2x_7-x_8$}};
\node (v2) at (12,2) {\scriptsize{$2l-x_2-x_3-x_6-x_7-x_8-x_9$}};
\node (v1_1_1_1) at (10,1) {\scriptsize{$x_1-x_4$}};
\node (v1_2) at (12,1) {\scriptsize{$x_2-x_1$}};
\node (v1_3) at (14,1) {\scriptsize{$x_3-x_2$}};
\node (v1_4) at (12,0) {\scriptsize{$l-x_2-x_3-x_5$}};
\draw  (v1_2) edge (v1_3);
\draw  (v1) edge (v1_1_1);
\draw  (v1) edge (v1_1);
\draw  (v2) edge (v1_2);
\draw  (v1_1_1_1) edge (v1_2);
\draw  (v1_4) edge (v1_2);
\draw  (v1_1_1) edge (v1_1);
\end{tikzpicture}
\end{array}
\end{align}\\

\begin{center}
\ubf{$\fg_C=\so(10)$}\,:
\end{center}

\begin{align}
\nn
\begin{array}{c}
\begin{tikzpicture}
\node (v2) at (14,1) {\scriptsize{$2l-x_2-x_3-x_6-x_7-x_8-x_9$}};
\node (v5) at (8,0) {\scriptsize{$x_4-x_5$}};
\node (v1_1_1_1) at (10,0) {\scriptsize{$x_1-x_4$}};
\node (v1_2) at (14,0) {\scriptsize{$x_2-x_1$}};
\node (v1_3) at (16,0) {\scriptsize{$x_3-x_2$}};
\node (v1_4) at (10,1) {\scriptsize{$l-x_1-x_2-x_3$}};
\draw  (v5) edge (v1_1_1_1);
\draw  (v1_1_1_1) edge (v1_2);
\draw  (v1_2) edge (v1_3);
\draw  (v1_1_1_1) edge (v1_4);
\draw  (v1_2) edge (v2);
\end{tikzpicture}
\end{array}
\end{align}\\

\begin{center}
\ubf{$\fg_C=\su(6)$}\,:
\end{center}

\begin{align}
\nn
\begin{array}{c}
\begin{tikzpicture}
\node (v1) at (0,0) {\scriptsize{$x_8-x_9$}};
\node (v2) at (2,0) {\scriptsize{$x_7-x_8$}};
\node (v3) at (4,0) {\scriptsize{$x_6-x_7$}};
\node (v4) at (6,0) {\scriptsize{$x_5-x_6$}};
\node (v5) at (8,0) {\scriptsize{$x_4-x_5$}};
\node (v1_4) at (4,2) {\scriptsize{$3l-x_1-x_2-x_3-2x_4-x_5-x_6-x_7-x_8$}};
\draw  (v1) edge (v2);
\draw  (v2) edge (v3);
\draw  (v3) edge (v4);
\draw  (v4) edge (v5);
\draw  (v1_4) edge (v1);
\draw  (v1_4) edge (v5);
\end{tikzpicture}
\end{array}
\end{align}\\

\begin{center}
\ubf{$\fg_C\oplus\fg_E=\so(8)\oplus\su(2)$}\,:
\end{center}

\begin{align}
\nn
\begin{array}{c}
\begin{tikzpicture}
\node (v1) at (9,-1.5) {\scriptsize{$x_8-x_9$}};
\node (v1_1_1) at (13.5,-1.5) {\scriptsize{$3l-x_1-x_2-x_3-x_4-x_5-x_6-x_7-2x_8$}};
\node (v2) at (12,2) {\scriptsize{$2l-x_2-x_3-x_6-x_7-x_8-x_9$}};
\node (v1_1_1_1) at (10,1) {\scriptsize{$x_1-x_4$}};
\node (v1_2) at (12,1) {\scriptsize{$x_2-x_1$}};
\node (v1_3) at (14,1) {\scriptsize{$x_3-x_2$}};
\node (v1_4) at (12,0) {\scriptsize{$l-x_2-x_3-x_5$}};
\draw  (v1_2) edge (v1_3);
\draw[double]  (v1) -- (v1_1_1);
\draw  (v2) edge (v1_2);
\draw  (v1_1_1_1) edge (v1_2);
\draw  (v1_4) edge (v1_2);
\end{tikzpicture}
\end{array}
\end{align}\\

\begin{center}
\ubf{$\fg_C\oplus\fg_E=\su(3)\oplus\su(3)$}\,:
\end{center}

\begin{align}
\nn
\begin{array}{c}
\begin{tikzpicture}
\node (v1) at (8,3.5) {\scriptsize{$x_8-x_9$}};
\node (v1_1) at (11,3.5) {\scriptsize{$x_7-x_8$}};
\node (v5) at (8,1.6) {\scriptsize{$x_4-x_5$}};
\node (v1_1_1_1) at (11,1.6) {\scriptsize{$x_1-x_4$}};
\draw  (v5) edge (v1_1_1_1);
\draw  (v1) edge (v1_1);
\node (v2) at (9.5,4.7) {\scriptsize{$3l-x_1-x_2-x_3-x_4-x_5-x_6-2x_7-x_8$}};
\node (v3) at (9.5,2.6) {\scriptsize{$3l-2x_1-x_2-x_3-x_4-x_6-2x_7-x_8-x_9$}};
\draw  (v2) edge (v1);
\draw  (v2) edge (v1_1);
\draw  (v3) edge (v5);
\draw  (v3) edge (v1_1_1_1);
\end{tikzpicture}
\end{array}
\end{align}\\

\begin{center}
\ubf{$\fg_C=\so(8)$}\,:
\end{center}

\begin{align}
\nn
\begin{array}{c}
\begin{tikzpicture}
\node (v2) at (12,2) {\scriptsize{$2l-x_2-x_3-x_6-x_7-x_8-x_9$}};
\node (v1_1_1_1) at (10,1) {\scriptsize{$x_1-x_4$}};
\node (v1_2) at (12,1) {\scriptsize{$x_2-x_1$}};
\node (v1_3) at (14,1) {\scriptsize{$x_3-x_2$}};
\node (v1_4) at (12,0) {\scriptsize{$l-x_2-x_3-x_5$}};
\draw  (v1_2) edge (v1_3);
\draw  (v2) edge (v1_2);
\draw  (v1_1_1_1) edge (v1_2);
\draw  (v1_4) edge (v1_2);
\end{tikzpicture}
\end{array}
\end{align}\\

\begin{center}
\ubf{$\fg_C=\su(5)$}\,:
\end{center}

\begin{align}
\nn
\begin{array}{c}
\begin{tikzpicture}
\node (v1) at (0,0) {\scriptsize{$x_8-x_9$}};
\node (v2) at (2,0) {\scriptsize{$x_7-x_8$}};
\node (v3) at (4,0) {\scriptsize{$x_6-x_7$}};
\node (v4) at (6,0) {\scriptsize{$x_5-x_6$}};
\node (v1_4) at (3,1.5) {\scriptsize{$3l-x_1-x_2-x_3-x_4-2x_5-x_6-x_7-x_8$}};
\draw  (v1) edge (v2);
\draw  (v2) edge (v3);
\draw  (v3) edge (v4);
\draw  (v1_4) edge (v1);
\draw  (v1_4) edge (v4);
\end{tikzpicture}
\end{array}
\end{align}\\

\begin{center}
\ubf{$\fg_C=\su(4)$}\,:
\end{center}

\begin{align}
\nn
\begin{array}{c}
\begin{tikzpicture}
\node (v1) at (10,-2) {\scriptsize{$x_8-x_9$}};
\node (v1_1) at (12,-3) {\scriptsize{$x_7-x_8$}};
\node (v1_1_1) at (12,-1) {\scriptsize{$3l-x_1-x_2-x_3-x_4-x_5-2x_6-x_7-x_8$}};
\node (v4) at (14,-2) {\scriptsize{$x_6-x_7$}};
\draw  (v1) edge (v1_1_1);
\draw  (v1) edge (v1_1);
\draw  (v1_1) edge (v4);
\draw  (v4) edge (v1_1_1);
\end{tikzpicture}
\end{array}
\end{align}\\

\begin{center}
\ubf{$\fg_C=\su(3)$}\,:
\end{center}

\begin{align}
\nn
\begin{array}{c}
\begin{tikzpicture}
\node (v1) at (10,-2) {\scriptsize{$x_8-x_9$}};
\node (v1_1) at (14,-2) {\scriptsize{$x_7-x_8$}};
\node (v1_1_1) at (12,-1) {\scriptsize{$3l-x_1-x_2-x_3-x_4-x_5-x_6-2x_7-x_8$}};
\draw  (v1) edge (v1_1_1);
\draw  (v1) edge (v1_1);
\draw  (v1_1_1) edge (v1_1);
\end{tikzpicture}
\end{array}
\end{align}\\

\begin{center}
\ubf{$\fg_C=\su(2)$}\,:
\end{center}

\begin{align}
\nn
\begin{array}{c}
\begin{tikzpicture}
\node (v1) at (9,-1.5) {\scriptsize{$x_8-x_9$}};
\node (v1_1_1) at (13.5,-1.5) {\scriptsize{$3l-x_1-x_2-x_3-x_4-x_5-x_6-x_7-2x_8$}};
\draw[double]  (v1) -- (v1_1_1);
\end{tikzpicture}
\end{array}
\end{align}

\subsubsection{Non-simply laced}

Now we move onto the other cases in Table \ref{table2} in which at least one of $\fg_C,\fg_E$ is non-simply laced. We have already found a set of gluing curves $D[f^i_C]$ for each simply laced $\fg_C\subseteq\fe_8$. Now consider a folding of $\fg_C$ giving rise to a non-simply laced algebra $\fh_C$ and let $C$ carry $\fh_C$. The action of folding can be represented as identifications of fibers $f^i_C\sim f^j_C$. We call the resulting fibers obtained after folding as $f'^{i'}_C=f^i_C\sim f^j_C$. Let us call the collection of surfaces living over $C$ associated to $\fh_C$ as $S'_C$ with the irreducible components being $S'^{i'}_C$. $S'^i_C$ is defined to be the surface carrying fiber $f'^i_C$. The curves gluing $S_D$ to $S'_C$ can then again be taken as the curves $D[f^i_C]$ that we discussed above. The only difference is that $D[f^i_C]$ and $D[f^j_C]$ each glues to a copy of $f'^i_C$. This automatically satisfies (\ref{glue1}) as well since $D[f^i_C]$ satisfy (\ref{glue1}).

One important difference compared to the simply laced case is that the constraint due to consistency of triple intersection numbers is slightly relaxed in this case since in transitioning from $S^i_C$ to $S'^{i'}_C$ the number of surfaces decrease but the number of gluing curves $D[f^i_C]$ remain the same. Whereas earlier we had to satisfy (\ref{EEE}), now we only need to satisfy
\be
\left(\sum_{i\to i'} D[f^i_C]\right)\cdot \left(\sum_{k\to k'} D[f^k_E]\right)=0 \label{EIV}
\ee
where the sum over $i\to i'$ means that we sum those $f^i_C$ which become a copy of $f'^{i'}_C$ under monodromy. This means that certain possibilities of $\fg_C\oplus\fg_E$ were not allowed due to the non-existence of gluing curves satisfying (\ref{EEE}), but their foldings are allowed. For example, $\fe_6\oplus\so(8)$ is not allowed but $\ff_4\oplus\fg_2$ is allowed, and this can be understood as the difference between the consistency conditions ({\ref{EEE}) vs. (\ref{EIV}) required of the gluing curves. 

Another small constraint that we have to keep in mind is that our choice of flop frames in Section \ref{flop-eq} for non-simply laced algebras was such that the surface $S^0$ glued to $B$ never participated in the monodromy. This imposes a constraint on our gluing curves below that the curve containing $x_9$ is not allowed to participate in folding. Let us provide the gluing curves now.

\begin{center}
\ubf{$\fg_C=\so(15)$}\,:
\end{center}

\begin{align}
\nn
\begin{array}{c}
\begin{tikzpicture}
\node (v1) at (0,0) {\scriptsize{$x_8-x_9$}};
\node (v2) at (2,0) {\scriptsize{$x_7-x_8$}};
\node (v3) at (4,0) {\scriptsize{$x_6-x_7$}};
\node (v4) at (6,0) {\scriptsize{$x_5-x_6$}};
\node (v5) at (8,0) {\scriptsize{$x_4-x_5$}};
\node (v1_1) at (10,0) {\scriptsize{$x_1-x_4$}};
\node (v1_2) at (13,0) {\scriptsize{$x_2-x_1$,\:\:$l-x_1-x_2-x_3$}};
\node (v1_3) at (2,1) {\scriptsize{$2l-x_1-x_2-x_4-x_5-x_6-x_7$}};
\draw  (v1) edge (v2);
\draw  (v2) edge (v3);
\draw  (v3) edge (v4);
\draw  (v4) edge (v5);
\draw  (v5) edge (v1_1);
\draw[double]  (v1_1) -- (v1_2);
\draw  (v1_3) edge (v2);
\end{tikzpicture}
\end{array}
\end{align}
which is simply a folded version of the graph for $\so(16)$, where we have folded the right end of the graph for $\so(16)$. Note that, as commented above, we could not have folded the left end of $\so(16)$ since it contains the curve containing $x_9$. The entry $x_2-x_1,l-x_1-x_2-x_3$ at the rightmost node of the graph means that we have two curves namely $x_2-x_1$ and $l-x_1-x_2-x_3$ gluing to different copies of the same fiber in the collection $S_C$. The double edge between $x_2-x_1,l-x_1-x_2-x_3$ and $x_1-x_4$ indicates the intersection of $x_1-x_4$ with the sum of $x_2-x_1$ and $l-x_1-x_2-x_3$. From now on, if carries multiple curves $\ell,\ell',\cdots$ then the edges to that node will always signify the total intersection with $\ell+\ell'+\cdots$.

\begin{center}
\ubf{$\fg_C\oplus\fg_E=\so(9)\oplus\so(7)$}\,:
\end{center}

\begin{align}
\nn
\begin{array}{c}
\begin{tikzpicture}
\node (v1) at (8,3.5) {\scriptsize{$x_8-x_9$}};
\node (v1_1) at (11,3.5) {\scriptsize{$x_7-x_8$}};
\node (v1_1_1) at (11,5) {\scriptsize{$2l-x_1-x_2-x_4-x_5-x_6-x_7$}};
\node (v2) at (11,2) {\scriptsize{$2l-x_1-x_2-x_6-x_7-x_8-x_9$}};
\node (v4) at (14,3.5) {\scriptsize{$x_6-x_7$}};
\node (v5) at (8,0.5) {\scriptsize{$x_4-x_5$}};
\node (v1_1_1_1) at (11,0.5) {\scriptsize{$x_1-x_4$}};
\node (v1_2) at (15,0.5) {\scriptsize{$x_2-x_1$,\:\:$l-x_1-x_2-x_3$}};
\node (v1_3) at (17.5,3.5) {\scriptsize{$x_5-x_6$,\:\:$l-x_3-x_5-x_6$}};
\draw  (v5) edge (v1_1_1_1);
\draw[double]  (v1_1_1_1) -- (v1_2);
\draw  (v1) edge (v1_1);
\draw  (v1_1) edge (v4);
\draw  (v1_1_1_1) edge (v2);
\draw  (v1_1_1) edge (v1_1);
\draw[double]  (v4) -- (v1_3);
\end{tikzpicture}
\end{array}
\end{align}
Notice that if we unfold the graphs, then we would obtain gluing curves for $\so(10)$ and $\so(8)$ but they will mutually intersect with each other thus violating (\ref{EEE}) and in fact $\so(10)\oplus\so(8)$ is not an allowed value for $\fg_C\oplus\fg_E$ as can be seen from Table \ref{table2}.

\begin{center}
\ubf{$\fg_C\oplus\fg_E=\so(13)\oplus\su(2)$}\,:
\end{center}

\begin{align}
\nn
\begin{array}{c}
\begin{tikzpicture}
\node (v1) at (0,0) {\scriptsize{$x_8-x_9$}};
\node (v2) at (2,0) {\scriptsize{$x_7-x_8$}};
\node (v3) at (4,0) {\scriptsize{$x_6-x_7$}};
\node (v4) at (6,0) {\scriptsize{$x_5-x_6$}};
\node (v5) at (8,0) {\scriptsize{$x_4-x_5$}};
\node (v1_1) at (11,0) {\scriptsize{$x_1-x_4$,\:\:$l-x_1-x_3-x_4$}};
\node (v1_3) at (2,1) {\scriptsize{$2l-x_1-x_2-x_4-x_5-x_6-x_7$}};
\draw  (v1) edge (v2);
\draw  (v2) edge (v3);
\draw  (v3) edge (v4);
\draw  (v4) edge (v5);
\draw[double]  (v5) -- (v1_1);
\draw  (v1_3) edge (v2);
\node (v6) at (2,-1.5) {\scriptsize{$x_2-x_1$}};
\node (v7) at (7,-1.5) {\scriptsize{$3l-2x_2-x_3-x_4-x_5-x_6-x_7-x_8-x_9$}};
\draw [double] (v6) -- (v7);
\end{tikzpicture}
\end{array}
\end{align}\\

\begin{center}
\ubf{$\fg_C\oplus\fg_E=\so(9)\oplus\su(4)$}\,:
\end{center}

\begin{align}
\nn
\begin{array}{c}
\begin{tikzpicture}
\node (v1) at (10.5,-2.5) {\scriptsize{$x_8-x_9$}};
\node (v1_1) at (12.5,-3.5) {\scriptsize{$x_7-x_8$}};
\node (v1_1_1) at (12.5,-1.5) {\scriptsize{$3l-x_1-x_2-x_3-x_4-x_5-2x_6-x_7-x_8$}};
\node (v2) at (14,1) {\scriptsize{$2l-x_2-x_3-x_6-x_7-x_8-x_9$}};
\node (v4) at (14.5,-2.5) {\scriptsize{$x_6-x_7$}};
\node (v5) at (9,0) {\scriptsize{$l-x_1-x_2-x_3$,\:\:$x_4-x_5$}};
\node (v1_1_1_1) at (12,0) {\scriptsize{$x_1-x_4$}};
\node (v1_2) at (14,0) {\scriptsize{$x_2-x_1$}};
\node (v1_3) at (16,0) {\scriptsize{$x_3-x_2$}};
\draw[double]  (v5) -- (v1_1_1_1);
\draw  (v1_1_1_1) edge (v1_2);
\draw  (v1_2) edge (v1_3);
\draw  (v1) edge (v1_1_1);
\draw  (v1) edge (v1_1);
\draw  (v2) edge (v1_2);
\draw  (v1_1) edge (v4);
\draw  (v4) edge (v1_1_1);
\end{tikzpicture}
\end{array}
\end{align}\\

\begin{center}
\ubf{$\fg_C=\so(13)$}\,:
\end{center}

\begin{align}
\nn
\begin{array}{c}
\begin{tikzpicture}
\node (v1) at (0,0) {\scriptsize{$x_8-x_9$}};
\node (v2) at (2,0) {\scriptsize{$x_7-x_8$}};
\node (v3) at (4,0) {\scriptsize{$x_6-x_7$}};
\node (v4) at (6,0) {\scriptsize{$x_5-x_6$}};
\node (v5) at (8,0) {\scriptsize{$x_4-x_5$}};
\node (v1_1) at (11,0) {\scriptsize{$x_1-x_4$,\:\:$l-x_1-x_3-x_4$}};
\node (v1_3) at (2,1) {\scriptsize{$2l-x_1-x_2-x_4-x_5-x_6-x_7$}};
\draw  (v1) edge (v2);
\draw  (v2) edge (v3);
\draw  (v3) edge (v4);
\draw  (v4) edge (v5);
\draw[double]  (v5) -- (v1_1);
\draw  (v1_3) edge (v2);
\end{tikzpicture}
\end{array}
\end{align}\\

\begin{center}
\ubf{$\fg_C\oplus\fg_e=\ff_4\oplus\fg_2$}\,:
\end{center}

\begin{align}
\nn
\begin{array}{c}
\begin{tikzpicture}
\node (v2) at (5.5,6.5) {\scriptsize{$l-x_7-x_8-x_9$}};
\node (v1_1_1_1) at (11,6.5) {\scriptsize{$x_1-x_4$}};
\node (v1_2) at (13.5,6.5) {\scriptsize{$x_2-x_1$,\:\:$x_4-x_5$}};
\node (v1_3) at (16.5,6.5) {\scriptsize{$x_3-x_2$,\:\:$x_5-x_6$}};
\node (v1_4) at (8.5,6.5) {\scriptsize{$l-x_1-x_2-x_3$}};
\draw [double] (v1_1_1_1) -- (v1_2);
\draw[double]  (v1_2) -- (v1_3);
\draw  (v1_1_1_1) edge (v1_4);
\draw  (v1_4) edge (v2);
\node (v1) at (6.5,5) {\scriptsize{$x_8-x_9$}};
\node (v1_1) at (8.5,5) {\scriptsize{$x_7-x_8$}};
\node (v4_1) at (14.1,5) {\scriptsize{$x_6-x_7$,\:\:$l-x_3-x_6-x_7$,\:\:$2l-x_1-x_2-x_4-x_5-x_6-x_7$}};
\draw  (v1) edge (v1_1);
\draw (9.08,5) -- (10.04,5);
\draw (9.08,4.96) -- (10.04,4.96);
\draw (9.08,5.04) -- (10.04,5.04);
\end{tikzpicture}
\end{array}
\end{align}\\

\begin{center}
\ubf{$\fg_C\oplus\fg_E=\ff_4\oplus\su(3)$}\,:
\end{center}

\begin{align}
\nn
\begin{array}{c}
\begin{tikzpicture}
\node (v1) at (8,-2.5) {\scriptsize{$x_8-x_9$}};
\node (v1_1) at (12,-2.5) {\scriptsize{$x_7-x_8$}};
\node (v1_1_1) at (10,-1.5) {\scriptsize{$3l-x_1-x_2-x_3-x_4-x_5-x_6-2x_7-x_8$}};
\node (v2) at (4.5,0) {\scriptsize{$l-x_7-x_8-x_9$}};
\node (v1_1_1_1) at (10,0) {\scriptsize{$x_1-x_4$}};
\node (v1_2) at (12.5,0) {\scriptsize{$x_2-x_1$,\:\:$x_4-x_5$}};
\node (v1_3) at (15.5,0) {\scriptsize{$x_3-x_2$,\:\:$x_5-x_6$}};
\node (v1_4) at (7.5,0) {\scriptsize{$l-x_1-x_2-x_3$}};
\draw [double] (v1_1_1_1) -- (v1_2);
\draw[double]  (v1_2) -- (v1_3);
\draw  (v1_1_1_1) edge (v1_4);
\draw  (v1_4) edge (v2);
\draw  (v1_1_1) -- (v1_1);
\draw  (v1) edge (v1_1_1);
\draw  (v1) edge (v1_1);
\end{tikzpicture}
\end{array}
\end{align}\\

\begin{center}
\ubf{$\fg_C\oplus\fg_E=\so(9)\oplus\fg_2$}\,:
\end{center}

\begin{align}
\nn
\begin{array}{c}
\begin{tikzpicture}
\node (v2) at (13,7.5) {\scriptsize{$2l-x_2-x_3-x_6-x_7-x_8-x_9$}};
\node (v5) at (8,6.5) {\scriptsize{$x_4-x_5$,\:\:$l-x_1-x_2-x_3$}};
\node (v1_1_1_1) at (11,6.5) {\scriptsize{$x_1-x_4$}};
\node (v1_2) at (13,6.5) {\scriptsize{$x_2-x_1$}};
\node (v1_3) at (15,6.5) {\scriptsize{$x_3-x_2$}};
\draw[double]  (v5) -- (v1_1_1_1);
\draw  (v1_1_1_1) edge (v1_2);
\draw  (v1_2) edge (v1_3);
\draw  (v2) edge (v1_2);
\node (v1) at (6.5,5) {\scriptsize{$x_8-x_9$}};
\node (v1_1) at (8.5,5) {\scriptsize{$x_7-x_8$}};
\node (v4_1) at (14.1,5) {\scriptsize{$x_6-x_7$,\:\:$l-x_3-x_6-x_7$,\:\:$2l-x_1-x_2-x_4-x_5-x_6-x_7$}};
\draw  (v1) edge (v1_1);
\draw (9.08,5) -- (10.04,5);
\draw (9.08,4.96) -- (10.04,4.96);
\draw (9.08,5.04) -- (10.04,5.04);
\end{tikzpicture}
\end{array}
\end{align}\\

\begin{center}
\ubf{$\fg_C\oplus\fg_E=\so(8)\oplus\fg_2$}\,:
\end{center}

\begin{align}
\nn
\begin{array}{c}
\begin{tikzpicture}
\node (v2) at (11,6.5) {\scriptsize{$2l-x_1-x_2-x_6-x_7-x_8-x_9$}};
\node (v5) at (8,8) {\scriptsize{$x_4-x_5$}};
\node (v1_1_1_1) at (11,8) {\scriptsize{$x_1-x_4$}};
\node (v1_2) at (14,8) {\scriptsize{$x_2-x_1$}};
\node (v1_4) at (11,9.5) {\scriptsize{$l-x_1-x_2-x_3$}};
\draw  (v5) edge (v1_1_1_1);
\draw  (v1_1_1_1) edge (v1_2);
\draw  (v1_1_1_1) edge (v1_4);
\draw  (v1_1_1_1) edge (v2);
\node (v1) at (6.5,5) {\scriptsize{$x_8-x_9$}};
\node (v1_1) at (8.5,5) {\scriptsize{$x_7-x_8$}};
\node (v4_1) at (13.75,5) {\scriptsize{$x_6-x_7$,\:\:$l-x_3-x_6-x_7$,\:\:$2l-x_1-x_2-x_4-x_5-x_6-x_7$}};
\draw  (v1) edge (v1_1);
\draw (9.08,5) -- (10.04,5);
\draw (9.08,4.96) -- (10.04,4.96);
\draw (9.08,5.04) -- (10.04,5.04);
\end{tikzpicture}
\end{array}
\end{align}\\

\begin{center}
\ubf{$\fg_C\oplus\fg_E=\so(7)\oplus\so(7)$}\,:
\end{center}

\begin{align}
\nn
\begin{array}{c}
\begin{tikzpicture}
\node (v1) at (8,3.5) {\scriptsize{$x_8-x_9$}};
\node (v1_1) at (11,3.5) {\scriptsize{$x_7-x_8$}};
\node (v1_1_1) at (11,5) {\scriptsize{$2l-x_1-x_2-x_4-x_5-x_6-x_7$}};
\node (v2) at (11,2) {\scriptsize{$2l-x_1-x_2-x_6-x_7-x_8-x_9$}};
\node (v4) at (14.5,3.5) {\scriptsize{$x_6-x_7$,\:\:$l-x_3-x_6-x_7$}};
\node (v5) at (8,0.5) {\scriptsize{$x_4-x_5$}};
\node (v1_1_1_1) at (11,0.5) {\scriptsize{$x_1-x_4$}};
\node (v1_2) at (14.5,0.5) {\scriptsize{$x_2-x_1$,\:\:$l-x_1-x_2-x_3$}};
\draw  (v5) edge (v1_1_1_1);
\draw[double]  (v1_1_1_1) -- (v1_2);
\draw  (v1) edge (v1_1);
\draw[double]  (v1_1) -- (v4);
\draw  (v1_1_1_1) edge (v2);
\draw  (v1_1_1) edge (v1_1);
\end{tikzpicture}
\end{array}
\end{align}\\

\begin{center}
\ubf{$\fg_C\oplus\fg_E=\so(8)\oplus\so(7)$}\,:
\end{center}

\begin{align}
\nn
\begin{array}{c}
\begin{tikzpicture}
\node (v1) at (8,3.5) {\scriptsize{$x_8-x_9$}};
\node (v1_1) at (11,3.5) {\scriptsize{$x_7-x_8$}};
\node (v1_1_1) at (11,5) {\scriptsize{$2l-x_1-x_2-x_4-x_5-x_6-x_7$}};
\node (v4) at (14.5,3.5) {\scriptsize{$x_6-x_7$,\:\:$l-x_3-x_6-x_7$}};
\node (v2) at (11,6.5) {\scriptsize{$2l-x_1-x_2-x_6-x_7-x_8-x_9$}};
\node (v5) at (8,8) {\scriptsize{$x_4-x_5$}};
\node (v1_1_1_1) at (11,8) {\scriptsize{$x_1-x_4$}};
\node (v1_2) at (14,8) {\scriptsize{$x_2-x_1$}};
\node (v1_4) at (11,9.5) {\scriptsize{$l-x_1-x_2-x_3$}};
\draw  (v5) edge (v1_1_1_1);
\draw  (v1_1_1_1) edge (v1_2);
\draw  (v1_1_1_1) edge (v1_4);
\draw  (v1_1_1_1) edge (v2);
\draw  (v1) edge (v1_1);
\draw[double]  (v1_1) -- (v4);
\draw  (v1_1_1) edge (v1_1);
\end{tikzpicture}
\end{array}
\end{align}\\

\begin{center}
\ubf{$\fg_C\oplus\fg_E=\so(11)\oplus\su(2)$}\,:
\end{center}

\begin{align}
\nn
\begin{array}{c}
\begin{tikzpicture}
\node (v1) at (5.5,-1.5) {\scriptsize{$x_8-x_9$}};
\node (v2) at (10.5,-1.5) {\scriptsize{$3l-x_1-x_2-x_3-x_4-x_5-x_6-x_7-2x_8$}};
\node (v3) at (4,0) {\scriptsize{$x_6-x_7$}};
\node (v4) at (6,0) {\scriptsize{$x_5-x_6$}};
\node (v5) at (8,0) {\scriptsize{$x_4-x_5$}};
\node (v1_1) at (10,0) {\scriptsize{$x_1-x_4$}};
\node (v1_2) at (13,0) {\scriptsize{$x_2-x_1$,\:\:$l-x_1-x_2-x_3$}};
\node (v1_3) at (6,1) {\scriptsize{$2l-x_1-x_2-x_4-x_5-x_8-x_9$}};
\draw  (v3) edge (v4);
\draw  (v4) edge (v5);
\draw  (v5) edge (v1_1);
\draw[double]  (v1_1) -- (v1_2);
\draw  (v1_3) edge (v4);
\draw[double]  (v2) -- (v1);
\end{tikzpicture}
\end{array}
\end{align}\\

\begin{center}
\ubf{$\fg_C\oplus\fg_E=\so(9)\oplus\su(3)$}\,:
\end{center}

\begin{align}
\nn
\begin{array}{c}
\begin{tikzpicture}
\node (v1) at (10.5,-2.5) {\scriptsize{$x_8-x_9$}};
\node (v1_1) at (14.5,-2.5) {\scriptsize{$x_7-x_8$}};
\node (v1_1_1) at (12.5,-1.5) {\scriptsize{$3l-x_1-x_2-x_3-x_4-x_5-x_6-2x_7-x_8$}};
\node (v2) at (14,1) {\scriptsize{$2l-x_2-x_3-x_6-x_7-x_8-x_9$}};
\node (v5) at (9,0) {\scriptsize{$l-x_1-x_2-x_3$,\:\:$x_4-x_5$}};
\node (v1_1_1_1) at (12,0) {\scriptsize{$x_1-x_4$}};
\node (v1_2) at (14,0) {\scriptsize{$x_2-x_1$}};
\node (v1_3) at (16,0) {\scriptsize{$x_3-x_2$}};
\draw [double] (v5) -- (v1_1_1_1);
\draw  (v1_1_1_1) edge (v1_2);
\draw  (v1_2) edge (v1_3);
\draw  (v1) edge (v1_1_1);
\draw  (v1) edge (v1_1);
\draw  (v2) edge (v1_2);
\draw  (v1_1_1) edge (v1_1);
\end{tikzpicture}
\end{array}
\end{align}\\

\begin{center}
\ubf{$\fg_C\oplus\fg_E=\so(7)\oplus\su(4)$}\,:
\end{center}

\begin{align}
\nn
\begin{array}{c}
\begin{tikzpicture}
\node (v1) at (10,-1.5) {\scriptsize{$x_8-x_9$}};
\node (v1_1) at (12,-2.5) {\scriptsize{$x_7-x_8$}};
\node (v1_1_1) at (12,-0.5) {\scriptsize{$3l-x_1-x_2-x_3-x_4-x_5-2x_6-x_7-x_8$}};
\node (v2) at (12,2) {\scriptsize{$2l-x_2-x_3-x_6-x_7-x_8-x_9$}};
\node (v4) at (14,-1.5) {\scriptsize{$x_6-x_7$}};
\node (v1_1_1_1) at (10,1) {\scriptsize{$x_1-x_4$}};
\node (v1_2) at (12,1) {\scriptsize{$x_2-x_1$}};
\node (v1_3) at (15,1) {\scriptsize{$x_3-x_2$,\:\:$l-x_2-x_3-x_5$}};
\draw [double] (v1_2) -- (v1_3);
\draw  (v1) edge (v1_1_1);
\draw  (v1) edge (v1_1);
\draw  (v2) edge (v1_2);
\draw  (v1_1) edge (v4);
\draw  (v4) edge (v1_1_1);
\draw  (v1_1_1_1) edge (v1_2);
\end{tikzpicture}
\end{array}
\end{align}\\

\begin{center}
\ubf{$\fg_C=\so(11)$}\,:
\end{center}

\begin{align}
\nn
\begin{array}{c}
\begin{tikzpicture}
\node (v3) at (4,0) {\scriptsize{$x_6-x_7$}};
\node (v4) at (6,0) {\scriptsize{$x_5-x_6$}};
\node (v5) at (8,0) {\scriptsize{$x_4-x_5$}};
\node (v1_1) at (10,0) {\scriptsize{$x_1-x_4$}};
\node (v1_2) at (13,0) {\scriptsize{$x_2-x_1$,\:\:$l-x_1-x_2-x_3$}};
\node (v1_3) at (6,1) {\scriptsize{$2l-x_1-x_2-x_4-x_5-x_8-x_9$}};
\draw  (v3) edge (v4);
\draw  (v4) edge (v5);
\draw  (v5) edge (v1_1);
\draw [double] (v1_1) -- (v1_2);
\draw  (v1_3) edge (v4);
\end{tikzpicture}
\end{array}
\end{align}\\

\begin{center}
\ubf{$\fg_C\oplus\fg_E=\ff_4\oplus\su(2)$}\,:
\end{center}

\begin{align}
\nn
\begin{array}{c}
\begin{tikzpicture}
\node (v1) at (7,-1.5) {\scriptsize{$x_8-x_9$}};
\node (v1_1_1) at (11.5,-1.5) {\scriptsize{$3l-x_1-x_2-x_3-x_4-x_5-x_6-x_7-2x_8$}};
\node (v2) at (4.5,0) {\scriptsize{$l-x_7-x_8-x_9$}};
\node (v1_1_1_1) at (10,0) {\scriptsize{$x_1-x_4$}};
\node (v1_2) at (12.5,0) {\scriptsize{$x_2-x_1$,\:\:$x_4-x_5$}};
\node (v1_3) at (15.5,0) {\scriptsize{$x_3-x_2$,\:\:$x_5-x_6$}};
\node (v1_4) at (7.5,0) {\scriptsize{$l-x_1-x_2-x_3$}};
\draw [double] (v1_1_1_1) -- (v1_2);
\draw[double]  (v1_2) -- (v1_3);
\draw  (v1_1_1_1) edge (v1_4);
\draw  (v1_4) edge (v2);
\draw [double] (v1) -- (v1_1_1);
\end{tikzpicture}
\end{array}
\end{align}\\

\begin{center}
\ubf{$\fg_C\oplus\fg_E=\so(7)\oplus\fg_2$}\,:
\end{center}

\begin{align}
\nn
\begin{array}{c}
\begin{tikzpicture}
\node (v5) at (7.5,6.5) {\scriptsize{$x_4-x_5$}};
\node (v1_1_1_1) at (10.5,6.5) {\scriptsize{$x_1-x_4$}};
\node (v1_2) at (14,6.5) {\scriptsize{$x_2-x_1$,\:\:$l-x_1-x_2-x_3$}};
\node (v1_4) at (10.5,8) {\scriptsize{$2l-x_1-x_2-x_6-x_7-x_8-x_9$}};
\draw  (v5) edge (v1_1_1_1);
\draw[double]  (v1_1_1_1) -- (v1_2);
\draw  (v1_1_1_1) edge (v1_4);
\node (v1) at (6.5,5) {\scriptsize{$x_8-x_9$}};
\node (v1_1) at (8.5,5) {\scriptsize{$x_7-x_8$}};
\node (v4_1) at (14.1,5) {\scriptsize{$x_6-x_7$,\:\:$l-x_3-x_6-x_7$,\:\:$2l-x_1-x_2-x_4-x_5-x_6-x_7$}};
\draw  (v1) edge (v1_1);
\draw (9.08,5) -- (10.04,5);
\draw (9.08,4.96) -- (10.04,4.96);
\draw (9.08,5.04) -- (10.04,5.04);
\end{tikzpicture}
\end{array}
\end{align}\\

\begin{center}
\ubf{$\fg_C\oplus\fg_E=\su(4)\oplus\fg_2$}\,:
\end{center}

\begin{align}
\nn
\begin{array}{c}
\begin{tikzpicture}
\node (v5) at (8,8) {\scriptsize{$x_4-x_5$}};
\node (v1_1_1_1) at (11,6.5) {\scriptsize{$x_1-x_4$}};
\node (v1_2) at (14,8) {\scriptsize{$x_2-x_1$}};
\node (v1_4) at (11,9.5) {\scriptsize{$3l-x_1-2x_2-x_3-x_4-x_6-x_7-x_8-x_9$}};
\draw  (v5) edge (v1_1_1_1);
\draw  (v1_1_1_1) edge (v1_2);
\node (v1) at (6.5,5) {\scriptsize{$x_8-x_9$}};
\node (v1_1) at (8.5,5) {\scriptsize{$x_7-x_8$}};
\node (v4_1) at (14.1,5) {\scriptsize{$x_6-x_7$,\:\:$l-x_3-x_6-x_7$,\:\:$2l-x_1-x_2-x_4-x_5-x_6-x_7$}};
\draw  (v1) edge (v1_1);
\draw (9.08,5) -- (10.04,5);
\draw (9.08,4.96) -- (10.04,4.96);
\draw (9.08,5.04) -- (10.04,5.04);
\draw  (v5) edge (v1_4);
\draw  (v1_4) edge (v1_2);
\end{tikzpicture}
\end{array}
\end{align}\\

\begin{center}
\ubf{$\fg_C\oplus\fg_E=\so(9)\oplus\su(2)$}\,:
\end{center}

\begin{align}
\nn
\begin{array}{c}
\begin{tikzpicture}
\node (v1) at (8.5,-1.5) {\scriptsize{$x_8-x_9$}};
\node (v1_1_1) at (13.5,-1.5) {\scriptsize{$3l-x_1-x_2-x_3-x_4-x_5-x_6-x_7-2x_8$}};
\node (v2) at (14,1) {\scriptsize{$2l-x_2-x_3-x_6-x_7-x_8-x_9$}};
\node (v5) at (9,0) {\scriptsize{$l-x_1-x_2-x_3$,\:\:$x_4-x_5$}};
\node (v1_1_1_1) at (12,0) {\scriptsize{$x_1-x_4$}};
\node (v1_2) at (14,0) {\scriptsize{$x_2-x_1$}};
\node (v1_3) at (16,0) {\scriptsize{$x_3-x_2$}};
\draw [double] (v5) -- (v1_1_1_1);
\draw  (v1_1_1_1) edge (v1_2);
\draw  (v1_2) edge (v1_3);
\draw  [double](v1) -- (v1_1_1);
\draw  (v2) edge (v1_2);
\end{tikzpicture}
\end{array}
\end{align}\\

\begin{center}
\ubf{$\fg_C\oplus\fg_E=\so(7)\oplus\su(3)$}\,:
\end{center}

\begin{align}
\nn
\begin{array}{c}
\begin{tikzpicture}
\node (v1) at (11,-1.5) {\scriptsize{$x_8-x_9$}};
\node (v1_1) at (15,-1.5) {\scriptsize{$x_7-x_8$}};
\node (v1_1_1) at (13,-0.5) {\scriptsize{$3l-x_1-x_2-x_3-x_4-x_5-x_6-2x_7-x_8$}};
\node (v2) at (12,2) {\scriptsize{$2l-x_2-x_3-x_6-x_7-x_8-x_9$}};
\node (v1_1_1_1) at (10,1) {\scriptsize{$x_1-x_4$}};
\node (v1_2) at (12,1) {\scriptsize{$x_2-x_1$}};
\node (v1_3) at (15,1) {\scriptsize{$x_3-x_2$,\:\:$l-x_2-x_3-x_5$}};
\draw [double] (v1_2) -- (v1_3);
\draw  (v1) edge (v1_1_1);
\draw  (v1) edge (v1_1);
\draw  (v2) edge (v1_2);
\draw  (v1_1_1_1) edge (v1_2);
\draw  (v1_1_1) edge (v1_1);
\end{tikzpicture}
\end{array}
\end{align}\\

\begin{center}
\ubf{$\fg_C=\so(9)$}\,:
\end{center}

\begin{align}
\nn
\begin{array}{c}
\begin{tikzpicture}
\node (v2) at (14,1) {\scriptsize{$2l-x_2-x_3-x_6-x_7-x_8-x_9$}};
\node (v5) at (9,0) {\scriptsize{$l-x_1-x_2-x_3$,\:\:$x_4-x_5$}};
\node (v1_1_1_1) at (12,0) {\scriptsize{$x_1-x_4$}};
\node (v1_2) at (14,0) {\scriptsize{$x_2-x_1$}};
\node (v1_3) at (16,0) {\scriptsize{$x_3-x_2$}};
\draw [double] (v5) -- (v1_1_1_1);
\draw  (v1_1_1_1) edge (v1_2);
\draw  (v1_2) edge (v1_3);
\draw  (v2) edge (v1_2);
\end{tikzpicture}
\end{array}
\end{align}\\

\begin{center}
\ubf{$\fg_C=\ff_4$}\,:
\end{center}

\begin{align}
\nn
\begin{array}{c}
\begin{tikzpicture}
\node (v2) at (4.5,0) {\scriptsize{$l-x_7-x_8-x_9$}};
\node (v1_1_1_1) at (10,0) {\scriptsize{$x_1-x_4$}};
\node (v1_2) at (12.5,0) {\scriptsize{$x_2-x_1$,\:\:$x_4-x_5$}};
\node (v1_3) at (15.5,0) {\scriptsize{$x_3-x_2$,\:\:$x_5-x_6$}};
\node (v1_4) at (7.5,0) {\scriptsize{$l-x_1-x_2-x_3$}};
\draw [double] (v1_1_1_1) -- (v1_2);
\draw[double]  (v1_2) -- (v1_3);
\draw  (v1_1_1_1) edge (v1_4);
\draw  (v1_4) edge (v2);
\end{tikzpicture}
\end{array}
\end{align}\\

\begin{center}
\ubf{$\fg_C\oplus\fg_E=\fg_2\oplus\fg_2$}\,:
\end{center}

\begin{align}
\nn
\begin{array}{c}
\begin{tikzpicture}
\node (v1_1_1_1) at (10.5,6.5) {\scriptsize{$x_1-x_4$}};
\node (v1_2) at (14.5,6.5) {\scriptsize{$x_2-x_1$,\:\:$x_4-x_5$,\:\:$l-x_1-x_2-x_3$}};
\node (v1_4) at (7,6.5) {\scriptsize{$2l-x_1-x_2-x_6-x_7-x_8-x_9$}};
\draw  (v1_1_1_1) edge (v1_4);
\node (v1) at (6.5,5) {\scriptsize{$x_8-x_9$}};
\node (v1_1) at (8.5,5) {\scriptsize{$x_7-x_8$}};
\node (v4_1) at (14.1,5) {\scriptsize{$x_6-x_7$,\:\:$l-x_3-x_6-x_7$,\:\:$2l-x_1-x_2-x_4-x_5-x_6-x_7$}};
\draw  (v1) edge (v1_1);
\draw (9.08,5) -- (10.04,5);
\draw (9.08,4.96) -- (10.04,4.96);
\draw (9.08,5.04) -- (10.04,5.04);
\begin{scope}[shift={(2.02,1.5)}]
\draw (9.08,5) -- (10.04,5);
\draw (9.08,4.96) -- (10.04,4.96);
\draw (9.08,5.04) -- (10.04,5.04);
\end{scope}
\end{tikzpicture}
\end{array}
\end{align}\\

\begin{center}
\ubf{$\fg_C\oplus\fg_E=\su(3)\oplus\fg_2$}\,:
\end{center}

\begin{align}
\nn
\begin{array}{c}
\begin{tikzpicture}
\node (v5) at (8,6.5) {\scriptsize{$x_4-x_5$}};
\node (v1_1_1_1) at (14,6.5) {\scriptsize{$x_1-x_4$}};
\node (v1_4) at (11,8) {\scriptsize{$3l-2x_1-x_2-x_3-x_4-x_6-x_7-x_8-x_9$}};
\draw  (v5) edge (v1_1_1_1);
\node (v1) at (6.5,5) {\scriptsize{$x_8-x_9$}};
\node (v1_1) at (8.5,5) {\scriptsize{$x_7-x_8$}};
\node (v4_1) at (14.1,5) {\scriptsize{$x_6-x_7$,\:\:$l-x_3-x_6-x_7$,\:\:$2l-x_1-x_2-x_4-x_5-x_6-x_7$}};
\draw  (v1) edge (v1_1);
\draw (9.08,5) -- (10.04,5);
\draw (9.08,4.96) -- (10.04,4.96);
\draw (9.08,5.04) -- (10.04,5.04);
\draw  (v5) edge (v1_4);
\draw  (v1_1_1_1) edge (v1_4);
\end{tikzpicture}
\end{array}
\end{align}\\

\begin{center}
\ubf{$\fg_C\oplus\fg_E=\so(7)\oplus\su(2)$}\,:
\end{center}

\begin{align}
\nn
\begin{array}{c}
\begin{tikzpicture}
\node (v1) at (9.5,-0.5) {\scriptsize{$x_8-x_9$}};
\node (v1_1_1) at (14.5,-0.5) {\scriptsize{$3l-x_1-x_2-x_3-x_4-x_5-x_6-x_7-2x_8$}};
\node (v2) at (12,2) {\scriptsize{$2l-x_2-x_3-x_6-x_7-x_8-x_9$}};
\node (v1_1_1_1) at (10,1) {\scriptsize{$x_1-x_4$}};
\node (v1_2) at (12,1) {\scriptsize{$x_2-x_1$}};
\node (v1_3) at (15,1) {\scriptsize{$x_3-x_2$,\:\:$l-x_2-x_3-x_5$}};
\draw [double] (v1_2) -- (v1_3);
\draw  [double](v1) -- (v1_1_1);
\draw  (v2) edge (v1_2);
\draw  (v1_1_1_1) edge (v1_2);
\end{tikzpicture}
\end{array}
\end{align}\\

\begin{center}
\ubf{$\fg_C=\so(7)$}\,:
\end{center}

\begin{align}
\nn
\begin{array}{c}
\begin{tikzpicture}
\node (v2) at (12,2) {\scriptsize{$2l-x_2-x_3-x_6-x_7-x_8-x_9$}};
\node (v1_1_1_1) at (10,1) {\scriptsize{$x_1-x_4$}};
\node (v1_2) at (12,1) {\scriptsize{$x_2-x_1$}};
\node (v1_3) at (15,1) {\scriptsize{$x_3-x_2$,\:\:$l-x_2-x_3-x_5$}};
\draw [double] (v1_2) -- (v1_3);
\draw  (v2) edge (v1_2);
\draw  (v1_1_1_1) edge (v1_2);
\end{tikzpicture}
\end{array}
\end{align}\\

\begin{center}
\ubf{$\fg_C\oplus\fg_E=\su(2)\oplus\fg_2$}\,:
\end{center}

\begin{align}
\nn
\begin{array}{c}
\begin{tikzpicture}
\node (v5) at (8,6.5) {\scriptsize{$x_4-x_5$}};
\node (v1_4) at (13,6.5) {\scriptsize{$3l-x_1-x_2-x_3-2x_4-x_6-x_7-x_8-x_9$}};
\node (v1) at (6.5,5) {\scriptsize{$x_8-x_9$}};
\node (v1_1) at (8.5,5) {\scriptsize{$x_7-x_8$}};
\node (v4_1) at (14.1,5) {\scriptsize{$x_6-x_7$,\:\:$l-x_3-x_6-x_7$,\:\:$2l-x_1-x_2-x_4-x_5-x_6-x_7$}};
\draw  (v1) edge (v1_1);
\draw (9.08,5) -- (10.04,5);
\draw (9.08,4.96) -- (10.04,4.96);
\draw (9.08,5.04) -- (10.04,5.04);
\draw[double]  (v5) -- (v1_4);
\end{tikzpicture}
\end{array}
\end{align}\\

\begin{center}
\ubf{$\fg_C=\fg_2$}\,:
\end{center}

\begin{align}
\nn
\begin{array}{c}
\begin{tikzpicture}
\node (v1) at (6.5,5) {\scriptsize{$x_8-x_9$}};
\node (v1_1) at (8.5,5) {\scriptsize{$x_7-x_8$}};
\node (v4_1) at (14.1,5) {\scriptsize{$x_6-x_7$,\:\:$l-x_3-x_6-x_7$,\:\:$2l-x_1-x_2-x_4-x_5-x_6-x_7$}};
\draw  (v1) edge (v1_1);
\draw (9.08,5) -- (10.04,5);
\draw (9.08,4.96) -- (10.04,4.96);
\draw (9.08,5.04) -- (10.04,5.04);
\end{tikzpicture}
\end{array}
\end{align}\\

We would like to end this section by commenting about the seemingly different configurations of gluing curves that glue $\sp(0)$ to some gauge algebra $\fg_C$. The different configurations can, for instance, be discovered by looking at the subgraph corresponding to gluing to $\fg_C$ in the graph proposed for gluing to $\fg_C\oplus\fg_E$ and comparing it to the graph proposed for gluing to $\fg_C$ alone. In many cases, these two graphs are different, and one might worry that the two different graphs lead to two different gluings. For example, one can compare the $\so(8)$ subgraphs proposed in the gluing to $\so(8)\oplus\so(8)$ to the graph proposed in the gluing to $\so(8)$. We conjecture that the different graphs should be related  by automorphisms of $dP_9$ and hence the corresponding gluings should be equivalent. We plan to return to this point in the near future. This is related to the first item in the list presented in Section \ref{conc}.

\section{Future work}\label{conc}
This work opens up many interesting directions for future research:
\bit
\item First of all, the description of Mori cone $\cM_\fT$ of the threefold $\tilde X_\fT$ in Section \ref{mori} was only implicit. It should be possible to flesh it out and give a more explicit description which concretely identifies the generators of $\cM_\fT$.
\item The above can be viewed as a preliminary step towards a more explicit description of the RG flows since they are implemented by $-1$ curves that are generators of $\cM_\fT$. Once the generators of $\cM_\fT$ are explicitly known, as discussed in Section \ref{RG}, blowing down and flopping the $-1$ curves among those generators would generate all the descendant Calabi-Yaus describing $5d$ SCFTs at the endpoints of the corresponding RG flows. In this way, it should be possible to perform a much more refined classification of $5d$ SCFTs than presented so far. This is a work currently in progress.
\item It should be possible to use the ideas and results described in this work to identify T-dual pairs of supersymmetric little string theories (LSTs) admitting an F-theory construction without frozen singularities \cite{Bhardwaj:2015oru}. LSTs are UV complete theories in $6d$ without dynamical gravity, and are expected to have the $R\to\frac{1}{R}$ duality when compactified on a circle of radius $R$. Using the methods of this paper, one can associate a Calabi-Yau threefold to each LST. The T-dual theories can then be identified by the criterion that their associated Calabi-Yaus should be flop equivalent.
\item In this paper, by using a duality to M-theory, we were able to associate a Calabi-yau threefold to \emph{untwisted} compactification of a $6d$ SCFT admitting an F-theory construction without frozen singularities. By similar dualities to M-theory, it should also be possible to associate a Calabi-Yau threefold to a \emph{twisted} compactification of the $6d$ SCFT. Using the ideas described in Section \ref{gen}, with possibly small extensions, it should then be possible to perform a sequence of flops and blowdowns to produce all $5d$ SCFTs that come from compactifications of $6d$ SCFTs admitting F-theory construction without frozen singularities.
\item There are $6d$ SCFTs that can only be constructed in F-theory by using frozen singularities \cite{Bhardwaj:2018jgp}. It would be interesting to further understand such setups and to see if their compactifications on circle admit a duality to M-theory. If they do, then the ideas discussed in this paper should presumably turn out to be useful again in classifying $5d$ SCFTs originating from such $6d$ SCFTs.
\eit

\section*{Acknowledgements}
The work of LB is supported by the NSF grant PHY-1719924. The work of PJ is supported by the Harvard University Graduate Prize Fellowship.

\appendix
\section{User's guide for Mathematica notebook \emph{Pushforward.nb}}\label{appendix}
The Mathematica notebook \emph{Pushfoward.nb} contains a module \texttt{push[]} designed to compute triple intersection numbers $(S_i \cdot S_j \cdot S_k)_{X_r}$ associated to a basis of fibral divisors $S_{i=0,\dots,r}$ in a resolved elliptically fibered Calabi-Yau 3-fold $X_r$. Here, we assume $X_r$ is realized by means of a Weierstrass model characterized by a collection of curves $\Sigma_i \subset B$ with prescribed intesection data $(\Sigma_i \cdot \Sigma_j)_B$ and with each curve carrying a particular choice of Kodaira singularity. We emphasize here that the 3-fold $X_r \subset Y_r$ is a hypersurface of a 4 dimensional projective bundle $Y_r \rightarrow B$. All classes and intersection products are initially defined in the intersection ring of $Y_r$.
\subsection*{Input}
The module \texttt{push[]} is designed to accept three inputs arranged sequentially in the form
\begin{align}
	\emph{output}~\texttt{=}~ \texttt{push[}\emph{generators,exceptional,fibral}\texttt{]}
\end{align}
where 
	\begin{enumerate}
		\item The object \emph{generators} is a two dimensional array of divisor classes for the generators of each blowup center. Denote \emph{generators} by $G$. Then, $G$ takes the form 
			\begin{align}
				G = \texttt{\{} G_1,G_2, \dots, G_r\},~~G_i = \{ G_{i1},G_{i2},\dots \}
			\end{align}
		where each $G_{ij}$ represents a divisor class. 
		\item The object \emph{exceptional} is a one dimensional array of divisor classes $[e_i], i = 0, \dots r$, i.e. the classes of the divisors $e_i=0$. Denote \emph{exceptional} by $E$. Then $E$ takes the form
			\begin{align}
				E = \{ [e_0] , [e_1], \dots, [e_r]\}.
			\end{align}
		Any ordering for $E$ is permissible so long as the ordering matches the ordering of the array \emph{fibral}. Moreover, note that the precise syntax
			\begin{align}
				\texttt{e[i]}
			\end{align}
		is used to denote $E_i$, namely the (total transform of the) exceptional divisor associated to a given blowup $f_i : Y_i \rightarrow Y_{i-1}$. We stress here that because an abuse of notation has been made, the classes $[e_i]$ are in general linear combinations of the divisor classes $E_i$. The symbol
			\begin{align}
					\texttt{s[i]}
			\end{align}
		must be used for the pullbacks of the divisor classes $\Sigma_i$ of the curves in the base.
		\item The object \emph{fibral} is a one dimensional array of the classes of the $[e_i]$ expanded in a basis of the fibral divisors $S_i$. Let $S$ denote \emph{fibral}. Then, $S$ takes the form:
			\begin{align}
					S= \{ [e_0], [e_1], \dots \}. 
			\end{align}	
		Note again that the ordering of $S$ should match that of $E$ such that the identification
			\begin{align}
				\emph{fibral} = \emph{exceptional}
			\end{align}
		is consistent sense component-wise. 
 	\end{enumerate}
	
\subsection*{Output}
The object \emph{output} represents the output of the module \texttt{push[]}. Denote \emph{output} by $O$. Then $O$ is a four component array
	\begin{align}
		O = \{ O_1, O_2,O_3, O_4 \}.
	\end{align}
 We now describe each component $O_i$; note that each component carries its own symbolic labeling in Mathematica to facilitate the interpretation of the output data:
 	\begin{enumerate}
 		\item The first component $O_1$ is a $(r+1) \times 2$ array where each entry takes the form
 			\begin{align}
 				O_{1i} = \{ \texttt{S}_{\texttt{i}}^3 , 8 - (S_i \cdot S_i \cdot S_i)_{X_r}\},~~i=0,\dots, r.
 			\end{align}
 		\item The second component $O_2$ is a matrix of classes 
 			\begin{align}
 				O_{2ij} = ( S_i \cdot S_j)_{X_r}.
 			\end{align}
 		For a given gauge algebra $\prod_i \mathfrak{g_i}$ where the $i$th component, arising due to a Kodaira singularity, is supported on a curve $\Sigma_i$, the matrix $O_2$ should be equal to 
 			\begin{align}
 				O_2 = -\bigoplus_i \Sigma_i h_{i},
 			\end{align}
 		where $h_i$ is the inverse quadratic form of the algebra $\mathfrak{g}_i$ and (by abuse of notation) $\Sigma_i$ is the pullback of the divisor class of the curve $\Sigma_i \subset B$.
 		\item The third component $Q_3$ is such that each component is of the form
 			\begin{align}
 			\begin{split}
 				O_{3j} &= \{ O_{3j1}, O_{3j2}\}\\
 				 O_{3j1} &= \{\dots, \texttt{S}_{\texttt{j}_1} \texttt{S}_{\texttt{j}_2}^2, \texttt{S}_{\texttt{j}_1}^2 \texttt{S}_{\texttt{j}_2},\dots\}\\
 				  Q_{3j2} &=\{\dots, (S_{j_1} \cdot S_{j_2}\cdot S_{j_2})_{X_r}, (S_{j_1} \cdot S_{j_1} \cdot S_{j_2})_{X_r},\dots\}.
 			\end{split}
 			\end{align}
 		Note that the maximum value of $j$ is determined by the total number of transverse intersections $S_i \cdot S_j$ which exist among the fibral divisors. These intersections are displayed in lexicographic order, with $j_1 < j_2$.
 		\item Finally, the fourth component $O_4$ is given by 
 			\begin{align}
 			\begin{split}
 				O_{4j} &= \{ O_{4j1},O_{4j2}\}\\
 				O_{4j1}&=  \{\dots,\texttt{S}_{\texttt{j}_1} \texttt{S}_{\texttt{j}_2} \texttt{S}_{\texttt{j}_3}, \dots\} \\
 				O_{4j2}&= \{\dots,(S_{j_1} \cdot S_{j_2} \cdot S_{j_3})_{X_r},\dots\}
 			\end{split}
 			\end{align}
 		Here, the maximum value of $j$ is determined by the total number of intersection points\footnote{These ``points'' can also be $-1$ curves.} which exist among triples of inequivalent $S_i$. These intersections are displayed in lexicographic order, with $j_1 < j_2 < j_3$. 
 	\end{enumerate}
 In order to fully specify the output $O$, one must supply the intersection data $(\Sigma_i \cdot \Sigma_j)_B$ at the end of the computation.

\bibliographystyle{ytphys}
\let\bbb\bibitem\def\bibitem{\itemsep4pt\bbb}
\bibliography{ref_multi}

\end{document}